%% file: Desiree.tex
\pdfoutput=1
\documentclass[a4paper,12pt,titlepage]{book}
\usepackage[pdftex]{graphicx}
\usepackage{fancyhdr}
\usepackage{plain}
\usepackage{epigraph}

\usepackage{url}
\usepackage{hyperref}
\usepackage{csquotes}
\usepackage{setspace}
\usepackage{amsmath, bm}
\usepackage{amssymb}
\usepackage{multirow}
\usepackage{color}

\usepackage[flushleft]{threeparttable}
\usepackage{verbatim}
\usepackage{tabularx}
\usepackage{array}
\usepackage{amsmath}
\usepackage{cases}
\usepackage{longtable}

\usepackage{threeparttablex}
\usepackage{longtable}
\usepackage{alltt}
\usepackage{upquote}
\usepackage{eurosym}

\usepackage{lmodern}
\usepackage{array}
\usepackage{mathabx}

\usepackage{epstopdf}
\usepackage{pdfpages}
\makeatletter
\renewcommand\part{%
  \if@openright
    \cleardoublepage
  \else
    \clearpage
  \fi
  \thispagestyle{empty}%
  \if@twocolumn
    \onecolumn
    \@tempswatrue
  \else
    \@tempswafalse
  \fi
  \null\vfil
  \secdef\@part\@spart}
\makeatother

\newcommand{\clearemptydoublepage}{\newpage{\pagestyle{empty}\cleardoublepage}}

\makeindex
\begin{document}
\pagestyle{plain}
\pagestyle{fancy}

\fancyhead{}
\fancyfoot{}
\fancyhead[RO, LE] {\thepage}
\fancyhead[LO]{\itshape\nouppercase{\rightmark}}
\fancyhead[RE]{\itshape\nouppercase{\leftmark}}

\input{preface}

\newpage
\clearemptydoublepage
\thispagestyle{empty}
%\large

\input{abstract}

\newpage
\clearemptydoublepage
\thispagestyle{empty}
\input{acknowledgement}

\clearemptydoublepage

%\linespread{1}
\pagenumbering{roman}
\tableofcontents
\clearemptydoublepage
\listoftables
\clearemptydoublepage
\listoffigures
\clearemptydoublepage
\pagestyle{fancy}

%\pagenumbering{num_style}
\pagenumbering{arabic}

%\input{intro.2}
%\input{interpretation}
%\input{concepts}
%\input{semantics}
%\input{semantics.2}

%Visio 2014, 64 bit product key: GCM2J-JGNKP-WFGJP-9KYQD-QRQB2
%\input{tool}
%\input{evaluation}
%\input{conclusion}
%\begin{comment}
\input{intro.2}
\input{related.work.2}
\input{baseline}

\input{interpretation}

\input{concepts}

\input{semantics}
\input{tool}
\input{evaluation}
\input{conclusion}
%\end{comment}

\clearemptydoublepage

%----

\thispagestyle{empty}
%----
\makeatletter
\addcontentsline{toc}{chapter}{Bibliography}
\bibliographystyle{habbrv}
\bibliography{Thesis}

\clearemptydoublepage

\appendix

\input{appendixs}

\end{document}

%% file: preface.tex
\newpage
\clearemptydoublepage
\thispagestyle{empty}
\begin{center}

\textbf{\large PhD Dissertation}\\

\ \hrulefill \\\

%\begin{figure}[h!]
%  \centerline{\psfig{file=./pics/logo.eps,width=0.7\textwidth}}
%\end{figure}

\begin{figure}[!h]
  \centering
  \includegraphics[width=0.7\textwidth]{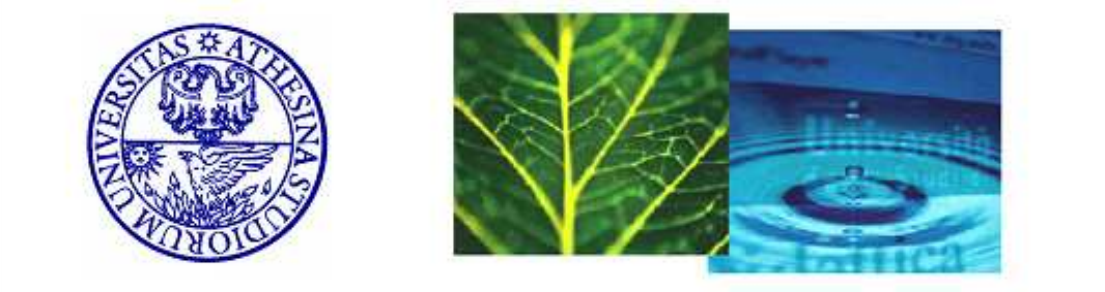}\\
\end{figure}

\textbf{\large International Doctorate School in Information
  and\\Communication Technologies}\\

\vspace{0.3 cm}

\LARGE DISI - University of Trento\\

\vspace{0.3 cm} \LARGE\textsc{Desiree: a Refinement Calculus for Requirements Engineering}

\vspace{0.3 cm}

\begin{center}
\begin{tabular}{l}
\Large Feng-Lin Li\\
\end{tabular}
\end{center}

\begin{flushleft}
\begin{tabular}{l}
{\large Advisor:}\\
\large Prof. John Mylopoulos\\
\large Universit\`a degli Studi di Trento\\
\end{tabular}
\end{flushleft}

\begin{comment}
\begin{flushleft}
\begin{tabular}{l}
{\large Co-Advisor:}\\
\large Prof. Lin Liu\\
\large Tsinghua University\\
\end{tabular}
\end{flushleft}
\end{comment}

\hrulefill

\normalsize
January $2016$
\end{center}

%\end{document}

%% file: abstract.tex
{\bf \Huge Abstract}

%\vspace{4cm}

\vspace{2cm}

%\emph{}

%\begin{normalsize}

\noindent \emph{The requirements elicited from stakeholders suffer from various afflictions, including informality, incompleteness, ambiguity, vagueness, inconsistencies, and more. It is the task of requirements engineering $(RE)$ processes to derive from these an eligible (formal, complete enough, unambiguous, consistent, measurable, satisfiable, modifiable and traceable) requirements specification that truly captures stakeholder needs.}
%consists of the functions and qualities that the system-to-be needs to operationalize, and
%(with various afflictions being addressed)

\emph{We propose \textbf{Desiree}, a refinement calculus for systematically transforming stakeholder requirements into an eligible specification. The core of the calculus is a rich set of requirements operators that iteratively transform stakeholder requirements by strengthening or weakening them, thereby reducing incompleteness, removing ambiguities and vagueness, eliminating unattainability and conflicts, turning them into an eligible specification. The framework also includes an ontology for modeling and classifying requirements, a description-based language for representing requirements, as well as a systematic method for applying the concepts and operators in order to engineer an eligible specification from stakeholder requirements. In addition, we define the semantics of the requirements concepts and operators, and develop a graphical modeling tool in support of the entire framework.}
%in order to engineer an eligible specification from stakeholder requirements

\emph{To evaluate our proposal, we have conducted a series of empirical evaluations, including an ontology evaluation by classifying a large public requirements set, a language evaluation by rewriting the large set of requirements using our description-based syntax, a method evaluation through a realistic case study, and an evaluation of the entire framework through three controlled experiments. The results of our evaluations show that our ontology, language, and method are adequate in capturing requirements in practice, and offer strong evidence that with sufficient training, our framework indeed helps people conduct more effective requirements engineering.}

%\vspace{1cm}
\vspace{1cm}
\noindent
{\bf Keywords}

\noindent
[Requirements Problem, Requirements Ontology, Description-based Syntax, Requirements Operators, Goal-oriented Requirements Engineering]
%\end{normalsize}

\begin{comment}
  The x abstract goes here.  The abstract should be selfcontained
  and:
  1. clearly state the problem dealt with by the thesis
  2. give a synthetic description of the proposed solution
  3. highlight the sense in which the proposed solution enhances the
  state of the art.
  The abstract must be limited to max. one page with no bibliographic
  references, nor external references on any kind.
\end{comment}

%% file: acknowledgement.tex
{\bf \Large Acknowledgement}

%\vspace{4cm}

\vspace{1cm}

%\emph{}
\begin{normalsize}

\noindent This thesis would not have been possible without the help of several important people \footnote{As for research funding, this research has been funded by the ERC advanced grant 267856 ``Lucretius: Foundations for Software Evolution'', unfolding during the period of April 2011 - March 2016. It has also been supported by the Key Project of National Natural Science Foundation of China (no. 61432020).}. I would also like to recognize and thank colleagues, friends and family for their help and efforts during these past four years.

To John Mylopoulos, my supervisor, thank you for your sagacious guidance on every piece of my work, for your great enthusiasm and support that has encouraged my research over the last four years. Thank you, Prof. John Mylopoulos, with a four-year research training, I gained a more wonderful view and a much deeper understanding of the things in the same world. In short, thank you for helping me get to where I am now.

To Lin Liu, my co-supervisor, thank you for encouraging me to pursue my Ph.D. degree, for your insightful guidance and constructive suggestions on my work, for your great help in the evaluation of our framework.

%To Alexander Borgida, Giancarlo Guizzardi, Jennifer Horkoff, Renata Guizzardi, and Lin Liu, my main colleagues, thank you for your large amount of efforts put on the discussions, and for your insightful and useful comments and suggestions. Especially, Prof. Alexander Borgida has helped a lot on the semantics of the framework, Prof. Giancarlo Guizzardi and Renata Guizzardi have helped a lot on the ontological foundation part, Jennifer Horkoff has helped a lot on the requirement operators, and Prof. Lin Liu has helped a lot on the evaluation.

To Alexander Borgida, Giancarlo Guizzardi, Renata Guizzardi, and Jennifer Horkoff, my main colleagues, thank you for your large amount of efforts put on the discussions, and for your insightful and useful comments and suggestions. Especially, Prof. Alexander Borgida has helped a lot on the semantics of the framework, Prof. Giancarlo Guizzardi and Renata Guizzardi have helped a lot on the ontological foundation part, Jennifer Horkoff has helped a lot on the requirement operators.

%, and Prof. Lin Liu has helped a lot on the evaluation.
%Each of them, has given very useful comments and and given many and many useful feed
To Paolo Giorgini (University of Trento), Barbara Pernici (Politecnico di Milano), and Alexander Borgida (Rutgers University), the professors of my thesis committee, thank you for accepting the invitation to participate in.

To Mauro Poggianella, thank you for your help in developing the prototype tool.

To my colleagues in the research group of Software Engineering and Formal Methods, University of Trento, thank you for helping me in conducting the controlled experiment in University of Trento; to my colleagues in the Institute of Information System and Engineering, School of Software, Tsinghua University, thank you for volunteering to participate in and for helping me to organize the two controlled experiments in Tsinghua University. I apologize for not listing your names.

%To our research secretary and department secretary, thank you for helping me a lot in handling various kinds of procedures.

Last, but definitely not least, to my wife and family. For your unconditional love and support, and for always taking
care of me.

To all of you, my many thanks, \emph{grazie mille} !

%\vspace{6pt}
%\hrulefill
%\noindent\rule[0.5ex]{\linewidth}{1pt}
%\begin{small}
%\end{small}
\end{normalsize}

%% file: intro.2.tex
%\vspace{4cm}
\chapter{Introduction}
\label{cha:intro}

\setlength\epigraphwidth{0.8\textwidth}
\setlength\epigraphrule{0pt}
\epigraph{\normalsize\itshape``The requirements for a system, in enough detail for its development, do not arise naturally; instead, they need to be engineered and have continuing review and revision.''}{--- \textup \itshape T.E. Bell and T.A. Thayer}

% This seems can be taken for granted at first sight.
%or software-intensive
%or profitable
%(-intensive)
%what conditions will the system be built on

In general, software systems are designed for solving real-world problems. For example, a meeting scheduler is designed for scheduling meetings, and probably also for managing resources (meeting rooms, room equipments); an online flight reservation system is designed for distributing airline tickets in an efficient way, also providing promotional services to customers. To make sure that a software system correctly solves a real-world problem, we need to clearly understand and define what things the system needs to achieve in the world (\textit{requirements}), under what assumptions it will operate (\textit{domain assumptions} or \textit{domain properties}), and what the system needs to do in order to meet the requirements (\textit{specification}).

As problems nowadays are increasingly complex, getting the problem right is not easy, and sometimes surprisingly difficult. For example, when designing an online flight reservation system, we need to consider various kinds of customers (e.g., first-class, business-class, economic-class, adults, kids, infants), different kinds of  nationalities (that may need special kinds of in-flight meals), special kinds of consignments (e.g., pets, baby carriages, wheelchairs), etc. It is the task of Requirements Engineering (RE) to discover, understand, analyze and formulate the requirements that the system we are designing needs to fulfill, the context in which the system will be used, and the functions and qualities that the system needs to operationalize. Put differently, Requirements Engineering bridges the gap between the initial recognition of a problem and the task of producing a specification for the system-to-be.
%building a software(-intensive) system to address that problem.

In this chapter, we review the role of Requirements Engineering in software development, and discuss the RE problem. At the end, we present an overview of the thesis.

%this thesis's proposal: \textbf{\textit{Desiree}}, a refinement calculus for addressing the RE problem (we refer to the calculus as \emph{requirements calculus} in the rest of the thesis, since the objects of refinement here are requirements).

%~\footnote{We refer to the calculus as \emph{requirements calculus} in the rest of this thesis, since the objects of refinement here are requirements.}
%we need to discover, understand, formulate, analyse and agree on \textit{what} problem should be solved, \textit{why} such a problem needs to be solved and \textit{who} should be involved in solving the problem. Broadly, this is what \textit{requirements engineering} (hereafter RE) is all about~\cite{van_lamsweerde_requirements_2009}.
%\vspace{-6pt}
\section{Requirements Engineering}
\label{sec:context}
Requirements Engineering (RE) was born in the middle of the '70s~\cite{mylopoulos_object-oriented_1999}, partly thanks to Douglas Ross and his SADT proposal (Structured Analysis and Design Technique~\cite{ross_structured_1977-1}, and partly thanks to others who established that the rumored `\textit{requirements problem}' was actually a reality~\cite{bell_software_1976}. In their seminar work~\cite{ross_structured_1977}, Ross and Schoman have defined `\textit{requirements definition}' (an early name for requirements engineering) as ``\textit{a careful assessment of the needs that a system is to fulfill}'', and pointed out that it must deal with ``\textit{context analysis}'' (why a system is needed, based on current or foreseen conditions), ``\textit{functional specification}'' (what the system is to be, in terms of functions that the system must accomplish), and ``\textit{design constraints}'' (how the functions are to be designed and implemented). Later work by Zave~\cite{zave_classification_1997} has further developed this idea and suggested that RE also needs to correlate the real-word (contextual) goals for, the functions of, and the constraints on a software system to precise specifications of software behavior, and concern their evolution over time.

%a language for communicating ideas),
%further developed this idea suggested that the evolution of specifications also needs to be considered.

%This early definition has already covered the main components of RE, namely requirements (needs), domain properties (current or foreseen conditions) and specifications (functions, design constraints), and suggests that RE is rather broad and covers a set of intertwined activities such as domain analysis, elicitation, negotiation and agreement, specification, verification and validation~\cite{van_lamsweerde_requirements_2000}. Later,

%the translation of informal requirements which are about real-word phenomena to formal specification languages (requirements analysis, specification derivation)
These definitions suggest that the subject of RE is inherently broad~\cite{van_lamsweerde_requirements_2000}: it needs to deal with the discovery of requirements (requirements elicitation), the analysis of domain assumptions/properties (domain analysis), the check of requirements deficiencies such as inadequacy, incompleteness and inconsistency (requirements analysis), documentation, the management of requirements and specifications, etc.

%as suggested by many researchers
Roughly, these activities can be categorized into two processes~\cite{naumann_determining_1980}\cite{pohl_requirements_2010}: (1) the process of eliciting stakeholder requirements/needs; and (2) the process of refining stakeholder requirements into a requirements specification that can serve as the basis for subsequent system development activities. In addition, evolution/change management also needs be seriously considered as stakeholder requirements/needs are ever-changing~\cite{li_requirements-driven_2013}.

%appear to be more changeable nowadays
%the validation of requirements and the verification of specifications,
%\vspace{-6pt}
\subsection{RE in Traditional Software Development}
\label{sec:RE_in_SE}

In traditional software development methodologies (e.g., waterfall model~\cite{royce_managing_1970}), requirements engineering is presented as a first phase, followed by others, such as design, implementation, testing and maintenance. In such methodologies, business analysts (or requirements engineers) have to mediate between customers and software engineers: they translate informal requirements elicited from customers into a formal specification that can be handed over to software engineers for downstream development.

%RE has been argued be to be of key importance to traditional software development processes for two main reasons: (1) the cost of fixing errors in later software development process is much higher that that in the RE process, e.g., Boehm and Papaccio~\cite{boehm_understanding_1988} reported that it costs 5 times more to detect and fix requirements defects during design, 10 times more during implementation, 20 times more during unit testing and up to 200 times more after system delivery; (2) poor requirements are the leading factor of project failure, e.g., the well-known CHAOS report of the Standish Group has reported a staggering software project success rate of 16.2\% in 1994, and identified poor requirements as the major source of problems (47\%)~\cite{group_chaos_1995}.

RE has been argued be to be of key importance to traditional software development processes for two main reasons: (1) the cost of fixing errors in later software development process is much higher that that in the RE process, e.g. Boehm and Papaccio~\cite{boehm_understanding_1988} reported that it costs 5 times more to detect and fix requirements defects during design, 10 times more during implementation, 20 times more during unit testing and up to 200 times more after system delivery; (2) poor requirements are the leading factor of project failure, as shown in the empirical evaluations and investigations below.

%e.g., the well-known CHAOS report of the Standish Group has reported a staggering software project success rate of 16.2\% in 1994, and identified poor requirements as the major source of problems (47\%)~\cite{group_chaos_1995}.

Software requirements have been recognized as a {{real}} problem since the '70s. In their early empirical study, Bell and Thayer~\cite{bell_software_1976} have reported that software requirements, of both small and large systems, are often incorrect, ambiguous, inconsistent, or incomplete. %They established that the rumored `\textit{requirements problems}' are a reality, and concluded that "\textit{the requirements for a system, in enough detail for its development, do not arise naturally; instead, they need to be engineered and have continuing review and revision}".
In the sequel, more studies have confirmed the requirements problems~\cite{group_chaos_1995}\cite{ibanez_european_1996}\cite{verner_requirements_2007}. One broadly circulated report is the survey over a total of 8380 projects in 350 US companies, which is conducted by the Standish Group in 1994~\cite{group_chaos_1995}. The survey revealed that 31.1\% of the investigated projects were failed, and 52.7\% of them were challenged (e.g., offering only partial functionalities, being over budget or over time).
%\footnote{These statistics were updated each year. In their manifesto~\cite{manifesto_think_2013}, they reported that in 2012, 18\% were failed, 43\% were challenged, 39\% were successful.}
Poor requirements, such as the lack of user involvement, requirements incompleteness, changing requirements, unrealistic expectations and unclear objectives, were identified as the major source of problems (47\%). Similarly, in a survey conducted with 3800 organizations over 17 countries in Europe, most software problems are in the area of requirements specifications (50\%) and requirements management (50\%)~\cite{ibanez_european_1996}.

\indent More recently, in a survey of 12 UK companies~\cite{hall_requirements_2002}, it was reported that requirements problems accounted for 48\% of all software development problems (2002). In 2014, the Project Management Institute (PMI) has conducted a comprehensive survey with more than 2,000 practitioners, and found that 47\% of unsuccessful projects failed to meet goals due to poor requirements management, including interpreting and clearly articulating requirements, aligning them to the strategic vision for projects, communicating with stakeholders, dealing with ambiguities, etc.~\cite{profession_requirements_2014}.

\subsection{RE in Agile Software Development}
\label{sec:RE_in_Agile}
% Verner et al.~\cite{verner_requirements_2007} reported that good requirements quality and management are the the best predictor of project success (93\% of projects were predicted correctly).
%~\cite{committee_ieee_1998}.

Agile software development is a collection of \textit{lightweight} software development methods (e.g., scrum~\cite{schwaber_scrum_1997}) evolved in the middle '90s in reaction to the \textit{heavyweight} waterfall-oriented methods~\cite{_agile_????}. As suggested in the \textit{Agile Manifesto}~\cite{_manifesto_????}, agile development encourages creating minimum amount of documentations needed to accurately guide developers and testers (note that it does not mean \textit{{no}} documentation). That is, one does not need to prepare a comprehensive requirements document when starting a software project.

Due to its lightweight nature, agile development has enjoyed much attention in recent years. By checking Standish Group's report in 2010 and 2012, one may notice that they have suggested ``\textit{agile process}" as a factor of success to small projects (with less than \$1 million in labor content)~\cite{manifesto_think_2013}. Also, a recent IT survey in 2014~\cite{ambler_2014_2014} revealed that 52\% of  the 231 respondents responded that their teams were using agile development.

%We can see that agile development is enjoying attentions in software engineering.
One may think that requirements engineering is not important any more in nowadays software development at first sight. However, this is not the case. First, agile development makes strong assumptions (e.g., all stakeholder roles, including customers and users, can be reduced to one single role) and is not applicable to all projects, especially mission-critical projects~\cite{van_lamsweerde_requirements_2009}. For example, we would obviously not like our air traffic control, transportation, power plant, medical operation or e-banking systems to be obtained through agile development of critical parts of the software~\cite{van_lamsweerde_requirements_2009}. Second, agile development has its own challenges, e.g., the very limited documentation would make a software system hard to maintain~\cite{nawrocki_extreme_2002}, the ignorance of non-functional requirements would result in major issues as the system matures and becomes ready for larger-scale deployment~\cite{cao_agile_2008}, etc.

% (the $21^st$ Working Conference on Requirements Engineering: Foundation for Software Quality)
More importantly, agile development also needs requirements engineering. Through a survey over 12,000 software projects between 1984 and 2003, Jones~\cite{jones_variations_2003} has shown that almost any project includes some RE activities, whatever its type and size. Specifically, Wiegers et al.~\cite{wiegers_software_2013} have pointed out that ``\textit{agile projects require fundamentally the same types of requirements activities as traditional development project}", and the difference is that ``\textit{detailed requirements are not documented all at once at the beginning of an agile project, and will be developed in small portions through the entire project}''.

In an industrial keynote ``\textit{Agile RE -- recipe for success or project wish-list?}''~\footnote{http://refsq.org/2015/files/2015/04/Rogers.pdf} at REFSQ 2015, a solution designer from {1\&1 Internet}, Germany has shared their experiences on adopting agile development (which is not successful). He pointed out that ``\textit{pure scrum is over-simplistic in complex, commercial environment}'', and RE is needed for several objectives, e.g., completeness, consistency, model-driven elicitation and analysis.

%Note that the statement ``\textit{working software over comprehensive documentation}'' in the manifesto for agile software development~\cite{_manifesto_????} does not mean no documentation. Instead, agile development encourages creating minimum amount of documentations needed to accurately guide developers and testers.
% for implementations

In agile development, requirements are often specified as user stories, which are allocated to specific iterations~\cite{wiegers_software_2013}. In each iteration, the allocated user stories need to be further analyzed, clarified, specified and verified. This new characteristic of agile development indicates that traditional RE needs to be adapted: to embrace agile processes, RE shall be able to produce small requirements specifications, which can be easily modified and integrated with requirements specifications produced in other iterations, before and after. That is, a requirements specification shall be easily modifiable (i.e., offering effective support for evolution/change management).
%This require the interrelations between requirements to be captured at a fine-granularity, instead of using a general link, as many current RE research and industrial practices have done.
%That is, a requirement need to be treated as a set of constituents, instead of a simple sentence (proposition).
%as they have observed a much higher success rate rate of small projects than that of large projects (with more than \$10 million in labor content)

%Improving requirements quality is thus critical. Moreover, according to the IEEE 830-1998 standard~\cite{committee_ieee_1998}, as stakeholder requirements are transformed into formal, complete, measurable, and attainable specifications, they need to be well organized such that the resulting specifications can be easily modifiable and traceable (i.e., easy to manage).

\section{The RE Problem}
\label{sec:problem}

%As suggested by the IEEE 830-1998 standard~\cite{committee_ieee_1998}, a good software requirements specification (SRS) shall be correct, complete, ambiguous, consistent, ranked, measurable, modifiable and traceable.
%Engineering high quality requirement specifications is thus critical.

Upon elicitation, requirements are typically mere informal approximations of stakeholder needs that the system-to-be must fulfill. It is the task of the requirements engineering process to transform these requirements into a specification that describes formally and precisely the functions and qualities of the system-to-be. However, this is complicated by the very nature of requirements elicited from stakeholders, which are often ambiguous, incomplete, unverifiable, unattainable/unsatisfiable, etc. In our earlier studies on the PROMISE requirements data set~\cite{menzies_promise_2012}, we found that 3.84\% of all the 625 (functional and non-functional) requirements are ambiguous~\cite{li_stakeholder_2015}, 25.22\% of the 370 non-functional requirements (NFRs) are vague, and 15.17\% of the NFRs are potentially unattainable (e.g., they implicitly or explicitly use universals like ``\emph{any}'' as in ``any time'')~\cite{li_non-functional_2014}.

%Moreover, as requirements are commonly written in natural language, the modifiability of requirements specifications is often poor:

%Moreover, as discussed in the Section~\ref{sec:RE_in_Agile}, requirements need to be well captured such that they can be easily modifiable.

%Typically, requirements elicited from stakeholders are mere informal, approximate statements of stakeholder needs within certain contexts. It is the task of Requirements Engineering to transform such requirements into a formal, complete, measurable, and consistent specification of the functions and qualities that the system-to-be needs to operationalize.
\subsection{Problem Definition}
\label{sec:intro_problem_def}
The problem of transforming informal stakeholder requirements into precise specifications (i.e., the requirements problem) was elegantly characterized by Jackson and Zave~\cite{jackson_deriving_1995} as finding the specification \textit{S} that for certain domain assumptions \textit{DA} satisfies the given requirements \textit{R}, and was formulated as in Eq.~\ref{core_p1}.
\begin{equation}\label{core_p1}
   DA, S \models R
\end{equation}

\indent This formulation determines what does it mean to successfully complete RE: once the specification \textit{S} is found such that Eq.~\ref{core_p1} holds, the requirements problem will be solved~\cite{jureta_revisiting_2008}. The transformation of requirements to specifications, or alternatively, the derivation of specifications from requirements, is thus of key importance to RE: the process will determine the quality of requirements (e.g., are they complete enough? are they measurable? etc.), and the way how requirements are specified will influence requirements management (e.g., if well structured, the resulting requirements specification could have good modifiability and traceability). In fact, this is the core problem of RE.

%\indent In RE, the process of finding of a specification \textit{S} is in fact a processing of transforming  stakeholder requirements into a requirements specification that describes formally and precisely the functions and qualities of the system-to-be, or, put differently, a process of incrementally improving requirements quality such that the resulting requirements specifications will be correct, complete, unambiguous, consistent, verifiable, attainable, etc.

%\cite{zave_four_1997}

Jackson and Zave have distinguished between a machine (a software system) and its environment, and stated that a machine can affect, or can be affected by the environment only when they have some shared phenomena, which are events or states located at the interface/boundary of the machine~\cite{jackson_deriving_1995}. Take a turnstile at the entry to a zoo for example, inserting a coin and pushing the barrier to its intermediate position are shared phenomena, entering the zoo is merely an environment phenomenon, and the lock and unlock of the turnstile are only machine phenomena.

Based on this, they distinguish between requirements and specifications as follows: a ``\textit{requirement}'' states desired property in the environment -- properties that will be brought about or maintained by the machine; a ``\textit{specification}'' for the machine is restricted to implementable behaviour of a machine that can ensure the satisfaction of the requirements, and can only be written in terms of shared phenomena. Further, they define ``\textit{domain assumption}'' as an indicative property of the environment irrespective of whether we build the machine or not~\cite{zave_four_1997}. As such, ideally, a specification shall not refer to private phenomena in the environment, because a machine has no access to them; it also shall not include private phenomena in the machine, which shall be in the design.

%a software ecosystem (or, alternatively, socio-technical system)
We favor this distinction, as it makes the requirements problem quite clear, and suggests the need of a systematic methodology for transforming stakeholder requirements into specifications. Note that this distinction between (stakeholder) requirements and (software) specifications is sometimes captured using other terms, e.g., user requirements (needs) vs. (software) system requirements as in Maiden~\cite{maiden_user_2008}. Moreover, we generalize a {machine} (a pure software system) in Jackson and Zave's characterization~\cite{jackson_deriving_1995} to a software ecosystem (or, alternatively, a socio-technical system), as the solution for a real-world problem often involves other agents (e.g., other software systems, or human beings) in addition to the software system we are designing. That is, by specification, we mean socio-technical system specification, in which a function will not necessarily be performed by the software system being designed. For example, a meeting scheduling socio-technical system may have a person collect timetables, feeding them to the software, which then generates a time slot and location for a meeting.
%software ecosystem specification
%For example, maintaining a software will be performed by software maintainers, who are external agents.

%Note that we do not restrict ourselves to software solutions only when addressing a real-world problem, and treat ecosystem. e.g., we allow a function to be performed by an external agent (relative to a software), e.g., a human-being.

%\footnote{Note that this distinction is not universally accepted in RE. For example, the IEEE 830-1998 standard~\cite{committee_ieee_1998} and the Volere template~\cite{robertson_mastering_2012} do not make such distinction.}

%\footnote{Note that this distinction is not universally accepted in RE. For example, Robertson et al.~\cite{robertson_mastering_2012} have provided the Volere template for documenting requirements as part of requirements specifications, and the IEEE 830-1998 standard~\cite{committee_ieee_1998} also does not make such distinction.}

%Fraser et al.~\cite{fraser_informal_1991} proposed guidelines for developing VDM specifications from Structural Analysis (mostly Data Flow Diagrams).

\subsection{Requirements Engineering in Practice}
\label{sec:intro_deficiencies}

\begin{comment}
Roughly, to handle this RE problem, we need to consider at least three major aspects:
\begin{enumerate}
  \item What are the right concepts for requirements modeling? What kinds of things can we say when engineering a requirements specification (e.g., functions, qualities and data)?
  \item What are the proper representation for capturing requirements? What can we say about a particular thing (e.g., the actor, object of a function), can the presentation provides methodological guidelines for refining requirements?
  \item What kinds of methods can we use to incrementally refine requirements such that we can derive a specification of high quality ?
\end{enumerate}
\end{comment}

Predictably, there has been much work on transforming informal stakeholder requirements to a formal specifications,
going back to the early '90s and before~\cite{cauvet_alecsi:_1991}\cite{rolland_natural_1992}\cite{fraser_informal_1991}\cite{dardenne_goal-directed_1993}. Some of this work exploited AI techniques such as expert systems~\cite{cauvet_alecsi:_1991} and natural language processing (NLP)~\cite{rolland_natural_1992}. Others proposed guidelines for developing formal specifications from informal requirements~\cite{fraser_informal_1991}. However, the RE problem has not been addressed effectively and has remained open, as attested by current requirements engineering practice, where word processors and spreadsheets continue to constitute the main tools for engineering requirements. For example, according to a webcast audience poll\footnote{http://www.blueprintsys.com/lp/the-business-impact-of-poor-requirements/} conducted by Blueprint Software System in 2014, more than 50\% of the participants said that they are using documents and spreadsheets for conducting requirements engineering. To address the poor support offered by such vanilla tools, there have been many RE tools (e.g., Rational DOORS~\cite{_ibm_????}) that support RE-specific activities, such as elicitation, specification and traceability management. However, the existing RE tools mainly focus on the management of individual requirements, and pay little attention to the transformation of requirements to specifications.

Generally speaking, current solutions have (some of) the following deficiencies:

 \begin{itemize}
   \item \textbf{Leaving out non-functional requirements}. In traditional structural analysis and object-oriented analysis, non-functional requirements (NFRs) are generally left outside~\cite{van_lamsweerde_object_2004}. Early work by Fraser et al.~\cite{fraser_informal_1991} have proposed guidelines for developing VDM specifications from Structural Analysis (mostly Data Flow Diagrams). More recently, Giese at al.~\cite{giese_informal_2004} have tried to relate informal requirements specified in UML use cases to formal specifications written in Object Constraint Language (OCL). Seater et al.~\cite{seater_requirement_2007} have discussed how to derive system specifications from Problem Frame descriptions through a series of incremental problem reduction. Such approaches focused on functional requirements (FRs), but left out non-functional ones (NFRs), a very important class of requirements in RE.

   \item \textbf{Lacking a unified language for representing both FRs and NFRs}. Although a multitude of languages have been proposed for specifying requirements, few of them offer a syntax for both functional and non-functional requirements. For example, EARS (the Easy Approach to Requirements Syntax)~\cite{mavin_easy_2009} is designed mainly for functional requirements (FRs), while Planguage~\cite{gilb_competitive_2005} is a keyword-driven language developed for writing measurable quality requirements. Similarly, the NFR Framework (NFR-F)~\cite{chung_non-functional_2000} has proposed to use ``\textit{softgoals}'' (goals without clear criteria for success) to capture NFRs, and offered a simple syntactic form for softgoals; however, it does not offer a companion syntax for goals, which are used to model functional requirements. Natural language is known to everyone and can be used to specify any kind of requirements, but there is much evidence that requirements written in natural language is inherently vague, incomplete, and ambiguous~\cite{fabbrini_linguistic_2001}\cite{berry_contract_2003}\cite{kordon_advances_2008}\cite{yang_automatic_2010}\cite{kamalrudin_automated_2009}.

       %Goal-oriented techniques are the first to treat non-functional requirements in depth~\cite{chung_non-functional_2009}. The NFR Framework (NFR-F), first proposed  in 1992~\cite{mylopoulos_representing_1992} and extended into a monograph~\cite{chung_non-functional_2000}, has proposed to use ``\textit{softgoals}'' (goals without clear criteria for success) to capture NFRs, and offered a simple syntactic form for softgoals. The framework offers contribution links for linking software design elements to softgoals, and several operators for decomposing softgoals. However, the NFR Framework uses natural language, instead of a companion syntax, for goals that are used to model functional requirements (FRs).

       %, and hence does not capture the dependency relation between NFRs and FRs~\cite{li_non-functional_2013} (e.g., in the requirement ``the processing time of the file search function shall take less than 30 seconds'', the quality ``processing time'' depends on the function ``file search'').
       %(e.g., the dependency relation between FRs and NFRs)
       %Lacking capture of interrelations based on requirements details

   \item \textbf{Missing important inter-dependencies between requirements}. Often, in current RE research (e.g., goal modeling approaches such as \textit{i}*~\cite{yu_modelling_2011}) and practice (e.g., RE tools such as Rational DOORS~\cite{_ibm_????}), individual requirements are treated as propositions, and traces between requirements are done on the basis of propositional (whole) requirements without considering requirements details. This treatment has several flaws: (1) the establishment and maintenance of traces is quite laborious and needs to be done manually~\cite{smialek_introducing_2007}, resulting in poor modifiability of requirement specifications and making change/evolution management harder~\footnote{Approaches for dynamic link generation have been proposed but they suffer from a certain level of imprecision, and user evaluation over the generated links is necessary~\cite{cleland-huang_goal-centric_2005}.}; (2) some kinds of relations have been missing in current RE techniques~\cite{li_non-functional_2013}, e.g., the existential dependency relation between qualities and functions as in ``\emph{the \underline{processing time} of \underline{file search} shall be less than 30 seconds}''; (3) requirements specification/modeling languages do not have explicit links to the domain vocabulary, hence the vocabulary is often used inconsistently throughout the whole specification, e.g., (``\emph{interest rate}'' vs. ``\emph{loan interest}'')~\cite{smialek_introducing_2007}.

       %``file search'' in the requirement ``the processing time of the file search function shall take less than 30 seconds''
       %First, the establishment of such trace links is difficult since it needs an analysis of all the individual requirements. Second, the dependency relation between FRs and NFRs (e.g., in the requirement ``the processing time of the file search function shall take less than 30 seconds'', the quality ``processing time'' depends on the function ``file search'') has been missing in current RE techniques, including the NFR Framework~\cite{li_non-functional_2013}. Third, the trace links do not capture the meaning of interrelations, possibly giving rise to imprecise identification of requirements to be affected in change impact analysis. For example, the change of a requirement may not affect all related requirements, but only some of them.

       %Cleand Huang : Traceability

       %, we may identify all the related requirements as the candidates of change impact. we can capture the relations between an entity (schema) and the functions referring to it, we can easily and precisely know the possibly affected functions when changing the entity (schema).
       %why requirements are interrelated
       %that needs to be defined in advance

       %For example, the refinement of the goal $G$ ``trip be scheduled'' into ``airline ticket be booked'' and ``hotel be booked'' is a strengthening, because $G$ can be refined to other alternatives, e.g., ``train ticket be booked'' and ``hostel be booked''.

       %ensuring that each refinement is strictly a strengthening:
   \item \textbf{Lacking support for weakening requirements}. Goal-oriented techniques, such as KAOS~\cite{dardenne_goal-directed_1993}, \textit{i}*~\cite{yu_modelling_2011}, Tropos~\cite{bresciani_tropos:_2004}, and Techne~\cite{jureta_techne:_2010}, have provided an elegant way for going from high-level strategic goals to low-level technical goals through AND/OR refinements, and finally tasks (aka functions) through operationalization. In these classic goal modeling frameworks, the AND/OR refinements are generally strengthening: the resulting sub-goal(s) of a refinement logically imply the refined goal, or, alternatively, the resulting set of sub-goal(s) has less solutions than the refined goal~\footnote{Strengthening can be interpreted as shrinking the solution space of a requirement: if a requirement $r$ is strengthened to a set of sub-goals (one or more), say $r_s$, then each solution for $r_s$ is also a solution for $r$, not vice versa. For example, the refinement of the goal $G$ ``\emph{trip be scheduled}'' into ``\emph{airline ticket be booked}'' and ``\emph{hotel be booked}'' is a strengthening, because $G$ can be refined to other alternatives, e.g., ``\emph{train ticket be booked}'' and ``\emph{hostel be booked}''.}. However, stakeholder requirements are sometimes practically unattainable and need to be relaxed. For example, is it possible to require every surveyed user to agree that the system under test is attractive? Is it possible to require a file search function to take less than 30 seconds at all times? How about if there are some exceptions (e.g., when having heavy workload)? In such cases, we need a refinement operator that weakens a requirement so it only needs to be fulfilled $x\%$ of the time ($0 < x < 100$). These kinds of weakening refinements are not supported by current goal-oriented techniques.

       %(reducing the choices for satisfying)
       %a weakening from 100\% of the individuals in a domain to $x\% \; ( 0 \textless x \textless 100)$ is needed

       %a weakening: the former logically imply the latter (or the former has less solutions than the latter)
       %Especially, KAOS~\cite{dardenne_goal-directed_1993} has used LTL (Linear Temporal Logic) to formalize stakeholder requirements, and offered patterns and rules for ensuring the completeness of refinements. That is, ensuring that if a goal $G$ is refined to a set of sub-goals (one or more), the resulting sub-goals should logically imply $G$.

       %That is, KAOS's formalism is able to ensure the completeness of formalized requirements, but has difficulties with completeness check of stakeholder requirements to be formalized~\cite{al-subaie_evaluating_2006}.

   \item \textbf{Lacking support for incremental formalization}. Formal languages have been advocated for specifying requirements specifications as they have a clear syntax and semantics, and support automatic analysis such as ambiguity detection, inconsistency check and verification. However, their limitations are also obvious: they are hard to write and understand for non-experts. To alleviate the pain of directly writing formal specifications, many controlled natural languages (CNLs) based (e.g., Fuchs et al.~\cite{fuchs_attempto_2006}, Konrad et al.~\cite{konrad_real-time_2005}) and natural language processing (NLP) based approaches (e.g., Fantechi et al.~\cite{fantechi_assisting_1994}, Gervasi et al.~\cite{gervasi_reasoning_2005}) have been suggested. In such approaches, NL requirements are first mapped to an intermediate representation, which is then further translated into targeted formal specification language(s).

       One important issue with these approaches is that they just deal with what stakeholders have stated and do not support refining either stakeholder requirements or formalized specification, probably giving rise to requirements inadequacy, incompleteness and incorrectness. Moreover, the NLP-based techniques inevitably suffer from the incompleteness issue (e.g., missing information from the original requirements documents), being challenged in the RE field~\cite{berry_case_2012}.

       %Note that although KAOS has offered refinement patterns and rules for refining formalized requirements, it also assumes that stakeholder requirements are in enough detail and can be directly formalized~\cite{al-subaie_evaluating_2006}. That is, KAOS assumes formalized requirements are complete enough, and what KAOS ensures is the completeness of subsequent refinements. As such, KAOS also does not support incrementally going from informal stakeholder requirements to complete (enough) formal requirements specification, which may elicit missing information in initial stakeholder requirements.

       %when dealing with safety, security or mission-critical systems
       %Especially, the NLP based techniques inevitably suffer from the incompleteness issue.
       %have been employed to extract useful information from requirements written in natural language, according to some parsing rules, and then construct some sorts of requirements/design models based on the extracted information (e.g., Ambriola et al.~\cite{ambriola_processing_1997}) or translate this intermediate result into a formal logic for some tasks such as  (e.g., Fantechi et al.~\cite{fantechi_assisting_1994}, Gervasi et al.~\cite{gervasi_reasoning_2005}).
 \end{itemize}

\section{Our Research Objective}
\label{sec:solution}
%We have reviewed the role of Requirements Engineering in nowadays software development in section~\ref{sec:context}, and discussed the core RE problem and efforts devoted to addressing this problem in section~\ref{sec:problem}. As we can see, although many efforts have been put on the core RE problem, it has not been addressed effectively and has remained open. We want to tackle this problem in this thesis.

We have reviewed the role of Requirements Engineering in state-of-the-art software development in section~\ref{sec:context}, and discussed the RE problem and the deficiencies of current solutions to it in section~\ref{sec:problem}. As we can see, although much research has focused on the RE problem, it has not been addressed effectively and remains largely an open problem. We want to tackle this problem in this thesis.

%many efforts have been put on
%has remained open
We specify our research objective as follows.

%We now specify our research objective and the research questions that this thesis intends to answer.
%In particular, we are interested in addressing issues of ambiguity (e.g.,``\textit{notify users with email}'', is ``\textit{email}'' a means of sending or an attribute of user?), incompleteness (e.g., ``\textit{sort a list of integers}'', ascending or descending?), unattainability (e.g., ``\textit{the system shall remain operational at all times}'', is it necessary for ``\textit{all the time}"?) and conflicts (e.g., ``\textit{high comfort}'' vs. ``\textit{low cost}'') in the process of transforming stakeholder requirements to specifications.

\begin{displayquote}
%\textbf{Research objective}: \textit{to develop a systematic framework for incrementally transforming informal, incomplete, ambiguous, inconsistent, vague, unattainable stakeholder requirements into a formal, complete, unambiguous, consistent, measurable, attainable specification. Moreover, the resulting specification shall be easily modifiable and traceable to support evolution management.}
\textbf{Research objective}: \textit{to develop a holistic~\footnote{The meaning of ``holistic'' is two fold: (1) the framework includes the necessary components for addressing the requirements problem (defined as in Eq.~\ref{core_p1}); (2) the framework aims to address a comprehensive set of requirements issues (to defined in Table~\ref{tab:req_desired_properties}).} framework for incrementally transforming stakeholder requirements into an eligible specification.}
\end{displayquote}

We define the desired characteristics of an eligible specification in Table~\ref{tab:req_desired_properties}, including formal, valid, sufficiently complete, unambiguous, consistent, measurable, satisfiable, modifiable and traceable, according to the IEEE standard~\cite{committee_ieee_1998}. As such, requirements issues such as informal, invalid, incomplete, ambiguous, inconsistent, unverifiable, unsatisfiable, unmodifiable, and untraceable can be defined by identifying the inverse of each desired characteristic.

%Note that we currently do not consider the ``ranked'' characteristic (i.e., requirements are prioritized) and will take it into consideration in our future work.
\begin{table}[!htbp]
  \caption {The characteristics of an eligible specification (adapted from~\cite{committee_ieee_1998})}
  \label{tab:req_desired_properties}
  \vspace {0.3 cm}
  \centering
  \small
  \setlength\tabcolsep{2pt}
  %\begin{tabular}{|c|p{0.8\textwidth}|}
  \begin{tabular}{|m{0.18\textwidth}|m{0.78\textwidth}|@{}m{0pt}@{}} %<{\centering}
  \hline
  \textbf{Characteristic} & \textbf{Definition} \\ \hline
  Formal &  A specification is formal, if, and only if, each requirement therein is stated using a language, which has a mathematically precise semantics\\ \hline
  Valid &  A specification is valid/correct if, and only if, every requirement therein is one that the software system shall meet (i.e., the stakeholders want). \\ \hline
  (Sufficiently) Complete & A specification is sufficiently complete if, and only if (both conditions need to be met): (1) every requirement has the necessary information for implementation; and (2) no significant requirement is missing \\ \hline
  Unambiguous & A specification is unambiguous if, and only if, every requirement therein has only one interpretation (e.g., ``\textit{notify user with email}'' is ambiguous because it has two interpretations: ``\textit{notify users through email}'' vs. ``\textit{notify users who have email}'') \\ \hline
  Consistent & A specification is (internally) consistent if, and only if, no subset of individual requirements described in it conflict. By conflict, it means: (1) the specified properties of real-world objects may conflict (e.g., \textit{the lights shall be green} vs. \textit{the lights shall be red}); or (2) logical or temporal conflict between two specified actions (e.g., ``\textit{event A follows B}'' vs. ``\textit{event B follows A}''); or (3) term inconsistencies (e.g., the use of ``\textit{loan rate}'' vs. ``\textit{interest rate}'' in the same specification)\\ \hline
  Measurable &  A specification is verifiable if, and only if, every requirement stated therein is measurable (e.g., ``\textit{low}'' is unverifiable since we can not measure it)\\ \hline
  Satisfiable \; \;  \; (Attainable) & A specification is practically satisfiable (attainable), if, and only if, every requirement therein is technically and economically feasible to fulfill\\  \hline
  Modifiable &  A specification is modifiable if, and only if, its structure and style are such that any changes to the requirements can be made easily, completely, and consistently. In general, modifiability requires a specification to: (1) be structurally organized; (2) not be redundant; (3) express each requirement separately rather than intermixing them  \\ \hline
  Traceable &  A specification is traceable if: (1) each requirements explicitly references to its source in earlier documents; (2) interrelations between requirements within the specification are well captured; (3) each requirement in the specification having a unique label for later reference \\ \hline
  \end{tabular}
\end{table}

We propose to address the research objective by tackling four research questions (RQs):
%decompose the research objective into

\vspace{6pt}
%\noindent \textbf{RQ1: What are functional (resp. non-functional) requirements?}
%\noindent
\textbf{RQ1: What are the proper concepts for modeling requirements?}
\vspace{6pt}

We follow the commonly accepted classification of requirements as functional (FRs) and non-functional (NFRs). We provide ontological interpretations for these two kinds of requirements, and accordingly proposing an ontological classification of all requirements. Specifically, we treat FRs as requirements referring to \textit{functions} and NFRs as requirements that refer to \textit{qualities}, and look to foundational ontologies (e.g., the Unified Foundational Ontology~\cite{guizzardi_ontological_2005}, UFO for short) to tell us precisely what functions and qualities are, respectively. Further, we propose new concepts to capture what have been traditionally called NFRs but do not refer to qualities.

\vspace{6pt}
%\noindent \textbf{RQ2: How to represent functional (resp. non-functional) requirements?}
%\noindent
\textbf{RQ2: What are the proper models for engineering requirements?}
\vspace{6pt}

Based on our ontological interpretation for requirements, and inspired by AI frames and Description Logics (DL)~\cite{baader_description_2003}, we propose to use a description-based representation for requirements, both functional and non-functional. We treat a requirement as a description consisting of a set of constituent parts rather than a simple propositional sentence. As such, we allow analysts/engineers to reference the already defined constituent parts of another requirement when specifying the current one, and to refine any constituent part of a requirement description by using proper requirements operators (introduced for addressing \textbf{RQ3}). We also provide precise semantics for this language by using set theory.

\vspace{6pt}
%\noindent \textbf{RQ3: Given a functional (resp. non-functional) requirement, how to incrementally transform it into a specification fragment?}
%\noindent
\textbf{RQ3: What are appropriate techniques and tools for transforming stakeholder requirements into eligible specifications?}
\vspace{6pt}

To address this question, we revise the classic ideas of AND/OR refinements and operationalization in existing goal modeling techniques, which allow us going from high-level goals to low-level goals, and finally functions (aka tasks). To make sure that resulting specification is eligible, we propose a set of new requirements operators based on our language for requirements, and offer a methodology for applying the operators to incrementally transform stakeholder requirements into an eligible specification.

\vspace{6pt}
%\noindent
\textbf{RQ4: How well does our proposed framework work in realistic settings?}
\vspace{6pt}

%\noindent
When an approach is proposed, it needs empirical evidence to assess its effectiveness. To assess our framework, we shall conduct thorough evaluation to verify whether it is helpful in conducting requirements engineering. Specifically, we shall evaluate whether (1) our classification of requirements (functional vs. non-functional) is adequate in covering requirements; (2) our syntax is sufficiently expressive in capturing requirements; (3) the methodology is effective in transforming stakeholder requirement to an eligible specification; and (4) the whole approach can indeed help people to perform better requirements analysis, i.e., identifying and addressing more requirements issues during requirements analysis.

\section{Overview and Contribution}
\label{sec:innovation}
%strengthening/weakening
We propose $Desiree$, a requirements calculus for systematically transforming stakeholder requirements into a specification. The framework includes six main building blocks: (1) an ontology for classifying requirements; (2) a language for representing requirements; (3) a set of requirements operators for transforming (refining or operationalizing) requirements; (4) a methodology for applying the operators on requirements concepts; (5) a prototype tool in support of the entire framework; (6) a set of empirical evaluations conducted to evaluate the framework. The relations between the building blocks and the contributed research questions are shown in Table~\ref{tab:bblocks}.
%, including the ontology, operators and methodology
\begin{table}[!htbp]
  \caption {The building blocks of the \textit{Desiree} framework}
  \label{tab:bblocks}
  \vspace {0.2 cm}
  \centering
  \small
  \begin{tabular}{|c|c|c|}
  \hline
  \textbf{Id} & \textbf{Building blocks} & \textbf{Contributed research questions} \\ \hline
  1 & An ontology for requirements & RQ1 \\ \hline
  2 & A language for requirements and semantics& RQ2 \\ \hline
  3 & A set of requirements operators and semantics& RQ3 \\ \hline
  4 & A methodology for applying operators & RQ3 \\ \hline
  5 & A prototype tool & RQ4 \\ \hline
  6 & A set of empirical evaluations & RQ4 \\
  \hline
  \end{tabular}
\end{table}

We overview now each building block, showing its contribution in advancing the state of the art.

%In our ontology for requirements, we have 9 kinds of requirements concepts in total, being classified into into four classes: (i) functional requirements, including ``\textit{functional goal}'' (FG), ``\textit{function}'' (F) and ``\textit{functional constraint}'' (FC); (ii) quality requirements, including ``\textit{quality goal}'' (QG) and ``\textit{quality constraint}'' (QC); (iii) content requirements, including ``\textit{content goal}'' (CTG) and ``\textit{state constraint}'' (SC); and (iv) ``\textit{domain assumption}'' (DA).

% is entirely about environment phenomena
% We provide an ontology for requirements. any natural language
%\begin{itemize}
%\item
\vspace {12pt}
\noindent \textbf{An Ontology for Requirements}. In general, we take a goal-oriented perspective and capture stakeholder requirements as \textit{goals} (G). We have 8 kinds of concepts except the most general \textit{goal}: 3 sub-kinds of goals (i.e., requirements), 4 sub-kinds of specification elements, and domain assumptions. Orthogonally, these 8 requirements concepts can be classified into into four categories: (1) function-related; (2) quality-related; (3) content-related; and (4) domain assumptions.

  \begin{itemize}
        %We treat a functional requirement (FR) as a requirement referring to a \textit{function}, a desired capability of its subject/bearer.
        %We treat an non-functional requirement (NFR) as a requirement referring to a \textit{quality}, and requires that quality to take quality value in an expected quality region.
      \item Function-related requirements. If a requirement specifies a desired state, and needs some functions to make it true (e.g., ``\emph{airline tickets be booked}''), it is a \textbf{{functional goal}} (FG); if a requirement specifies what the software ecosystem shall do (e.g., ``\emph{the system shall allow users to book airline tickets}''), it is a  \textbf{{function}} (F); if a requirement constrains the situation of a function (e.g., ``\emph{only managers are allowed to activate pre-paid cards}''), it is a \textbf{functional constraint} (FC).
      \item Quality-related requirements. If a requirement refers to a quality and specifies a vague quality region (e.g., ``\emph{the file search function shall be fast}''), it is a  \textbf{{quality goal}} (QG); if a requirement refers to a quality but has a measurable quality region (e.g., ``\emph{the file search function shall take less than 30 sec.}''), it is a  \textbf{{quality constraint}} (QC).
      \item Content-related requirements. We use a \textbf{{content goal}} (CTG) to specify a set of properties of a real world entity (e.g., ``\emph{a student shall have Id, name and GPA}''), and a \textbf{{state constraint}} (SC) specify the system state that represents the desired word entity/sate, capturing conventional data requirements (e.g., the student record table shall include three columns: Id, name and GPA).
      \item Domain assumptions. We use ``\textbf{domain assumption}'' (DA) to describe the environment of the system-to-be (e.g., ``\emph{the system shall have a functioning power supply}'') or domain knowledge (e.g., ``\emph{Tomcat is a web server}'').
  \end{itemize}

  %\textit{Addressed research questions}: \textbf{RQ1}.\\
  %(see the practical implication discussed in~\cite{guizzardi_ontological_2014})
  %operational guidance for distinguishing between them, and provides support for designing requirements specification language

  \textit{{Contribution in advancing the state of the art}}. Our ontological interpretation for requirements clarifies the concepts of functional and non-functional requirements, enabling us to (1) effectively distinguish between FRs and NFRs; (2) address the inconsistency between quality classifications; (3) reasonably measure the satisfaction of ``good-enough'' NFRs. It also offers support for designing requirements specification language. In our ontology, we have revised the concept ``\textit{functional goal}'' in the Core Ontology for Requirements Engineering (aka CORE)~\cite{jureta_revisiting_2008}\cite{juretaa_core_2009}, the state of the art requirements ontology, and enriched this CORE ontology with new requirements concepts ``\textit{functional constraint}'', ``\textit{content goal}'' and ``\textit{state constraint}''. Moreover, we have validated our ontology for requirements by classifying all the 625 requirements in the PROMISE requirements set~\cite{menzies_promise_2012}. To the best of our knowledge, our evaluation is the first attempt of applying an ontology for requirements to realistic requirements data.

    %(e.g., ``${Backup <actor: {the\_system}> <object: Data> <when: Weekday>}$'')
    %(e.g., ``\textit{$Airline\_ticket \, :< Booked$}'')
    %, and further suggest guidelines for refining requirements inspired by the description-based representation
    %We capture the interrelations between requirements through shared slot-description pairs (SlotDs) or descriptions.
    %and identify possibly missing requirements (e.g., if ``\textit{user}'' is an information entity, what kinds of attributes will be used to characterize it?).
%\item

\vspace {12pt}
\noindent \textbf{A Description-based Language for Requirements}. We propose a description-based language for representing requirements, both functional and non-functional. We treat a FR/NFR as a description consisting of slot-description pairs, and capture the intention of an entity (e.g., a physical object, a software function) to be in certain state/situation by using an assertion-like syntax. We define the semantics for the syntax using set theory, and capture part of the meaning by \textit{{automatically}} translating the syntax into DL~\cite{baader_description_2003}. With the support of off-the-shelf reasoners (e.g., Hermit~\cite{shearer_hermit:_2008}), we are able to query the interrelations between requirements (e.g., what kinds of requirements will be affected if we change a function?), detect inconsistencies (e.g., ``\textit{user}'' as a physical entity in the real word vs. ``\textit{user}'' as an information entity stored in a system), and ``what-if'' analysis (e.g., what kinds of elements in a model will be affected if some Function, FCs, or QCs are fulfilled while others are not).

      Descriptions, inspired by AI frames and DL, have the general form ``$Concept <slot_1: D_1> ... <slot_n: D_n>$'' where $D_i$ is a description that restricts $slot_i$. For example, the requirement $R_1$ ``\emph{the system shall be able to backup data at weekdays}'' can be captured as ``${Backup <actor: \{the\_system\}> <object: Data> <when: Weekday>}$''. This form offers intuitive ways to strengthen or weaken\footnote{Weakening a requirement means that one will have more choices to satisfy that requirement, while strengthening means the opposite. In this sense, traditional AND- and OR-refinements can be generally taken as strengthening. In fact, current goal modeling techniques rarely allow modelers to weaken a requirement, which is a distinguishing feature of this work. This will be further discussed when defining the semantics of operators.} requirements. For instance, $R_1$ can be strengthened into ``\textit{$Backup <actor: \{the\_system\}> <object: Data> <when: \{Mon, \, Wed, \, Fri\}>$}'', or weakened into ``$Backup <actor: \{the\_system\}> <object: Data> <when: Weekday \lor \{Sat\}>$''. With this syntax, the interrelations between qualities and functions can be captured by writing ``$Quality \, <inheres\_in: Function>$'', or more elegantly, ``$Quality \, (Function)$''. For example, the requirement ``\emph{the file search function shall take less than 30 sec.}'' can be denoted as ``$Processing\_time \, (File\_search) :: \, \leq 30 \, (sec.)$'', where ``$::$'' means taking value in the region ``$[0, 30 \, (sec.)]$''. The syntax for each requirement concept is sketched as follows.

      %For example, the requirement ``\textit{the system shall be able to backup data at weekdays}'' can be captured as ``${Backup <actor: {the\_system}> <object: Data> <when: Weekday>}$''.
      %(e.g., one can specialize $<object: Airline\_ticket>$ to $<object: Economic\_airline\_ticket>$)
      %Both \textbf{Functional goals} (FGs) and \textbf{functional constraints} (FCs) have an assertion-like syntax ``\textit{$ E_t :< S_s $}'', where ``\textit{$E_t$}'' is an entity (e.g., a physical object, a software function), ``\textit{$S_s$}'' is an intended state/situation, ``\textit{$:<$}''
      %where ``\textit{Q}'' is a quality name (type), ``\textit{SubjT}'' is the type of the subject that a quality (of type ``\textit{Q}'') inheres in, ``\textit{QRG}'' is the desired quality region that the quality shall take its value in
  \begin{itemize}
    \item \textbf{Functions} (Fs) have the syntactic form ``\textit{$F := FName <slot: D>^*$}'', where ``\textit{FName}'' is the function name that describes the desired capability, ``*'' means zero or more. For example, the requirement ``\emph{the system shall allow users to book airline tickets}'' can be captured as ``\textit{$F_1 := Book <actor: User> <object: Airline\_ticket> $}''. Both \textbf{functional goals} (FGs) and \textbf{functional constraints} (FCs) have an assertion-like syntax. For example, ``\emph{airline tickets be booked}'' will be captured as an FG ``\textit{$FG_1 := Airline\_ticket :< Booked$}''; ``\emph{only managers are able to activate pre-paid cards}'' can be captured as an FC ``\textit{$FC_1 := Activate <object: Prepaid\_card> \, :< \, <actor: ONLY Manager>$}''.
    \item \textbf{Quality goals} (QGs) have the form ``\textit{$QG := Q \, (SubjT) :: QRG$}'', meaning that each quality (of type ``\textit{Q}'') shall map its subject/bearer (of type ``\textit{SubjT}''), if exist, to a quality value that is in the desired region ``\textit{QRG}''. For example,  the requirement ``\emph{the file search function shall be fast}'' can be captured as ``\textit{$QG_1 := Processing\_time \, (File\_search) :: \, Fast$}''. Here ``\textit{File\_search}'' is a function that is defined elsewhere. \textbf{Quality constraints} (QCs) have the same syntax as QGs, but with measurable quality regions. For example, ``\textit{$QC_1 := Processing\_time \, (File\_search) :: \, \leq 30 \, (sec.)$}''.
    \item \textbf{Content goals} (CTGs) and \textbf{state constraints} (SCs) share an assertion-like syntax. For example, ``$CTG_1 := Student \, :< \, <has\_id: ID> <has\_name: Name> <has\_gpa: GPA>$'' and ``$SC_2 := Student\_record :< \, <ID: String> <name: String> <GPA: Float>$''.
    \item \textbf{Domain assumptions} (DAs) have a syntax similar to that for FCs. For instance, ``\textit{$DA_1 := \{the\_system\} :< \, <has\_power: Power>$}'' or  ``$DA_2$ := Tomcat $:<$ Web\_server''.
        %The difference is that an FC requires a subject to possess certain properties (e.g., in certain situations), while a DA assumes that a subject will have certain properties.
\end{itemize}

  %\textit{Addressed research questions}: \textbf{RQ2}.\\
  %(2) it is relatively easy to write, understand and communicate, and can be automatically translated to other formal specifications (e.g., DL~\cite{baader_description_2003}), hence it does not require high expertise as those mathematical specification languages do (e.g., Linear Temporal Language, Z and Software Cost Reduction, see a summary and discussion in~\cite{van_lamsweerde_building_2001});
  %We propose a description-based representation for requirements.
  %(1) it is able to specify not only functional, but also non-functional requirements;
  %(2) it is able to disambiguate requirements~\cite{li_stakeholder_2015}, and capture requirements even if they are informal or incomplete (unlike KAOS~\cite{dardenne_goal-directed_1993} and formal Tropos~\cite{fuxman_formal_2003}, which assume stakeholder requirements to be in enough detail and can be directly formalized);
  \textit{{Contribution in advancing the state of the art}}. Our description-based representation for requirements makes several contributions: (1) unlike many formal languages that are able to capture only functional requirements (e.g., Linear Temporal Language, Z and Software Cost Reduction, see a summary and discussion in~\cite{van_lamsweerde_building_2001}), our language is able to represent both functional and non-functional requirements; (2) our description-based syntax facilitates the identification of requirements issues such as incompleteness, ambiguity and unverifiability because it drives analysts/engineers to think about the properties (slots) of a function capability (e.g., actor, object, means, etc.), the cardinality of the description of a slot, the quantification of a vague quality region, etc. (see the analysis in Section~\ref{sec:expr_discussion} for more detail); (3) our syntax suggests operational guidelines for refining requirements, e.g., one can specialize the description of a slot in a function (F) description via an ``\textit{IsA}'' hierarchy, or refine a QG/QC via standard quality hierarchies like the ISO/IEC 25010 standard~\cite{iso/iec_25010_systems_2011}; (4) it captures the interrelation between requirements based on requirements details, especially the existential dependency relation between qualities and functions, which has been missing in current RE techniques~\cite{li_non-functional_2013}; moreover, this allows us to systematically identify the requirements to be affected when changes occur to a requirement specification (i.e., requirements specifications will have better modifiability).
  %to make it more precise

%  \item
\vspace {12pt}
\label{block:operator}
\noindent \textbf{Requirements Operators}. We introduce a set of of 8 operators for refining or operationalizing requirements, mainly inspired by our representation for requirements. For refinement, we have ``\textit{Reduce}''(\bm{$R_d$}),``\textit{Interpret}''(\bm{$I$}), ``\textit{Focus}''(\bm{$F_k$}), ``\textit{deUniversalize}''(\bm{$U$}), ``\textit{Scale}''(\bm{$G$}), and ``\textit{Resolve}'' (\bm{$R_s$}); for operationalization, we have ``\textit{Operationalize}''(\bm{$O_p$}) and ``\textit{Observe}''(\bm{$O_b$}). The \bm{$R_d$} and \bm{$O_p$} operators are adapted from existing goal modeling techniques, while the other 6 operators are newly proposed in this thesis. These 8 operators can be applied as follows.

  % ``\textit{refinement}'' and ``\textit{operationalization}'' operators

  %\item The ``\textit{Reduce}'' (\bm{$R_d$}) operator is used for refining a higher-level modeling element {$E_R$} to lower-level elements of the same type. Here the element {$E_R$} can be a goal (G), functional goal (FG), function (F), functional constraint (FC), and domain assumption (DA). That is, we allow reducing a goal to goals, a FG to FG(s), an F to F(s), etc. For example, the goal G1 ``\textit{collect real time traffic info}'' can be reduced to G2 ``\textit{traffic info be collected}'' and G3 ``\textit{collected traffic info be in real time}''.
  %\item The ``\textit{Interpret}'' (\bm{$I$}) operator is used for disambiguating a requirement (if needed), and encoding a goal written in natural language to an element $E_I$ specified using our syntax. Here the element type $E_I$ can be FG, QG, CTG, F, FC, QC and SC. For example, the ambiguous goal G1 ``\textit{notify users with email}'' can be disambiguated and encoded as F2 ``$Notify <object: User> <means: Email>$''.
  \begin{itemize}
  \item The ``\textit{Reduce}'' (\bm{$R_d$}) operator is used for refining a higher-level modeling element to lower-level elements of the same type. That is, we allow reducing a goal to goals, a FG to FG(s), an F to F(s), etc. For example, the goal $G_1$ ``\textit{collect real time traffic info}'' can be reduced to $G_2$ ``\textit{traffic info be collected}'' and $G_3$ ``\textit{collected traffic info be in real time}''.
  \item The ``\textit{Interpret}'' (\bm{$I$}) operator is used for disambiguating a requirement (if needed), and encoding a goal written in natural language using our syntax. For example, the ambiguous goal $G_1$ ``\textit{notify users with email}'' can be disambiguated and encoded as $F_2$ ``$Notify <object: User> <means: Email>$'' (another possible interpretation is $Notify <object: User \; <has\_email: Email>>$).
  \item ``\textit{Focus}'' (\bm{$F_k$}), ``\textit{de-Universalize}'' (\bm{$U$}) and ``\textit{Scale}'' (\bm{$G$}) are used for refining QGs/QCs: \bm{$F_k$} refines QGs/QCs via quality type or subject type (e.g., for the quality goal ``${Security \, (\{the\_system\}):: Good}$'', one can focus the quality type ``\textit{Security}'' to its sub-dimensions like ``\textit{Integrity}'', or focus the subject ``\textit{\{the\_system\}}'' into its parts, e.g., ``\textit{\{the\_data\_storage\}}''); \bm{$U$} is used to relax a QG/QC such that it is no longer expected to hold for 100\% of the individuals in the subject domain  (e.g., ``\textit{the file search function take less than 30 sec. all the time }'' to ``\textit{(at least) $80\%$ of the time}''); \bm{$G$} is used to enlarge or shrink the quality region of a QG/QC (e.g., enlarging ``$[0, \, 30 \, (sec.)]$'' to ``$[0, \, 35 \, (sec.)]$'' or shrinking it to ``$[0, \, 25 \, (sec.)]$''). In addition, the \bm{$O_b$} operator is used to operationalize subjective QGs such as ``\textit{the interface shall be simple}'' into QCs by asking obervers, in order to make them measurable.
  \item The ``\textit{Operationalize}''(\bm{$O_p$}) is used to operationalize a FG as F(s) and/or FC(s), e.g., the operationalization of the FG ``${User\_access :< Controlled}$'' as functions ``$Authenticate <object:: User>$'' and ``$Authorize <object:: User>$''; operationalize a QG as QC(s) to make it measurable, or operationalize the QG as F(s) and/or FC(s) to make it implementable; operationalize a CTG as SC(s). Note that any goal (G, FG, QG, or CTG) can be simply operationalized as a domain assumption, i.e., assuming it to be true.
  \item The ``$Resolve$'' (\bm{$R_s$}) operator is used to solve conflicts between requirements, which are captured by using the ``\textit{conflict}'' relation. Note that by conflict we do not necessarily mean logical inconsistency, but can also be other kinds such as normative conflict or infeasibility.
\end{itemize}
    %(2)transform requirements to a specification;
    %We propose a set of 8 operators for refining and operationalizing requirements based on their representations: the \bm{$R_d$} and \bm{$O_p$} operators are adapted from existing goal modeling techniques, the other 6 operators are newly proposed in this thesis.

    \textit{{Contribution in advancing state of the art}}. Our set of 8 operators enables us to: (1) go from informal to formal through disambiguating natural language requirements, and translating them into the \textit{Desiree} syntax; (2) repeatedly refine requirements and specification-elements, which are written in \textit{Desiree} syntax, making them complete enough, measurable, and consistent; (3) waken requirements if they are too hard or costly to attain, making them practically satisfiable. The incremental transition from informal to formal and the weakening of requirements are lacking in current RE techniques.

    %In general, a transformation process would include two important transitions: (1) from stakeholder requirements (i.e., sub-kinds of goals, FGs, QGs, or CTGs) written in natural language to \textit{Desiree} requirements specified in our syntax; (2) from \textit{Desiree} requirements to \textit{Desiree} specification elements (F, FC, QC, SC).
    %We use \bm{$I$} and \bm{$O_p$} for the first and second transition, respectively. In addition, we can repeatedly use other operators to strengthen or weaken stakeholder requirements, \textit{Desiree} requirements, and specification elements, in order to obtain an eligible specification at the end.
    %That is, specification elements, derived from requirements, can still be refined.
    %\item \textit{The first transition}. We use \bm{$I$} to interpret each atomic stakeholder requirement (i.e., a requirement with a single concern) as a \textit{Desiree} requirement specified in our syntax. We also allow directly interpreting stakeholder requirements as specification elements (if the given requirements can be classified as such).
    %\item \textit{The smithing phase}. This phase covers the second transition. This means that we would allow refining both \textit{Desiree} requirements and specification elements (i.e., before and after operationalization). Specifically,
%\item
\vspace {12pt}
\noindent \textbf{Transformation Methodology}. We provide a three-staged methodology for applying the operators to transform stakeholder requirements into an eligible specification. We outline the important steps as follows.
      \begin{itemize}
        \item \textit{The informal phase}. In this phase, we identify the concern of each stakeholder requirement, and use \bm{$R_d$} to separate its concerns if a requirement has multiple ones (i.e., it is complex). Meanwhile, if conflicts are found, we capture the conflicting requirements by using the ``\textit{Conflict}'' relation.
        \item \textit{The interpretation phase}. In this phase, we use the \bm{$I$} operator to disambiguate stakeholder requirements (if needed), classify and encode each atomic stakeholder requirement (i.e., a requirement with a single concern) by using our description-based syntax. In this phase, we allow interpreting stakeholder requirements as sub-kinds of goals (FG, QG, CTG) or directly as specification elements (F, FC, QC, SC). %(if the given requirements can be classified as such).
        \item \textit{The smithing phase}. In this phase, we incrementally refine the structurally specified goals (FG, QG, and CTG) and operationalize them as specification elements. Moreover, if needed, we keep refining specification elements (F, FC, QC, and SC). Specifically,
            \begin{itemize}
              \item FGs, FCs, CTGs, SCs can be refined by using the \bm{$R_d$} operator.
              \item QGs and QCs can be refined by using \bm{$F_k$}, relaxed using \bm{$U$} and \bm{$G$}.
              \item Conflicts can be resolved by using the \bm{$R_s$} operator.
              \item FGs, QGs, CTGs can be operationalized as corresponding specification elements by using the \bm{$O_p$} operator (as discussed in the requirements operators sub-section above).
            \end{itemize}
        %\item \textit{The second transition}. We use the \bm{$O_p$} operator to operationalize FGs, QGs, CTGs to corresponding specification elements (as discussed in the requirements operators sub-section above).
      \end{itemize}
    %is interweaved with

  \textit{{Contribution in advancing state of the art}}. The methodology tells analysts/engineers when to use which operator, in order to remove ambiguities and vagueness, reducing incompleteness, eliminating unattainability and conflict, turning informal requirements into an eligible specification. The existing goal-oriented frameworks often lack such a methodology, especially guidelines for going from informal to formal and weakening requirements during requirements analysis.

    %drag and drop nodes and edit their
%  \item
\vspace {12pt}
\noindent \textbf{Prototype Tool}. We have developed a prototype tool in support of the \textit{Desiree} framework, including all 9 requirements concepts, the 8 operators and the transformation process. The tool includes three key components: (1) a textual editor, which allows analysts to write requirements using our language; (2) a graphical editor, which allows analysts to draw models through a graphic interface; (3) a reasoning component, which (partially~\footnote{The translation does not support expressions with nested \bm$U$ operators.}) translates requirement specifications, either texts or models, to DL ontologies, and make use of existing reasoners (e.g., Hermit~\cite{shearer_hermit:_2008}) to perform reasoning tasks such as interrelations query, inconsistency check, and ``what-if'' analysis.

    \textit{{Contribution in advancing state of the art}}. The prototype tool supports the entire framework, and provides hints for applying requirements operators. It automatically translates the \textit{Desiree} syntax to DL~\cite{baader_description_2003}, allowing us to do some interesting reasoning (see Section~\ref{sec:eval_method} for more detail).
    %Li et al.~\cite{li_stakeholder_2015} for more detail on the types of reasoning questions that our framework can support
  %\item

\vspace {12pt}
\noindent \textbf{Empirical Evaluations}. We have conducted a set of empirical evaluations to assess our proposal, including case studies and controlled experiments. We use the PROMISE requirements set~\cite{menzies_promise_2012} as our requirements and projects repository.
      \begin{itemize}
        \item We evaluated the coverage of our ontology for requirements by classifying all the 625 requirements, functional and non-functional, in this repository.
        \item We evaluated the expressiveness of our language by rewriting the 625 requirements in the data set.
        \item We evaluated the effectiveness of our methodology by conducting a realistic ``\textit{Meeting Scheduler}'' case study chosen from the data set. % and ``\textit{Nursing Scheduler}''.
        \item We evaluated the effectiveness of the entire \textit{Desiree} framework by conducting a series of three controlled experiments, using two testing projects chosen from the data set: ``\emph{Meeting Scheduler}'' and ``\emph{Realtor Buddy}''.
      \end{itemize}

    \textit{{Contribution in advancing state of the art}}. The evaluations show that our ontology and language are adequate in capturing requirements, our methodology is effective in engineering requirements specifications from stakeholder requirements. The controlled experiments provide strong evidence that \emph{Desiree} can indeed help people to conduct better requirements analysis (i.e., identifying and addressing more issues when transforming requirements into specifications). These empirical evaluations also allow us to identify the limitations of our language, and give useful feedback on improving the methodology and tool.

\section{Structure of the Thesis}
\label{sec:structure}
The remainder of the thesis is structured as follows:

%First, the baseline for our proposal, including GORE and the ontological foundation for function and quality. Second, related work on the treatment of FRs and NFRs, transforming stakeholder requirements to specifications, requirements elaboration (writing good requirements), empirical evaluation about languages and methods in RE.
%the treatment of FRs and NFRs, requirements elaboration (writing good requirements), transformation of stakeholder requirements to specifications, empirical evaluations about RE languages and methods.

%We first discuss the ontological foundations of functions and qualities. Based on the ontological foundations, we then

\begin{itemize}
  \item Chapter 2 summarizes the state of the art related to our work. We mainly review the related work on requirements ontologies/classifications, requirements specification and modeling, and requirements transformation (i.e., deriving specifications from requirements).
  \item Chapter 3 presents the research baseline of this thesis, including (1) goal-oriented requirements engineering; and (2) the ontological foundations of \emph{functions} and \emph{qualities}.
  \item Chapter 4 introduces the ontological interpretations of requirements. We first provide an ontological interpretation for functional and non-functional requirements based on the ontological meaning of functions and qualities, and then accordingly propose a syntax for representing both of them.
  \item Chapter 5 presents the \emph{Desiree} framework, which includes a set of requirements concepts, a set of requirement operators, a description-based syntax for representing these concepts and operators, and a systematic methodology for transforming stakeholder requirements into an eligible requirements specification.
  \item Chapter 6 presents the semantics of the \emph{Desiree} framework. We first define the semantics of requirements concepts and operators, and then discuss the partial translation of \emph{Desiree} syntax to DL ontologies.
  \item Chapter 7 introduces a prototype tool. In this chapter, we introduce the key components and features of our prototype tool. This tool supports the entire \emph{Desiree} framework, including concepts, operators, methodology and reasoning.
  \item Chapter 8 presents a set of empirical evaluations, which are conducted in order to evaluate: (1) the coverage of our requirements ontology; (2) the expressiveness of our language; (3)  the effectiveness of our methodology; (4) the effectiveness of the entire framework.
  \item Chapter 9 concludes the thesis by listing the contributions and limitations of the \emph{Desiree} framework, and sketches directions for further research.
\end{itemize}

\begin{comment}
\item Chapter 3 introduces the requirements concepts for modeling functional requirements (FRs). First, we provide an ontological interpretation and introduce a description-based syntax for FRs. Second, we define the semantics for the syntax using set theory, and discuss its translation to DL.
  \item Chapter 4 introduces the requirements concepts for modeling non-functional requirements (NFRs). First, we provide an ontological interpretation and introduce a description-based syntax for NFRs. Second, we define the semantics for the syntax using set theory, and discuss its translation to DL.
  \item Chapter 5 presents a set of requirements operators used for refining or operationalizing requirements. First, we introduce the signatures of the operators using examples. Second, we define the semantics for these operators and discuss their translation to DL.
  \item Chapter 6 offers a methodology for incrementally transforming stakeholder requirements to an eligible specification. We discuss the main steps involved in this process, and provide guidelines for applying the operators.
\end{comment}

\section{Published Papers}
\label{sec:papers}
%We list here published work related to this thesis. They are split into refereed and un-refereed, and and ordered by date of publication.
We list here published work related to this thesis, ordered by date of publication.
%\subsection{Refereed}
%\label{sec:papers_refereed}

\begin{itemize}
  \item  Feng-Lin Li, Jennifer Horkoff, John Mylopoulos, Lin Liu, and Alexander Borgida. Non-Functional Requirements Revisited. iStar'13, pp. 109-114, 2013.
  \item Renata Guizzardi, Feng-Lin Li, Alexander Borgida, Giancarlo Guizzardi, Jennifer Horkoff, and John Mylopoulos. An ontological interpretation of non-functional requirements. FOIS'14, pp. 344--347, 2014.
  \item Feng-Lin Li, Jennifer Horkoff, John Mylopoulos, Renata SS Guizzardi, Giancarlo Guizzardi, Alexander Borgida, and Lin Liu. Non-functional requirements as qualities, with a spice of ontology. RE'14, pp. 293--302, 2014.
  \item Feng-Lin Li, Jennifer Horkoff, Alexander Borgida, Giancarlo Guizzardi, Lin Liu, and John Mylopoulos. From Stakeholder Requirements to Formal Specifications Through Refinement. REFSQ'15, pp. 164--180, 2015.
  \item Jennifer Horkoff, Fatma Baak Aydemir, Feng-Lin Li, Tong Li, and John Mylopoulos. Evaluating Modeling Languages: An Example from the Requirements Domain. ER'14, pp. 260--274, 2014.
  \item Jennifer Horkoff, Tong Li, Feng-Lin Li, Mattia Salnitri, Elsa Cardoso, Paolo Giorgini,John Mylopoulos, and Joo Pimentel. Taking goal models downstream: A systematic roadmap. RCIS'14, pp. 1--12, 2014.
  \item Jennifer Horkoff, Tong Li, Feng-Lin Li, Mattia Salnitri, Evellin Cardoso, Paolo Giorgini, and John Mylopoulos. Using Goal Models Downstream: A Systematic Roadmap and Literature Review. IJISMD 6(2):1--42, 2015.
  \item Feng-Lin Li, Jennifer Horkoff, Lin Liu, Alexander Borgida, Giancarlo Guizzardi, and John Mylopoulos. Engineering Requirements with \emph{Desiree} : An Empirical Evaluation. Submitted to CAiSE'16.
  \item Feng-Lin Li, Alexander Borgida, Giancarlo Guizzardi, Jennifer Horkoff, Lin Liu, and John Mylopoulos. \emph{Desiree}: a Requirements Calculus for the Requirements Problem. To be submitted to RE'16.
\end{itemize}

%% file: related.work.2.tex
%\vspace{4cm}
\chapter{State of the Art}
\label{cha:state_art}
The state of the art in our research area can be generally classified into three categories: (1) requirements ontologies/classifications; (2) requirements specification and modeling; (3) requirements transformation (i.e., deriving specifications from requirements). We review these sub-areas and assess their adequacy in addressing the RE problem.

%\section{Requirements Concepts and Languages}
%\label{sec:art_fr_nfr}
\section{Requirements Ontology}
\label{sec:art_ontologies}

An initial conceptualization for RE was offered by Jackson and Zave~\cite{jackson_deriving_1995} nearly two decades ago, founded on three basic concepts: \textit{requirement}, \textit{specification} and \textit{domain assumption}. Based on this characterization, the classical requirements problem is defined as finding the specification \textit{S} consistent with the given domain assumptions \textit{DA} to satisfy the given requirements \textit{R}. For example, to satisfy the requirement ``\emph{make online payment}'' (\textit{R}), a software or service needs to support a function ``\emph{pay with credit card}'' (\textit{S}) under the (implicit) domain assumption of ``\emph{having a credit card with available credits}'' (\textit{DA}).

%follows: given a set of requirements \textit{R}, and a set of domain assumptions \textit{DA}, find a specification \textit{S} consistent with \textit{DA} such that $DA, S \models R$.
%In RE, it is widely accepted that requirements can be classified as functional and non-functional.

%Jackson and Zave do not distinct the taxonomic dimension for the ``requirement'' concept -- the distinction between functional and non-functional requirements, which is widely accepted in the RE field.

%This example indicates two deficiencies: (1) the definition for ``function'' is narrow as it is restricted to ``what a software system shall do'' only; (2) other agents (e.g., human beings) involved in the software ecosystem are not taken into consideration (in fact, this is related to the ``WHO'' issues as pointed by van Lamsweerde in~\cite{van_lamsweerde_building_2001}).

%does not allow partial fulfillment of some requirements and
On observing that this initial characterization leaves out important notions such as non-functional requirements, Jureta et al.~\cite{jureta_revisiting_2008}\cite{juretaa_core_2009} have proposed the CORE ontology based on goal-oriented requirements engineering (GORE), which is founded on the premise that requirements are stakeholder goals. Based on the understanding that functional requirements (FRs) describe what the system should do, whereas non-functional ones (NFRs) describe how well the system should perform its functions ~\cite{paech_non-functional_2004}, the CORE ontology distinguishes between non-functional and functional requirements using \textit{qualities} introduced in the DOLCE foundational ontology~\cite{masolo_ontology_2003}: (1) a requirement is non-functional if it refers to a quality; further, this requirement will be a \textit{softgoal} if it is vague for agreed success, or, alternatively, a \textit{quality constraint} if it is clear for success; (2) a requirement is a \textit{functional goal} if it refers to a perdurant (something accumulates parts over time, e.g., a process, an event) and does not refer to a quality.

Since its proposal in 2008, the CORE ontology has enjoyed considerable attention, and has served as the baseline of new research directions in RE~\cite{jureta_techne:_2010}\cite{liaskos_integrating_2010}. However, in our experience, it has several deficiencies in classifying requirements. First, it is ineffective on capturing requirements that refer to neither qualities nor perdurants, but endurants (whose proper parts are wholly present at any time, e.g., physical objects). For example, ``\emph{the user interface shall have menu buttons for navigation}'', where ``\emph{menu buttons}'' are endurants. We could classify this requirement as a general \textit{goal}, but can not further specialize it since it is neither functional nor non-functional according to CORE. This is insufficient if we want to transform such requirements into a specification. Second, it is difficult to capture requirements that are vague for success but do not refer to qualities. For example, requirements such as ``\emph{attract customers}'' and ``\emph{increase sales}'' refer to perdurants (processes/events) rather than qualities, and are accordingly classified as functional goals. However, this conclusion contradicts Jureta et al.~\cite{juretaa_core_2009}'s claim that ``functional goals are Boolean, i.e., true or false'' since these examples, like softgoals, have no clear-cut criteria for success. Third, requirements could refer to both qualities and functions. For example, although we can classify the requirement ``\emph{the system shall collect real-time traffic information}'' as a softgoal according to CORE (``\emph{real-time}'', i.e., timeliness, is a quality of traffic information), we are still left with the question ``is it only an NFR?''. It seems to be a combination of functional and non-functional concerns, which should be refined into distinct sub-goals.

Also, ontologies of specific domains, for which requirements are desired, have been employed in RE mainly for activities (e.g., requirements elicitation~\cite{kaiya_using_2006} and evolution~\cite{machado_using_2011}) or processes (e.g.,~\cite{falbo_evolving_2008}). These efforts, however, are not proposals for an ontological analysis of requirements notions. In fact, few researchers have attempted to ontologically analyze requirements. Our goal here is in the ontological classification and conceptual clarification of different requirement kinds.

\subsection{Functional Requirements (FRs)}
\label{sec:art_ontology_fr}
% According to Glinz's survey~\cite{glinz_non-functional_2007},

In RE, there is a broad consensus on how to define ``\textit{functional requirements}'' (FRs)~\cite{glinz_non-functional_2007}: the existing definitions focus on either functions (what a system shall do)~\cite{eee_standard_1990}\cite{sommerville_software_2006}\cite{robertson_mastering_2012} or behaviors (inputs, outputs, and the relations between them)~\cite{davis_software_1993}\cite{wiegers_software_2013}. For example, Robertson et al.~\cite{robertson_mastering_2012} define an FR as a requirement that describes an action a (software) product must take; while Davis~\cite{davis_software_1993} defines FRs as ``those requirements that specify the inputs (stimuli) to the system, the outputs (responses) from the system, and behavioral relationships between them''. As a (software) function often includes input and output~\cite{committee_ieee_1998}\cite{wiley_incose_2015}, these two threads of definitions (``what kinds of functions a software system shall have'' and ``what kinds of behavior a software system shall exhibit'') are compatible: a function can be treated as an input-output relation, and be specified using input, output, and maybe other things such as pre-, post- and trigger conditions as in Letier et al.~\cite{letier_deriving_2002}.

There are two points to be noted about these definitions for functional requirements. First, they focus narrowly on the software part of the system to be designed, and do not consider its environment. For example, how can one classify the requirement ``\emph{the product shall be supported by the corporate support center}'', which specifies a function (``\textit{support}'') that will not be performed by the system but by an external agent (``\textit{the corporate support center}'') in a system-to-be ecosystem?  Second, they do not acknowledge that a requirement can be a mix of concerns (e.g., having more than one function/quality, or including both functions and qualities). For instance, how to classify the requirement ``\emph{the system shall have standard menu buttons for navigation}''? This one has several concerns: (1) it refers to a (software) system function ``$navigate \, (users)$''; (2) it restricts the way to implement the function to ``\textit{use menu buttons}''; (3) it requires the menu buttons to be ``\textit{standard}'', which is about qualities (e.g., size, color).

\subsection{Non-functional Requirements (NFRs)}
\label{sec:art_ontology_nfr}
Unlike FRs, ``\textit{non-functional requirements}'' (NFRs) are highly controversial in RE: despite the wide acceptance of their importance and the many efforts devoted to them, there is still no ultimate consensus on what NFRs are and how to capture them~\cite{glinz_non-functional_2007}.

\vspace{12pt}
\noindent {{\textbf{NFRs Definition}}}. Recently, there are two important reviews on NFRs by Martin Glinz~\cite{glinz_non-functional_2007} and Chung et al.~\cite{chung_non-functional_2009}. Glinz~\cite{glinz_non-functional_2007} surveyed 13 NFR definitions, and suggested his own on the basis of attributes and constraints. However, his definition focuses on only (software) system NFRs, and does not consider project and process requirements (e.g., development, deployment, maintenance, etc.). Moreover, his proposal does not offer methodological guidance for designing language and method for capturing NFRs.

%In our framework, such processes will serve as subjects, allowing us to specify different kinds of desired qualities of them.

After analyzing a list of NFR definitions, Chung et al.~\cite{chung_non-functional_2009} defined FRs as mathematical functions of the form $f: I \rightarrow O$ ($I$ and $O$ represent the input and output of $f$), and NFRs as anything that concern characteristics of $f$, $I$, $O$ or relationships between $I$ and $O$. This definition, accordingly, will treat constraints over functions (FCs) as NFRs, e.g., ``\emph{updates of the database can only be performed by managers}'' is a functional constraint (it requires the system to check who is updating, and is not about the quality of updating), but an NFR according to Chung et al.~\cite{chung_non-functional_2009}'s definition.

\vspace{12pt}
\noindent \textbf{NFRs Classification}. NFRs are often classified into sub-categories. For example, the IEEE Std 830~\cite{committee_ieee_1998} classifies NFRs into interface requirements, performance requirements, attributes and design constraints, and documents them in a separate section from FRs using natural language. Similarly, the Volere template~\cite{robertson_mastering_2012} categories NFRs as look and feel, usability and humanity, performance, operational and environmental, maintainability and support, security, cultural and political, and legal requirements, and captures requirements (functional and non-functional) sentences as informal text using a structured template which includes attributes like ID, requirements type, description, rationale, fit criterion, etc.

%as shown in our examples chosen from the PROMISE requirements set~\cite{menzies_promise_2012}
There are several deficiencies in such approaches. First, they classify requirements strictly as functional or non-functional, leading to difficulties when classifying requirements that mix both functional and quality concerns (as shown in our evaluation in Section~\ref{sec:eval_ontology}, also observed in the work of Svensson et al.~\cite{svensson_investigation_2013}). Second, the sub-categories of NFRs in these approaches are open-ended, and would probably change as time, domain, or researcher varies. Third, such approaches document FRs and NFRs separately, hence do not capture the interrelations between them and do not reflect the cross-cutting concerns of NFRs (i.e., an NFR could refer to multiple system functions, parts or artifacts).

%(as they are documented separately in requirements documents)

%For example, the requirement ``the processing time of file search shall be acceptable'' specifies a quality ``processing time'' of the function ``file search'', which will be documented elsewhere according to the IEEE recommended practice for software requirements specifications (i.e., the 830-1998 standard)~\cite{committee_ieee_1998}. That is, the dependency relations between qualities and functions are missing~\cite{li_non-functional_2013}.

%Moreover, they do not reflect the cross-cutting concerns of NFRs, i.e., an NFR can refer to multiple parts of a system (e.g., different system functions or modules), and do not well capture the interrelations between FRs and NFRs (e.g., the requirement ``the processing time of file search shall be acceptable'' specifies a quality ``processing time'' of the function ``file search'', which will be documented elsewhere in the IEEE recommended practice for software requirements specifications).

\vspace{12pt}
\noindent \textbf{NFRs as Qualities}. \textit{Quality} is the most popular term adopted to specify NFRs. Many efforts have been made towards modeling and classifying qualities, resulting in fruitful quality models. The well-known models include McCall et al.~\cite{company_factors_1977}, Boehm et al.~\cite{boehm_characteristics_1978}, ISO/IEC 9126-1~\cite{iso/iec_9126-1_software_2001}, ISO/IEC 25010~\cite{iso/iec_25010_systems_2011}, etc., in which qualities and their interdependencies are usually organized in a hierarchical structure. One issue with these quality models is that they, even the well-known ones, are neither terminologically nor categorically consistent with each other~\cite{chung_non-functional_2009}. For example, ``understandability'' is a sub-quality of ``usability'' in ISO/IEC 9126-1~\cite{iso/iec_9126-1_software_2001}, but is a sub-class of ``maintainability'' in Boehm et al.~\cite{boehm_characteristics_1978}. The inconsistency in the classifications of software qualities has been also observed in comparative studies over quality models, e.g.,~\cite{al-badareen_software_2011}.

%To address this issue, Chung et al.~\cite{chung_non-functional_2009} have suggested that a software practitioner shall be aware of some of the well-known classification schemes, and consider one or more to adopt with some tailoring.

Although Chung et al.~\cite{chung_non-functional_2009} have suggested be aware of some of the well-known classification schemes and consider one or more to adopt with some tailoring, we are still left with the question: why these quality models differ from, and sometimes even conflict with each other?

\vspace{12pt}
\noindent \textbf{NFRs as Softgoals}. In goal-oriented approaches, the NFR framework~\cite{chung_non-functional_2000} and \textit{i}*~\cite{yu_modelling_2011} have used ``\emph{softgoals}'' (goals without clear criteria for success) to model non-functional requirements. However, goals lacking a clear criterion of satisfaction (i.e., softgoals) turn out to be not always NFRs -- most early requirements, as elicited from stakeholders are also ``\emph{soft}''. For instance, when a stakeholder says ``\emph{Upon request, the system shall schedule a meeting}'', this is also vague and needs to be made more firm: do we allow requests for any time (e.g., weekends)? Should the system notify participants about the scheduled meeting? Should it account for contingencies (e.g., power outage)? etc. Similarly, NFRs are not necessarily vague, e.g., the ``\emph{the processing time of file search shall be less than 30 seconds}'' is an NFR but is clear for success.

So, softgoals constitute a useful abstraction for early (vague) requirements, both functional and non-functional, rather than just non-functional ones. However, this conclusion begs the next question: what then are non-functional requirements, how do we model them and how do we use these models in RE processes?
%(and later)
%\end{itemize}

%such as UFO  and DOLCE~\cite{masolo_ontology_2003}
%to tell us precisely what qualities are.
%: we begin with treating NFRs as ``qualities''
%As we have discussed in Guizzardi et al.~\cite{guizzardi_ontological_2014}, NFRs are not necessarily vague for success (e.g., the ``the processing time of file search shall be less than 30 seconds'' is an NFR but is clear for success), and softgoals are not necessarily non-functional (e.g., ``increase profits'' is a softgoal since it is not clear how much )

\section{Requirements Specification and Modeling Languages}
\label{sec:art_specificaiton_and_modeling}
%To capture functional requirements, which define the fundamental actions that a software system must do, many modeling/specification languages were proposed.

In RE, great efforts have been placed to designing effective language for capturing requirements, resulting in various kinds of requirement specification and modeling techniques, including (1) natural language (NL); (2) requirements specification (RS) templates, e.g., the IEEE Std 830~\cite{committee_ieee_1998}, the Volere template~\cite{robertson_mastering_2012}; (3) writing guidelines, e.g., Alexander et al.~\cite{alexander_writing_2002}, Wiegers et al.~\cite{wiegers_software_2013}; (4) structured languages, e.g., EARS (the Easy Approach to Requirements Syntax)~\cite{mavin_easy_2009}, Planguage~\cite{gilb_competitive_2005}; (5) controlled natural languages, e.g., ACE~\cite{fuchs_attempto_2006}, Gervasi et al.~\cite{gervasi_reasoning_2005}; (6) visual modeling languages, e.g., $SADT^{TM}$ (Structured Analysis and Design Technique~\cite{ross_structured_1977-1}, UML~\cite{rumbaugh_unified_2004}, KAOS~\cite{dardenne_goal-directed_1993}, the NFR Framework (NFR-F)~\cite{chung_non-functional_2000}, \textit{i}*~\cite{yu_modelling_2011}; (7) formal languages, e.g., RML~\cite{greenspan_capturing_1982}\cite{greenspan_formal_1994}, VDM (Vienna Development Method)~\cite{fitzgerald_validated_2005}, Linear Temporal Logic~\cite{dardenne_goal-directed_1993}, and SCR (Software Cost Reduction) table~\cite{heitmeyer_automated_1996}.

% language style (textual and graphical) and
We classify these techniques according to their formalities (informal, semi-formal and formal), and show the classification result in Table~\ref{tab:classification}. There are two points to be noted. First, we take the viewpoint that modeling is broader than specification: a specification describes the behavior of software system, which is a particular type of model -- behavior model; in addition to behavior model, the models of a software system include many other kinds, e.g., architecture model, deployment model, responsibility model, etc. Second, writing guidelines by themselves are not languages, but empirical rules that help people to use languages. So, we do not classify them into any of the formality categories.

%classification over language style is just for the purpose of structuring the state-of-the-art, and there could be exceptions. For example, an SCR specification~\cite{heitmeyer_automated_1996} would include tables, use cases in a UML model can be expressed using natural language or tables.

%and is not complete and restrict. For example, a graphical language could use a textual language (natural or artificial) to specify its constructs. For example, UML~\cite{rumbaugh_unified_2004} often uses natural language for describing use cases, NFR-F~\cite{chung_non-functional_2000} has proposed a simple syntax (``$type[topic]$'', e.g., ``$accuracy[account]$'') for its key concept -- softgoal.

%A natural language or an artificial language for a particular application can be embedded into this graphical framework.
%There are two points to be noted. First, a graphical specification language  we treat graphical requirements models as a kind of requirements specification.

\begin{table}[!htbp]
  \caption {The classification of requirements specification languages}
  \label{tab:classification}
  \vspace {0.2 cm}
  \centering
  \small
  \begin{tabular}{|c|c|c|l|}
  \hline
  \textbf{Type} & \textbf{Kind} & \textbf{Formality} & \textbf{Examples} \\ \hline
  Guideline & Writing Guidelines & -- & Alexander et al.~\cite{alexander_writing_2002}, Wiegers et al.~\cite{wiegers_software_2013}\\ \hline
  \multirow{6}{*}{Specification} & Natural Language & Informal & English \\
  & RS Template & Semi-formal & IEEE Std 830~\cite{committee_ieee_1998}, Volere~\cite{robertson_mastering_2012} \\
  & Structured Language & Semi-formal & EARS~\cite{mavin_easy_2009}, Planguage~\cite{gilb_competitive_2005}\\
  & Controlled NL & Semi-formal & ACE~\cite{fuchs_attempto_2006}, Gervasi et al.~\cite{gervasi_reasoning_2005} \\
  & Formal Language & Formal & RML~\cite{greenspan_capturing_1982}, SCR~\cite{heitmeyer_automated_1996}, VDM~\cite{fitzgerald_validated_2005}\\ \hline
  \multirow{3}{*}{Modeling} & Structural Analysis & Semi-formal & $SADT^{TM}$~\cite{ross_structured_1977-1}\\
  & Object Orientation & Semi-formal & UML~\cite{rumbaugh_unified_2004}\\
  & Goal Orientation & Semi-formal & KAOS~\cite{dardenne_goal-directed_1993}, NFR-F~\cite{chung_non-functional_2000}, \textit{i}*~\cite{yu_modelling_2011}\\
  \hline
  \end{tabular}
\end{table}

%ER (Entity-Relation) diagram~\cite{chen_entity-relationship_1976},
%Z~\cite{potter_introduction_1996},
%In practice, requirements are usually expressed in natural language as it is known to everyone and can serve as the common ground of all the people involved in software development processes (e.g., customers, analysts, engineers, etc.).
%Natural language (NL) is a common way used for writing requirements in practice.
\subsection{Requirements Specification Languages}
\label{sec:art_specification}

%missing information or requirements,
When considering the use of a specification language for representing requirements, there are several candidates: (1) natural language, which is known to every one but error prone, (2) formal languages, which are precise but hard to understand for non-experts; (3) semi-formal/structured languages, which are proposed with the purpose of mediating natural and formal languages.

\vspace{12pt}
\noindent \textbf{Natural Language}. A common way to express requirements is to use natural language (NL). According to a survey conducted in 2004 by Luisa et al.~\cite{luisa_market_2004}, 79\% of the 151 respondents have reported that their previous requirements documents were written in natural language, while only 16\% and 5\% of them reported the use of structured and formal languages, respectively. Natural language is easy to use and understand, and can serve as the common ground of everyone involved in software development processes, e.g., customers, analysts, engineers, etc.

%many experiences have shown
%various kinds of issues
%~\cite{menzies_promise_2012}
However, there is much evidence that requirements written in natural language are inherently vague and error prone~\cite{fabbrini_linguistic_2001}, leading to undesirable properties such as ambiguity, incompleteness and inconsistency~\cite{berry_contract_2003}\cite{kordon_advances_2008}\cite{yang_automatic_2010}\cite{kamalrudin_automated_2009}. As mentioned earlier, we have found that 3.84\% of all the 625 requirements (functional and non-functional) in the PROMISE requirements set are ambiguous~\cite{li_stakeholder_2015}, 25.22\% of the 370 non-functional requirements (NFRs) are vague, and 15.17\% of the NFRs are potentially unattainable~\cite{li_non-functional_2014}. For example, the requirement ``\emph{the system shall be able to send meeting notification via Email and SMS}'' is ambiguous because it can be interpreted in two ways: (1) a meeting notification need to be sent via both Email and SMS; (2) the system needs to support these two means, but a meeting notification only needs to be sent by one of them. Moreover, this requirement is incomplete: it is unclear who can send meeting notifications and to whom the notifications will be sent since such information is missing.

\vspace{12pt}
\noindent \textbf{Requirements Specification (RS) Templates}. Templates such as the IEEE Recommended Practice for Software Requirements Specifications~\cite{committee_ieee_1998} and the Volere Requirements Specification Template~\cite{robertson_mastering_2012} represent the most basic types of tool used for requirements engineering. Using such RS templates, requirements are usually represented as requirements sentences (e.g., a set of ``\textit{the system shall}'' statements), and are listed under different sections of a requirements document. For example, in the IEEE Std 830 template~\cite{committee_ieee_1998}, section 3.2 is suggested for specifying functional requirements, while the rest of section 3 is used to document different types of NFRs. Further, the Volere template~\cite{robertson_mastering_2012} has provided a structure around requirements sentences, and use a set of attributes such as ID, requirements type, description, rationale, and fit criterion to specify individual requirements.

RS templates are useful in classifying and documenting individual requirements, but they offer very limited support for requirements management (e.g., FRs and NFRs are documented separately, the interrelations between them are hence scattered in RS documents and are difficult to trace).

% for writing requirements
\vspace{12pt}
\noindent \textbf{Writing Guidelines}. To help people write better requirements, writing guidelines have been suggested~\cite{alexander_writing_2002}\cite{wiegers_more_2005}\cite{wiegers_software_2013}. These approaches usually use a set of properties of good requirements (e.g., the IEEE Std 830~\cite{committee_ieee_1998}, or the semiotic quality framework~\cite{krogstie_semiotic_2002}) as criteria, and provide a set of operational guidelines. For example, Alexander et al.~\cite{alexander_writing_2002} have offered a list of guidelines for writing good requirements (e.g., use simple direct sentences, use a limited vocabulary) and a set of ``DO NOT'' rules (e.g., do not ramble, do not make mixed requirements). Similarly, after discussing each characteristic in the IEEE Std 830~\cite{committee_ieee_1998}, Wiegers et al.~\cite{wiegers_software_2013} have provided a set of guidelines for reducing or eliminating requirements issues. For example, the authors listed a set of terms that may lead to ambiguity (e.g., ``and'', ``or''), and suggested a symmetry strategy for reducing incompleteness (e.g., for an ``if'' statement, we need to look for the corresponding ``else'' statement). In general, these techniques are often informal, and lack a systematic methodology and tool support.

%and offer very limited support for requirements management (e.g., FRs and NFRs are documented separately, their interrelations are scattered in requirements documents and are difficult to trace).

%Structured languages, e.g., EARS (the Easy Approach to Requirements Syntax)~\cite{mavin_easy_2009} and Planguage~\cite{gilb_competitive_2005}, offer a structured syntax for requirements, intending to reduce or eliminate some kinds of requirements issues such as ambiguity and vagueness.

\vspace{12pt}
\noindent \textbf{Structured Languages}. Structured languages such as EARS~\cite{mavin_easy_2009} and Planguage~\cite{gilb_competitive_2005}, have been proposed based on practical experiences, intending to reduce or eliminate certain kinds of requirements issues (e.g., ambiguity and vagueness).

A generic structure for requirements from the EARS approach is shown in Eq.~\ref{syntax_ears}, where ``[]'' means optional, ``$\langle \rangle$'' indicates parameters to be instantiated. EARS also includes additional syntax constructs for event-driven, unwanted behavior, state-driven, optional and complex requirements. For example, the requirement ``\emph{If the computed airspeed fault flag is set, then the control system shall use modelled airspeed}'' is an example of ``unwanted behavior'' requirements. EARS is developed mainly for specifying functional requirements and has been reported effective on reducing ambiguity, vagueness and wordiness~\cite{mavin_easy_2009}.
\begin{equation}\label{syntax_ears}
            \lbrack \textit{pre-condition}\rbrack \, \lbrack \textit{trigger}\rbrack \,\, \textit{the} \,\, \langle system \, name \rangle \,\, \textit{shall} \,\, \langle \textit{system response} \rangle
\end{equation}

The planning language (i.e., Planguage~\cite{gilb_competitive_2005}), proposed by Tom Glib [38] and widely used in industry, uses a set of keywords such as $tag$ (a unique, persistent identifier), $scale$ (the scale of measure used to quantify the requirement, e.g., time, temperature, speed), $meter$ (the process or device used to establish location on a scale, e.g., watch, thermometer, speedometer) and $must$ (the minimum level required to avoid failure) to write measurable quality requirements. Following is an example use of the Planguage for specifying the NFR ``\emph{the processing of order shall be fast}''~\footnote{http://www.iaria.org/conferences2012/filesICCGI12/Tutorial\%20Specifying\%20Effective\%20Non-func.pdf}.

        \begin{table}[!htbp]
        \label{tab:example_planguage}
        \centering
        \begin{tabular}{|rp{0.6\textwidth}|}
        \hline
            Tag: & Order Processing Time \\
            %Ambition: & Do not make the users wait too long for order processing \\
            Scale: & Time \\
            Meter: & Measured from the click on the ``Submit Order'' button to the display of the ``Order Complete'' message \\
            Must: & 5 seconds \\
        \hline
        \end{tabular}
        \end{table}

%have shown
%, and are not for both of them
There is empirical evidence~\cite{mavin_easy_2009}\cite{jacobs_introducing_1999} that these approaches are effective for their intended purpose. However, they are designed exclusively for only FRs or NFRs. The interrelations between functional and non-functional requirements (e.g., NFRs could introduce FRs~\cite{glinz_non-functional_2007}, NFRs could depend on FRs~\cite{li_non-functional_2013}) have been also missing.

%The transition from informal ideas to formal specifications is an inescapable part of software development because stakeholder needs, which are inherently informal, must be transformed into software, which is inherently formal~\cite{kordon_advances_2008}. Because of this, formal specification languages have been advocated for specifying requirements specifications~\cite{bowen_seven_1995}\cite{bowen_ten_1995}\cite{bowen_ten_2006}. Example formal requirements specification languages include RML~\cite{greenspan_capturing_1982}\cite{greenspan_formal_1994}, VDM (Vienna Development Method)~\cite{fitzgerald_validated_2005}, SCR (Software Cost Reduction) table~\cite{heitmeyer_automated_1996}, KAOS~\cite{dardenne_goal-directed_1993} (uses Linear Temporal Logic as its formalism, also does Formal Tropos~\cite{fuxman_formal_2003}), etc.

\vspace{12pt}
\noindent \textbf{Formal Languages}. In software development, stakeholder needs, which are inherently informal, will be finally transformed into software, which is inherently formal~\cite{kordon_advances_2008}. Since the transition from informal ideas to formal specifications is inescapable, formal languages have been advocated for specifying requirements specifications~\cite{kordon_advances_2008}. Example formal requirements specification languages include RML~\cite{greenspan_capturing_1982}\cite{greenspan_formal_1994}, VDM (Vienna Development Method)~\cite{fitzgerald_validated_2005}, SCR (Software Cost Reduction) table~\cite{heitmeyer_automated_1996}, KAOS~\cite{dardenne_goal-directed_1993} (with Linear Temporal Logic as its formalism, also does Formal Tropos~\cite{fuxman_formal_2003}), etc.

Formal specification languages have been advocated because they have a clear syntax and semantics, and promise much more sophisticated analysis such as ambiguity detection~\cite{fuchs_attempto_2006}, inconsistency check~\cite{gervasi_reasoning_2005}, animation of a specification, and validation of a specification against certain properties~\cite{fuxman_formal_2003}. However, they suffer from major shortcomings~\cite{van_lamsweerde_building_2001}: (1) they require substantial expertise, are hard to write, understand and communicate, and are not really accessible to practitioners and customers; (2) they mainly focus on functional aspects, and leave out non-functional ones, a very important class of requirements in RE. Interested readers can refer to van Lamsweerde~\cite{van_lamsweerde_building_2001}\cite{van_lamsweerde_requirements_2009} for a detailed review of (semi-formal and) formal approaches used for requirements specifications.

%it is intuitive to design some kind of
\vspace{12pt}
\noindent \textbf{Controlled Natural Languages}. To bridge the gap between natural and formal specification languages, ``intermediate'' languages, which are relatively easy to use and understand, and can be automatically translated into a formal language, have been suggested. In this direction, controlled natural languages (CNLs), which are sub-sets of natural languages and are obtained by restricting the grammar and vocabulary of NLs in order to reduce or eliminate ambiguity and complexity~\cite{_controlled_????}, have enjoyed much attention~\cite{gervasi_reasoning_2005}\cite{fuchs_attempto_2006}~\cite{konrad_real-time_2005}\cite{mavin_easy_2009}\cite{gilb_competitive_2005}.

%Controlled natural languages (CNLs) are sub-sets of natural languages, obtained by restricting the grammar and vocabulary in order to reduce or eliminate ambiguity and complexity~\cite{_controlled_????}. In general, CNLs fall into two categories: (1) the first type of CNLs (often called ``simplified'' or ``technical'') are designed to improve human readability and are used to increase the quality of technical documentation; (2) the second type of CNLs are designed to enable automatic semantic analysis: have a formal syntax and semantics, and can be mapped to an existing formal logic~\cite{_controlled_????}.

%Logic-based CNLs have been favored because they can integrate the familiarity of natural language and the rigor of formal specification languages.

%In the literature,

Some of these approaches require users to write requirements using controlled grammars and vocabularies (e.g., Fuchs et al.~\cite{fuchs_attempto_2006}, Konrad et al.~\cite{konrad_real-time_2005}); some of them employ natural language processing (NLP) techniques to extract useful information from NL requirements according to some grammar rules and vocabularies (e.g., Fantechi et al.~\cite{fantechi_assisting_1994}, Gervasi et al.~\cite{gervasi_reasoning_2005}). In either way, full NL requirements are first mapped to a intermediate (restricted) representation, at which point ambiguities are resolved. This intermediate result is then further mapped into targeted formal specification language(s) for subsequent automatic analysis.

%In either way, once requirements are expressed in restricted forms, they usually can be automatically translated to some formal representations for later on automatic analysis.

Fuchs et al.~\cite{fuchs_attempto_2006} have proposed ACE (Attempto Controlled English), a CNL with a domain-specific vocabulary and a restricted grammar, aiming to reduce ambiguity and vagueness inherent in full natural language software system specifications. For example, ACE syntax disambiguates the statement ``\emph{The driver stops the train with the defect brake}'' as either ``The driver stops the train with the defect brake'' (according to their interpretation rule, ``with the defect brake'' modifies the verb ``stop'') or ``The driver stops the train that has a defect brake''. With a supporting tool, their ACE specifications can be translated to discourse representation structures (DRS) -- a syntactical variant of full first-order predicate logic, allowing some reasoning tasks such as query answering and abductive reasoning.
%(i.e., the Attempto system)

Konrad et al.~\cite{konrad_real-time_2005} have offered a structured English grammar in support of real-time requirements specification patterns. Requirements expressed in their syntax are mapped to three commonly used real-time temporal logics with a tool support: metric temporal logic (MTL), timed computational tree logic (TCTL) and real-time graphical interval logic (RTGIL), enabling quantitative reasoning about time.

Fantechi et al.~\cite{fantechi_assisting_1994} have proposed NL2ACTL, a prototype tool for automatic translation of NL requirements sentences into formulae of the action-based temporal logic ACTL, allowing the behavioural and logical properties of reactive systems to be checked. The translation process includes two phases: requirements sentence analysis and formula generation. In the first, useful pieces of information units are extracted from input sentences according to their grammar rules; the generation phase composes these pieces into ACTL formulae. During the translation, ambiguities are reflected in the possible generation of more than one ACTL formula. %a not unique

Similarly, Ambriola et al.~\cite{ambriola_processing_1997} have presented \textit{Cico}, a tool that supports the automatic construction of (semi-)formal models (e.g., data-flow diagrams, entity-relation diagrams) through extracting information from NL requirements text. More recently, Gervasi et al.~\cite{gervasi_reasoning_2005} have employed the \textit{Cico} algorithm to automatically transform NL requirements into propositional logic formulae, enabling inconsistency check.
% and the consistency check of these models.

The key idea of the \textit{Cico} algorithm is the parsing rules, which consist of three parts: a template/pattern $M$ to be matched in input NL requirements sentences, an action $A$ that records the intended semantics of the matched fragment (as a node in a NLP parse tree), and a substitution $S$ that replaces the matched fragment in the input sentence. Using the parsing rules, the extracted information will be in a representation specified by the template/pattern $M$, and can be used to construct (semi-)formal models~\cite{ambriola_processing_1997} or will be translated to propositional logic formulae~\cite{gervasi_reasoning_2005}.

In general, CNLs combine the advantages of both natural and formal languages: on one hand they are practically accessible to engineers/analysts and customers, on the other hand they can be (automatically) mapped to formal specification language(s) for certain kinds of automatic analysis. However, these approaches do not support refining stakeholder requirements: they assume stakeholders know what they need, and what they said are in enough detail and can be directly specified. This assumption is not always true, as pointed out by Robertson et al.~\cite{robertson_mastering_2012}: ``\emph{your customers will not always give you the right answer; sometimes it is impossible for customers to know what is right, and sometimes they just do not know what they need}''.

%This means that CNLs are useful if stakeholder requirements have been elaborated.

%which have been challenged when dealing with safety, security or mission-critical systems
One issue with the NLP-based techniques is that they can not ensure completeness (e.g., does a constructed model includes all the elements specified in NL requirements? does the approach detect all the ambiguites/inconsistencies?), which has been challenged in RE~\cite{berry_case_2012}: ``{It is important to understand the limitations of NLP-based tools for RE, because although good but imperfect performance is often helpful to the analyst, in certain circumstances it is of no help to the analyst at all. It may even make his or her job harder}''.
%\end{itemize}

%by Berry et al.

\subsection{Requirements Modeling Languages}
\label{sec:art_modeling}
%Another way to mediate natural and formal specification languages is to develop some modeling languages -- known as semi-formal approaches, which use graphical notations to capture requirements at different levels of abstraction.
%By graphical specification languages, we actually mean graphical requirements modeling languages~\footnote{Requirements modeling languages are often graphic, but can also be textual (e.g., RML~\cite{greenspan_capturing_1982}).}, which use graphical notations to capture requirements at different levels of abstraction.

Requirements modeling languages usually use graphical notations to capture requirements at different levels of abstraction~\footnote{Requirements modeling languages are often graphic, but can also be textual (e.g., RML~\cite{greenspan_capturing_1982}).}. Representative requirements modeling techniques include ($SADT^{TM}$)~\cite{ross_structured_1977-1}, UML~\cite{rumbaugh_unified_2004}, and Goal models~\cite{dardenne_goal-directed_1993}\cite{yu_modelling_2011}. Clearly, modeling languages have several advantages: (1) graphical notations are easy to understand and communicate; (2) the capability of modeling requirements at intermediate levels of abstraction contributes to better managing requirements, especially when there is a large amount of requirements; (3) different diagrams can provide complementary and interrelated views of the same system~\cite{van_lamsweerde_building_2001}\cite{robertson_models_????}.
%ER (Entity-Relation) diagrams~\cite{chen_entity-relationship_1976},
%These strengthes are probably the main reasons for the popularity of UML, which has been adopted as the industrial standard for object-oriented modeling~\cite{luisa_market_2004}.

%\begin{itemize}
%    \item
\vspace{12pt}
\noindent \textbf{Structured Analysis}. Structured Analysis and Design Technique ($SADT^{TM}$)~\cite{ross_structured_1977-1} -- a language for communicating ideas, is probably the first graphical language used for modeling and communicating requirements. SADT assumes that the world consists of activities and data, both of which are represented as boxes and arrows. According to SADT, each activity consumes some data (input), represented by an incoming arrow at the left-hand side of the activity box, produces some data (output), represented by an outgoing arrow at the right-hand side. In addition, each activity has associated constraints (e.g., purpose, viewpoint) controls its execution (control, represented by an incoming arrow on the upper part of the box) and some external agent that executes it (mechanism, represented by an incoming arrow on the bottom part). Data is modeled in a dual fashion, i.e., having activities as its input, output, control and mechanism~\footnote{Here ``mechanism'' will be interpreted as devices for storage, representation, implementation, etc.}. SADT has served as the starting point of other structured analysis techniques, e.g., the popular data flow diagrams~\cite{demarco_structured_1979}. According to Mylopoulos et al.~\cite{mylopoulos_object-oriented_1999}, $SADT^{TM}$ and other proposals~\cite{bell_software_1976} that established the ``\emph{requirements problem}'' have played an important role in establishing the field of RE in the 70's.

%instituted
%Note that ``any idea'', not just a functional specification, as been made more clearer by RML ``capturing more wold knowledge'', this is where conceptual modeling shall comes in.
        %Among them, the Unified Modeling Language (UML)~\cite{rumbaugh_unified_2004}, which employs a set of structural and behavior diagrams to model software systems, constitutes a landmark OOA technique.

%\item
\vspace{12pt}
\noindent \textbf{Object Orientation}. With the programming paradigm being shifted from structured programming to object orientation in the 90's, object-oriented techniques such as Object-Oriented Systems Analysis (OOSA)~\cite{mellor_object-oriented_1989}, Object-Oriented Analysis~\cite{coad_object_1991} and the Object-Oriented Modeling Technique (OMT)~\cite{rumbaugh_object-oriented_1991} have been proposed. These approaches were consolidated into UML~\cite{rumbaugh_unified_2004}.

Use case, a UML concept that represents the externally visible functionalities of the system-to-be, has been widely used for representing requirements. Use cases capture functional/behavior requirements by describing sequences of interactions (including the mainline sequences, different variations on normal behavior, the exceptional conditions that can occur with such behavior, together with the desired responses) between external actors and the system under consideration~\cite{rumbaugh_unified_2004}. Use case diagrams, a kind of UML behavior diagrams, structure use cases through three kinds of relations: \textit{generalization}, \textit{include}, and \textit{extend}~\cite{rumbaugh_unified_2004}.

%Use cases have been shown effective in capturing functional requirements as it is

Use cases have been widely used in practice~\cite{luisa_market_2004}. However, it is worth pointing out that: (1) use cases capture only part of the requirements (i.e., functional requirements), and leave out non-functional ones (e.g., user interface requirements, data requirements, quality requirements)~\cite{cockburn_writing_1999}; (2) use cases have their own deficiencies (e.g., UML is insufficient in capturing the structure between use cases)~\cite{glinz_problems_2000}.

%use case model cannot specify interaction requirements where the system shall initiate an interaction between the system and an external actor

%In order to capture non-functional requirements,
%Considering its popularity, researchers have tried to use UML for capturing also non-functional requirements, e.g.,~\cite{cysneiros_using_2001}\cite{supakkul_integrating_2005}.

Many efforts have been made to use or extend use cases for capturing non-functional requirements, e.g., combining use case with the NFR framework (NFR-F)~\cite{cysneiros_using_2001}\cite{supakkul_integrating_2005}, using misuse case to elicit quality requirements~\cite{herrmann_quality_2005}\cite{herrmann_moqare:_2008}. Cysneiros et al.~\cite{cysneiros_using_2001} model NFRs as softgoals, and use the NFR framework~\cite{chung_non-functional_2000} to refine softgoals and finally operationalize them to attributes or functions in UML classes, aiming to ensure that the designed conceptual models reflect the desired NFRs. Similarly, Supakkul et al.~\cite{supakkul_integrating_2005} use the NFR framework to model and refine NFRs, and associate NFRs with use cases through a set of association points and propagation rules. These approaches combine two modeling techniques, namely UML and NFR-F. Our proposal, instead, intend to offer a unified language for representing both functional and non-functional requirements. Moreover, as these approaches have used NFR-F to handle NFRs, they suffer from the same problems as NFR-F does, e.g., softgoals are not the proper concept for modeling NFRs.
%the interrelations between FRs and NFRs are not captured at a fine granularity.

Herrmann et al.~\cite{herrmann_quality_2005}\cite{herrmann_moqare:_2008} have proposed MOQARE (misuse-oriented quality requirements engineering) for eliciting quality requirements. Their approach is based on misuse cases -- inverted use cases to denote functions that a system should not allow~\cite{sindre_templates_2001}. They use the ISO 9126-1 quality standard~\cite{iso/iec_9126-1_software_2001} as a check list to identify a core set of quality attributes (QAs) that has essential influence on business goals, and derive quality goals from these QAs by adding affected assets. They then describe misuse cases that could threats the derived quality goals, and define countermeasures, some of which are quality goals, for each misuse case. They have noticed that quality attributes depend on assets, but they do not offer a syntax for quality goals. Also, they focused on eliciting, instead of refining quality requirements.

%They are more focused on how NFRs will be reflected and implemented in the software design phase, and do not address the interdependency issue in requirements themselves.

%In practice, use cases usually represent very high-level user requirements and need to be further refined to derive system requirements. When refining a use case, how to relate NFRs to each derived system requirements is still a problem (as mentioned at the beginning).

Constantine and Lockwood~\cite{constantine_software_1999} have proposed essential use cases (EUC) to overcome the limitation of use case (UC) on user interface design. The key difference between EUC and UC is that EUC captures user intentions and system responsibilities, instead of lower-level user actions and system responses. For example, to withdraw money from an ATM, a use case will describe a sequence of interactions ``\emph{insert card}; \emph{enter PIN}; \emph{enter amount}; \emph{take card}; \emph{take cash}'', while an EUC only include an abstract interaction sequence ``\emph{identify self}; \emph{make selection}; \emph{take cash}''.

Kamalrudin et al.~\cite{kamalrudin_improving_2011} have discussed about improving requirements quality based on EUC interaction patterns. They first translate NL requirements into a set of abstract interaction sequences, based on which EUC models are constructed. They then compare these extracted EUC models to the ``best practice'' examples of EUC interaction pattern templates, in order to identify incompleteness, inconsistencies, and incorrectness. To be practically useful, this approach needs a large set of standard EUC interaction patterns (the authors have 30 templates when they published their work in 2011).

%They use a large database of EUC abstract interaction patterns to identify phrases in the NL requirements text, which map onto an EUC abstract interaction concept.

%\item
\vspace{12pt}
\noindent \textbf{Goal Orientation}. Realizing the limitations of OOA techniques, e.g., focusing on the software alone and lacking support for reasoning about the software ecosystem made of software and its environment, leaving out non-functional requirements, lacking of rationale capture, etc., van Lamsweerde and his colleagues~\cite{van_lamsweerde_object_2004} have suggested to use the concept ``\textit{goal}'', which represents the objective a system under consideration should achieve, to capture requirements, and accordingly proposed KAOS~\cite{dardenne_goal-directed_1993}, which stands for ``\emph{Knowledge Acquisition in autOmated Specification}'',  or, alternatively, ``\emph{Keep All Objectives Satisfied}''.

KAOS~\cite{dardenne_goal-directed_1993} constitutes the landmark goal-oriented technique for RE. It offers four complementary and interrelated views on the whole system, not just the software part of it: (1) goal model, which captures stakeholder goals/needs and the derived requirements needed to achieve them; (2) responsibility model, which describes the assignment of requirements and expectations to each agent (human being or automated agents); (3) object model, which defines the concepts and their relations in the application domain; (4) operation model, which describes the behaviors that agents need to exhibit in order to fulfill their requirements.

At the same period, Mylopoulos et al.~\cite{mylopoulos_representing_1992} have proposed the NFR framework (NFR-F), which uses ``\textit{softgoals}'' (goals without a clear-cut criteria for success) to capture non-functional requirements. They also provided a simple syntactic form for softgoals: ``$type \, [topic]$'' (e.g., ``Processing\_time \, [File\_search]'', where ``Processing\_time'' is a type and ``File\_search'' is a topic), and several operators for decomposing softgoals. These ideas are further extended into a monograph~\cite{chung_non-functional_2000}.

%(e.g., "$accuracy [account]$", where ``accuracy'' is a type and ``account'' is a topic)

These two frameworks pioneered in promoting goal-oriented requirements engineering (GORE). For example, KAOS has been employed in many other proposals for deriving functional specification from stakeholder goals (e.g., Hassan et al.~\cite{hassan_goal-oriented_2008}, Aziz et al.~\cite{aziz_goal-oriented_2009}), and NFR-F has been widely adopted for capturing non-functional requirements (e.g., \textit{i}*~\cite{yu_modelling_2011}, Tropos~\cite{bresciani_tropos:_2004}, and Techne~\cite{jureta_techne:_2010}).

%Note that the syntax for softgoals is often surprisingly neglected in these subsequent frameworks.
%to support them

%(2) Goal AND-refinement provides a natural mechanism for structuring complex requirements documents for increased readability; (3) Goal OR-refinement allows reasoning about alternatives.

Note that goal oriented analysis and OOA are complementary. As pointed out by Mylopoulos et al.~\cite{mylopoulos_object-oriented_1999}, ``(goal oriented techniques) focus on the early stages of requirements analysis and on the rationalization of development processes, (object oriented techniques) on late stages of requirements analysis (and design, as they define all the objects and activities mentioned in the detailed requirements for the new system)''. The KAOS methodology gives an excellent sample of how these two types of analyses complement each other: OOA (e.g., object model) starts with where goal-oriented requirements elaboration process (goal model) ends~\cite{van_lamsweerde_object_2004}.

Goal-oriented requirements engineering (GORE) is advocated for multiple reasons~\cite{van_lamsweerde_goal-oriented_2001}: (1) goals drive the elaboration of requirements, justifying requirements (i.e., providing a rationale for requirements) and providing a criterion for requirements completeness (a requirement specification is complete if all the goals can be achieved from the specification); (2) goal models allowing refining from high-level strategic goals to low-level technical goals through AND/OR refinements, providing a natural mechanism for structuring requirements and allowing reasoning about alternatives. In addition, goal-oriented techniques treat NFRs in depth~\cite{chung_non-functional_2009}.

%However, when employing existing goal oriented techniques to address the RE problem, we found some deficiencies.
We agree with these claims about the advantages of GORE. However, goal oriented techniques also have some of the deficiencies we have discussed in Section~\ref{sec:intro_deficiencies}. We present here the two flaws of current GORE techniques at the language level (flaws at the method level will be discussed in Section~\ref{sec:art_gore}). First, goal oriented techniques lack a unified language for representing both functional and non-functional requirements, except natural language that is error prone. Second, they treat individual requirements as propositions (wholes) and trace between requirements based on propositional (whole) requirements without considering requirements details, hence missing some kinds of interrelation between requirements (e.g., the existential dependency relation between qualities and functions) and making the requirements change/evolution management (e.g., how to precisely identify the possibly affected requirements for requirements changes) harder.

\vspace {-12pt}
\section{Requirements Transformation}
\label{sec:art_trans}
The nature of software or software intensive systems is to resolve real-world problems (from stakeholders). In software development, problems are represented as the requirements for and the accompanying domain assumptions about the environment of the system to be designed. To resolve a problem, the engineering of a specification that it is able to satisfy the given requirements under the given domain assumptions is of key importance. Often, the engineering of a specification from requirements is done through requirements transformation, which includes decomposition, refinement, operationalization, etc.

%Problems, especially complex ones, need to be decomposed in to smaller ones in order to gain more insights, and subsequently, a better understanding of stakeholder needs.
%Accordingly, the refinement of requirements and domain assumptions is of key importance to the the engineering of a specification that it is able to satisfy the given requirements under the given domain assumptions.
%resolving the real world problems.
\subsection{Structural and Object-Oriented Decomposition}
\label{sec:art_general}
%As Douglas T. Ross has eloquently declared ``Synthesis is composition, analysis is decomposition'', (Structured Analysis and Design Technique)
The early $SADT^{TM}$ proposal~\cite{ross_structured_1977-1} has already suggested useful structural decomposition mechanisms for decomposing a software system into activities and data. In SADT, structural decomposition is done by following a top-down manner, along strictly either the activity or the data dimension. That is, a software system will be decomposed into a SA activity (resp. data) diagram consisting of six or fewer activities (resp. data boxes), each of which will further become an activity (resp. data) diagram in its own right with its own internal structure. This decomposition process is recursive, and ends when a proper level of detail is achieved. Note that SADT forces six or fewer pieces of things (activities or data boxes) and does not allow leaving nothing out at any decomposition stage. That is, if a subject is broken into six or fewer pieces, every single thing of the subject must go into exactly one of those (non-overlapping) pieces.

%The use of \textit{functional decomposition} and the resulting of hierarchical diagrams is one of the key features of structured analysis~\cite{sage_handbook_2009}, which has its roots in SADT.
These ideas of structural decomposition are further developed into \textit{functional decomposition}, one of the key features of structured analysis that has its roots in $SADT^{TM}$~\cite{sage_handbook_2009}. Starting with a verb or verb phrase that describes the function of the system, a first-level decomposition would separate the top-level function into first-level sub-functions, which are mutually exclusive and could be totaly exhaustive. Each of these functions can be decomposed into level-two functions that are part of it, and so forth~\cite{sage_handbook_2009}.

In object-oriented development (OOD), the decomposition of a system is based on objects. This is fundamentally different from traditional functional decomposition: instead of decomposing a system into modules that denote functions, object-oriented method structures the system around the objects in the model of reality~\cite{booch_object-oriented_1986}. Specifically, the important steps of OO development includes: (1) identify objects and their attributes; (2) identify the operations suffered and required for each object; (3) establish the visibility of each object in relation to other objects; (4) establish the interface of each object; (5) implement each object. That is, in OO development, a module in a system denotes an object or class of objects from the problem space, rather than a function~\cite{booch_object-oriented_1986}.

On one hand, structural decomposition allows us to decompose functions following a process-oriented manner. On the other hand, object-oriented decomposition enables us to effectively analyze real-world entities and their interrelations. However, as pointed out by van Lamsweerde~\cite{van_lamsweerde_building_2001}, both approaches have limited scope: they focus on the software system alone (i.e., WHAT a software system shall do), and do not consider its environment (e.g., WHY does the system need to achieve a specific requirement, WHO is responsible for satisfying certain requirements). Moreover, they do not support capturing non-functional requirements and reasoning about alternatives.

%for which the primary criteria for decomposition is that each module in the system represents a major step in the overall process.
%Domain Analysis. Feature Model. \\

\subsection{Goal-Oriented Refinement and Operationalization}
\label{sec:art_gore}
%Goal-oriented requirements engineering (GORE) techniques have proposed to deal with requirements at a higher-level of abstraction --  using ``goals'', which drive the elaboration and justify the completeness of requirements, instead of ``objects'' or ``functions'', to capture stakeholder requirements.

With the proposal of ``\textit{goals}'', which drive the elaboration and justify the completeness of requirements, for capturing stakeholder requirements in the '90s, goal-oriented requirements engineering (GORE) has been playing a key role in tackling the RE problem. Goal-oriented techniques, such as KAOS~\cite{dardenne_goal-directed_1993}, NFR-F~\cite{chung_non-functional_2000}, \textit{i}*~\cite{yu_modelling_2011}, Tropos~\cite{bresciani_tropos:_2004}, and Techne~\cite{jureta_techne:_2010}, capture stakeholder requirements as goals, use AND/OR refinement to refine high-level (strategic) goals into low-level (operational) goals, and use operationalization to operationalize low-level goals as tasks (aka functions)~\cite{lapouchnian_goal-oriented_2005}.

%For example, with AND-refinement, we can decompose a problem into smaller ones, e.g., ``\textit{trip be scheduled}'' can be refined to ``\textit{accommodation be booked}'' and ``\textit{ticket be booked}''; with OR-refinement, we are able to capture alternatives, e.g., ``\textit{accommodation be booked}'' can be refined to ``\textit{hotel be booked}'' or ``\textit{hostel be booked}''; with operationalization, low-level goals will be operationalized as functions to be performed by the system-to-be, e.g., ``\textit{book hotel through credit card}''. %These techniques provide an elegant way for going from high-level goals to low-level specifications. Moreover, using an accompanying label propagation algorithm, analysts/engineers can check which alternative is better to satisfy the high-level goals for the system-to-be~\cite{liaskos_integrating_2010}.

%However, they do not support incrementally going from informal to formal and weakening of requirements.
%However, they do not support incrementally going from informal to formal (e.g., \textit{i}*~\cite{yu_modelling_2011}, Techne~\cite{jureta_techne:_2010}), or lack effective support for dealing with non-functional requirements (e.g., KAOS~\cite{dardenne_goal-directed_1993},  Tropos~\cite{bresciani_tropos:_2004}).

%and does not address ontological considerations for requirements.
%\begin{itemize}
%  \item
\vspace{12pt}
\noindent \textbf{KAOS}. As a pioneering goal-oriented modeling framework, KAOS~\cite{dardenne_goal-directed_1993} has proposed to use goal models for capturing the rationale for requirements, and responsibility models for defining the assignment of individual requirements to different agents in a software ecosystem. In KAOS, goals elicited from stakeholders are formalized using LTL (Linear Temporal Logic), and can be refined to sub-goals through a set of refinement patterns, and operationalized as specifications of system operations (pre-, post- and trigger conditions) by following a set of formal derivation rules~\cite{letier_deriving_2002}. Due to the support for formal specifications, this transformation process has been extended by many other researchers for deriving formal system specifications (e.g., the B language) from KAOS goal models, e.g.,~\cite{aziz_goal-oriented_2009}\cite{hassan_goal-oriented_2008}.

There are three points to be noted. First, KAOS does facilitate the derivation of functional system specification from stakeholder goals, but it does not offer effective support for specifying and refining NFRs (in fact, KAOS adopts ``\emph{softgoals}'' from the NFR framework~\cite{chung_non-functional_2000} to capture NFRs). Second, KAOS assumes that stakeholder requirements are in enough detail and can be directly formalized~\cite{al-subaie_evaluating_2006}. That is, KAOS does not support incrementally going from informal stakeholder requirements (which can be incomplete) to (complete enough) formal requirements specification. Third, the refinements in KAOS are strengthening, and weakening of requirements is not supported. For example, KAOS does not allow relaxing ``\emph{the file search function shall take less than 30 seconds} (\emph{all of the time})'' to ``(\emph{at least}) \emph{80\% of the time}''.

%Here the former logically implies the latter, which is the reverse of a strengthening~\footnote{A strengthening means that if a goal $G$ is refined to a set of sub-goals (one or more), the resulting sub-goals should logically imply $G$.}.

%Third, what KAOS ensures is the completeness of the refinements of requirements: if a goal $G$ is refined to a set of sub-goals (one or more), the resulting sub-goals should logically imply $G$. That is, the refinements in KAOS are strengthening, and weakening of requirements is not supported, e.g., relaxing ``the file search function shall take less than 30 seconds (all of the time)'' to ``(at least) 80\% of the time''.

%(\textit{Help}: +; \textit{Hurt}: $-$; \textit{Make}: ++; \textit{Break}: $- -$)
%\item
%to satisfice softgoals (i.e., operationalization)

\vspace{12pt}
\noindent \textbf{The NFR Framework}. The NFR framework (NFR-F), proposed  in 1992~\cite{mylopoulos_representing_1992} and extended into a monograph~\cite{chung_non-functional_2000}, was the first proposal to treat NFRs in depth. NFR-F uses ``\emph{softgoals}'' to capture NFRs and offers a simple syntactic form for softgoals: ``$type \,[topic]$''. The framework offers operators for decomposing softgoals along the ``type'' and ``topic'' dimension (AND/OR), and contribution links (\textit{Help}, \textit{Hurt}, \textit{Make}, \textit{Break}) for linking software design elements or lower-level techniques to softgoals. For example, the softgoal ``Security [Account]'' can be decomposed into ``Confidentiality [Account]'' and ``Integrity [Account]'' along the type dimension (``Security''); or, alternatively, ``Security [Golden\_account]'' and ``Security [Silver\_account]'' along the topic dimension (``Account'').

%(positive/negative)
%(\textit{Help}, \textit{Hurt}, \textit{Make}, \textit{Break})
Further, NFR-F provides ``\textit{softgoal interdependency graph}'' (SIG), which graphically structures softgoals, softgoal refinements (AND/OR), softgoal contributions, softgoal operationalization (operationalizing softgoals as lower-level design elements or techniques for satisficing them) and claims (design rationale for softgoal refinements). Using an accompanying label propagation algorithm, analysts/engineers can check which alternative is better to satisfy the non-functional requirements for the system-to-be~\cite{liaskos_integrating_2010}.

The use of ``\emph{softgoals}'' for modeling NFRs has been adopted by many other proposals, e.g., \textit{i}*~\cite{yu_modelling_2011}, Tropos~\cite{bresciani_tropos:_2004}, and Techne~\cite{jureta_techne:_2010}. However, in our observation, this treatment has some flaws: (1) ``\emph{softgoal}'' is not the proper concept for modeling NFRs: softgoals can be used to model also early vague FRs, while NFRs are not necessarily ``\emph{soft}'' (vague), as we have discussed in section~\ref{sec:art_ontology_nfr}; (2) NFR-F acknowledges relations between ``type'' and ``topic'', but does not go into depth in analyzing the interrelations between FRs and NFRs, e.g., the topic ``Account'' in the softgoal ``Security [Account]'' could be the object operated by a function; (3) NFR-F is a process-oriented approach for dealing with NFRs and focuses on rationalizing the development process in terms of NFRs (e.g., justify design decisions)~\cite{mylopoulos_representing_1992}, but does not push for measurable and testable NFR specifications. %and does not offer a companion syntax for goals, which are used to capture FRs.

% The \textit{i}* framework centers on the notion of ``intentional actor'' and ``intentional dependency'' (goal, softgoal, task and resource dependencies), and
%SR process elements are structured using two types of links
%RE processes in terms of process elements, such as goals, softgoals, tasks, and resources, and their relations.
%\item
\vspace{12pt}
\noindent \textbf{\textit{i}*/Tropos}. The \textit{i}*~\cite{yu_modelling_2011} framework proposes an agent-oriented approach for RE centering on the intentional properties (e.g., goals, abilities, beliefs, and commitments) of agents/actors. The framework offers two kinds of models: the Strategic Dependency (SD) model and the Strategic Rationale (SR) model. An SD model captures the dependencies among actors in an organizational context; an SR model describes the the rationale behind such dependencies, and provides information about how actors achieve their goals and softgoals. SR models use three types of links to structure modeling elements, such as goals, softgoals, tasks, and resources: (1) \textit{decomposition} (AND-refinement), a goal/task can be decomposed into sub-goals, sub-tasks, etc.; (2) \textit{means-ends} (OR-refinement), a goal can be operationalized as alternative tasks; (3) \emph{contribution} (\textit{Help}, \textit{Hurt}, \textit{Make}, \textit{Break}), a softgoal can be satisficed by other goal(s), softgoal(s), or task(s). In addition, \textit{i}* also uses softgoals as criteria for choosing better alternatives.

The \textit{i}* framework has served as the basis of Tropos~\cite{bresciani_tropos:_2004}, an agent-oriented software development methodology. Tropos guides the development of agent-based software system from early and late requirements analysis through architecture and detailed design to implementation. Tropos uses the \textit{i}* framework to represent and reason about requirements and architecture styles, and maps \textit{i}* concepts to the Belief-Desire-Intention (BDI) agent architecture for implementation (e.g., actor as agent, resource as belief, goal and softgoal as desire, task as intention)~\cite{castro_towards_2002}. Formal Tropos (FT)~\cite{fuxman_formal_2003}, the formal counterpart of Tropos, has offered an intermediate language for specifying \textit{i}* models. Further, FT translates the intermediate representation into LTL, supporting automatic verification of early requirements, including animation, consistency check, possibility checks (if some expected scenarios are excluded)~\cite{fuxman_formal_2003}.

The \textit{i}* framework adopts ``\emph{softgoals}'' from the NFR framework for capturing NFRs~\footnote{Note that \textit{i}* does not acknowledge the dependency relation between types and topics, and simply use natural language to describe softgoals.''}, suffering from similar problems as the NFR framework does. Like KAOS, Formal Tropos also lacks support for incrementally going from informal to formal. Both \textit{i}* and Tropos do not support weakening requirements.

%The methodology includes five phases: early requirements analysis (identifying stakeholders and their goals), late requirements analysis (focusing on the system-to-be within its environments), architectural design (defining the system's global architecture in terms of sub-systems (actors) interconnected through data and control flows (dependencies)), detailed design (specifying agents' goals, beliefs, and capabilities, as well as communication among agents), and implementation (using JACK Intelligent Agent platform). Tropos has two novel features: (1) the use of agent and mentalistic notions in all phases of software development; (2) the importance placed on early requirements analysis.
%\item
\vspace{12pt}
\noindent \textbf{GBRAM}. The goal-based requirements analysis method (GBRAM)~\cite{anton_goal-based_1996} includes two important activities: \textit{goal analysis} and \textit{goal evolution}. Goal analysis is about the identification of goals from various information sources such as diagrams, textual statements, interview transcripts, and followed by classification. Goal evolution concerns the elaboration, refinement, and operationalization of the identified goals into operational requirements specifications. GBRAM provides a set of heuristics for goal evolution: (1) refine: eliminating redundancies, reconciling/merging synonymous goals (e.g., ``\emph{Meeting arranged}'' and ``\emph{Meeting scheduled}'' are synonymous and can be reconciled); (2) elaborate:  identifying goal obstacles, and analyzing scenarios and constraints, in order to uncover hidden goals and requirements; (3) operationalize: translating goal information into a requirements specification.

%\item
\vspace{12pt}
\noindent \textbf{Techne}. Jureta et al.~\cite{jureta_revisiting_2008} interpreted softgoals as goas referring to qualities but are vague for success, and used quality constraints (QCs), which refer to qualities but are clear for success, to quantify softgoals. These ideas were introduced into Techne~\cite{jureta_techne:_2010}, a requirements modeling language for handling preferences and conflict.

The abstract Techne syntax, as introduced in~\cite{jureta_techne:_2010}, consists of several requirements concepts: \emph{goals} ($g$), \emph{softgoals} ($s$), \emph{tasks} ($t$), \emph{domain assumptions} ($k$), and \emph{quality constraints} ($q$). The framework provides three relations between elements: \emph{inference} ($I$), \emph{conflict} ($C$), and \emph{priority} ($P$). If a premise element ($g$, $s$, $q$, or $t$), e.g., $e_1$, infers a conclusion element, $e$, this means that the achievement of $e$ can be inferred from the achievement of $e_1$.  Multiple premise elements, e.g., $e_1$ ... $e_n$, can individually infer the same conclusion element, $e$, and this is treated as a (non-exclusive) OR, where achievement of any $e_1$ ... $e_n$ means $e$ is achieved. Multiple premise elements can also be aggregated together to infer a conclusion element, $e$. The aggregation is described using functions of arbitrary complexity captured by associated domain assumptions ($k$). The most common aggregation is AND, meaning $e_1$ ... $e_n$ must all be achieved for the $e$ to be achieved. For simplicity,~\cite{jureta_techne:_2010} suggests that a concrete syntax may be used to represent OR and aggregation via AND (i.e., OR and AND inferences). Conflict ($C$) and priority ($P$) relations map a single element to a single element. Relating elements via conflict means that these elements cannot be satisfied simultaneously. Priorities between elements mean that one element has a higher priority than another.

%Techne uses \emph{inference} ($I$) to capture traditional AND/OR refinement.
Techne has restricted non-functional requirements to quality requirements (softgoals and quality constraints), partially addressing the question ``what are the proper concept for modeling NFRs?''. The issue is only partially addressed because Techne does not answer ``what are the requirements that are traditionally not functional but do not refer to qualities?''. In addition, Techne treats its modeling elements as propositions, lacking support for capture of interrelation between requirements based on requirements details.
%\end{itemize}

To sum up, goal-oriented techniques provide an elegant way for going from high-level goals to low-level goals through AND/OR refinement, and finally functions (i.e., tasks) through operationalization. Some of the approaches have used formal languages to formalize requirements specifications (e.g., KAOS, Formal Tropos), enabling certain automatic requirements analysis. However, these existing goal modeling techniques have some common deficiencies at the method level: (1) they do not support incrementally improving requirements quality and going from informal to formal; (2) they do not support weakening of requirements.

% no unified language

%formalized

%which may elicit missing information in initial stakeholder requirements.
%Especially, KAOS~\cite{dardenne_goal-directed_1993} has used LTL (Linear Temporal Logic) to formalize stakeholder requirements, and offered patterns and rules for ensuring the completeness of refinements. That is, ensuring that if a goal $G$ is refined to a set of sub-goals (one or more), the resulting sub-goals should logically imply $G$.

%This idea of using softgoals to capture NFRs has been adopted in many other approaches, including KAOS, i*, Tropos and Techne.

%Softgoals are not the proper concepts for NFRs (Jureta).

%Built on these ideas on softgoals and quality constraints, our previous work on NFRs~\cite{li_non-functional_2014} has proposed the syntax ``Q (SubjT) :: QRG'' for capturing, and introduced some operators for refining quality requirements (QRs). For example, we have used ``gradability'' for the degree of fulfillment of QGs/QCs, and ``agreement'' for addressing subjectivity of QRs. In this paper we do not con-sider partial fulfillment. Instead, we allow QGs/QCs to tolerate some deviations of quality values through scaling their QRGs. The previous "agreement" operator was a mix of observation and relaxation, we divide it into ``Observe'' and ``deUniversalize''.

%Feature models offer a similar way for decomposing system.
\subsection{Problem-Oriented Decomposition and Reduction}
\label{sec:art_pore}
The problem frames (PF) approach concretizes many of Jackson's ideas on the characterization of requirements engineering (e.g., requirements $R$, specifications $S$, domain assumptions $DA$, and their interrelations)~\cite{jackson_software_1995}, and offers a framework for representing, classifying and analyzing software problems. In this approach, a software problem is regarded as a requirement within a real-world context (i.e., a problem consists of two parts: the requirement and the context)~\footnote{The requirement (what is desired, i.e., optative properties of the environment) and the context (what is given, i.e., indicative properties of the environment) are separated and kept distinct in problem frames.}, and the solution to the problem is to develop a machine -- a software running in a computer -- that ensures the satisfaction of the requirement in the given context~\cite{cox_roadmap_2005}.

Problem frames begin with understanding the problem to be solved through the development of a \textit{context diagram}, which identifies relevant problem domains in the real world (the world is decomposed into many problem domains~\cite{jackson_problem_1999}) and the machine to be built, and the interconnections thereof. The problem will be captured in a \textit{problem diagram}, which is developed from the context diagram by associating a requirement with (a subset of) the domains. The end-product of problem analysis is a specification that is able to satisfy the given requirement in the given context~\cite{cox_roadmap_2005}.

\begin{figure}[!htbp]
  \centering
  \vspace {-0.2 cm}
  \includegraphics[width=\textwidth]{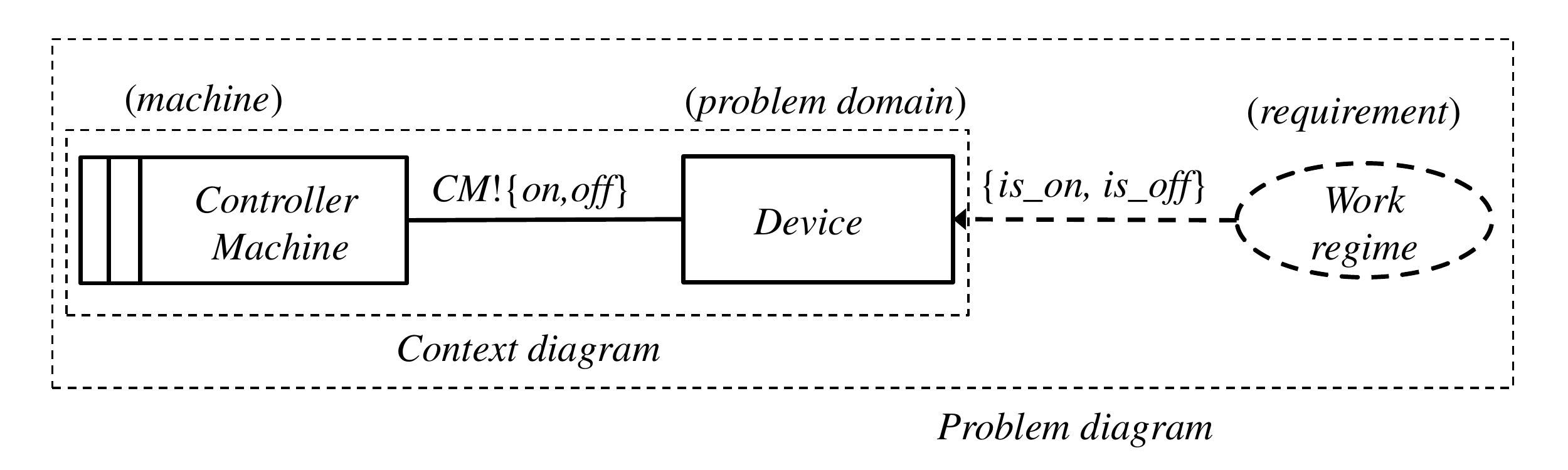}\\
  \vspace {-0.3 cm}
  \caption{A simple problem diagram (adapted from~\cite{cox_roadmap_2005})}\label{fig:pore}
\end{figure}

Figure~\ref{fig:pore} represents a simple problem (adapted from~\cite{cox_roadmap_2005}) that needs a machine (the solution) to control a device (the problem domain/context) such that a certain work regime (the requirement) is satisfied. The link between the controller machine (CM) and the device indicates the shared phenomena between these two domains. Here the shared phenomena are commands (switch ``\emph{on}'' and ``\emph{off}'') that are issued by the controller machine (indicated by the ``!'' symbol) and observed by the device. The arrowhead link indicates the states of the device (i.e., problem world properties) that are constrained by the requirement (depicted as a dotted oval). The domain descriptions and the statements of requirement, which can be described using un-prescribed notations (e.g., natural, structured or formal languages, any of them), are not included in the program diagram. Another important notion in problem frames is \textit{adequacy argument}, which argues that the derived specification satisfies the stated requirement in the given context.

%In problem frames, a problem is captured in a \textit{problem diagram}, which defines the context of a problem and a requirement to be satisfied in such a context. A problem context is characterized by a \textit{context diagram}, which includes various relevant problem domains in the real-world and the machine to be built, and interconnections thereof. A problem diagram is formed by associating a requirement with (a subset of) the domains~\cite{cox_roadmap_2005}.
%Note that the context of a problem can be arbitrarily complex,
Indeed, many real-world problems are complex, being reflected by complex context diagrams, where requirements are initially abstract and far from any machine phenomena~\cite{cox_roadmap_2005}\cite{rapanotti_problem_2006} (recall that a requirement states desired properties in the context of the machine, while a specification describes the behavior of a machine at its interfaces with the problem world, i.e., the connected problem domains)~\cite{hall_problem_2008}). In such cases, the initial problem needs to be transformed, such that the derived requirements (of new problems) become closer to a machine specification. To support this, there have been two forms of problem transformation: problem decomposition and problem reduction/pregression~\cite{cox_roadmap_2005}.

%There are several points to be noted. First, a link between two domains indicate the shared phenomena of them.
%bounds the scope of the problem and identifies the in-scope domains. A context diagram typically
%when they are complex and can not be directly matched to known problem frames

%\begin{itemize}
%\item
\vspace{12pt}
\noindent \textbf{Problem decomposition}. Problem decomposition allows the transformation of a complex problem into smaller and simpler ones, the solution of each would contribute to the original problem. The most basic type of decomposition is to match the problem under consideration to already known problem frames~\cite{jackson_problem_2001} (e.g., the Workpieces frame, the Required Behaviour frame, the Commanded Behaviour frame, the Information Display frame and the Transformation frame). Jackson~\cite{jackson_problem_2001} has offered a number of decomposition heuristics for decomposing problems, with problem \textit{projection} being one of them.

Problem projection is very similar to data table projection in relational databases~\cite{haley_using_2004}: a projection of a relational database table is a new table consisting of a subset of columns, a projection of a problem is a sub-problem with a subset of the problem domains. That is, the context of the sub-problem is a projection of the context of the original problem, limiting the domains and/or phenomena needed to describe the sub-problem.

Rapanotti et al.~\cite{rapanotti_architecture-driven_2004} have proposed ``\emph{Architecture Frame}'' (AFrame) for decomposing problems. In their approach, an AFrame is a Problem Frame (characterising a problem class) that provides a collection of template diagrams for sub-problem decomposition. They offered two AFrames, namely ``\emph{the Pipe-and-Filter Transformation Frame}'' and ``\emph{the Active Store Transformation Frame}'' for decomposing the ``\emph{Key Word in Context}'' (KWIC) problem.

Closely related to problem decomposition is the composition of the solutions to sub-problems. In the composition process, interactions between sub-problems (e.g., consistency, precedence, interference and scheduling~\cite{jackson_problem_2001}) could arise if they share a problem domain or the machine domain. This issue is akin to ``\emph{feature interaction}''\cite{calder_feature_2003}, and has been addressed to some extent~\cite{cox_roadmap_2005}.

%\item
\vspace{12pt}
\noindent \textbf{Problem reduction/progression}. Problem reduction/progression allows to simplify the context of a problem by removing some (one or more) of the domains and re-express the requirement using the phenomena in the remaining domains~\cite{cox_roadmap_2005}. By iterative reduction/progression, a requirement gradually moves closer toward the machine phenomena. Early work by Jackson and Zave~\cite{jackson_deriving_1995} has discussed some principled elements of a method for deriving a specification from requirements. In line with this, Rapanotti et al.~\cite{rapanotti_problem_2006}\cite{rapanotti_deriving_2006} have offered a set of systematic transformation rules for achieving the same purpose.

Problem reduction/pregression shares a similar idea with other approaches for deriving specifications from requirements, e.g., goal-oriented approaches. In goal models, high-level strategic goals can be treated as expressions of requirements about the world, while low-level technical goals represent technical requirements for a software system to satisfy~\cite{rapanotti_problem_2006}. In this sense, the shift from essential use cases (EUCs)~\cite{constantine_software_1999}, which capture use intentions and system responsibilities, to use cases~\cite{rumbaugh_unified_2004}, which capture user actions and system responses, is also in line with this view.

%\end{itemize}

The problem frames approach offers a clear characterization for the key concepts in requirements engineering, namely requirements, specifications, and domain assumptions (context or environment). Nevertheless, this approach does not distinguish the taxonomic dimension for the ``requirement'' concept -- the distinction between functional and non-functional requirements, which is widely accepted in the RE field.
Indeed, this approach is more generally seen as an conceptual framework in which many other techniques (e.g., goal-oriented approaches) may work together~\cite{cox_roadmap_2005}.

\subsection{Others Aspects}
\label{sec:art_others}

We briefly summarize some other techniques related to requirements transformation in this subsection, including feature models, quality quantification, and empirical evaluation.

%\begin{itemize}
%\item
\vspace{12pt}
\noindent \textbf{Feature models}. In the original FODA (Feature Oriented Domain Analysis) proposal, \emph{feature} is defined as a prominent or distinctive user-visible aspect, quality, or characteristic of a software system~\cite{kang_feature-oriented_1990}. Other researchers treat a feature as either a unit of functionality~\cite{felty_feature_2003} or a cohesive set of requirements~\cite{heisel_heuristic_2001}. After investigating the existing definitions,  Classen et al.~\cite{classen_whats_2008} have defined the concept ``\emph{feature}'' from a RE perspective: a feature is a triplet $f = (R, W, S)$, where $R$ represents the requirements the feature $f$ satisfies, $W$ the assumptions the feature takes about its environment, and $S$ its specification.

% that represents all valid system configurations of features

\emph{Feature models} (FMs) are widely used to model the variability of a software system in terms of features (depicted as boxes) and feature dependencies (depicted as directed links)~\cite{schobbens_generic_2007}. FMs structure features into multiple levels of increasing detail, often in the form of tree or directed acrylic graph (DAG), where the root feature represents a complete system~\cite{kang_feature-oriented_1990}. When decomposing a feature in to sub-features, several kinds of decomposition operators can be used: \emph{mandatory} (sub-feature is required), \emph{optional} (sub-feature is optional), \emph{or} (at least one of the sub-features need to be selected) and \emph{xor} (exactly one of the sub-features must be selected). Two kinds of feature dependencies, \emph{requires} and \emph{excludes}, are also commonly used. There are many variants of FMs, interested readers can refer to~\cite{schobbens_feature_2006}\cite{schobbens_generic_2007} for a survey.

Feature models offer a rich set of operators for decomposing software systems, functionalities, entities, qualities, etc. These ideas, e.g., ``inclusive or'' vs. ``exclusive or'', ``mandatory'' vs. ``optional'', can be also applied to goal models~\cite{jureta_techne:_2010}. Note that FMs have a similar issue with goal models: feature specifications focus on the functional/behavior aspect (e.g.,~\cite{heisel_heuristic_2001}\cite{felty_feature_2003}\cite{shaker_feature-oriented_2012}), but leave out non-functional concerns.

%than goal models (e.g., goal models often do not explicitly distinguish between ``inclusive or'' and ``exclusive or'').

%Feature-oriented requirements modeling languages have been also proposed~\cite{shaker_feature-oriented_2012}.

%\cite{van_gurp_notion_2001}. A feature model defines features and their correlations, typically in the form of a feature diagram and left-over constraints.
%There are four types of features: mandatory, optional, or and alternative. Also, typical constraints include inclusion and exclusion.
%\emph{Feature} has been extensively studied in telecommunication and software engineering.
%and many efforts have been made towards defining the notion precisely.

%\item
\vspace{12pt}
\noindent \textbf{Quality quantification}. Quality quantification is an important step in deriving measurable and testable quality requirements. It is similar to goal refinement and operationalization: a quality is often decomposed to several sub-qualities and then quantified using metrics~\cite{iso/iec_tr_9126-2_software_2003}. For example, the quality ``usability'' can be decomposed into ``learnability'', ``operability'', ``accessibility'', etc., according to ISO/IEC Std 9126-1~\cite{iso/iec_9126-1_software_2001}; further, ``learnability'' can be measured by using metrics such as ``learning time'' and ``help frequency'' according to ISO/IEC Std 9126-2~\cite{iso/iec_tr_9126-2_software_2003}.

An issue associated with quality quantification is how to elicit meaningful numbers. Commonly, numbers can be obtained by asking stakeholders or investigating competing products. However, in our empirical evaluation of Techne~\cite{horkoff_evaluating_2014}, we found that stakeholders had some difficulty in quantifying softgoals into quality constraints. Often stakeholders responded with ``\emph{it depends}'', meaning that techniques capturing domain context (e.g., subjects/bearers) of qualities are needed (this is also applicable when investigating competitors).
%(i.e., quantify qualities)

Moreover, non-functional or quality requirements are known as ``good enough'', i.e., their satisfaction is not make-or-break. For example, if a customer states that the cost of trip is low if takes less than 500 Euros, how about if the trip takes 520 Euros finally? Is it still ``low''? Fuzzy logic has been employed to capture such partial satisfaction of  quality requirements~\cite{baresi_fuzzy_2010}; however, the fuzzy membership functions are either difficult to elicit~\cite{yen_systematic_1997} or constructed by inventing made-up numbers~\cite{baresi_fuzzy_2010}.

%The planning language (i.e., Planguage), proposed by Tom Glib [38] and widely used in industry, is a typical example along this line: it uses a set of keywords such as scale, meter must and plan, and a syntax that captures fuzzy concepts, quantifiers and collections to express and quantify quality requirements. However, Planguage is informal and textual, and does not allow compositional notions for specifying the subjects of concerned qualities as well as the degree of fulfillment of other requirements. Moreover, it does not offer a methodology for refining informal stakeholder goals to unambiguous, satisfiable and measurable requirements, only providing a language to capture the results of such a process.

%\item

\vspace{12pt}
\noindent \textbf{Empirical evaluation}. In RE, many empirical evaluations have been conducted to assess the utility of some languages or methods, but mainly on their expressiveness and effectiveness~\cite{estrada_empirical_2006}\cite{horkoff_evaluating_2014}. For example, Estrada et al.~\cite{estrada_empirical_2006} have evaluated the \emph{i}* framework by using three industrial case studies, and reported good expressiveness and domain applicability, but poor modularity and scalability of \emph{i}*. Horkoff et al.~\cite{horkoff_evaluating_2014} have used three studies to evaluate Techne, and reported some challenges related to both expressiveness and effectiveness (e.g., it is difficult to capture contributions between softgoals).

Al-Subaie et al.~\cite{al-subaie_evaluating_2006} have used a realistic case study to evaluate KAOS and its supporting tool, Objectiver, with regarding to a set of properties of requirements, introduced in Davis et al.~\cite{davis_software_1993} and the IEEE Std 830-1998~\cite{committee_ieee_1998}. They reported that KAOS is helpful in detecting ambiguity and capture traceability. However, they also pointed out that the formalism of KAOS is only applicable to goals that are in enough detail and can be directly formalized. %In this sense, KAOS does not support going from informal to formal, and leave out softgoals and functional goals which are not formalisable.

Work by Matulevicius et al.~\cite{matulevicius_comparing_2007} is quite relevant. In their evaluation, the authors have compared two goal modeling languages, namely \emph{i}* and KAOS. Beside the quality of languages themselves, they also compared the models generated by using the two framework with regarding to a set of qualitative properties in the semiotic quality framework~\cite{krogstie_semiotic_2002}. Their findings indicate a higher quality of the KAOS language (not significant), but a higher quality of the \emph{i}* models (note that the participants are not required to write formal specifications with KAOS). They also found that the goal models produced by both frameworks are evaluated low at several aspects, including verifiability, completeness and ambiguity.

These evaluations show that stakeholder requirements initially captured in goal models are of low quality and error prone. The quality of such requirements models needs to be improved, no matter the models will be formalized or not later. That is, techniques for improving the quality of requirements captured in traditional goal models, and incrementally transforming stakeholder requirements to formal specifications are needed.

%Previous work has evaluated RE languages and methods with regarding to desired properties of a requirements specification. For instance, Al-Subaie et al. [31] have evaluated KAOS and its supporting tool (i.e., Objectiver) through a realistic case study, using used a set of properties introduced in Davis et al. [32] and the IEEE standard [2]. They reported that KAOS is helpful in detecting ambiguity and capturing traceability; however, they also pointed out that KAOS does not support going from informal to formal, hence leaving out softgoals and functional goals which are not easily formalizable. Matulevicius et al. [33] have compared KAOS with i* at two aspects: the quality of languages and the quality of generated models. They used the properties introduced in the semiotic quality framework [24] as criteria. Their findings indicate a higher quality of the KAOS language (not significant), but a higher quality of i* models. They also found that the goal models produced by both frameworks have poor evaluations in several aspects, including verifiability, completeness and ambiguity. These evaluations confirm the need for techniques to incrementally go from informal to formal, increasing the quality of requirements initially captured in traditional goal-oriented frameworks.

%\end{itemize}

\section{Chapter Summary}
\label{sec:art_summary}
In this chapter, we have reviewed the state of the art in our chosen research area, including requirements ontologies, requirements specification and modeling, and requirements transformation. We have discussed their strengthens and limitations, assessing their adequacy in addressing the RE problem. We also justified the need of techniques for addressing the common deficiencies of current techniques (for this RE problem), such as lacking a unified language for representing both FRs and NFRs, lacking capture of interrelations based on requirements details, lacking support for weakening requirements, and lacking support for incremental formalization.

%% file: baseline.tex
\chapter{Research Baseline}
\label{cha:baseline}
This thesis builds on two important threads of research: (1) goal-oriented requirements engineering; and (2) foundational ontologies. We adopt some of the important concepts (e.g., functional goal, softgoal, quality constraint) from GORE techniques, re-interpret them based on fundamental concepts such as \emph{function} and \emph{quality} that are defined in the unified foundational ontology (UFO), and accordingly propose an ontology for classifying and a description-based syntax for representing requirements. We also adapt some conventional goal refinement operators (e.g., AND/OR), revise them and propose several new operators in order to offer effective support for transforming requirements.

% them the ontological foundations of functions and qualities.
% We review the important concepts of these two aspects in this section.
\section{Goal-Oriented Requirements Engineering}
\label{sec:baseline_gore}

The rise of goal orientation as a research paradigm for requirements engineering (GORE) is founded on the premise that requirements can be modeled and analyzed as stakeholder goals. According to this view, functional requirements are modeled as hard goals with a clear-cut criterion for fulfillment, while non-functional requirements (NFRs) are modeled as soft goals (aka softgoals) with no such criterion~\cite{mylopoulos_representing_1992}, hence their name. This paradigm served as research baseline for \emph{i}*~\cite{yu_modelling_2011} and Tropos~\cite{bresciani_tropos:_2004}, and has also enjoyed much broader attention within the RE community during the past decades.

\begin{figure}[!htbp]
  \centering
  %\vspace {-0.3 cm}
  \includegraphics[width=\textwidth]{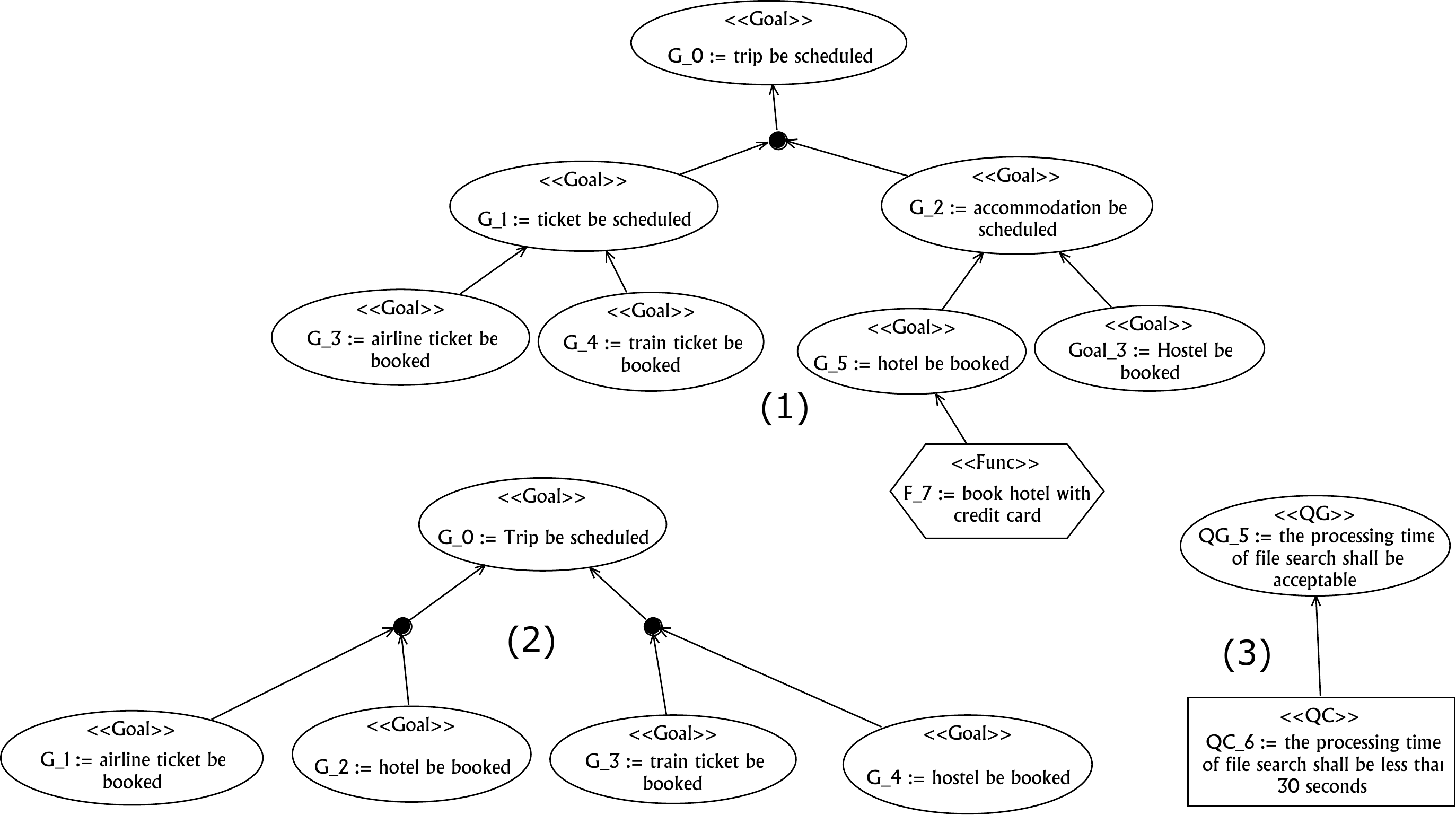}\\
  %\vspace {-0.2 cm}
  \caption{A simple example of the \emph{Techne} syntax}\label{fig:techne_syntax}
\end{figure}

%Goal-Oriented Requirements Engineering (GORE) is
Goal oriented approaches also offer useful operators for transforming requirements, e.g., AND/OR refinement for refining higher-level goals to lower-level goals, and operationalization for operationalizing goals as functions. For example, with AND-refinement, we can decompose a problem into smaller ones, e.g., ``\textit{trip be scheduled}'' can be refined to ``\textit{accommodation be booked}'' and ``\textit{ticket be booked}''; with OR-refinement, we are able to capture alternatives, e.g., ``\textit{accommodation be booked}'' can be refined to ``\textit{hotel be booked}'' or ``\textit{hostel be booked}''; with operationalization (note that current GORE techniques do not distinguish between AND and OR structures on operationalization), low-level goals will be operationalized as functions to be performed by the system-to-be, e.g., ``\textit{book hotel through credit card}''. We capture this simple example in the first fragment of Fig.~\ref{fig:techne_syntax} using \emph{Techne}~\cite{jureta_techne:_2010}, the state of the art goal-oriented modeling framework.

Techne has used the ``\emph{inference}'' link to model both traditional AND/OR refinements (and also operationalization): if multiple premise elements (\emph{goals}, \emph{softgoals}, \emph{tasks}, \emph{quality constraints}), e.g., $e_1$ ... $e_n$, can individually infer the same conclusion element, $e$, then it is a (non-exclusive) OR; if multiple premise elements aggregated together can infer a conclusion element, $e$, it is an AND. As in Fig.~\ref{fig:techne_syntax}, a directed arrow represents an ``\emph{inference}'' link, a black circle indicates an aggregation. The use of ``\emph{inference}'' in Techne is more expressive than traditional AND/OR refinement. For example, we could have either $G_1$ ``\emph{airline ticket be booked}'' and $G_2$ ``\emph{hotel be booked}'' (\{$G_1$, $G_2$\}) or $G_3$ `\emph{`train ticket be booked}'' and $G_4$ ``\emph{hostel be booked}'' (\{$G_3$, $G_4$\}) inferring $G$ ``\emph{trip be scheduled}'', as shown in the second fragment of Fig.~\ref{fig:techne_syntax}. Traditionally, one needs to first OR-decompose $G$ into two (fake) intermediate goals, with each of them being further AND-decomposed.

%if multiple premise elements, e.g., $e_1$ ... $e_n$, can individually infer the same conclusion element, $e$, then it is a (non-exclusive) OR; if multiple premise elements aggregated together can infer a conclusion element, $e$, it is an AND. This treatment is able to capture more than one AND-refinement for a goal, being more expressive than traditional AND.

%Techne~\cite{jureta_techne:_2010} has further developed these concepts and operators.

Techne~\cite{jureta_techne:_2010} has further developed the concepts. In Techne, \emph{softgoals} (we rename this concept as \emph{quality goal}, and use $QG$ for a shorthand) are restricted to goals that are vague and refer to qualities, e.g., ``\emph{the processing time of file search shall be acceptable}''; \emph{quality constraints} ($QC$), which are goals that refer to qualities and are clear for success, e.g., ``\emph{the processing time of file search shall be less than 30 seconds}'', are used to operationalize softgoals. For example, as in the third fragment of Fig.~\ref{fig:techne_syntax}, the inference link between $QC_6$ and $QG_5$ means that the former quantifies the latter. In addition, Techne offers two relations, ``\emph{conflict}'' and ``\emph{priority}'', for capturing inconsistency between, and preferences over requirements.

%\emph{hard goals} are specialized into \emph{functional goals}, which are boolean and refer to perdurants (something accumulates parts over time, e.g., a process, an event), e.g., ``the system shall allow users to search products''.

This work adapts and revises important concepts (goals, tasks, softgoals, and quality constraints) and relations (inferences, conflict) in Techne. Based on these, we propose a requirements ontology, offer a unified syntax for representing and a set of operators for transforming the requirements concepts in the proposed ontology.
%, in order to transform informal stakeholder requirements into a formal specification.

\section{Ontological Foundations}
\label{sec:basline_ontology}

%In general, we take FRs as requirements that refer to functions, and NFRs as requirements on qualities. We then go deeper to capture the ontological meaning of function and quality, and use it to interpret functional and non-functional requirements. For this purpose, we need to review some of the concepts defined in the Unified Foundational Ontology (UFO)~\cite{guizzardi_ontological_2005}, the adopted foundational ontology in this work.

The fundamental concepts, ``\emph{function}'' and ``\emph{quality}'', are of key importance to our requirements ontology, language and operators. In this section, we look to the Unified Foundational Ontology (UFO)~\cite{guizzardi_ontological_2005} to tell us precisely what functions and qualities are~\footnote{We have also considered DOLCE~\cite{masolo_ontology_2003}, another foundational ontology that aims at capturing the ontological categories underlying natural language and human common sense. We choose UFO because: (1) UFO is compatible with DOLCE, at least in the ontology fragment relevant for this thesis; (2) UFO offers a more complete set of categories to cover some important aspects of the domain we target, especially regarding the analysis of quality spaces, situations and goals.}.

Over the years, UFO has been successfully employed to provide ontological seman-tics and methodological guidelines, as well as for analyzing and redesigning modeling languages, standards and reference models in domains ranging from Software Engineering, Enterprise Modeling, Telecommunications, Bio-informatics, among others~\footnote{See http://nemo.inf.ufes.br/en/publications for publications on the different UFO applications.}.

\begin{figure}[!htbp]
  \centering
   \vspace {-0.3 cm}
  \includegraphics[width=0.98\textwidth]{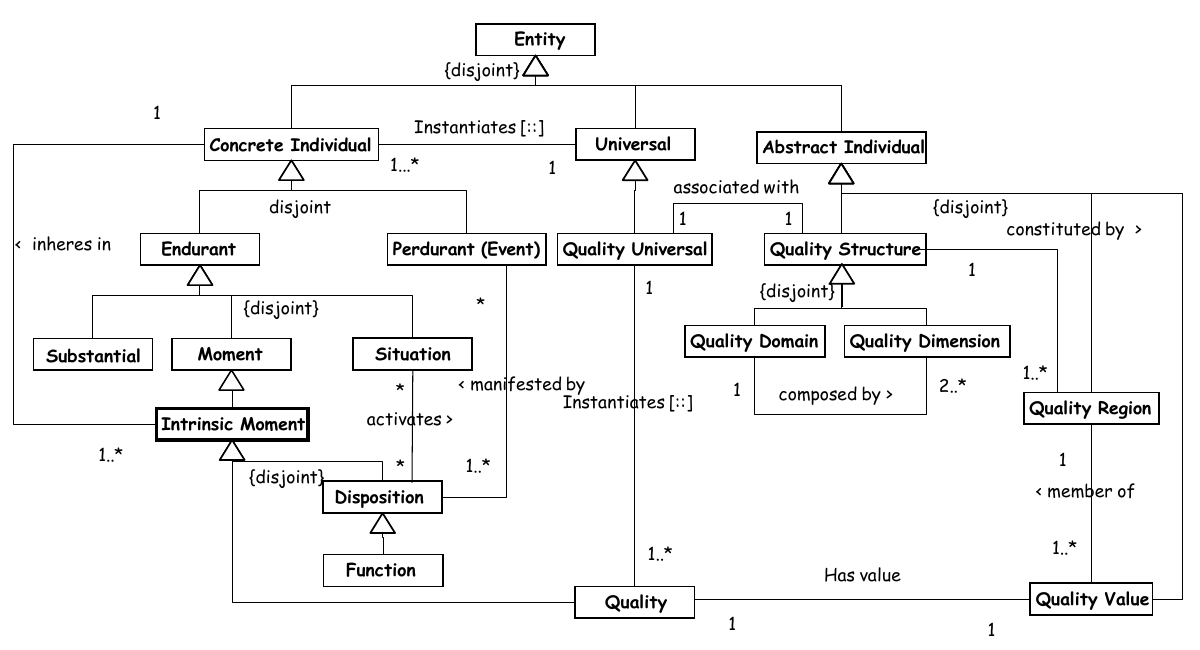}\\
  \vspace {-0.2 cm}
  \caption{A fragment of UFO representing function and quality related categories}\label{fig:ufo_frag}
\end{figure}

We present here (Figure~\ref{fig:ufo_frag}) only a fragment of the UFO ontology containing the categories that are related to function and quality. We illustrate these categories and some relevant relations with UML diagrams. These diagrams express typed relations (represented by lines with a reading direction pointed by ``$<$'' or ``$>$'') connecting categories (represented as rectangles), cardinality constraints for these relations, subsumption constraints (represented by open-headed arrows connecting a sub-category to its subsuming super-category), as well as disjointness constraints relating sub-categories with the same super-category, meaning that these sub-categories do not have common instances. These diagrams are used here primarily for visualization. The reader interested in an in-depth discussion and formal characterization of UFO is referred to~\cite{guizzardi_ontological_2005}.

UFO distinguishs between individuals and universals. \emph{Individuals} are entities that exist in reality possessing a unique identity, while \emph{universals} are patterns of features that are repeatable in a number of different individuals. A \emph{concrete individual} can be either an endurant or a perdurant. \emph{Endurants}~\footnote{By convention, if the word ``universal'' is not part of a term, then the term is assumed to refer to a particular.} do not have temporal parts, and persist in time while keeping their identity (e.g. a person) while \emph{perdurants} (also referred to as events) are composed of temporal parts (e.g. an execution of a software function). \emph{{Substantials}} are existentially independent endurants (e.g. a person or a system). \emph{{Moments}}, in contrast, are existentially dependent on other individuals, inhering in these individuals (e.g. the skill of a person, the performance of a system). \emph{Inherence} (symbolized as ``\emph{inh}'') is a type of non-reflexive, asymmetric and anti-transitive existential dependent relation connecting a moment to its bearer/subject. We focus here on \emph{intrinsic moments}, i.e., moments that are dependent on one single individual (e.g., a skill of a person, a capability of a system).

Note that most distinctions made for individuals, mutatis mutandis, also apply to universals; thus, we have the counterparts: \emph{substantial universal}, \emph{moment universal} and \emph{intrinsic moment universal}. As shown in Figure~\ref{fig:ufo_frag}, a \emph{quality universal} is defined as an intrinsic moment universal that is associated to a \emph{quality structure}, which can be either a \emph{quality dimension} or a \emph{quality domain} (a set of integral quality dimensions).

%For example, the product search function of an e-commerce web-site would depend on the specific system.

%\begin{itemize}
%\item
\vspace{12pt}
\noindent \textbf{Function}. UFO categories \emph{functions} as a sub-category of intrinsic moments, i.e., existentially dependent entities, and considers functions as particular types of dispositions (capacities, capabilities) exhibited by an entity~\cite{guizzardi_towards_2013}. For example, the disposition of a magnet \emph{m} to attract metallic material would depend on and be exhibited by that specific magnet \emph{m}. Functions (or more generally, dispositions) are potential (realizable) property instances manifested through the occurrence of an event that happens if a specific situation/state obtains. For example, given a situation in which \emph{m} is in the presence of a particular metallic object (at a certain distance, of a certain mass), the disposition of the magnet will be manifested by the movement of the object towards the magnet. If the activating situation does not hold, functions may never be manifested. For example, the disposition of the magnet \emph{m} to attract metallic material will not be manifested if it is never close to any magnetic material. The occurrence of the event, in turn, brings about a certain situation in the real world~\cite{hoehndorf_contributions_2009}.

Moreover, in UFO, most perceived events are \emph{polygenic}, i.e., when an event is occurring, there are a number of dispositions of different participants being manifested at the same time. For example, a manifestation event above will involve the capacities of both the magnet and the metallic object.

%For example, a manifestation event (i.e., a run) of the product search function will involve the capacities of both the system and a user.

%The understanding of manifesting events of functions as polygenic enables us to systematically elaborate requirements. For example, in software development, we can design the capacities of the system (a search function), but make assumptions about the capacities of the user (e.g., the user is not visually impaired, the user masters a certain language).

%These kinds of requirements will be captured as functional goals and domain assumptions in our requirements ontology, respectively.

%\item
\vspace{12pt}
\noindent \textbf{Quality}. In UFO (also DOLCE~\cite{masolo_ontology_2003}), a \emph{quality} is defined as a basic perceivable or measurable characteristic that inheres in and existentially depends on its bearer/subject. As shown in figure~\ref{fig:ufo_frag}, a \emph{quality} instantiates a \emph{quality universal} and has a \emph{quality value} in a \emph{quality structure} (either a \emph{quality domain} or a \emph{quality dimension}) associated with that quality universal. Moreover, as an intrinsic moment, a quality inheres in individuals.
%, while a quality domain is composed of a set of integral quality dimensions

UFO's notions of \emph{quality structure}, \emph{quality dimension} and \emph{quality domain} are based on the work of Gardenfors~\cite{gardenfors_conceptual_2004}\cite{gardenfors_how_2004}. According to this work, for all perceivable or conceivable quality universal, there is an associated quality structure in human cognition. For example, height, mass, and response time are associated with one-dimensional structures; other quality universals such as color, taste, and usability are represented by several dimensions. For instance, color can be represented in terms of the dimensions of hue, saturation and brightness; usability in RE is composed of learnability, operability, accessibility, among other dimensions. And, a set of integral dimensions that are separable from all other dimensions constitute a \emph{quality domain}~\footnote{Gardenfors~\cite{gardenfors_conceptual_2004} differentiates integral and separable quality dimensions: ``certain quality dimensions are integral in the sense that one cannot assign an object a value on one dimension without giving it a value on the other. For example, an object cannot be given a hue without giving it a brightness value. Dimensions that are not integral are said to be separable, as for example the size and hue dimensions.''}~\cite{gardenfors_conceptual_2004}.

In pace with DOLCE~\cite{masolo_ontology_2003}, if a quality universal is associated to a quality domain, its instances bear sub-qualities that take values in each of the dimensions of that domain. That is, UFO allows qualities to inhere in other qualities. For instance, the color of an individual apple is itself a bearer for individual qualities of hue, saturation and brightness (e.g., the hue of the color of the apple).

A quality region is a convex region \emph{C} of a quality structure (i.e. either a dimension or a domain); \emph{C} is convex if and only if: for all pairs of points $(x, \, y)$ in \emph{C}, all points between $x$ and $y$ are also in C~\cite{gardenfors_conceptual_2004}. The value of a quality individual can be represented as a point in a quality domain. UFO names this point a quality value (which DOLCE calls ``\emph{quale}''~\cite{masolo_ontology_2003}). For example, a color quality \emph{c} of an apple \emph{a} takes its value in a three-dimensional quality domain constituted of the quality dimensions hue, saturation and brightness. It is relevant to highlight that in UFO both physical (e.g., color, height, shape) and nominal quality types (e.g., social security number, the economic value of an asset) are sorts of quality universals and, hence, are associated with quality structures.

%\end{itemize}

%In UFO, an \emph{agent} is a substantial that creates actions, perceives events and to which we can ascribe mental states (i.e., \emph{intentional moments}). Intentionality in UFO is intended in a broader sense than ``intending something''. Rather, it refers to the capacity of some properties of individuals to refer to possible situations of reality. Thus, ``intending something'' is a specific type of intentionality termed intention in UFO. \emph{Intentions}  are  intentional  moments  that  represent  an  internal  commitment  of  the  agent  to  act towards  that  will. A \emph{goal} is a proposition, and more specifically, the propositional content of an intention. Furthermore, a goal is satisfied by a situation if and only if the situation makes true the proposition expressed by that goal.

\section{Chapter Summary}
\label{sec:basline_summary}
In this chapter, we briefly reviewed the baseline of our research, including: (1) Techne, the state of the art goal-oriented modeling framework, based on which we developed our requirements concepts and operators; (2) the ontological meaning of the fundamental concepts such as function and quality, which are of key importance for understanding, defining, classifying, representing and transforming requirements.

%% file: interpretation.tex
%\vspace{4cm}
%\vspace{-12pt}
\chapter{Ontological Interpretation of Requirements}
\label{cha:frs}
%\vspace{-12pt}
%In this chapter, we define ``\emph{functional requirements}'' according to the ontological meaning of ``\emph{function}'', and design a description-based syntax for representing them. We further formalize the semantics of the proposed syntax using set theory, and translate the syntax into Description Logics in support of reasoning tasks such as query and inconsistency detection.
We have seen that the traditional definitions of functional requirements focus narrowly on the software part of the system to be designed and do not consider its environment, and the classifications of non-functional requirements differ from, sometimes even conflict with each other. These observations beg some basic questions: what are functional and non-functional requirements? how to classify (and represent) them?

In this section, we try to answer these two questions through ontologically interpreting functional and non-functional requirements. Based on the ontological interpretation of requirements, we then accordingly providing a requirements ontology for classifying and a syntax for representing requirements in practice.

%through presenting an ontological interpretation for functional and non-functional requirements
%Further, we offer a description-based syntax for representing the requirements concepts in our ontology, covering both functional and non-functional requirements.
%\section{Research Baseline}
%\label{sec:interp_baseline}
%\vspace{-6pt}
\section{Requirements}
\label{sec:interp_requirments}
%In general, we are in line with Jureta et al.~\cite{jureta_revisiting_2008}\cite{juretaa_core_2009} that NFRs are requirements refer to ``qualities''. We go deeper to capture the ontological meaning of quality through looking to the foundational ontologies such as UFO~\cite{guizzardi_ontological_2005}, and use it to interpret and capture non-functional requirements.
In general, we are in line with Jackson and Zave~\cite{jackson_deriving_1995} that a ``\emph{requirement}'' states a desired property in the environment -- properties that will be brought about or maintained by the system to be designed, and model requirements as goals as in goal-oriented requirements engineering (GORE). We follow the common categorization of requirements as functional and non-functional (i.e., FRs and NFRs). Specifically, we take FRs as requirements that refer to functions, and NFRs as requirements on qualities. We then go deeper to capture the ontological meaning of function and quality, and use it to interpret functional and non-functional requirements.
%For this purpose, we need to review some of the concepts defined in the Unified Foundational Ontology (UFO)~\cite{guizzardi_ontological_2005}, the adopted foundational ontology in this work.
%\vspace{-6pt}
\section{Functional Requirements (FRs)}
\label{sec:interpret_fr}
%We take that a functional requirement (FR) refers to a function (capability or capacity) that has the potential to manifest certain behavior in particular situations. That is, an FR requires a certain entity to bear a function, which being a disposition is realizable through the occurrence of processes/events (or, perdurants, in ontology terms) of a given type. For example, the ``keyword search'' function of an online shop will be manifested by a process (perdurant) of matching between an input key-word and the list of keywords in the system in a particular situation (when the keyword is given and the search button is clicked by users) and brings about a certain effect (the matched product will be displayed).

We take that a functional requirement (FR) refers to a function (capability or capacity) that has the potential to manifest certain behavior in a particular situation and bring about a certain effect in the real world. In other words, an FR requires a certain entity, often the system-to-be in software engineering, but can also be other agents in the software ecosystem, to bear a function of a given type.

\begin{figure}[!htbp]
  \centering
  \vspace {-0.2 cm}
  \includegraphics[width=0.9\textwidth]{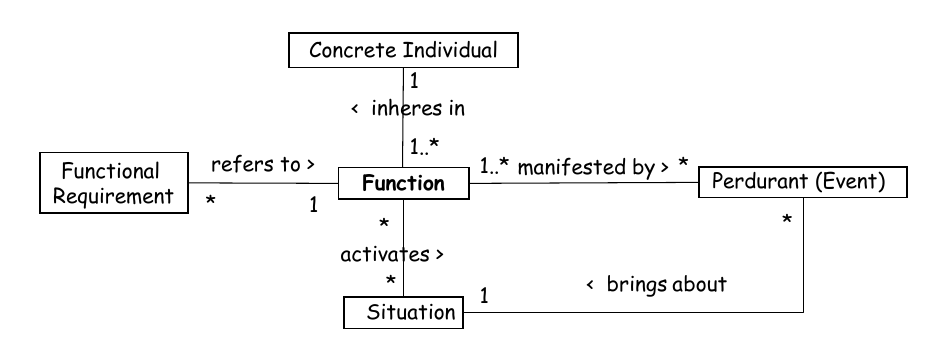}\\
  \vspace {-0.2 cm}
  \caption{The ontological meaning of function (adopted from~\cite{guizzardi_ontological_2005})}\label{fig:function_meaning}
\end{figure}

As shown in Fig.~\ref{fig:function_meaning}, ontologically, a \emph{function} would come with the following associated information~\cite{guizzardi_ontological_2005}\cite{guizzardi_towards_2013}: (1) \emph{function} -- the nature of the required capability; (2) \emph{situation} (state of affairs) -- the conditions under which the function can be activated; often this includes pre-conditions (characterizations of the situation), triggers (the event that brings about that situation), but also actors (agents), objects, targets, etc.; (3) \emph{event} -- the manifestations or occurrences of the function; (4) \emph{effect} (post-conditions) -- situations that are brought about after the execution of the function; and (5) \emph{subject} -- the individual that the function inheres in~\footnote{An subject is often an individual, but can also be a set of individuals in the case that a function can not exist without any of them.}. For example, in the requirement ``\emph{the system shall notify the realtor in a timely fashion when a seller or buyer responds to an appointment request}'', the ``notify'' function, which inheres in ``the system'', will be activated by the situation ``when a seller or buyer responds to an appointment request''; and, its manifestation, a notification event, will bring about the effect ``realtors be notified''; moreover, the notification event is required to occur in a timely fashion (a quality of the notification).
%(note this is not an effect, but a quality of the notification).

%We explain here each of them and discuss their influences on establishing our requirements ontology and language.
%That is, we can not talk about a function without mentioning its subject/bearer.

There are five points to be noted about the ontological meaning of function. First, a function is ontologically a disposition (capability or capacity), an existentially dependent characteristic (specifically, an intrinsic moment, see Fig~\ref{fig:ufo_frag}) that can only exist by inhering in its subject/bearer. For example, the ``keyword search'' function of an online shop would depend on that specific system. In practice, the subject of a function is defaulted as the system to be designed, and often omitted. This restricts the scope of functional requirements to software system functions, and leaves out FRs that are related to software systems life-cycle (e.g., development, deployment, or maintenance requirements) and software system environment. Explicitly considering the subject of a function allows us to incorporate such functional requirements. For example, the requirement ``\emph{the product shall be supported by the corporate support center}'' is also an FR since it specifies a ``support'' function that inheres in ``the corporate support center'', an external agent in a system-to-be ecosystem.

%related to software development life-cycles and software system environment.
%according to our ontological interpretation,
%For example, as we have discussed in Section~\ref{sec:art_ontology_fr}, traditional definitions of FRs would be questionable when used to classify the requirement ``the product shall be supported by the corporate support center'', which specifies a ``\textit{support}'' function that inheres in ``\textit{the corporate support center}'', an external agent in a system-to-be ecosystem. Using our ontological interpretation, this requirement will be reasonably classified as a functional requirement.

%a function is ontologically a disposition (capability or capacity)
Second, being a disposition (capability or capacity), a function can be manifested by the occurrence of perdurants (e.g., events, processes) of a given type. That is, a function by itself is not a set of events or processes, but is realizable through the occurrence of a set of events or processes. So, contra Jureta et al.~\cite{juretaa_core_2009}, we take that FRs refer to perdurants (events or processes) only indirectly, i.e., by referring to a function, which being a disposition is realizable through the occurrence of perdurants of a given type. For example, the ``keyword search'' function of an online shop will be manifested by a process of matching an input key-word with the list of keywords in the system in a particular situation (when the keyword is given and the search button is clicked by a user) and brings about a certain effect (the matched product will be displayed to the user).

%(perdurant)

Third, a function can be manifested only in particular situations, and can also fail to be manifested. For example, the requirement ``\emph{only managers are able to activate pre-paid card}'' states that the ``activate pre-paid card'' function can be activated only if ``the actors are managers'', and would fail to be activated in other cases. When manifested, a function is manifested through the occurrence of events, which in turn would bring about a certain situation (state) in the world. For example, the execution of the ``activate pre-paid card'' would bring about the state ``pre-paid card be activated'' in the real world.

Forth, in UFO~\cite{guizzardi_ontological_2005}, most perceived events are \emph{polygenic}, i.e., when an event is occurring, there are a number of dispositions of different participants being manifested at the same time. For example, a manifestation event (i.e., an execution) of the product search function will involve the capacities of both the system and a user. The understanding of manifesting events of functions as polygenic enables us to systematically elaborate requirements. For example, in software development, we can design the capacities of the system (a search function), but make assumptions about the capacities of the user (e.g., the user is not visually impaired, the user masters a certain language).

\begin{figure}[!htbp]
  \centering
  %\vspace {-0.2 cm}
  \includegraphics[width=0.8\textwidth]{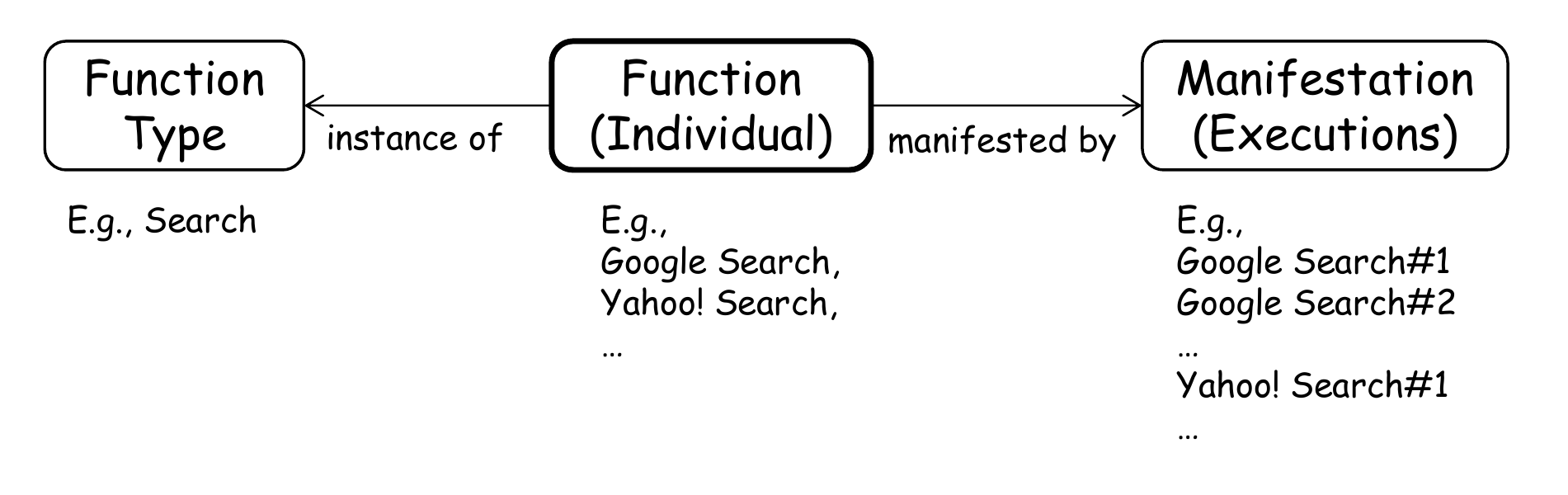}\\
  \vspace {-0.3 cm}
  \caption{Function individual, function type and function manifestation}\label{fig:function_type}
\end{figure}

Fifth, a function is an individual (instance, in object-oriented terms), not a type (class). For example, when mentioning ``Google search'', we are refereing to a particular search function that has the Google web-site as its subject/bearer. Apparently, ``Yahoo! search'', another function individual, is different from ``Google search'' although they share the same function type ``Search''. To make it easier to understand, we show the relations among function type, function (individual), and function manifestation in Fig.~\ref{fig:function_type}.
%as shown in Fig.~\ref{fig:function_meaning}, ``Google Search'' is a function individual, which instantiates the function type ``Search'', and can be manifested by a set of search executions conducted by users.

Our view of functions as individuals does not contradict with the common view of functions as ``action schemas'' in the programming world: a function individual is of a particular function type, and can be manifested by a set of executions. Moreover, the distinction between function (capability) and its manifestations allows us to talk about the function (capability) itself, increasing the expressiveness of our language. For example, we can differentiate ``the search capability of Google'' from ``the search capability of Yahoo!'', and can define qualities such as ``usability'' and ``learnability'' of ``the search capability of Google'', so on and so forth.

%(1) ``function'' is a capability, not what a system does; is an individual, not a type.
%(2) It has a subject, and has executions/manifestation

%(3) Its manifestation needs certain situations to hold -- function constraints
%(4) When manifested, it is polygenic, has many associated attributes: subject (narrow scope) vs. actor.
%(5) functional goal
%(5) ``refer to'' can be specialized: (i) functional goal, which needs a functions to be satisfied; (ii) function; (iii) functional constraints

% of a given type. So, contra Jureta et al. [4], we take that FRs refer to perdurants only indirectly, i.e., by referring to a function,

%In practice, a functional requirement (a requirement that refers to a function) can be further specialized: (1) an FR could specify a desired function (i.e., capability), e.g., ``the system shall allow users to query students' personal information''; (2) an FR could describe a desired state (situation) in the world, e.g., ``students be managed''; (3) an FR could constrain the situation, under which a function can be activated, e.g., ``only managers are able to query students' personal information''. Based on these phenomena, we accordingly propose three concepts to capture functional requirements:
%phenomena

In practice, a functional requirement (a requirement that refers to a function) often specifies only some of the constituting components. For example, an FR may just specify a desired function (i.e., capability), or only describe a desired state (situation) in the world, or merely constrain the situation under which a function can be activated. Based on our observations, we accordingly specialize functional requirements into three sub-categories:
%necessary information such as
\begin{enumerate}
  \item \emph{Functional goal} (\textbf{FG}): a function goal specifies a desired state in the real world that is fulfilled through one or more functions. For example, the functional goal ``\emph{student records shall be managed}'' describes a desired state ``student records be managed'', and will be operationalized by functions such as ``add'', ``update'' and ``remove'' student records.
  \item \emph{Function} (\textbf{F}): a function specifies a desired capability, and necessary information for completing its manifestation. For example, to execute the function ``\emph{the system shall be able to send meeting notification}'', we need to know ``who will send'', ``send to whom'', and ``how many notifications will be sent at a time''. Often, a function implies a functional goal, i.e., a desired state that will be brought about through its manifestation (``meeting notifications be sent'', in this case).
  \item \emph{Functional constraint} (\textbf{FC}): a functional constraint constrains the situation in which a function can be manifested. That is, an FC is usually stated as a restriction on the situation of a function. For example, in the function ``\emph{users shall be able to update the time schedule}'', one may impose a constraint ``\emph{only managers are allowed to perform the update}''.
\end{enumerate}

%As we can see in these examples, FGs, Fs and FCs cannot be simply taken as propositions, as some goal modeling techniques (e.g., \emph{i}*~\cite{yu_modelling_2011}, Techne~\cite{jureta_techne:_2010}) have it. Rather, they are descriptions. Inspired by this observation, we use a ``\emph{slot}: \emph{restriction}'' language to capture them. In general, we specify situations (states), functions and entities using the syntactic form ``[\emph{Name}] $< s : D >^*$''~\footnote{To be specific, it shall be ``\emph{Name} $< s : D >^*$'' or ``$< s : D >^+$'' (``+'' means more than 1). } (``[ ]'' mean optional, ``*'' indicates 0 or more), and capture the intention of something to be in a situation (state) by using the symbol ``$:<$'' (mechanically, it will be understood as description subsumption). So the above examples can be captured as $FG_1$, $F_2$ and $FC_3$ in Eq.~\ref{eq:eq_fr_ex}.

\subsection{Representing FRs}
\label{sec:interpret_representing_frs}
As we can see in these examples, FGs, Fs and FCs cannot be simply taken as propositions, as some goal modeling techniques (e.g., \emph{i}*~\cite{yu_modelling_2011}, Techne~\cite{jureta_techne:_2010}) have it. Rather, they are descriptions. Inspired by this observation, we use ``$<slot_i: Description_i>$'' pairs (SlotDs) as in Eq.~\ref{eq:eq_fr_syntax}, where a description $D_i$ restricts a slot $s_i$, to capture them. Similarly, we specify intended situations using ``$<s: D>$'' pairs, and capture the intention of something to be in a situation (state) using ``$:<$'' (mechanically, this symbol is taken as description subsumption). In addition, we assign names to expressions using ``:='' for later reference.\\
\begin{equation}\label{eq:eq_fr_syntax}
    \begin{aligned}
        \bm{Concept <slot_1: Description_1> ... <slot_n: Description_n>}
    \end{aligned}
\end{equation}

%``:='' means an assignment,
Using this syntax, the ``student records be managed'' example is modeled as $FG_1$ in Eq.~\ref{eq:eq_fr_fg}, where ``Student\_record'' indicates a set of student records, and ``Managed'' is the desired state and refers to an associated set of individuals that are in this specific state.\\
\begin{equation}\label{eq:eq_fr_fg}
    \begin{aligned}
            FG_1 \; := \; & Student\_records \; :< \; Managed; \\
            %FC_2 \; := \; & Update <object: Time\_schedule> \; :< \; <actor: Manager>;
    \end{aligned}
\end{equation}

Functional constraints and functional goals are similar: both of them require certain entities (e.g., objects, functions) to be in some intended situations. The difference is that a functional constraint constrains a situation under which a function will be manifested, while a function goal represents a desired state that will be brought about through the manifestation of a function. As above, we use a similar syntax to capture ``\emph{only managers are allowed to perform the update}'' as $FC_2$ in Eq.~\ref{eq:eq_fr_fc}, which means that the executions of the update function is subsumed by things that only have managers as their actors.\\
\begin{equation}\label{eq:eq_fr_fc}
    \begin{aligned}
            %FG_1 \; := \; & Student\_records \; :< \; Managed; \\
            FC_2 \; := \;   & Update :< \; <actor: ONLY \; Manager>;
    \end{aligned}
\end{equation}

%In general, we specify situations (states), functions and entities using the syntactic form ``[\emph{Name}] $< s : D >^*$''~\footnote{To be specific, it shall be ``\emph{Name} $< s : D >^*$'' or ``$< s : D >^+$'' (``+'' means more than 1). } (``[ ]'' mean optional, ``*'' indicates 0 or more), and capture the intention of something to be in a situation (state) by using the symbol ``$:<$'' (mechanically, it will be understood as description subsumption). So the above examples can be captured as $FG_1$, $F_2$ and $FC_3$ in Eq.~\ref{eq:eq_fr_ex}.
%Note that the curly brackets indicate a singleton, and ``$:<$'' denotes description subsumption (e.g., FC\_3 says that the update function is subsumed by things that only have managers as their actors).
In practice, when a function is manifested (i.e., executed), it would have properties of different participants being manifested at the same time~\cite{guizzardi_ontological_2005}. For example, an execution of ``send meeting notification'' will involve participants like the system, the sender, the receivers, and meeting notifications. This means that in addition to the desired capability, many pieces of information (e.g., descriptions of its actor, object, and trigger) can be associated with the description of a function. For example, ``\emph{the system shall be able to send meeting notification}'' can be captured as $F_3$ in Eq.~\ref{eq:eq_fr_f}.\\
\begin{equation}\label{eq:eq_fr_f}
    \begin{aligned}
            F_3 \; := \; & Send <subject: \{the\_system\}> <actor: Organizer> \\ & <object: Notification>;\\
    \end{aligned}
\end{equation}

%\; \geq 1 \;

We distinguish a function (i.e., an individual) from its manifestation (i.e., a set of executions) by using curly brackets: we use ``\{F\}'' to indicate a function individual, and ``F'' to represent a set of its executions. In the ``send meeting notification'' example, $F_3$ indicates a set of executions of the function, and ``\{$F_3$\}'' means the capability of sending meeting notification. Similarly,``\{the\_system\}'' indicates a singleton. In addition, we also allow the description of a slot to be numeric. For example, the expression ``Student $<$age: $\geq 20$ $>$'' represents a set of students who are older than 20.
%($<$\emph{slot}: \emph{restriction}$>$, SlotR for short, where ``\emph{descriptor}'' \emph{restricts} ``slot'')
\subsection{Refining Functions}
\label{sec:interpret_refining_functions}
Our description-based syntax offers intuitive ways for refining requirements. In general, there are four kinds of basic operations: (1) adding a ``$<s: D>$'' pair (SlotD); (2) removing a ``$<s: D>$'' pair; (3) refining the description of a slot; (4) refining a slot. For example, the requirement ``\emph{the system shall allow users to search meeting rooms}'', captured as a function $F_{4-1}$ in Eq.~\ref{eq:eq_refining_f}, can be refined to $F_{4-2}$ by adding a SlotD ``$<parameter: Room\_capacity>$'', or be refined to $F_{4-3}$ by specializing the description of the slot ``actor'' from ``User'' to ``Organizer''. An example of refining a slot is to refine ``dependency'' to ``goal\_dependency'' or ``task\_dependency'' as in \emph{i}*~\cite{yu_modelling_2011}.
\begin{equation}\label{eq:eq_refining_f}
    %\small
    \begin{aligned}
         F_{4-1} :=  Search & <subject: \{the\_system\}><actor: User>\\ & <object: Meeting\_room> \\
         F_{4-2} :=  Search & <subject: \{the\_system\}><actor: User> \\ & <object: Meeting\_room> <parameter: Room\_capacity> \\
         F_{4-3} :=  Search & <subject: \{the\_system\}><actor: Organizer> \\ & <object: Meeting\_room>
    \end{aligned}
\end{equation}

%In general, the refinement of a restriction is closely related to specializing or generalizing an entity, while that of a slot is akin to specializing or generalizing a relation. Moreover, the specialization of a restriction or a slot is a

When refining requirements we distinguish between strengthening and weakening: a strengthening shrinks the solution space of a requirement (i.e., reducing the choices for satisfying a requirement), while weakening is the opposite. For instance, the requirement ``\emph{the system shall be able to back up data at weekdays}'', captured as a function ``$F_{B1}$ := Backup $<$actor: \{the\_system\}$>$ $<$object: Data$>$ $<$when: Weekday$>$'', can be strengthened into ``$F_{B2}$ := Backup $<$actor: \{the\_system\}$>$ $<$object: Data$>$ $<$when: \{Mon, Wed, Fri\}$>$'' (it is a strengthening because $F_{B2}$ excludes the solutions that are able to backup data at Tuesday or Thursday, thus having less solutions), or weakened into ``$F_{B3}$ := Backup $<$actor: \{the\_system\}$>$ $<$object: Data$>$ $<$when: Weekday $\lor$ \{Sat\}$>$'' (it is a weakening because $F_{B3}$ includes extra solutions that are able to backup data at Saturday, thus having more solutions). In addition, slot-description (\emph{SlotD}) pairs ``$<s: \; D>$'' allow nesting, hence ``$<$object: Data$>$'' can be strengthened to ``$<$object: Data $<$associated\_with: Student$>$$>$''. In general, a requirement can be strengthened by adding slot-description pair(s), or by strengthening the description of a slot. Weakening is the converse of strengthening.

\section{Non-functional Requirements (NFRs)}
\label{sec:interpret_nfr}
Conversely to FRs, we treat NFRs as requirements referring to qualities, i.e., an NFR requires a certain entity to bear a quality or exemplify a quality of a given type. To be more specific, we treat NFRs as requirements that require qualities to take values in particular quality regions in their corresponding quality spaces. In order to distinguish requirements that refer to qualities from traditional NFRs, we call them \emph{quality requirements} (QRs). We will cover what have been traditionally called NFRs but do not directly refer to qualities in Section~\ref{sec:concepts}.

%We will discuss about what have been traditionally called NFRs but do not directly refer to qualities later in this section.

%For instance, the requirement ``the system shall have an intuitive interface'' requires the particular quality ``understandability'' of ``the interface'' to have a quality value ``intuitive''.

\begin{figure}[!htbp]
  \centering
  %\vspace {-0.2 cm}
  \includegraphics[width=\textwidth]{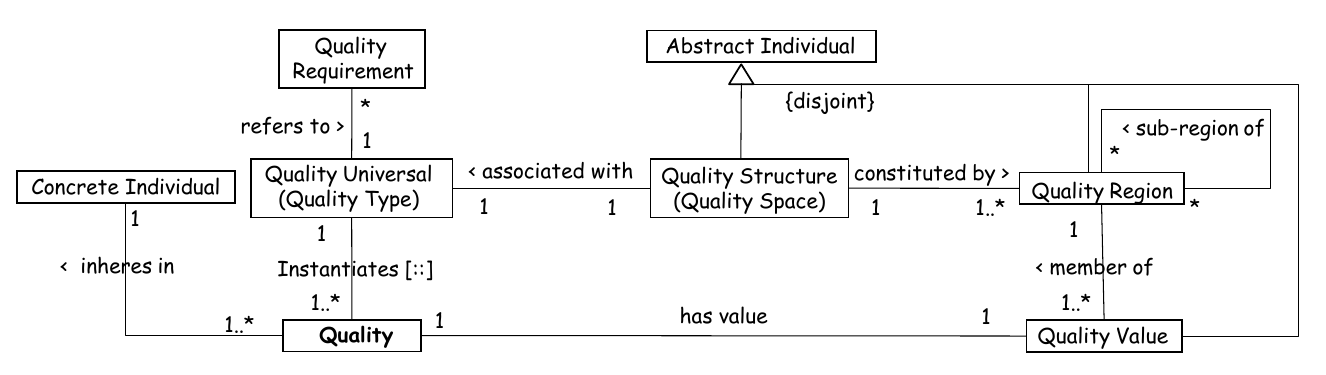}\\
  \vspace {-0.2 cm}
  \caption{The ontological meaning of quality (adopted from~\cite{guizzardi_ontological_2005})}\label{fig:quality_meaning}
\end{figure}

As shown in Fig.~\ref{fig:quality_meaning}, a \emph{quality} is ontologically defined as a basic perceivable or measurable characteristic that inheres in and existentially depends on its subject (a \emph{concrete individual})~\cite{guizzardi_ontological_2005}. The subject can be an object, process, action/task, goal, as well as collectives~\footnote{We can treat a set of individuals as a whole, i.e., a special individual.} of objects, processes, and so on. Like \emph{function}, quality is also an individual (i.e., instance), e.g. ``$cost\#1$'' represents the cost of a specific trip. Each quality has a \emph{{quality type}} {QT} (alternatively, \emph{quality universal}, e.g., ``\emph{Cost}''), which has an associated \emph{{quality space}} {QS} (alternatively, \emph{quality structure}, e.g., ``\emph{EuroValues}'')~\footnote{A quality type could have more than one quality space, e.g., ``\emph{Cost}'' could have ``\emph{EuroValues}'', ``\emph{USDValues}'' and ``\emph{RMBValues}'' as associated spaces. In such cases, a vague region ``Low'' can be mapped to different (math) regions in each space, e.g., ``[0, 1000 \euro]'', ``[0, 1200 \$]'', or ``[0, 7000 \yen]''. For simplicity, we consider only one space for each quality type.}. UFO also differentiates a quality, e.g. ``$cost\#1$'', from its value, e.g., ``1000 \euro'', which is a point or region in a region of the corresponding quality space. In the rest of this thesis, we use the terms ``\emph{quality type}'' and ``\emph{quality space}'', which are more familiar and acceptable to RE audiences, rather than ``\emph{quality universal}'' and ``\emph{quality structure}''.

%\footnote{In the rest of this thesis, we use ``quality type'' rather than ``quality universal'' as the former is more familiar and acceptable to RE audiences.}

The notion of quality space, adopted from UFO, is based on the notion of ``\emph{Conceptual Space}'' put forth by Gardenfors~\cite{gardenfors_conceptual_2004}. In this theory, quality spaces should be understood literally, given that these structures are endowed with geometrical and topological properties. For instance, associated with the quality type ``\emph{Cost}'' we can have a ``\emph{EuroValues}'' space, an one-dimensional structure isomorphic to the positive half-line of 2-place decimal numbers; other quality types such as ``\emph{Color}'', ``\emph{Security}'' and ``\emph{Usability}'' are associated with multi-dimensional spaces, with proper constraints on the constituting dimensions (reflecting the geometry of the space at hand). Moreover, UFO views a quality space as consisting of regions with sub-regions. So a trip might have ``\emph{Low}'' cost, with ``\emph{Very Low}'' being a sub-region.

%This theory can be adapted or extended to address a number of relevant conceptual phenomena, from context-dependent, non-monotonic and analogical reasoning~\cite{gardenfors_conceptual_2004} to graded membership in vague regions~\cite{guizzardi_ontological_2014}.

\subsection{NFRs as Quality Requirements}
\label{sec:interpret_nfr_as_qualities}
We simplify the rich quality theory by treating a quality \emph{Q} (be ontologically correct, \emph{Q} is a quality type) as a mapping (mathematical function) that takes an individual subject \emph{subj} of type \emph{SubjT}, to a quality value (point or region) in \emph{Q}'s codomain (quality space). For example, as a mapping, the quality ``{usability}'' takes its subject, say a software system ``{the E-movie manager}'', to a region ``{good}'' in its quality space. We refer interested readers to Appendix~\ref{cha:appendixes_qtheory} for the mathematic detail about quality mapping.

%In software engineering, qualities are also often applied to entire subject types. For example, the quality ``processing time'' in ``the product shall return (file) search results in an acceptable time'' applies to all possible runs of the system.

\vspace{12pt}
\noindent \textbf{Representing QRs.} Adopting a qualities-as-mappings perspective, we capture an NFR as a quality requirement (QR) that constrains a quality mapping $Q$ to take values in a desired region $QRG$ of its quality space for its subject type $SubjT$. As quality regions can be either vague (e.g., ``\emph{fast}'', ``\emph{low}'') or measurable (e.g., [0, 5 (\emph{Sec.})]), hence QRs can be accordingly vague or measurable. We accordingly specialize QRs into two sub-categories:

\begin{enumerate}
  \item \emph{Quality goals} (\emph{\textbf{QGs}}), whose quality regions are vague.
  \item \emph{Quality constraints} (\emph{\textbf{QCs}}), whose quality regions are clearly specified.
\end{enumerate}

As QRs are requirements referring to qualities, one must understand which quality it is and in which individual it inheres. Take, for instance, the requirement ``\emph{The user interface must have a standard format}''. The quality in this case is ``{format}'', while the subject/bearer is ``{the user interface}''; ``{standard}'' is a particular region in the interface format's quality space. Sometimes, the quality may not be explicit, e.g. ``\emph{The product should conform to the American Disabilities Act}'', in which case the quality is ``{regulatory compliance}'' and the subject is ``{the product}''.

%where the symbol ``:='' is used to assign labels to expressions for later reference, the expression ``$Q (SubjT) :: QRG$''
We capture QRs through using the notation in Eq.~\ref{eq:eq_qr_syntax}, which is an abbreviation of ``$\forall x. instance\_of \; (x, SubjT) \rightarrow subregion\_of \; (Q(x), QRG)$'', meaning that for each individual subject \emph{x} of type $SubjT$, the value of $Q(x)$ should be a sub-region of (including a point in) $QRG$.
\begin{equation}\label{eq:eq_qr_syntax}
    \begin{aligned}
       \bm{Q \; (SubjT) :: QRG}
    \end{aligned}
\end{equation}

%QR\_Label :=
%Using the syntax shown in Eq.~\ref{eq:eq_qr_syntax},
%In addition, we use ``:='' to assign names to expressions, for later reference.
%Using this syntax,
Using this syntax, the QR ``The product shall return (file) search results in an acceptable time'' can be modeled as $QG_1$ in Eq.~\ref{eq:eq_examples_qr}. QCs use the same syntax, but must involve measurable regions. For example, a corresponding quality constraint for $QG_5$ is shown in the same example, as $QC_5$.
\begin{equation}\label{eq:eq_examples_qr}
    %\small
    \begin{aligned}
        QG_5 & := \; Processing\_time (File\_search) :: Acceptable \\
        QC_5 & := \; Processing\_time (File\_search) :: \; \leq 8 \; (Sec.) \\
        %QC_1 & \; is-operationalization-of \; QG_1
    \end{aligned}
\end{equation}

\begin{comment}
\begin{table}[!htbp]
    \centering
    \small
    \begin{tabular}{|rp{0.75\textwidth}|}
        \hline
        \multicolumn{2}{l}{|\textbf{Example 1}|} \\ \hline
        QG1:= & Processing\_time (File\_search) :: Acceptable \\
        QC1:= & Processing\_time (File\_search) :: $\leq$ 8 (Sec.) \\
        \multicolumn{2}{|l|}{QC1 is-operationalization-of QG1} \\
        \hline
    \end{tabular}
\end{table}
\end{comment}
QRs can be defined over both subject types and individual subjects. In this example above, the subject ``file search'' is a type, not an individual (in object-oriented terms, a class, not an instance); here we refer to a set of its instantiations, i.e., a set of file searches. The expression of $QG_5$ ($QC_5$) implies a set of QGs (QCs), each of which requires a specific run ``\emph{file\_search}\#'' to take a processing time value in the acceptable region ($\leq$ 8 Sec.). Therefore, $QG_5$ ($QC_5$) is interpreted as ``\emph{for each file search, its processing time shall be acceptable} ($\leq$ 8 Sec.)''.

Consider another QR: ``The interface shall be intuitive''. In this case, the subject of the requirement is an individual subject, a singleton: ``Understandability (\{the\_interface\}) :: Intuitive'', where ``Understandability'' is a quality (type), ``Intuitive'' is the desired quality region in the its associated quality space where the ease of understanding is intuitive.

Often, a subject can be restricted by qualifiers, acting as relative clauses, e.g., going from ``activate (a pre-paid card) within 5 sec.'' to ``activate $<$a prepaid card$>$ $<$by Administrator$>$ $<$via the Administration section$>$ within 5 sec.''. We extend the basic syntax introduced in Eq.~\ref{eq:eq_qr_syntax} by allowing its subject type \emph{SubjT} to be restricted by qualifiers that consist of ``$<slot: description>$'' pairs referring to \emph{SubjT} (or descriptions, when nested). As such, we are able to define particular sets of individual subjects, over which we can talk about concerned qualities. For example, ``activate'', the subject of the requirement ``activate (a pre-paid card) within 5 sec.'', is a software function and can be qualified by the attributes ``object'', ``actor'' and ``means'', as in Eq.~\ref{eq:eq_examples_complex_qrs}. It represents the set of activations performed by administrators through the admin section (one can think that an activation starts when an administrator issue a request, and ends when necessary actions have been taken).\\ % (past or future).
\begin{equation}\label{eq:eq_examples_complex_qrs}
    \small
    \begin{aligned}
    Activate\_pcard' := & Activate <object: Prepaid\_card> \\
                        & <actor: Administrator> <means: Administration\_section>\\
    QG_6 := & Processing\_time (Activate\_pcard') :: \leq 5 \; (Sec.) \\
    \end{aligned}
\end{equation}
\begin{comment}
\begin{table}[!htbp]
    \centering
    \small
    \begin{tabular}{|rp{0.75\textwidth}|}
        \hline
        \multicolumn{2}{l}{|\textbf{Example 2}|} \\ \hline
        Activate\_pcard' := & Activate $<$object: Pre-paid\_card$>$ $<$actor: Administrator$>$ $<$means: Administration\_section$>$. \\
        QG2 := & Processing\_time (Activate\_pcard') :: $\leq$ 5 (Sec.) \\
        \hline
    \end{tabular}
\end{table}
\end{comment}

\vspace{12pt}
\noindent \textbf{Quality domains and codomains.} The concept of quality in UFO~\cite{guizzardi_ontological_2005} refers to a broad category of intrinsic properties of entities that can be projected on a quality space (roughly, the basis of a measurement structure that becomes the codomain of the associated quality mapping~\cite{albuquerque_ontological_2013}). Examples can be found in every domain, including color, shape, length, atomic number, electric charge, etc.

\begin{figure}[!htbp]
  \centering
  \vspace {-0.2 cm}
  \includegraphics[width=0.85\textwidth]{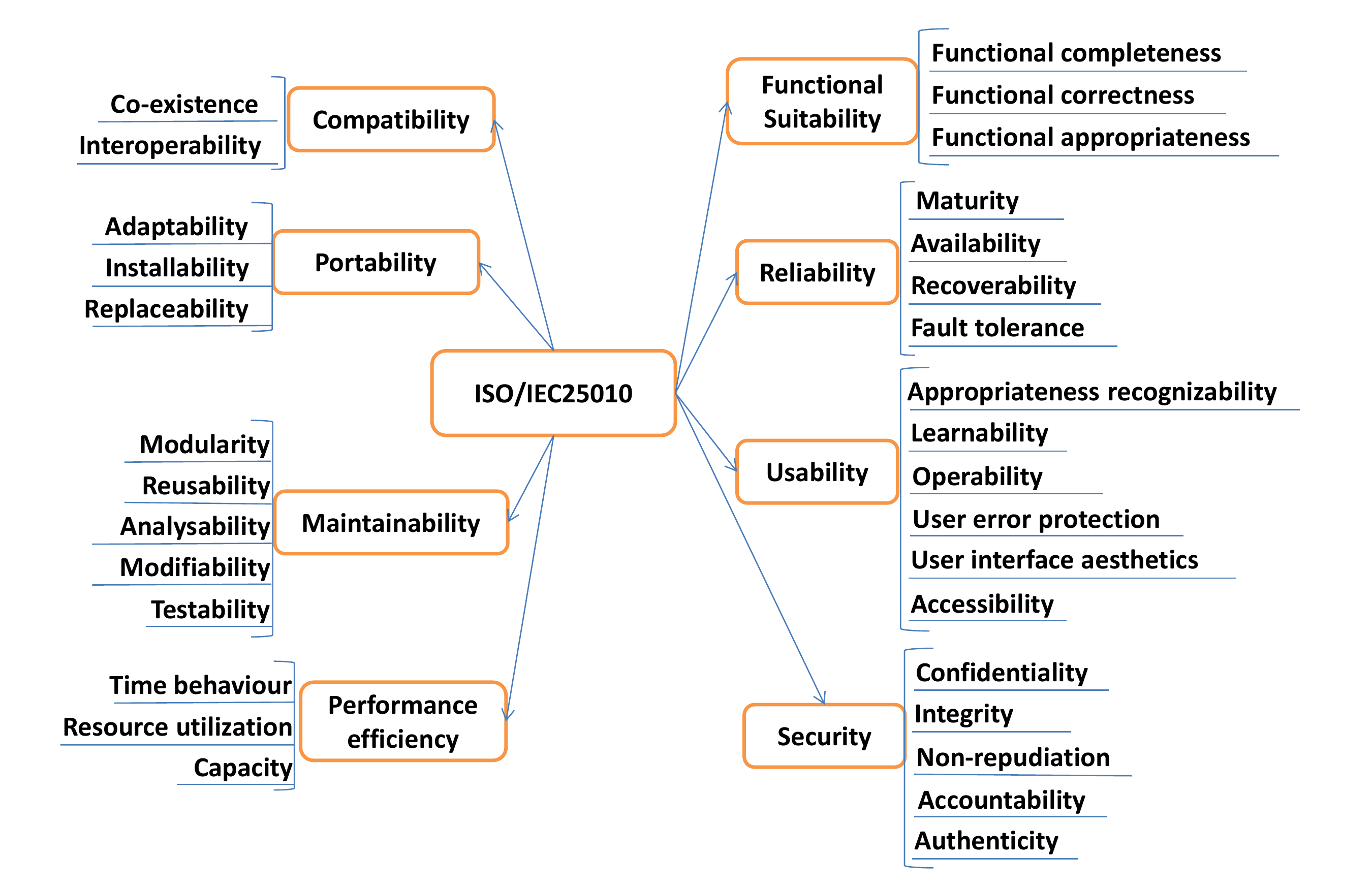}\\
  \vspace {-0.2 cm}
  \caption{The eight product qualities in ISO/IEC 25010 (with refinements)}\label{fig:quality_isoiec_25010}
\end{figure}

For our purposes, we adopt the quality model proposed by ISO/IEC 25010~\cite{iso/iec_25010_systems_2011} as our reference. This standard distinguishes two categories of qualities: ``\emph{qualities in use}'' and ``\emph{product qualities}'', with five and eight qualities, respectively. Fig.~\ref{fig:quality_isoiec_25010} shows the eight product qualities and their refinements. For example, ``Usability'' is refined into ``Learnability'', ``Operability'', ``Accessibility'', etc.

Domains and codomains of qualities are key components in the specification of an NFR. For a specific quality, the set of subject types that it can be applied to constitutes its domain, and the union of all the possible values will form its codomain (quality space).

On one hand, the domain of a software quality can be any aspect of a software system, including its constituents (code, architecture, requirements, etc.), the software processes that created it, its runtime environment, and the like. On the other hand, standards such as ISO/IEC 9126-1~\cite{iso/iec_9126-1_software_2001} and 25010~\cite{iso/iec_25010_systems_2011} are helpful, up to a point, in defining a codomain for qualities. For example, ``Availability'' is defined as ``\emph{degree to which a system, product or component is operational and accessible when required for use}''. Hence it will be associated with a codomain that is a scale ranging from 0\% to 100\%.

We show in Table~\ref{tab:quality_domain_codomain} possible domains and codomains of 10 frequently used qualities in our evaluation (more details can be found in Section~\ref{sec:eval_ontology}).

\begin{table}[!htbp]
    %\vspace {-0.2 cm}
    \caption {The domain and codomain of 10 frequent qualities in our evaluation}
    \label{tab:quality_domain_codomain}
    \small
    \vspace {0.3 cm}
    \centering
    \setlength\tabcolsep{2pt}
    \begin{tabular}{|c|c|p{0.4\textwidth}<{\centering}|} %<{\raggedright}
    \hline
    \textbf{Quality (Type)} & \textbf{Domain} & \textbf{Codomain}\\ \hline
    Operability & \{a system\} & \begin{tabular}{@{}c@{}}\{time to operate\} \\ \{ease of operating: easy, hard ...\}\end{tabular} \\ \hline
    Availability & \{a system\} & \{0\% $~$ 100\%\}\\ \hline
    Processing/Response time & \{functions/tasks\} & \begin{tabular}{@{}c@{}}\{time interval\} \\ \{slow, ... fast, ...\}\end{tabular}\\ \hline
    Scalability & \{a system\} & \{simultaneous transactions\}\\ \hline
    Learnability & \{a system\} & \begin{tabular}{@{}c@{}}\{time to learn\} \\ \{ease of learning\} \end{tabular} \\ \hline
    Frequency & \{functions/tasks\} & \{numbers per time unit\}\\ \hline
    Understandability & \{a system\} & \{ease of understanding\}\\ \hline
    Modifiability & \{a system\} & \{time to modify\}\\ \hline
    Look and feel & \{a system\} & \{degree of preferences\}\\ \hline
    \end{tabular}
    %\vspace{-0.3 cm}
\end{table}

The structure of the codomains of some qualities may be complex, and can differ depending on their subjects. For example, according to the ISO/IEC 25010 standard~\cite{iso/iec_25010_systems_2011}, the codomain of ``usability'' is a six-dimensional space, with each of its sub-qualities being one dimension. Of course, stakeholders may only be concerned with some of these sub-qualities, in which case a ``usability'' QG should be refined accordingly. For example, if only ``learnability'', ``operability'', and ``accessibility'' are of concern, then the codomain of usability becomes three-dimensional.

By differentiating a quality (type) from the quality spaces it can be projected on, we can account for the possibility of having multiple quality spaces (with measurement structures derived from them) for the same quality (type)~\cite{albuquerque_ontological_2013}). Thus such a quality (type) could map an individual subject to different quality values in their respective quality spaces. For example, as shown in Table~\ref{tab:quality_domain_codomain}, ``Learnability'' can map ``a system'' to either ``easy/good'' or ``\emph{x} minutes of training'' in different quality spaces.

%qualities vs. attributes

%It is important to highlight that our qualities cannot be equated with attributes in the tradition of conceptual modeling~\cite{guizzardi_ontological_2005}. In general, attributes are conventional ascriptions of property values to individuals. Qualities, in contrast, inhere in their bearers, i.e., there is something intrinsic (in the bearers) that makes true a certain property ascription to these bearers. That is, attributes are properties assigned to objects while qualities are properties intrinsic to them~\cite{iso/iec_iso/iec_2007} (ones we have to design for a system). For example, ``release date'' and ``serial number'' are merely conventional attributions of certain values to a software system. In contrast, when we state that a system has 50K LOCs or high reliability, there is something in the system that makes these statements true.

\subsection{QRs vs. Softgoals}
\label{sec:interpret_qgs_vs_softgoals}
In our view, the distinction between FRs and NFRs is orthogonal to that of hardgoals and softgoals. Traditionally, hardgoals and softgoals are informally differentiated depending on whether they have clear-cut criteria for success (the former) or not (the latter)~\cite{yu_modelling_2011}. Here, we take the following stance on these concepts, capturing their distinction as follows: a hardgoal can be objectively satisfied by a given set of situations (states of affairs). In contrast, a softgoal is an initial and temporary vague expression of intention before the goal at hand is properly refined. As such, we are not able to determine a priori the set of situations that satisfy a softgoal, i.e., its truth-making conditions.

For example, ``\emph{design the system's main menu}'' is a high-level goal and it can be considered vague (and thus modeled as a softgoal). In addition to capturing high-level vague goals, softgoals are also useful when capturing and analyzing vague QRs. As previously mentioned, a QR is a requirement (goal) referring to a quality (type). In this case, a softgoal refers to a vague quality region, meaning that although we are aware that such region exists in the quality space, we do not know where the boundaries of that region exactly are. For instance, consider that the aforementioned requirement is now refined into a QR: ``\emph{The menu buttons must have standard format}''. At first, the system's stakeholders and analyst may have difficulties in mapping standard to a specific region in the interface format quality space. As the analysis moves forward, vague QRs are continuously refined and operationalized, and hence such vagueness generally disappears.
%length and width

The NFR framework~\cite{chung_non-functional_2000} models NFRs as softgoals that are not clear for success, and on the other side, the CORE ontology~\cite{juretaa_core_2009} treats FRs as hardgoals i.e. goals whose satisfaction have a determinate truth-value. However, we claim that these definitions of FR/NFR and hardgoal/softgoal are, in fact, orthogonal, allowing us to identify FRs that are softgoals as well as NFRs that are in fact hardgoals. Moreover, the categories of FR and NFR are not disjoint, indicating that a requirement can fall into both categories. See Section~\ref{sec:interpret_benifits_fr_vs_nfr} for some interesting examples that illustrate the usefulness of this orthogonality principle.

%For example, the requirement ``The transportation system shall collect real-time traffic information'' specifies a desired function ``collect traffic information'' but also refers to a quality (timeliness) of ``collecting traffic information''.
%See Section 4 for some interesting examples that illustrate the usefulness of this orthogonality principle.

\subsection{Refining QRs}
\label{sec:interpret_refine_nfrs}
%Making QRs measurable often involves refinement operations, where a requirement \emph{r} is refined into \emph{r}' such that \emph{r}' is more precise and/or measurable. Often, refinement consists in conceptually deconstructing the QR's referred qualities. Our characterization of qualities offers several ways to refine quality requirements:

Making QRs measurable often involves refinement operations, which make the resulting requirement(s) more precise and/or measurable than the original one. Often, the refinement of a QR consists in conceptually deconstructing the QR's referred qualities. Our characterization of qualities offers several ways to refine a quality requirement:

%For example, a security quality can be refined in its sub-qualities confidentiality, integrity and availability~\cite{iso/iec_25010_systems_2011}.

\begin{itemize}
  \item One way to do this is to identify the sub-qualities of the quality associated with the QR. As an example of refining a QR ``\emph{The menu buttons must have standard format}'' by decomposing its quality, ``Format'', with respect to buttons may, for instance, be decomposed into ``Size'', ``Shape'', and ``Color''. Considering ``Size'', it may be further decomposed into ``Height'' and ``Width''. Hence, a further refinement could be ``\emph{The menu buttons must have standard height and width}''.
  \item A second way is to identify whether the quality associated with the QR is a resultant quality, which can be conceptually reduced to qualities of parts of the subject/bearer. For instance, we could refine the QR ``\emph{The user interface must have standard format}'' by reducing the quality at hand in terms of qualities of parts of the subject (``the interface''). Since the interface is usually composed of buttons, fields, icons etc., ``\emph{The menu buttons must have standard format}'' illustrates a possible refinement of the requirement.
  \item A third way is to refine, shrink or enlarge, the expected quality region. For example, we can shrink the expected quality region ``good'' in the QR ``the system shall have a good usability'' to ``very good'', or enlarge the region ``[0, 3 (\emph{Sec.})]'' in ``the file search function shall take less than 3 sec.'' to ``[0, 5 (\emph{Sec.})]''.
\end{itemize}

Keep in mind that if we refine a quality requirement by following the first and second way, the refinement is usually a weakening. This is because we often care about only some sub-qualities of the original quality or some parts of the original subject when refining a QR. For example, the refinement from ``\emph{The menu buttons must have standard format}'' to ``\emph{The menu buttons must have standard size}'' is a weakening since we only care about ``standard size'' and leave out ``standard shape'' and ``standard color''. In the case we care about all the sub-qualities or all the parts (e.g., we also need to consider ``standard shape'' and ``standard color''~\footnote{We assume that ``Format'' only has three sub-qualities: ``Size'', ``Shape'', and ``Color''.}), the refinement will be an equating (i.e., the solution space does not change). When following the third way, an enlarging of the desired quality region is a weakening while a shrinking is a strengthening: intuitively, it is easier (resp. harder) for a quality value to be located in a larger (resp. smaller) region.

%For example, we can enlarge the expected quality region ``low'' in the QR ``the cost of trip shall be low'' to ``relative low'', or shrink the region ``good'' in ``the system shall have a good usability'' to ``very good''.

\subsection{Operationalizing QRs}
\label{sec:interpret_operationalizing_nfrs}
To make QRs measurable, we need to operationalize them by constraining the referred qualities so that these qualities take values in measurable quality regions. That is, operationalizing QRs as quality constraints (QCs).

We may operationalize the QR ``\emph{The menu buttons must have standard length and width}'' by defining the quality constraint ``\emph{The menu buttons must have height 0.75 cm and width 1.75 cm}''. While in this example, qualities are constrained to have specific quality values, in other cases, operationalization of a QR may concern a region, as in ``\emph{The search functionality must be efficient}'', operationalized by ``\emph{The search results shall be returned within 30 seconds after the user enters search criteria}''. In our framework, the value ``efficient'' here is associated to a region in the time quality dimension, comprehending quality values from 0 to 30 seconds.

Note that terms such as ``\emph{efficient}'' and ``\emph{low}'' may refer to different quality regions, depending on the type of the subject. For instance, take the requirement ``\emph{Learning to operate the login function must be efficient}''. This QR may be operationalized by ``\emph{The user should learn how to operate the login function within 3 minutes}''. Thus ``\emph{efficient}'' for learning the login function and for returning search results (previous example) may map to different regions in the time quality dimension.

%Further, as our syntax allows to define subject types by restricting existing types (e.g., going from ``Trip'' to ``Trip $<$departure: A$>$$<$destination: B$>$$<$period: T$>$'') or using sets (e.g., we can define ``Common\_Relational\_DBMS'' using a set ``\{Oracle, SQLServer, MySQL\}''), we are also able to define vague quality regions like ``low'' for particular sets of individuals.

%In the case one want to define ``low'' for specific individuals, we can first define a corresponding subject type, and then

\subsection{Satisficing QRs}
\label{sec:interpret_satisficing_nfrs}
Consider the satisfaction of a quality constraint (QC) as a math function, which results in ``1'' (if the QC is satisfied) and ``0'' (if unsatisfied). The key point to determine the satisfaction of a QC is to understand if the measured or perceived quality value is a member of the region to which the QC is associated. If yes, the satisfaction function returns ``1'' and otherwise, it returns ``0''. For example, considering ``\emph{The search results shall be returned within 30 seconds after the user enters search criteria}'' (constraint defined region: [0, $\leq$ 30 (\emph{Sec.})]), if the runtime measurement of a search duration results in 25 seconds, the QC is satisfied; if the result is 32 seconds, then it is not.

However, this may be too strong a condition. Perhaps a 32 second response is ``good enough''. In many cases, ``good enough'' performance is sufficient, i.e., degree of fulfillment of a QC is what matters, rather than black-or-white fulfillment. Thus, in order to capture the intended semantics of many QRs communicated by stakeholders, the satisfaction function should not be a binary function but should instead return a graded satisfaction value in the interval between ``0'' and ``1''.

\begin{figure}[!htbp]
  \centering
  %\vspace {-0.2 cm}
  \includegraphics[width=0.9\textwidth]{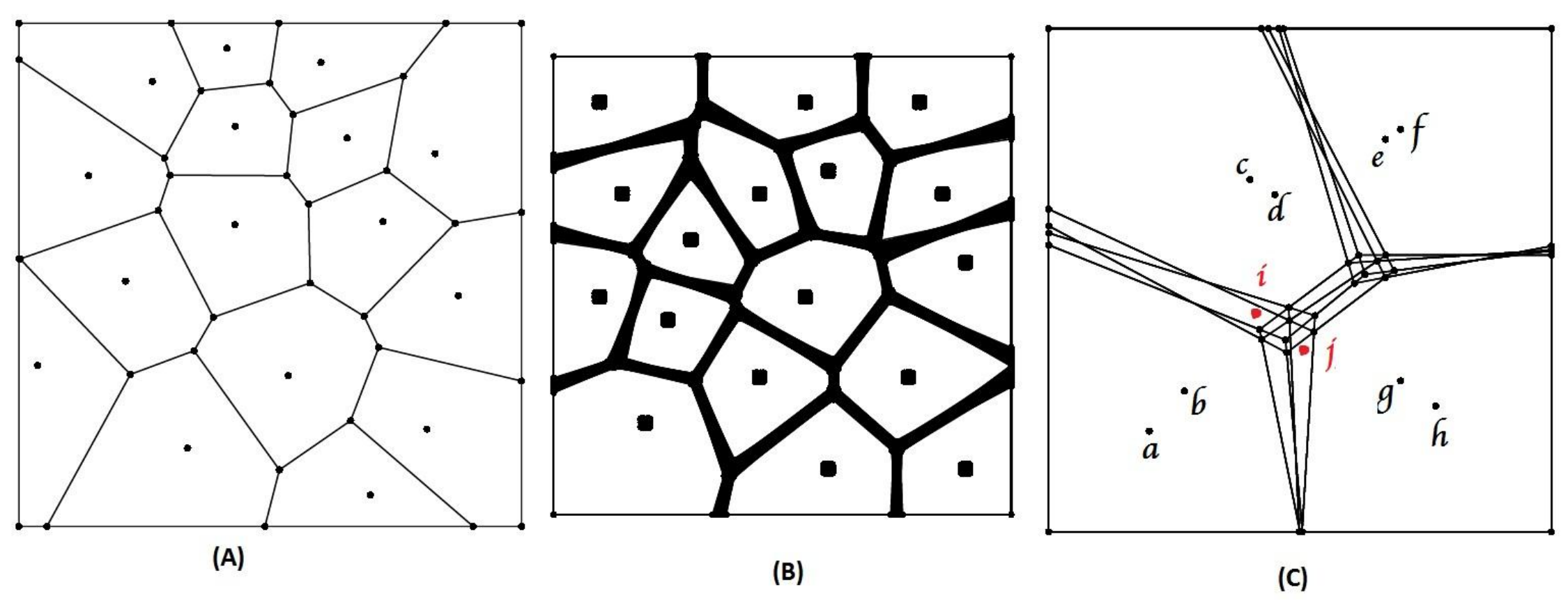}\\
  \vspace {-0.2 cm}
  \caption{Two-dimensional Voronoi and collated Voronoi diagrams (adapted from~\cite{douven_vagueness:_2013})}\label{fig:voronoi}
\end{figure}

%and some of its recent extensions of the original theory as proposed in [19][20].
To account for such phenomena, we use the ``\emph{graded membership}'' theory, proposed by Decock et al.~\cite{decock_what_2014} and based on the Gardenfors's conceptual space theory~\cite{gardenfors_conceptual_2004}. In Gardenfors~\cite{gardenfors_conceptual_2004}, the definition of quality region in the quality space is based on a combination of \emph{prototype theory}~\cite{rosch_cognitive_1975} and the mathematical technique of \emph{Voronoi} diagrams~\cite{aurenhammer_voronoi_1991}. Prototype theory claims that some instances of a concept are more representative or typical than others (thus termed prototypes). Thus, the prototype of a quality is nothing other than a point in its quality space. Creating Voronoi diagrams is a very general technique and can be applied to any metrical space in order to divide the space into cells. Each cell has a center, and contains all and only those points that lie no closer to the center of any other cell than to its own (see Fig.~\ref{fig:voronoi} (A) for an illustration). Combining prototype theory and this technique consists in defining \emph{Voronoi} diagrams by using quality prototypes as their central points.

%\caption{Two dimension (A) Voronoi Diagram and (B) Collated Voronoi Diagram (adapted from~\cite{douven_vagueness:_2013})}\label{fig:voronoi}
%(a vector of prototypical points coming from different regions)
To overcome the limitations in dealing with gradable concepts (e.g., ``\emph{low}'', ``\emph{relatively low}'', ``\emph{adequately low}'', and ``\emph{very low}''), Douven et al.~\cite{douven_vagueness:_2013} extend this approach by assuming that conceptual spaces may contain prototypical regions rather than isolated prototypical points. Using these prototypical regions, they develop a technique to generate what they call \emph{Collated Voronoi Diagrams}. This technique starts by considering the set of all possible selections of exactly one prototypical point from each prototypical region. Each selection can be used to generate a different diagram of the quality space \emph{QS}. Let us call the set of all these diagrams $V_{QS}$. The \emph{Collated Voronoi Diagram} can be generated by projecting all elements in $V_{QS}$ onto each other (thus overlaying the resulting diagrams). Fig.~\ref{fig:voronoi} (B) depicts this idea. In the resulting diagram, the regions created by the tessellation have thicker borders than in the traditional \emph{Voronoi} diagrams.

Decock and Douven~\cite{decock_what_2014} point out that Fig.~\ref{fig:voronoi} (B) is misleading in making one think that the transition from a crisp region (i.e., one of the white polygons/polyhedrons in it) to a borderline region is itself sharp, and provide a method for making the transition smooth. Their main idea is illustrated with the following example: suppose we have four prototypical regions, each consisting of two points \{\emph{a}, \emph{b}\}, \{\emph{c}, \emph{d}\}, \{\emph{e}, \emph{f}\} and \{\emph{g}, \emph{h}\}, and representing a prototype region (i.e., a concept such as ``\emph{low}'', ``\emph{medium}'', and ``\emph{high}''). Now, suppose we generate the $V_{QS}$  in the manner previously explained (i.e., $V_{QS}$  would contain $2 \times 2 \times 2 \times 2 = 16$ members in this case). The graded membership of a particular point \emph{p} in the quality space \emph{QS} to a concept \emph{X} (with degree \emph{D}) is defined as: the number of members of $V_{QS}$ that locate \emph{p} in the cell associated with \emph{X} divided by the total number of members of $V_{QS}$. For example, as in Fig.~\ref{fig:voronoi} (C), the point \emph{i} belongs to the concept associated with \{\emph{a}, \emph{b}\} with a degree 0.5, since 8 of the 16 members of $V_{QS}$ locate \emph{i} in the cell associated with \{\emph{a}, \emph{b}\}. Similarly, \emph{j}, belongs to that concept with a degree of 0.25.

By adopting the ``graded membership'' view, the satisfaction of a QG is defined by the membership of the observed quality value in the expected region \emph{QRG}. With this technique, what analysts need to do is to ask stakeholders to elicit prototype/typical values for concerned concepts, and use the elicited prototype values to derive membership functions for those concepts. These derived membership functions can be used to characterize the satisfaction of corresponding QGs for given quality values. We refer interested readers to Appendix~\ref{cha:appendixes_membership} for the technical details.
%(e.g., ``what do you think to be \emph{\textbf{low}} for a trip from A to B at period T?'')

%By adopting the ``graded membership'' view, the satisfaction of a QC for a subject/bearer \emph{subj} is defined by the membership of the observed quality value of \emph{subj} in the region \emph{QRG}. For the details about how to calculate graded membership, we refer interested readers to Appendix~\ref{cha:appendixes_membership}.

%By adopting the ``graded membership'' view, we define QCs as follows: a QC is an NFR that specifies a measurable region \emph{QRG} in a quality space \emph{QS}. As such, the satisfaction of a QC for a subject/bearer \emph{subj} is defined by the membership of the proper quality value of \emph{subj} in the region \emph{QRG}. For the details about calculating the graded membership, we refer interested readers to Appendix~\ref{cha:appendixes_membership}.
%As such, the satisfaction of an NFR to a certain degree is defined by the graded membership of the proper quality value of \emph{B} in region \emph{R}.

\section{Practical Implications}
\label{sec:practical_implication}

\subsection{Distinguishing between FRs and NFRs}
\label{sec:interpret_benifits_fr_vs_nfr}

In RE, there are two general criteria for distinguishing between functional and non-functional requirements, both of which are based on the premise that functional requirements (FRs) specify what a software system shall do (i.e., its functions): one takes the stance that non-functional requirements (NFRs) specify how well the system should perform its functions (i.e., qualities)~\cite{paech_non-functional_2004}, the other is to treat everything that is not an FR (i.e., not related to what a system shall do) as an NFR (i.e., as a sort of dispersive class defined by negation)~\cite{chung_non-functional_2000}. However, when put into practice, both criteria are deficient. For instance, how does one classify Ex.1 below, which specifies a function (``support'') that will not be performed by the system but by an external agent (``the corporate support center'')? One may treat it as an NFR by following the second criterion, but this is conceptually incorrect. In fact, Ex.1 will be classified as an FR in our proposal (because it requires a function of an entity in the system-to-be ecosystem).

%\cite{guizzardi_ontological_2014}

\begin{table}[!htbp]
    \centering
    %\small
    \begin{tabular}{|rp{0.8\textwidth}|}
        \hline
        %\multicolumn{2}{l}{|\textbf{Example 2}|} \\ \hline
        Ex.1: & \emph{The product shall be supported using the corporate support center}. \\
        Ex.2: & \emph{The system shall have a standard navigation button for navigation}. \\
        Ex.3: & \emph{The system shall help administrators to analyze failures/exceptions}. \\
        Ex.4: & \emph{The transportation system shall collect real-time traffic information}. \\
        \hline
    \end{tabular}
\end{table}

Jureta et al.~\cite{juretaa_core_2009} have made the first step in grounding this distinction on qualities in the foundational ontology DOLCE~\cite{masolo_ontology_2003}. As we have discussed in Section~\ref{sec:art_ontologies}, their requirements ontology still has deficiencies: it is not able to categorize requirements like Ex.2 (referring to neither qualities nor perdurants), Ex.3 (referring to a perdurant but being vague for success) and Ex.4 (referring to both perdurants and qualities).

Our guideline for distinguishing between FRs and NFRs: \emph{\textbf{if a requirement refers to a quality type}}\textbf{,} \emph{\textbf{then it is non-functional}}\textbf{;} \emph{\textbf{if it refers to a function}} (\emph{\textbf{in the ontological sense}})\textbf{,} \emph{\textbf{then it is functional}}\textbf{.}

Adopting this guideline, we can easily classify Ex.2 as functional, because it concerns a ``navigation'' function that is exhibited by the navigation button. Note that the distinction between FR and NFR is orthogonal to the one between hardgoal and softgoal. Hence, an FR can be vague while an NFR can also have a clear satisfaction criterion. For instance, Ex.3 is an FR but one that has a subjective criterion of satisfaction. Moreover, the classes of FRs and NFRs are not mutually exclusive. For example, Ex.4 specifies a desired function ``collect traffic information'' but also refers to a quality (timeliness) of ``collecting traffic information''. This is also the case for Ex.3: in addition to the ``navigation'' function, it also constrains the way of navigating users through navigation buttons (an FC), and require the navigation button to be standard (a QR).

%In fact, this requirement is a mix of concerns: it also constrains the way of navigating users through navigation buttons (an FC), and require the navigation button to be standard (a QR)

We have evaluated the guideline by applying it to the PROMISE requirements dataset ~\cite{menzies_promise_2012}, which includes 625 requirements (255 FRs and 370 NFRs) crossing 15 software projects. We found that most of the original security NFRs are often functional. For example, ``\emph{The website shall prevent the input of malicious data}'', originally labeled as a security NFR, should actually be a FR since it refers to a ``prevent'' function. The result suggests that our ontological interpretation is effective in distinguishing between FRs and NFRs in practice. For more details on the evaluation, interested readers can refer to Section~\ref{sec:eval_ontology}.

%For example, how can one classify the requirement ``\textit{the product shall be supported by the corporate support center}'', which specifies a function (``\textit{support}'') that will not be performed by the system but by an external agent (``\textit{the corporate support center}'') in a system-to-be ecosystem? This example indicates that the definition for ``function'' focuses on only the software

\subsection{Addressing Inconsistency between Quality Models}
\label{sec:interpret_benifits_addressing_inconsistency}
%Our characterization of qualities includes three key components: \emph{quality}, \emph{subject} and \emph{quality value}.

In the literature, when mentioning a quality, existing quality standards such as ISO/IEC 25010~\cite{iso/iec_25010_systems_2011} usually do not consider its subject/bearer; that is, we do not know whose quality it is or the quality of what. In our observation, the missing of subjects (of qualities) can be an important factor that leads to inconsistency when classifying qualities. For example, the ``understandability'' quality can be associated with many different aspects of a product, e.g., terminologies, user interface, architecture, code or documents. When taking a user's perspective (e.g., the understandability of the user interface), ``understandability'' will be classified as a sub-quality of ``usability'' as in ISO/IEC 9126-1~\cite{iso/iec_9126-1_software_2001}; however, when taking a developer's perspective (e.g., the understandability of the code or the architecture of a software system), it will be a sub-class of ``maintainability'' as in Boehm et al.~\cite{boehm_characteristics_1978}.

As we can see, the consideration of the subject of a quality contributes to addressing the inconsistency between quality hierachies/models. Note that the quality standards do classify qualities into categories, e.g.,``\emph{product quality}'' or ``\emph{quality in use}'' as in ISO/IEC 25010~\cite{iso/iec_25010_systems_2011}. However, categories are not the subjects/bearers of qualities: a quality existentially depends on particular things such as entities, processes, events. etc.

%We believe that it is important to take qualities' subjects into consideration when defining, specifying and classifying qualities.

\subsection{The Satisfaction of QRs}
\label{sec:interpret_benifits_qr_satisfaction}
Specifying QRs can be quite useful in practice since, in many cases (as exemplified in Section~\ref{sec:interpret_satisficing_nfrs}, it may be enough to ``almost'' reach the satisfaction of an QR. For instance, suppose that the associated quality value region ``\emph{low}'' of the QR ``\emph{the cost of trip from A to B at time period T shall be low}'' is represented by two prototype values 500\euro \; and 700\euro. Similarly, we can use 800\euro \; and 1000\euro \;, and 1200\euro \; and 1500\euro \; to represent the region ``\emph{medium}'' and ``\emph{high}''. Given the three prototype regions, the $V_{QS}$ will include 8 simple diagrams. Now if we have a cost value as 740\euro \;, then we will have 6 out of 8 diagrams classify it to the region ``low''. Thus, that QR is satisfied to a degree of 0.75. Interested readers can refer to the calculation details at Appendix~\ref{cha:appendixes_membership}.

The graded membership technique we adopted is more cognitively reasonable than fuzzy logic~\cite{baresi_fuzzy_2010}: with graded membership, we can use prototype/typical values to represent a region, and then use (collated) \emph{Voronoi} diagrams to derive the membership function (i.e., derive the boundary); however, with fuzzy logic, we need to explicitly specify boundary values, which are often uncertain or even made-up.

%The interesting point here is that we can use prototype values to represent a region, and then adopt (collated) \emph{Voronoi} diagrams to reason about the graded membership. without the need of inventing made-up numbers as that in fuzzy logic~\cite{baresi_fuzzy_2010}. Th

\section{Chapter Summary}
\label{sec:interpret_summary}
In this chapter, we have discussed about the ontological meaning of functional and non-functional requirements, and accordingly sketch a description-based syntax for their representation. Our ontological interpretation of requirements provides conceptual clarification, and enables us to clearly distinguish between functional and non-functional requirements. Also, our characterization of qualities contributes to addressing the inconsistency between quality hierarchies/models. Moreover, as we have seen, it offers support for reasoning about the satisfaction of quality requirements, designing requirement modeling languages, and refining/operationalizing requirements.

%In addition, our characterization of qualities contributes to address the classification inconsistency of NFRs/QRs in the literature: in our proposal, the categorization of qualities will be transformed into two open questions, ``what kinds of subjects exist (e.g., objects, artifacts, or functions)?'' and ``what kinds of qualities are related to them?''. To answer these questions, we could develop ontologies containing a number of upper-level categories showing what the qualities and subjects can be, and let stakeholders decide the specific quality hierarchies depending on system domains.

%% file: concepts.tex
%\vspace{4cm}
\chapter{The \emph{Desiree} Framework}
\label{cha:desiree_framework}
In the previous chapter, we have provided an ontological interpretation for functional and non-functional requirements, and sketched a syntax for their representation. Based on this, in this chapter, we present the \emph{Desiree} framework, including a set of requirement concepts, a set of requirements operators, a description-based syntax for representing these concepts and operators, and a systematic methodology for applying the concepts and operators in order to transform stakeholder requirements into a requirements specification.

\section{Requirements Concepts}
\label{sec:concepts}
%(derived from our ontological interpretation in the previous chapter and our experiences on examining the large PROMISE~\cite{menzies_promise_2012} requirements set)
%These concepts are derived from our experiences on examining the large PROMISE~\cite{menzies_promise_2012} requirements set.
% classified according to the RE problem (requirement \emph{R}, specification \emph{S}, and domain assumption \emph{DA}),

\begin{figure}[!htbp]
  \centering
  %\vspace {-0.5 cm}
  \includegraphics[width=1.0\textwidth]{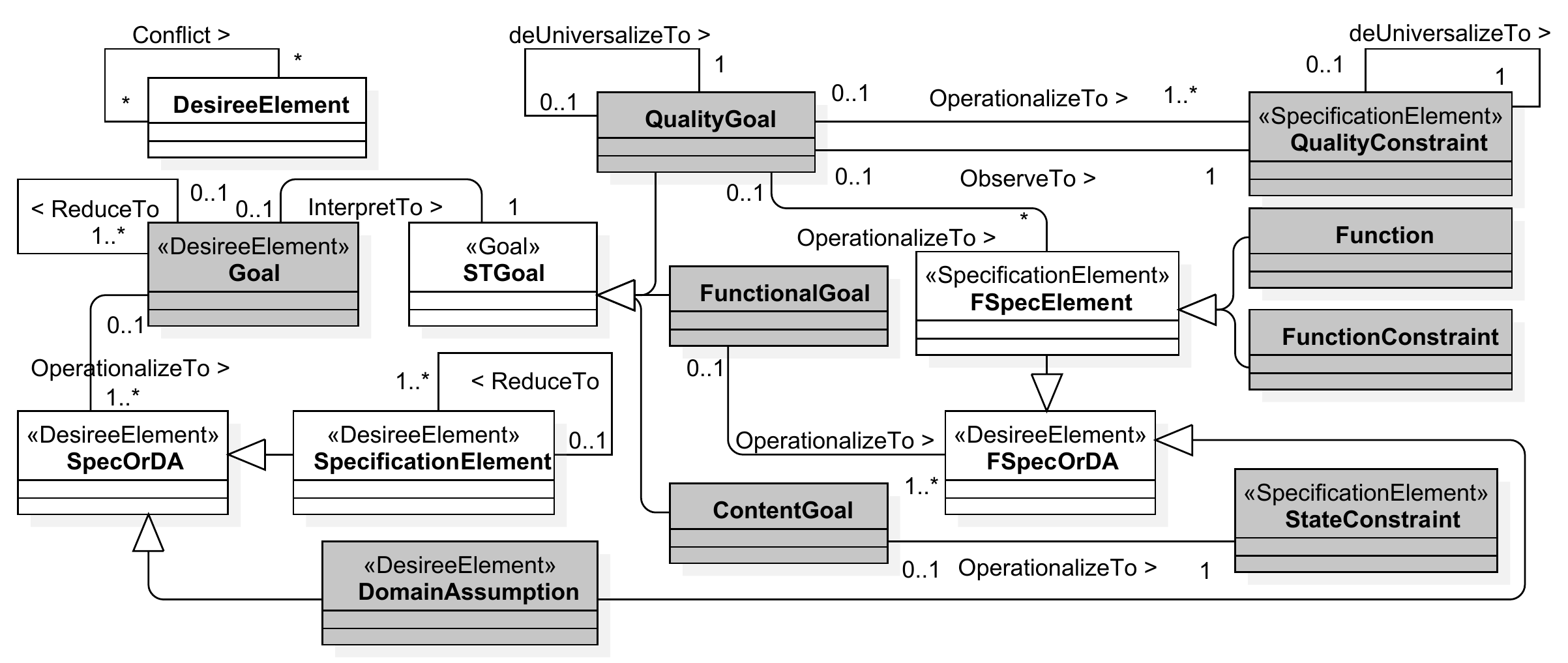}\\
  %\vspace {-0.6 cm}
  \caption{The requirements ontology}\label{fig:desiree_concepts}
\end{figure}

The core notions of \emph{Desiree} are shown in Fig.~\ref{fig:desiree_concepts}. As in goal-oriented RE, we capture stakeholder requirements as goals. We have 3 sub-kinds of goals, 4 sub-kinds of specification elements (those with stereotype ``Specification Element''), and domain assumptions, all of which are subclasses of ``Desiree Element''. These concepts are derived from our ontological interpretation in the previous chapter and our experiences on examining the large PROMISE requirements set~\cite{menzies_promise_2012}. In this chapter, we use examples from this set to illustrate each of these concepts and relations (see Table~\ref{tab:appendix_desiree_syntax} in Appendix~\ref{cha:appendixes_syntax} for the full syntax).

%(adapted from~\cite{li_stakeholder_2015})

There are three points to be noted. First, `$>$' and `$<$' are used to indicate the reading directions of the relations. Second, the relations (except ``Conflict'') are derived from applications of requirements operators (to be discussed in Section~\ref{sec:opeartors}): if an operator is applied to an X object then the result will be $n..m$ Y objects; if that operator is not applied, then there is no relation (thus we have $0..1$ as the lower bound). Third, ``STGoal'', ``SpecOrDA'', ``FSpecOrDA'' and ``FSpecElement'' are artificial concepts, used to represent the union of their sub-classes, e.g., ``STGoal'' represents the union of ``FunctionalGoal'', ``QualityGoal'' and ``ContentGoal''. These classes are added to overcome the limitations of UML in representing inclusive (e.g., an operationalization of a functional goal) and exclusive (e.g., an interpretation of a goal) OR.

\vspace{6pt}
\noindent \emph{\textbf{Functional Goal, Function and Functional Constraint.}} A functional goal (\textbf{FG}) states a desired state, and is operationalized to one or more functions (\textbf{F}). For example, the goal ``student records be managed'' specifies the desired state ``managed''. As mentioned in Section \ref{sec:interpret_representing_frs}, we capture the intention of something to be in a certain state (situation) by using the symbol ``$:<$''. So this example is interpreted as a functional goal ``$FG_1$ := Student\_record $:<$ Managed'' (here ``Managed'' refers to an associated set of individuals that are in this specific state). This FG will be operationalized using functions such as ``add'', ``update'' and ``remove'' on student records.

%In practice, when a function is manifested (i.e., executed), it would have properties of different participants being manifested at the same time~\cite{guizzardi_ontological_2005}. For example, an execution of ``product search'' will involve participants like the system, a user, and product info. This means that in addition to the desired capability, many pieces of information (e.g. restrictions on its actor, object, and trigger) can be associated with the description of a function (\textbf{F}).
As most perceived events in UFO~\cite{guizzardi_ontological_2005} are \emph{polygenic}, i.e., when an event is occurring, there are a number of dispositions of different participants being manifested at the same time, many pieces of information (e.g. actor, object, and trigger) can be associated with the desired capability when specifying a function (\textbf{F}). For example, an execution of ``product search'' will involve participants like the system, a user, and product info. That is, ``the system shall allow users to search products'' can be operationalized as a function ``$F_1$ := Search $<$subject: \{the\_system\}$>$$<$actor: User$>$$<$object: Product$>$''. Moreover, we can further add search parameters by adding a slot-description pair ``$<parameter: Product\_name>$''.

A functional constraint (\textbf{FC}) constrains the situation under which a function can be manifested. As above, we specify intended situations using ``$<s: D>$'' pairs and constrain a function or an entity involved in a function description to be in such a situation using ``$:<$''. For example, ``\emph{only managers are able to activate debit cards}'' can be captured as ``$FC_1$ := Active $<$object: Debit\_card$>$ $:<$ $<$actor: ONLY Manager$>$''.

%~\cite{guizzardi_ontological_2014}
\vspace{6pt}
\noindent \emph{\textbf{Quality Goal and Quality Constraint.}} We treat a quality as a mapping function that maps its subject to its value. Quality goal (\textbf{QG}) and quality constraint (\textbf{QC}) are requirements that require a quality to have value(s) in a desired quality region (QRG). In general, a QG/QC has the form ``Q (SubjT) :: QRG''. For instance, ``\emph{the file search function shall be fast}'' will be captured as a QG ``Processing\_time (File\_search) :: Fast''.

There are three points to be noted. First, QGs and QCs have the same syntax, but different kinds of QRGs: regions of QGs are vague (e.g., ``\emph{low}'') while those of QCs are often measurable (e.g., ``[0, 30 (\emph{Sec.})]'', {but see Example 5 in Section~\ref{sec:opeartors} for a more complex expression). Second, a quality name indicates a quality type, not a quality instance. By applying a quality type (e.g., ``Processing\_time'') to an individual subject \emph{x} of type \emph{SubjT} (e.g., a run of search, say \emph{search}\#1), we first get a particular quality \emph{q\#} (e.g., \emph{processing\_time}\#1), and then the associated quality value of \emph{q\#}. Third, when the subject is represented as individuals, we use curly brackets to indicate a set, e.g., ``\{the\_system\}''.

\vspace{6pt}
\noindent \emph{\textbf{Content Goal and State Constraint.}} A content goal (\textbf{CTG}) often specifies a set of properties of an entity in the real word, including both attributes and qualities, and these properties need to be represented by a system-to-be. To satisfy a CTG, a system needs to be in a certain state, which represents the desired world state. That is, concerned properties of real-world entities should be captured as data in the system. We use a state constraint (\textbf{SC}) to specify such desired system state.

For example, to satisfy the CTG ``A student shall have Id, name and GPA'', the student record database table of the system must include three columns: Id, name and GPA. This example can be captured as a content goal ``$CTG_1$ := Student $:<$ $<$has\_id: ID$>$ $<$has\_name: Name$>$ $<$has\_gpa: GPA$>$'' and a state constraint, ``$SC_2$ := Student\_record :$<$ $<$ID: String$>$ $<$Name: String$>$ $<$GPA: Float$>$''.

\vspace{6pt}
\noindent \emph{\textbf{Domain Assumption.}} A domain assumption (\textbf{DA}) is an assumption about the operational environment of a system. For instance, ``\emph{the system will have a functioning power supply}'', which will be captured as ``$DA_{1}$ := \{the\_system\} :$<$ $<$has\_power: Power$>$'' using our language. Note that the syntax for DAs is similar to that for FCs. The difference is that an FC requires a subject to possess certain properties (e.g., in certain situations), while a DA assumes that a subject will have certain properties. In addition, DAs in our framework are also used to capture domain knowledge, e.g., ``\emph{Tomcat is a web server}'' will be captured as ``$DA_{2}$ := Tomcat $:<$ Web\_server''.

According to Zave and Jackson's categorization of \emph{requirements} (R), \emph{specification} (S) and \emph{domain assumptions} (DA), these requirements concepts can be accordingly classified as: (1) (stakeholder) requirements, including ``goal'' (G), ``functional goal'' (FG), ``quality goal'' (QG), and ``content goal'' (CTG); (2) (requirements) specification, including ``function'' (F), ``functional constraint'' (FC), ``quality constraint'' (QC), and ``state constraint'' (SC); and (3) ``domain assumption'' (DA). Moreover, in our ontology, traditional non-functional requirements will be distributed into the following five kinds: QGs, QCs, CTGs, SCs or FCs. That is, what have been traditionally called NFRs but do not refer to qualities can be CTGs, SCs or FCs.

\section{Requirements Operators}
\label{sec:opeartors}
In this section, we introduce the set of requirements operators used for transforming requirements, which are inspired by traditional goal modeling techniques, and the syntactic form and semantics of our language. For example, since a QG/QC has the form ``Q (SubjT) :: QRG'', there can be different ways of refining it, based on whether Q, SubjT, or QRG is adjusted. In addition, since the semantics of such formulas have the form ``$\forall$ \emph{x}/\emph{SubjT}'', we also need to consider de-Universalizing them. Moreover, as qualities are measurable or observable properties of entities, we should also be able to add information about the observers who observe the quality.

In general, \emph{Desiree} includes two groups of operators (8 kinds in total): \emph{refinement} and \emph{operationalization}. An overview of these operators is shown in Table~\ref{tab:desiree_operators}, where ``\#'' means cardinality, ``\emph{m}'' and ``\emph{n}'' are positive integers (\emph{m} $\geq$ 0, \emph{n} $\geq$ 2). As shown, ``Reduce'', ``Interpret'', ``de-Universalize'', ``Scale'', ``Focus'' and ``Resolve'' are sub-kinds of refinement; ``Operationalize'' and ``Observe'' are sub-kinds of operationalization.

In the \emph{Desiree} framework, refinement operators are applied in the same category of elements: they refine goals to goals, or specification elements to specification elements. Operationalization operators map from goals to specification elements. Note that we do not support refinements from specifications to goals (i.e., requirements).

%Note that the ``Interpret'' operator is special: we allow interpreting a goal (requirement) as a sub-kind of goal or a specification element. In addition, we do not support refinements from specifications to goals (i.e., requirements).

\begin{table}[!htbp]
  \caption {An overview of the requirements operators }
  \label{tab:desiree_operators}
  \vspace {0.3 cm}
  \centering
  %\small
  \setlength\tabcolsep{2pt}
  %\begin{tabular}{|c|p{0.8\textwidth}|}
  %\begin{tabular}{|m{0.15\textwidth}|m{0.85\textwidth}|@{}m{0pt}@{}} %<{\centering}
  \begin{tabular}{|c|c|c|c|}
  \hline
  \multicolumn{2}{|c|}{\textbf{Requirements Operators}} & \textbf{\#InputSet} & \textbf{\#OutputSet} \\ \hline
  \multirow{6}{*}{\emph{Refinement}} & Reduce ($R_d$) & 1 & $1 ... m$ \\
  & Interpret (\emph{{I}}) & 1 & 1 \\
  & de-Universalize (\emph{{U}}) & 1 & 1 \\
  & Scale (\emph{{G}}) & 1 & 1 \\
  & Focus ($F_k$) & 1 & $1 ... m$ \\
  & Resolve ($R_s$) & $2 ... n$ & $0 ... m$ \\ \hline
  \multirow{2}{*}{\emph{Operationalization}}  & Operationalize ($O_p$) & 1 & $1...m$ \\
  & Observe ($O_b$) & 1 & 1 \\ \hline
  \end{tabular}
\end{table}

\vspace{6pt}
\noindent \textbf{\textbf{Reduce}} ($\bm{R_d}$). ``Reduce'' is used to refine a composite element (goal or specification element) to simple element(s), a high-level element to low-level element(s), or an under-specified element to sufficiently complete element(s). For instance, the composite goal $G_1$ ``\emph{collect real time traffic info}'' can be reduced to $G_2$ ``\emph{traffic info be collected}'' and $G_3$ ``\emph{collected traffic info be in real time}''.

The signature of operator ``${R_d}$'' is shown in Eq.~\ref{eq:reduce}, where $E'$ is a goal (e.g., goal, FG, QG, CTG) or a specification element (e.g., F, FC, QC, SC). It takes as input an element \emph{$E'$} and outputs a non-empty set (indicated by $\wp_1$, where $\wp$ represents power-set) of elements that are exactly of the same kind (with optional DAs). That is, we only allow reducing from goal to goal (not its sub-kind), FG to FG, F to F, etc; we also allow making explicit domain assumptions when applying the ``$R_d$'' operator. For example, when reducing $G_1$ ``\emph{pay for the book online}'' to $G_2$ ``\emph{pay with credit card}'', one needs to assume $DA_3$ ``\emph{having a credit card with enough credits}''. This refinement can be captured as ``$\bm{R_d}$ ($G_1$) = \{$G_2$, $DA_3$\}''.\\
\begin{equation}\label{eq:reduce}
  {\bm{R_d}: \; E' \rightarrow \bm{\wp_1} (E' \cup DA) }%\cup
\end{equation}

The ``$R_d$'' operator allows us to refine an element (a goal or a specification element) to several sub elements; hence it captures AND-refinement in traditional goal modeling techniques. To capture OR-refinement, we can apply the reduce operators several times, according to the different ways that a goal can be refined. For example, we will have two refinements ``$R_d$ ($G_1$) = \{$G_2$\}'' and ``$R_d$ ($G_1$) = \{$G_3$\}'' when reducing $G_1$ ``\emph{search products}'' to $G_2$ ``\emph{search by product name}'', and to $G_3$ ``\emph{search by product number}'', separately. As a result, ``$R_d$'' gives rise to a relation, not a (math) function.

%rephrase an underspecified requirement to make it precise,
\vspace{6pt}
\noindent \textbf{Interpret} ($\bm{I}$). The ``$\bm{I}$'' operator allows us to disambiguate a requirement by choosing the intended meaning, classify and encode a natural language requirement using our description-based syntax. For example, a goal $G_1$ ``\emph{notify users with email}'' can be interpreted as $FG_2$ ``User $:<$ Notified $<$means: Email$>$''. Note that $G_1$ is ambiguous since it has another interpretation $FG_3$ ``User $<$ has\_email: Email$>$ $:<$ Notified''. In such situation, analysts/engineers have to communicate with stakeholders in order to choose the intended interpretation.

We show the signature of ``$\bm{I}$'' as in Eq.~\ref{eq:interpret}, where $E$ is a \emph{Desiree} element (a goal, a specification element or a domain assumption) stated in natural language (NL), $E'$ is a structured \emph{Desiree} element, and is a subclass of $E$ or of the same type of $E$. Using this syntax, the ``notify user'' example will be written as ``$\bm{I}$ ($G_1$) = $FG_2$'' (suppose that $FG_2$ is the intended meaning).\\
\begin{equation}\label{eq:interpret}
  \bm{I}: \; E \rightarrow E'
\end{equation}

%\noindent \textbf{Interpret} ($\bm{I}$). The ``$\bm{I}$'' operator allows us to disambiguate a requirement by choosing the intended meaning, classify and encode a natural language requirement using our description-based syntax. For example, a goal $G_1$ ``\emph{notify users with email}'' can be interpreted as $F_2$ ``Notify $<$object: User$>$ $<$means: Email$>$''. Note that $G_1$ is ambiguous since it has another interpretation $F_3$ ``Notify $<$object: User $<$ has\_email: Email$>>$. In such situation, analysts/engineers have to communicate with stakeholders in order to choose the intended interpretation.

%We show the signature of ``$\bm{I}$'' as a partial function in Eq.~\ref{eq:interpret}. Here G is a goal, FG, QG and CTG are its subclasses, SE is a specification element, which can be F, FC, QC, or SC. Using this syntax, the ``notify user'' example will be written as ``$\bm{I}$ ($G_1$) = $F_2$'' (suppose that ``$F_2$'' is the intended meaning).\\
%\begin{equation}\label{eq:interpret}
%  \bm{I}: \; G  \rightarrow (FG \cup QG \cup CTG \cup SE)
%\end{equation}

%\vspace{6pt}
\noindent \textbf{Focus} ($\bm{F_k}$). The $\bm{F_k}$ operator is a special kind of ``Reduce'', and is used for refining a QGC~\footnote{A QGC is a QG or a QC, and models a QR (quality requirement); that is, a QGC is the structured version of a QR, which is in natural language.} (QG/QC) to sub-QGCs, following special hierarchies of its quality type or subject, e.g., dimension-of, part-of. For instance, for $QG_1$ ``Security (\{the\_system\}) :: Good'', the quality type ``Security'' can be focused to its sub-dimensions, e.g., ``Confidentiality'', the subject ``the system'' can be replaced by some of its parts, e.g., ``the data storage module''. The key point of applying $F_k$ is that if a quality goal $QG_1$ is focused to $QG_2$, then $QG_1$ would logically imply $QG_2$. For instance, if the system as a whole is secure, then its data storage module is secure, too.

The $\bm{F_k}$ operator has the signature as in Eq.~\ref{eq:focus}. In general, it replaces the quality type (resp. subject type) of a QGC with given Qs (resp., SubjTs), and returns new QGCs. Using this syntax, the focus of $QG_1$ to a sub-goal $QG_3$ ``Security (\{data\_storage\}) :: Good'' can be obtained as ``$\bm{F_k}$ ($QG_1$, \{data\_storage\}) = \{$QG_3$\}''.\\
\begin{equation}\label{eq:focus}
    \begin{aligned}
        \bm{F_k}: \; QGC \times \bm{\wp_1} (Q) \rightarrow \bm{\wp_1} (QGC) \\
        \bm{F_k}: \; QGC \times \bm{\wp_1}(SubjT) \rightarrow \bm{\wp_1} (QGC)
    \end{aligned}
\end{equation}

\vspace{6pt}
\noindent \textbf{Scale} ($\bm{G}$). In general, the $\bm{G}$ operator is used to enlarge the boundary of the quality region (QRG) of a QG/QC, in order to tolerate some deviations of quality values (i.e., relaxing the QG/QC to some degree). For instance, we can relax ``\emph{fast}'' to ``\emph{nearly fast}'', or ``\emph{in 30 seconds}'' to ``\emph{in 30 seconds with a scaling factor 1.2}''. The $\bm{G}$ operator can be also used to shrink the region of a QG/QC, strengthening the quality requirement. For example, we can replace a region ``\emph{good}'' with a sub-region ``\emph{very good}'', or more cleanly, replace ``[0, 30 (\emph{Sec.})]'' with ``[0, 20 (\emph{Sec.})]''. Based on the two cases, we specialize the ``\emph{scale}'' operator into ``\emph{scale down}'' ($\bm{G_d}$, relaxing through enlarging a QRG) and ``\emph{scale up}'' ($\bm{G_u}$, strengthening through shrinking a QRG) accordingly.

The two scaling operators have the syntax as in Eq.~\ref{eq:scale}. They take as input a QG (resp. QC), a qualitative (resp. quantitative) factor and return another QG (resp. QC).\\
\begin{equation}\label{eq:scale}
    \begin{aligned}
        \bm{G}: QG \times QualitativeFactor \rightarrow QG \\
        \bm{G}: QC \times QuantitativeFactor \rightarrow QC	
    \end{aligned}
\end{equation}

Using this syntax, the relaxation of ``\emph{fast}'' to ``\emph{nearly fast}'' can be captured as in Example 1, and the enlarging of ``[0, 30 (Sec.)]'' to ``[0, 36 (Sec.)]'' through a pair of scaling factors ``(1, 1.2)'' can be written as in Example 2.

\begin{table}[!htbp]
    \centering
    %\small
    \begin{tabular}{|rp{0.75\textwidth}|}
        \hline
        \multicolumn{2}{l}{|\textbf{Example 1}|} \\ \hline
        $QG_{1-1}$ := & Processing\_time (File\_search) :: Fast \\
        $QG_{1-2}$ := & $\bm{G_d}$ ($QG_{1-1}$, Nearly) \\
                = & Processing\_time (File\_search) :: Nearly\_fast \\
        \hline
        \multicolumn{2}{l}{|\textbf{Example 2}|} \\ \hline
        $QC_{2-1}$ := & Processing\_time(File\_search) :: [0, 30 (Sec.)] \\
        $QC_{2-2}$ := & $\bm{G_d}$ ($QC_{2-1}$, (1, 1.2)) \\
                    = & Processing\_time(File\_search) :: [0, 36 (sec.)] \\
        \hline
    \end{tabular}
\end{table}

Qualitative factors can be used to either strengthen (e.g., ``\emph{very}'') or weaken (e.g., ``\emph{nearly}'', ``\emph{almost}'') QGs, scaling their regions up and down, respectively. Quantitative factors are real numbers. We restrict ourselves to single-dimensional regions, which are intervals of the form $[Bound_{low} \; ... \; Bound_{high}]$. When enlarging a quality region, the scaling factor for $Bound_{low}$ shall be less than or equal to 1.0, and the factor for $Bound_{high}$ shall be greater than or equal to 1.0; when shrinking a region, opposite constraints must hold. Note that a region can be enlarged or shrunk, but not can be shifted. For example, we do not allow the change from ``[10, 20]'' to ``[15, 25]'', which is a shift. This is because we want to ensure the subsumption relation between regions when scaling them.

\vspace{6pt}
\noindent \textbf{de-Universalize} ($\bm{U}$). $\bm{U}$ applies to QGs and QCs to relax quality requirements, such that it is no longer expected to hold ``universally'', i.e., not to hold for 100\% of the individuals in a domain. For example, going from $QG_{3-1}$ ``(\emph{all}) \emph{file searches shall be fast}'' to $QG_{3-2}$ ``(\emph{at least}) \emph{80\% of the searches shall be fast}'' in Example 3.

\begin{table}[!htbp]
    \centering
    %\small
    \begin{tabular}{|rp{0.75\textwidth}|}
        \hline
        \multicolumn{2}{l}{|\textbf{Example 3}|} \\ \hline
        $QG_{3-1}$  := & Processing\_time (File\_search) :: Fast \\
        $QG_{3-2}$  := & $\bm{U}$ (?X, $QG_{3-1}$, $<$inheres\_in: ?X$>$, 80\%) \\
        \hline
    \end{tabular}
\end{table}

Note that a QGC (QG/QC) description has a built-in slot ``\textbf{inheres\_in}'', relating a quality to the subject to which it applies. For example, the right-hand side (RHS) of $QG_{3-1}$, ``Processing\_time (File\_search)'', is actually ``Processing\_time $<$inheres\_in: File\_search$>$''. The syntax of $\bm{U}$ refers to a subset ``?X'' of the subject concept by pattern matching, using the SlotD ``$<$inheres\_in: ?X$>$''. Hence the above relaxation will be captured as in Example 3, where ``?X'' represents a sub-set of ``File\_search''.

The general signature of the U operator is given in Eq.~\ref{eq:deUniversalize}.

\begin{equation}\label{eq:deUniversalize}
        \bm{U}: \; varId \times QGC \times SlotD \times [0\%...100\%] \rightarrow QGC
\end{equation}

Sometimes we want to capture relaxations of requirements over requirements, for example, ``\emph{system functions shall be fast at 90\% of the time}'', relaxed to ``(\emph{at least}) \emph{80\% of the system functions shall be fast $($at least$)$ 90\% of the time}''. For this, we use nested $\bm{U}$s, as in Example 4 below. Here, ``?F'' is a sub-set of system functions (i.e., ``?F'' matches to ``System\_function''), and ``?Y'' is a sub-set of executions of a function in ``?F'' (i.e., ``?Y'' matches to the description ``Run $<$run\_of: ?F$>$'').

\begin{table}[!htbp]
    \centering
    %\small
    \begin{tabular}{|rp{0.75\textwidth}|}
        \hline
        \multicolumn{2}{l}{|\textbf{Example 4}|} \\ \hline
        $QG_{4-1}$  := & Processing\_time (Run $<$run\_of: System\_function$>$) :: Fast \\
        $QG_{4-2}$ := & $\bm{U}$ (?F, $QG_{4-1}$, $<$inheres\_in: $<$run\_of:  ?F$>$$>$, 80\%) \\
        $QG_{4-3}$ := & $\bm{U}$ (?Y, $QG_{4-2}$, $<$inheres\_in: ?Y$>$, 90\%) \\
        \hline
    \end{tabular}
    \vspace{-0.3cm}
\end{table}

$\bm{U}$ applies to only QGs and QCs. If one wants to specify the success rate of a function, one can first define a QC, e.g., ``Success (File\_search) :: True'', and then apply $\bm{U}$.

\vspace{6pt}
\noindent \textbf{Resolve} ($\bm{R_s}$). In practice, it could be the case that some requirements can stand by themselves, but will be conflicting when put together, since they cannot be satisfied simultaneously. Note that by conflict, we do not necessarily mean logical inconsistency, but can also be other kinds like normative conflict (e.g., a requirement could conflict with a regulation rule) or unfeasibility given the state of technology. For example, the goal $G_1$ ``\emph{use digital certificate}'' would conflict with $G_2$ ``\emph{good usability}'' in a mobile payment scenario. In \emph{Desiree}, we use a ``\emph{conflict}'' relation to capture this phenomenon, and denote it as ``\emph{Conflict} (\{$G_1$, $G_2$\})''.

The ``$\bm{R_s}$'' operator is introduced to help deal with such conflicting requirements: it takes as input a set of conflicting requirements (more than one), while the output captures a set of compromise (conflicting-free) requirements, determined by the analyst. In the example above, we can replace $G_2$ by $G_2'$ ``acceptable usability'' or drop $G_1$.

The signature of $\bm{R_s}$ is shown in Eq.~\ref{eq:resolve}. Here we do not impose cardinality constraints on the output set, allowing stakeholders to totally drop the conflicting requirements when it is really necessary. Using this, we can write the resolution of this conflict as ``$\bm{R_s}$ (\{$G_1$, $G_2$\}) = \{$G_1$, $G_2'$\}'' or ``$\bm{R_s}$ (\{$G_1$, $G_2$\}) = \{$G_2$\}''.\\
\begin{equation}\label{eq:resolve}
    \bm{R_s}:  \bm{\wp_1} (E) \rightarrow \bm{\wp} (E)
\end{equation}

There are two points to be noted. First, analysts/engineers may declare known conflicts as DA axioms. For example, one can capture the conflict ``a user can not be both authorized and unauthorized'' as ``$DA_1$ := Authorized ($\cap$) Unauthorized $:<$ Nothing''. Second, an application of the $\bm{R_s}$ operator will not physically delete any ``dropped'' requirement from a \emph{Desiree} model. For example, in the case of ``$\bm{R_s}$ (\{$G_1$, $G_2$\}) = \{$G_2$\}'', $G_1$ will still be kept in the model, but will not be considered during fulfillment reasoning (i.e., we do not consider if $G_1$ can be fulfilled, but do consider the remaining $G_2$). This is the same for $G_2$ in the case of ``$\bm{R_s}$ (\{$G_1$, $G_2$\}) = \{$G_1$, $G_2'$\}''.

\vspace{6pt}
\noindent \textbf{Operationalize} ($\bm{O_p}$). The $\bm{O_p}$ operator is used to operationalize goals into specification elements. In general, $\bm{O_p}$ takes as input one goal, and outputs one/more specification elements with optional domain assumptions. For instance, the operationalization of $FG_1$ ``Products :$<$ Paid'' as $F_2$ ``Pay $<$object: Product$>$ $<$means: Credit\_card$>$'' and $DA_3$ ``Credit\_card :$<$ Having\_enough\_credit'' will be written as ``$\bm{O_p}$ ($G_1$) = \{$F_2$, $DA_3$\}''.

The generalized syntax of $\bm{O_p}$ is shown in Eq.~\ref{eq:operationalize}.\\
\begin{equation}\label{eq:operationalize}
    \begin{aligned}
        \bm{O_p}: \; FG \rightarrow \bm{\wp} (F \cup FC \cup DA)	\\
        \bm{O_p}: \; QG \rightarrow \bm{\wp} (QC \cup F \cup FC \cup DA)	\\
        \bm{O_p}: \; CTG \rightarrow \bm{\wp} (SC \cup DA)	\\
        \bm{O_p}: \; Goal \rightarrow \bm{\wp} (DA)	\\
    \end{aligned}
\end{equation}

Note that one can use $\bm{O_p}$ to operationalize a QG as QC(s) to make is measurable, as Fs and/or FCs to make it implementable, or simply by connecting it to DA(s), assuming the QG to be true.

\vspace{6pt}
\noindent \textbf{Observe} ($\bm{O_b}$). The $\bm{O_b}$ operator is employed to specify the means, measurement instruments or human used to measure the satisfaction of QGs/QCs, as the value of slot ``\emph{\textbf{observed\_by}}''. For instance, we can evaluate ``\emph{be within 30 seconds}'' by assigning a stopwatch or assess a subjective QG ``\emph{the interface shall be simple}'' by asking a set of observers. The $\bm{O_b}$ operator has the signature shown in Eq.~\ref{eq:observe}.\\
\begin{equation}\label{eq:observe}
        \bm{O_b}: \; (QG \cup QC) \times Observers \rightarrow QC
\end{equation}

Consider now the requirement ``(\emph{at least}) \emph{80\% of the surveyed users shall report the interface is simple}'', which operationalizes and relaxes the ``\emph{the interface shall be simple}''. The original goal will be expressed as $QG_{5-1}$ in Example 5. To capture the relaxation, we first use $\bm{O_b}$, asking a set of surveyed users to observe $QG_{5-1}$, and then use $\bm{U}$, to require (at least) 80\% of the users to agree that $QG_{5-1}$ hold. Here, the set variable ``?S'' represents a subset of surveyed users.

\begin{table}[!htbp]
    \centering
    %\small
    \begin{tabular}{|rp{0.75\textwidth}|}
        \hline
        \multicolumn{2}{l}{|\textbf{Example 5}|} \\ \hline
        $QG_{5-1}$ := & Style (\{the\_interface\}) :: Simple \\
        $QC_{5-2}$ := & $\bm{O_b}$ ($QG_{5-1}$, Surveyed\_user) \\
               = & $QG_{5-1}$ $<$observed\_by: Surveyed\_user$>$ \\
        $QC_{5-3}$ := & $\bm{U}$ (?S, $QC_{5-2}$, $<$observed\_by: ?S$>$, 80\%)\\
        \hline
    \end{tabular}
\end{table}

\begin{comment}
When the vague region of a QG is made measurable by specifying a QC, the QC is still not operational until the measuring instrument is specified. For example, if we have derived QC2-1 as in Example 2, we will need to specify a measuring device as in Example 6.

\begin{table}[!htbp]
    \centering
    %\small
    \begin{tabular}{|rp{0.75\textwidth}|}
        \hline
        \multicolumn{2}{l}{|\textbf{Example 6}|} \\ \hline
        QC6-1 := & $\bm{O_b}$ (QC2-1, Stopwatch) \\
        = & Processing\_time(File\_search) :: [0, 30 (sec.)] $<$observed\_by: Stopwatch$>$ \\
        \hline
    \end{tabular}
\end{table}
\end{comment}

\vspace{6pt}
\noindent \textbf{ReferTo}. We use the ``\emph{ReferTo}'' relation to capture the interrelations between Fs, FCs QGs, QCs, CTGs and SCs. In general, an F could refer to some CTGs or SCs, e.g., ``$F_1$ := Search $<$object: Product\_info$>$'' (``product info'' indicates a CTG); a QG/QC can take an F as its subject, containing its executions to be in certain time limitation, e.g., ``Processing\_time ($F_1$) :: [0, 30 (\emph{Sec.})]''; an FC could constrain an entity or a function involved in a function description, e.g., ``$F_1$ :$<$ $<$actor: ONLY Registered\_user$>$'' (only registered users can search); a CTG could refer to other SCs, e.g., when defining an attribute ``has\_product\_parameter'', we will need another SC ``Product\_parameter''.

The capture of such interrelations contributes to improving the modifiability and traceability of a requirements specification. For example, by translating a requirements specification with our syntax into a DL specification, we are able to performs queries such as ``what functions does a quality refer to?'', ``what qualities are of concern for a function?'', and ``what functions will an entity be involved in?''. That is, we can easily know what kinds of qualities will be affected if we change a function, what kinds of functions will be affected if we change the schema of an entity (i.e., an SC), etc. We will discuss this in detail in Section~\ref{sec:eval_method}.

%For example, by translating a requirements specification with our syntax into a DL specification (see our preliminary work on this topic~\cite{li_stakeholder_2015}), we can easily know what kinds of qualities will be affected if we change a function, what kinds of functions will be affected if we change the schema of an entity (i.e., an SC), etc.

\section{A Transformation Methodology}
\label{sec:transform_process}
The \emph{Desiree} approach takes as input informal stakeholder requirements, and outputs an eligible specification through incremental applications of the requirements operators. In general, the method consists of three major phases, as captured in Fig.~\ref{fig:desiree_methodology} using BPMN 2.0 notations~\footnote{http://www.omg.org/spec/BPMN/2.0/}: (1) an informal phase, where composite requirements are broken down into goals representing single requirements and high-level requirements are reduced to low-level ones; (2) an interpretation phase, where each informal goal is structurally specified, along with its relationships to other goals; (3) a smithing phase, where requirements operators are iteratively applied on structurally specified goals to derive an eligible specification. Note that a box with a ``$+$'' symbol indicates a collapsed sub-process.
%(complete enough, unambiguous, measurable, mutually consistent, satisfiable, modifiable, traceable)

\begin{figure}[!htbp]
  \centering
  \vspace {-0.2 cm}
  \includegraphics[width=1.0\textwidth]{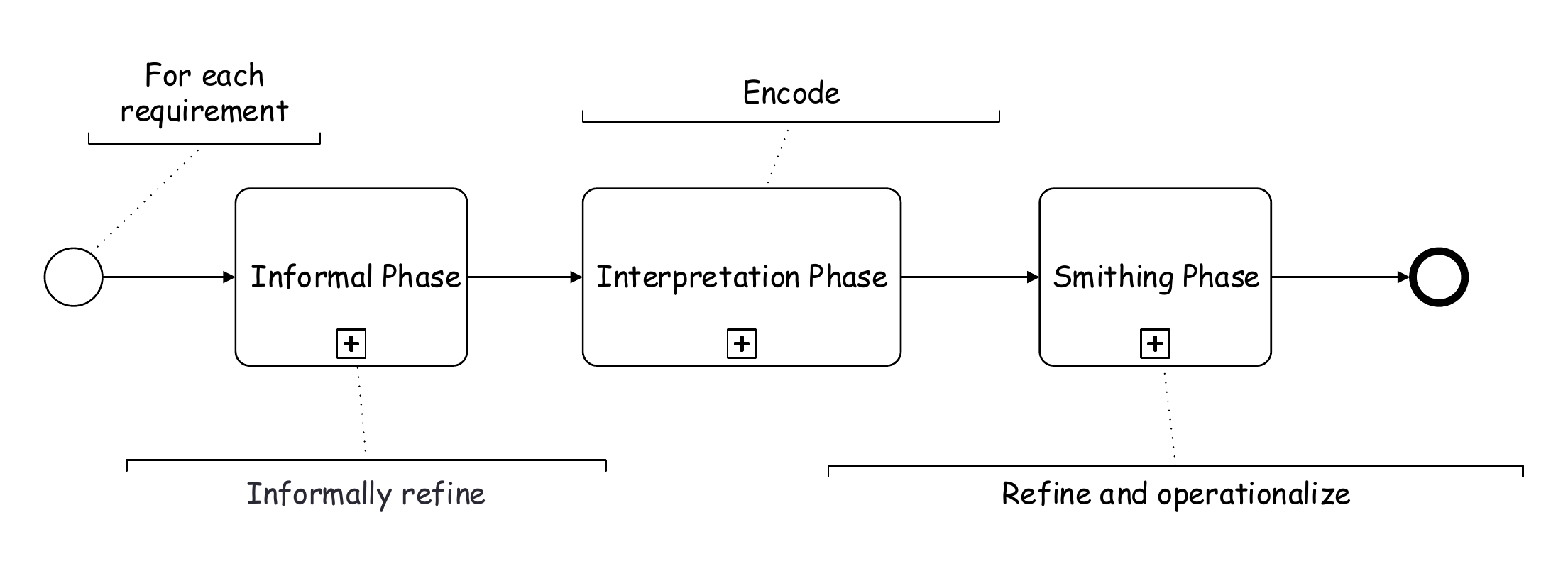}\\
  \vspace {-0.2 cm}
  \caption{The overview of the three-staged methodology}\label{fig:desiree_methodology}
\end{figure}
%(adapted from~\cite{li_stakeholder_2015})

%can be modeled and refined using existing goal modeling techniques (e.g., Techne~\cite{jureta_techne:_2010}).
%\emph{\textbf{Informal Phase.}}
%In this phase, each requirement is simply treated as a proposition and captured as a goal (G).

\subsection{Informal Phase}
\label{sec:desiree_informal_phase}
In this phase, each stakeholder requirement is simply captured as a goal (G) stated in natural language (NL). As shown in Fig.~\ref{fig:desiree_method_phase1}, the main tasks of requirement engineers in this phase are to: (1) identify key stakeholder concerns and determine their classifications according to the requirements ontology of Fig.~\ref{fig:desiree_concepts}; (2) de-couple composite concerns to make them atomic; (3) refine high-level requirements to low-level ones in the spirit of goal-oriented refinement techniques; and (4) capture conflicts if identified.

\begin{figure}[!htbp]
  \centering
  \vspace {-0.2 cm}
  \includegraphics[width=\textwidth]{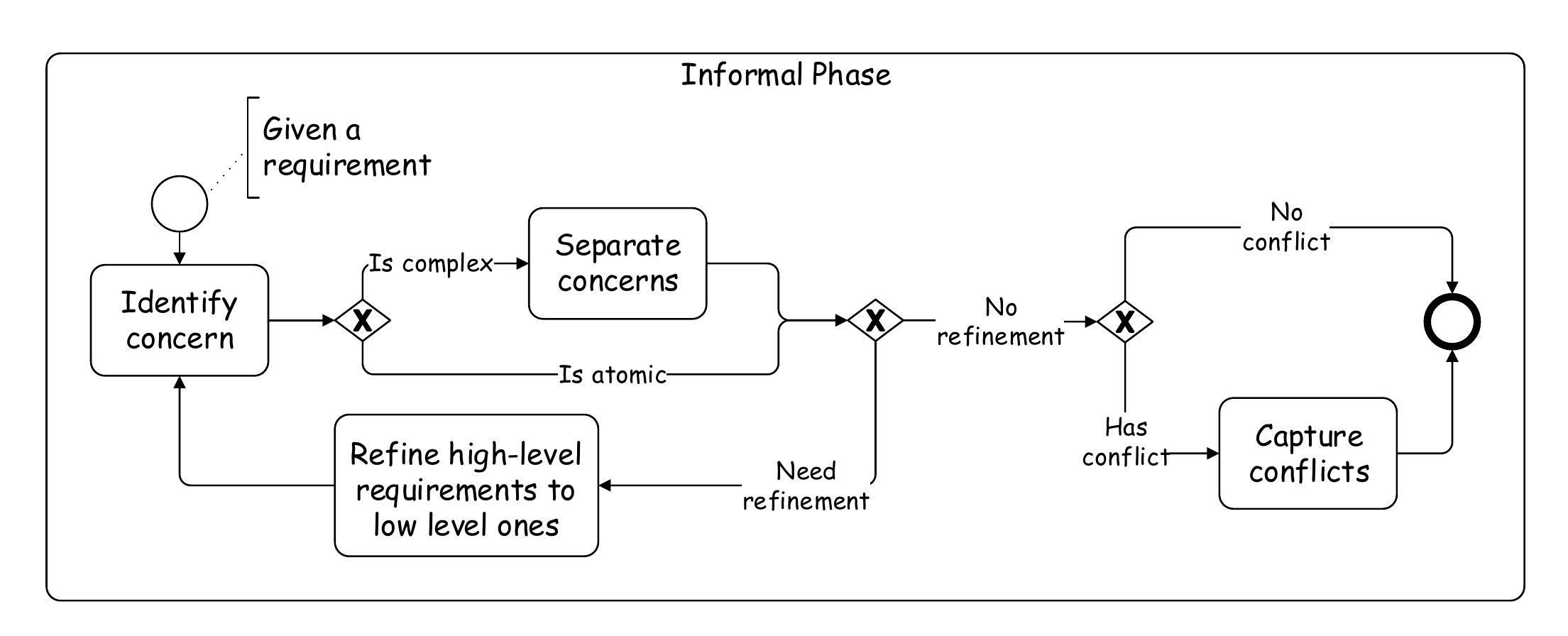}\\
  \vspace {-0.2 cm}
  \caption{The informal phase}\label{fig:desiree_method_phase1}
\end{figure}
%(adapted from~\cite{li_stakeholder_2015})

Step 1: \emph{Identify key concerns and determinate classifications}. We ask the question ``what does a goal (requirement) \emph{G} concern?'' to determine its classification, and provide some operational guidelines as follows.

%, which is more acceptable in practice
% Note that an FG differs from an FC in that an FG is operationalized as function(s) to be newly designed while an FC constrains certain function(s) already exist in a design.
%Implement certain extra mechanism in an already existing function.
\begin{itemize}
  \item If \emph{G} refers to several concerns, e.g., multiple functions, multiple qualities, or a mix of functions and qualities/content, then it is a \emph{composite} \textbf{goal} and needs to be decoupled. For example, ``\emph{the system shall be able to interface with most DBMSs}'' is composite since it refers to a function ``interface with'' and a universality quality ``most'' over the set of DBMSs.
  \item If \emph{G} requires an entity to be in a certain state, and will be fulfilled through one or more functions, then it is a functional goal \textbf{FG}. For example, ``\emph{order be paid}'' is an FG since it requires the entity ``order'' to be in a state, and could be operationalized as functions such as ``\emph{pay with credit card}'' and ``\emph{pay with debit card}''.
  \item If \emph{G} refers to only function(s), then it is a function \textbf{F}. For convenience, we treat a functional requirement as a function (F) rather than a functional goal (FG). For example, the requirement ``\emph{the system shall be able to send meeting notification}'', which states an implicit functional goal ``\emph{meeting notification be sent}'', will be preferably classified as a function.
  \item If \emph{G} constrains the situation of a function (e.g., actor, object, pre-condition, etc.), then it is an functional constraint \textbf{FC}. For example, ``\emph{the students added to a course shall be registered}''. Note that an FG differs from an FC in that an FG is operationalized as function(s) to be newly designed (or, alternatively, represents a desired state that will be brought about through the manifestation of a function) while an FC constrains certain function(s) already exist in a design (or, alternatively, constrains a situation under which a function will be manifested).
  \item If \emph{G} refers to only quality(-ies), then it is a quality requirement (QR). Further, if it has vague success criteria, then it is a quality goal \textbf{QG}; if it has clear success criteria, then it is a quality constraint \textbf{QC}.
   \item If \emph{G} describes the attributes a real-world entity shall possess, then it is a content goal \textbf{CTG}. For example, ``\emph{a meeting room} (\emph{in the real world}) \emph{has room number, room type, capacity}''.
  \item If \emph{G} defines the columns a data table shall have or the attributes of an information entity shall exhibit, then it is a state constraint \textbf{SC}. For example, ``\emph{a meeting room record} (\emph{an information entity}) \emph{shall include room number, room type, capacity}''.
  \item If \emph{G} makes an assumption about the environment of a system or describes domain knowledge, then it is a domain assumption \textbf{DA}. For example, ``\emph{the product will be used in an office environment}'', or ``\emph{MVC is a design pattern}''.
\end{itemize}

Step 2: \emph{Separate Concerns.} In case a goal (requirement) \emph{G} is a combination of concerns, its concerns need to be separated by using the ``Reduce'' ($\bm{R_d}$) operator.

\begin{itemize}
  \item If \emph{G} is a mix of function and quality, it can be reduced into two goals $G_1$ and $G_2$, with $G_1$ concerning the function while $G_2$ concerning the quality. For example, the DBMS example shall be decomposed into $G_1$ ``\emph{the system shall be able to interface with DBMSs}'' and $G_2$ ``\emph{most of the DBMSs}''.
  \item If \emph{G} refers to sibling functions or qualities, it shall be separated such that each resulting goal concerns one function/quality. For example, ``\emph{the system shall allow entering, storing and modifying product formulas}'' shall be decomposed into $G_1$ ``\emph{the system shall allow entering} ...'', $G_2$ ``... \emph{storing} ...'', and $G_3$ ``... \emph{modifying} ...''.
  \item If \emph{G} refers to nested qualities, we decouple them starting from the innermost layer. For example, the goal ``\emph{at least 90\% of the tasks shall be completed within 5 seconds}'' can be decoupled into two goals: $G_1$ ``\emph{processing time within 5 seconds}'', and $G_2$ ``$G_1$ \emph{shall be fulfilled for more than 90\% of tasks}''.
      %These two goals will be interpreted as two quality constraints QCs in the later interpretation phase.
  \item If \emph{G} is a mix of function and content, we derive two goals: a goal that describes the content using corresponding attributes, and a goal that operates on the content. For example, to capture the requirement ``display date and time'', we will have $G_1$ that describes an entity ``\emph{calendar}'' using attributes date and time, and $G_2$ ``\emph{display calendar}''.
  \item If \emph{G} includes purposes or means, it shall be decomposed to different goals. For example, for ``the product shall create an exception log of product problems for analysis'', we will have a goal $G_1$ ``analyze product problems'' being refined to $G_2$ ``create an exception log''.
\end{itemize}

Step 3: \emph{Refine Requirements.} In this step, we refine high-level goals to low-level ones by utilizing the $\bm{R_d}$ operator. For example, the goal $G_0$ ``{trip be scheduled}'' can be reduced to $G_1$ ``{accommodation be booked}'' and $G_2$ ``{ticket be booked}'' (``$\bm{R_d}$ ($G_0$) = \{$G_1$, $G_2$\}''); At the same time, if conflicts are found, we capture the conflicting requirements by using the ``\emph{Conflict}'' relation.

%or specification elements
In this informal phase, the type of requirement concepts will not be changed since only the ``Reduce'' operator has been employed. That is, the resulting elements of step 2 and 3 are still natural language goals (i.e., have not been specialized into sub-kinds of goals) if the starting point is a goal. For each of the resulting goals, we just keep in mind its classification, and will classify and structure it later in the interpretation phase.

For convenience, we also allow starting with specific kinds. For example, for a given requirement ``notify users with email'', an experienced analyst can directly capture it as a function ``$F_1$ := notify users with email'' or even ``$F_2$ := Notify $<$object: User$>$$<$means: Email$>$ ($F_1$ can be encoded as $F_2$ later in the interpretation phase), instead of merely modeling it as an early goal ``$G_1$ := notify users with email''. In this case, the resulting elements of the informal phase will be of specific kinds of the \emph{Desiree} elements.
\subsection{Interpretation Phase}
\label{sec:desiree_interpretation_phase}
%\emph{\textbf{Interpretation Phase.}}
%(9 in total, including Goal)
In this phase we encode each goal in accordance with its classification. If requirements issues (e.g., ambiguity) are identified, we need to communicate with stakeholders to elicit necessary information for resolution. This phase is captured in Fig.~\ref{fig:desiree_method_phase2}.

\begin{figure}[!htbp]
  \centering
  \vspace {-0.3 cm}
  \includegraphics[width=0.75\textwidth]{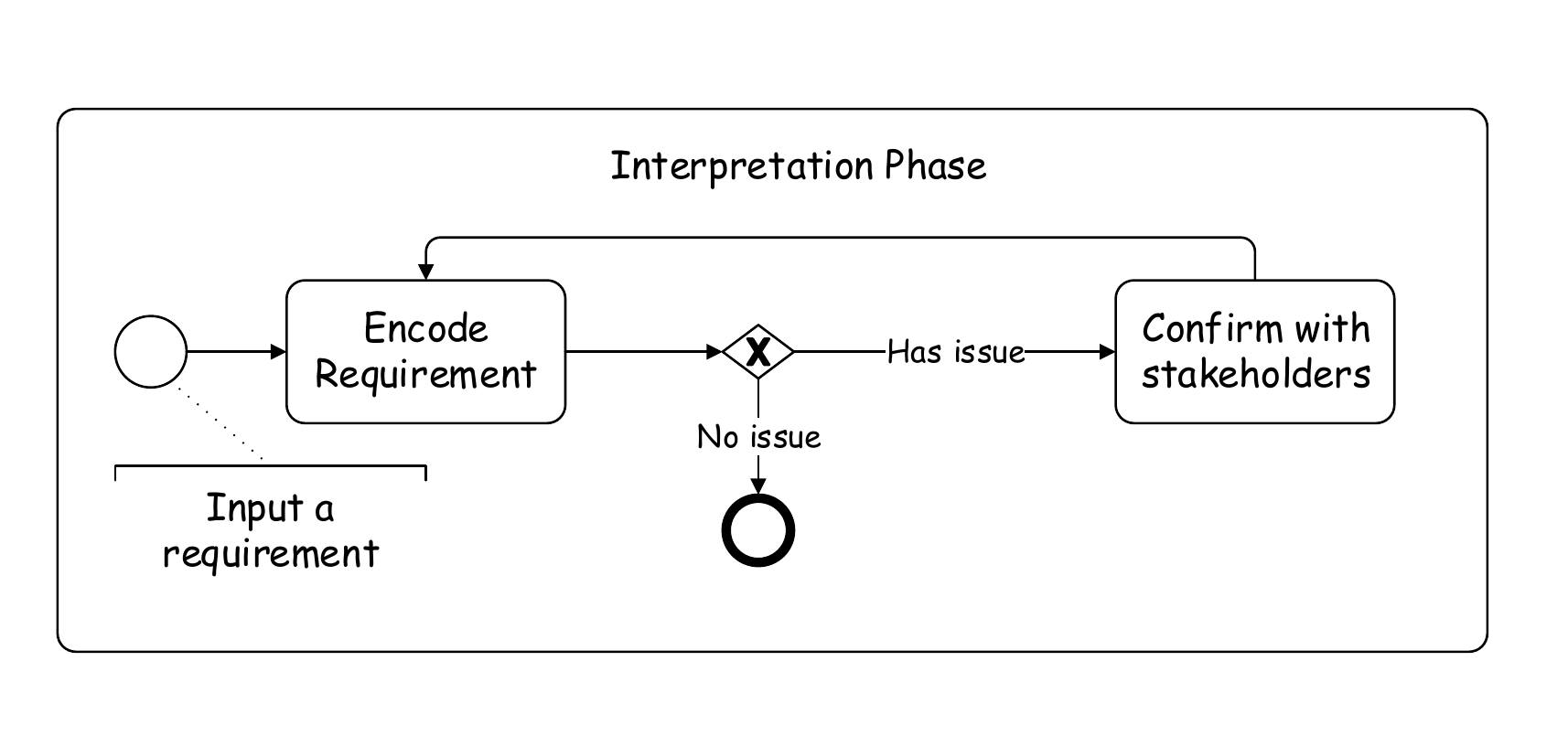}\\
  \vspace {-0.6 cm}
  \caption{The interpretation phase}\label{fig:desiree_method_phase2}
\end{figure}

In general, the syntactic forms for the 8 kinds of concepts (FG, F, FC, QG, QC, CTG, SC, and DA) can be classified into three categories: (1) ``FName $<s: D>^*$'' (F), where ``FName'' is the function name, ``$<s: D>^*$'' indicates zero or more slot-description pairs; (2) ``Q (SubjT) :: QRG'' (QG/QC), where ``Q'' is a quality type, ``SubjT'' is the subject type that a quality type refers to, and ``QRG'' is the desired quality region; (3) ``Subsumee $:<$ Subsumer'' (FG, FC, CTG, SC, DA), where ``Subsumee'' refers to a subject that is of concern, and ``Subsumer'' is a desired state, situation or property.

%In the discussion below we focus on FGs, QGs and CTGs. As for other elements such as FCs and DAs, readers can refer to our syntax introduction in Section~\ref{sec:concepts}.

\begin{itemize}
  \item \emph{Functions}. To specify a function, we often need to find out its actor, object, and sometimes its target, pre-, post- and trigger conditions. For example, ``\emph{the product shall be able to send meeting notifications}'' can be captured as ``$F_1$ := Send $<$subject: the\_system$>$$<$object: $\geq 1$ Meeting\_notification$>$''. Note that this requirement is incomplete since necessary information like ``who will send'' is missing, and will be further refined in the later smithing phase.
  \item \emph{Quality Goals and Quality Constraints}. The three key elements of a QG/QC include quality, subject and desired quality region. Note that the subject can be either a bare function/entity or a complex description. For example, for ``\emph{90\% of the maintainers shall be able to integrate new functionality into the product in 2 work days'}', there are two qualities: ``operating time'' for an integration process and ``universality'' for the set of maintainers. We thus define two QCs: ``$QC_1$ := Operating\_time (Integrate $<$actor: Maintainer$>$ $<$object: New\_functionality$>$ $<$target: \{the\_product\}$>$) :: $\leq$ 2 (work\_day)'', and ``$QC_2$ := $\bm{U}$ (?M, $QC_1$, $<$inhere\_in: Integrate $<$actor: ?M$>>$, 90\%)''.
  \item \emph{Functional Goals and Functional Constraints}. For an FG, we need to identify the entity that is of concern and the desired state. For example, for the functional requirement `\emph{`The product shall have a customizable look and feel}'', the concerned entity is ``the interface'', and the desired state is ``customizable''. This requirement can be accordingly captured as ``$FG_1$ := \{the\_interface\} $:<$ Customizable''. An FC can be structured in a similar way. For example, for ``\emph{when a conference room is reserved, the scheduler shall be updated}'', we can write ``$FC_2$ := Update $<$object: Scheduler$>$ $:<$ $<$trigger: Reserve $<$object: Room$>$$>$''. That is, by ``$FC_2$'', we constrain the ``Update'' function to be in a state ``having room reservations as its trigger''.
    %\item \emph{Functional Constraints}. To encode an FC, we need to identify the situation of a function that is to be constrained, and the desired property. For example, for ``when a conference room is reserved, the scheduler shall be updated'', we can write ``$FC_0$ := Update $<$object: Scheduler$>$ $:<$ $<$trigger: Reserve $<$object: Room$>$$>$''. That is, by ``$FC_0$'', we constrain the ``Update'' function to be in a state ``having room reservations as the trigger''.
  \item \emph{Content Goals and State Constraints}. CTGs and SCs share a similar syntax. For example, ``\emph{the system shall display date and time}'' will be captured as a CTG ``Calendar :$<$ $<$has\_date: Date$>$ $<$has\_time: Time$>$'' and an F ``Display $<$actor: \{the\_system\}$>$ $<$object: Calendar$>$''.
  \item \emph{Domain Assumption}. A DA can also be specified using description subsumption, like that for an FC. For example, ``\emph{the system will be used in an Windows environment}'' can be captured as ``\{the\_system\} $:<$ $<$has\_environment: Windows$>$''.
\end{itemize}
%The difference between an DA and a FC is that the former assumes that an entity have certain properties while the latter requires an entity to have certain properties.

%The interpretation process facilitates the detection and resolution of ambiguity: if there is more than one way to encode a requirement, then there is ambiguity. For example, ``notify users with email'' is ambiguous since it can be mapped into ``Notify $<$object: User$>$ $<$means: Email$>$'' or ``Notify $<$object: User $<$has\_email: Email$>$''. In such situation, stakeholders have to identify the intended meaning(s).

%The interpretation process facilitates the detection and resolution of ambiguity: if there is more than one way to encode a requirement, then there is ambiguity. For example, ``notify users with email'' is ambiguous since it can be mapped into ``Notify $<$object: User$>$ $<$means: Email$>$'' or ``Notify $<$object: User $<$has\_email: Email$>$''. In such situation, stakeholders have to identify the intended meaning(s).

The interpretation process facilitates the identification and resolution of requirements issues such as incompleteness, ambiguity and unverifiability because it drives analysts/ engineers to think about the properties (slots) of a function capability (e.g., actor, object, means, etc.), the cardinality of the description of a slot, the quantification of a vague quality region, etc. For example, a trained analyst/engineer could find that ``\emph{download contact information for client}'' is ambiguous since it can be mapped into ``Download $<$object: Client\_contact\_information $>$'' or ``Download $<$object: Contact\_information$>$ $<$beneficiary: Client$>$''. In such situation, stakeholders have to identify the intended meaning(s). We refer interested readers to our evaluation in Section~\ref{sec:eval_framework} for more detail.

%We suggest capture a requirement as it is in this phase, and refine it (e.g., add missing information) in the later simithing phase if needed.

\subsection{Smithing Phase}
\label{sec:desiree_smithing_phase}
%\emph{\textbf{Interpretation Phase.}}
%a complete (enough), unambiguous, measurable, satisfiable, and consistent
In this phase, structured goals (e.g., FG, QG, CTG) and specification elements (e.g., F, QC) are incrementally refined (structured goals are operationalized at proper time) using the requirements operators, to derive an eligible requirements specifications. During the process, if requirements issues are identified, stakeholders need to participate in and provide necessary information for resolving the identified issues.
%In this phase, structured goals (e.g., FG, QG, CTG) and specification elements (e.g., F, QC) are incrementally refined (structured goals are operationalized at proper time) using the requirements operators, to derive an eligible requirements specifications.

%In general, the requirements operators will be applied according to the properties of each requirement such as kind (e.g., FG, F, QG, QC, etc.), subjectivity and vagueness.
We sketch the guidelines for applying the operators in Fig.~\ref{fig:desiree_method_phase3}. For each input requirement, we can handle it by following the corresponding paths described therein. For example, for an FG, we first check whether it needs to be reduced; if not, we can directly operationalize it as Fs, FCs, DAs or combinations thereof. If these derived specification elements do not need further reduce, and no conflict has been identified when adding them to the target specification, then the smithing phase for this FG will end.

There are two points to be noted. First, a box with a circular arrow indicates a loop. That is, the ``Reduce'' and ``Focus'' operator can be used iteratively. For example, we can reduce a function $F_0$ to $F_1$, and then $F_1$ to $F_2$, so on and so forth. Second, all the operators shown in Fig.~\ref{fig:desiree_method_phase3} can be applied to the same requirement more than one time. For example, we can de-Universalize 100\% (for a QG) to 85\%, and 100\% (for the same QG) to 80\%, but not from 85\% to 80\%. That is, ``de-Universalize'' can not be iteratively applied to the same requirement (hence not represented as a loop task in Fig.~\ref{fig:desiree_method_phase3}), preventing the derivation of trivial requirements after some iterations.

%Once goals have been structured, we iteratively apply refinement operators ``Reduce", ``Focus'', relax (``de-Universalize'' or ``scale''), operationalization (``operationalize'' or ``observe''), ``resolve'', to derive complete (enough), unambiguous, measurable, satisfiable, and consistent requirements specifications:

\begin{figure}[!htbp]
  \centering
  %\vspace {-0.3 cm}
  \includegraphics[width=\textwidth]{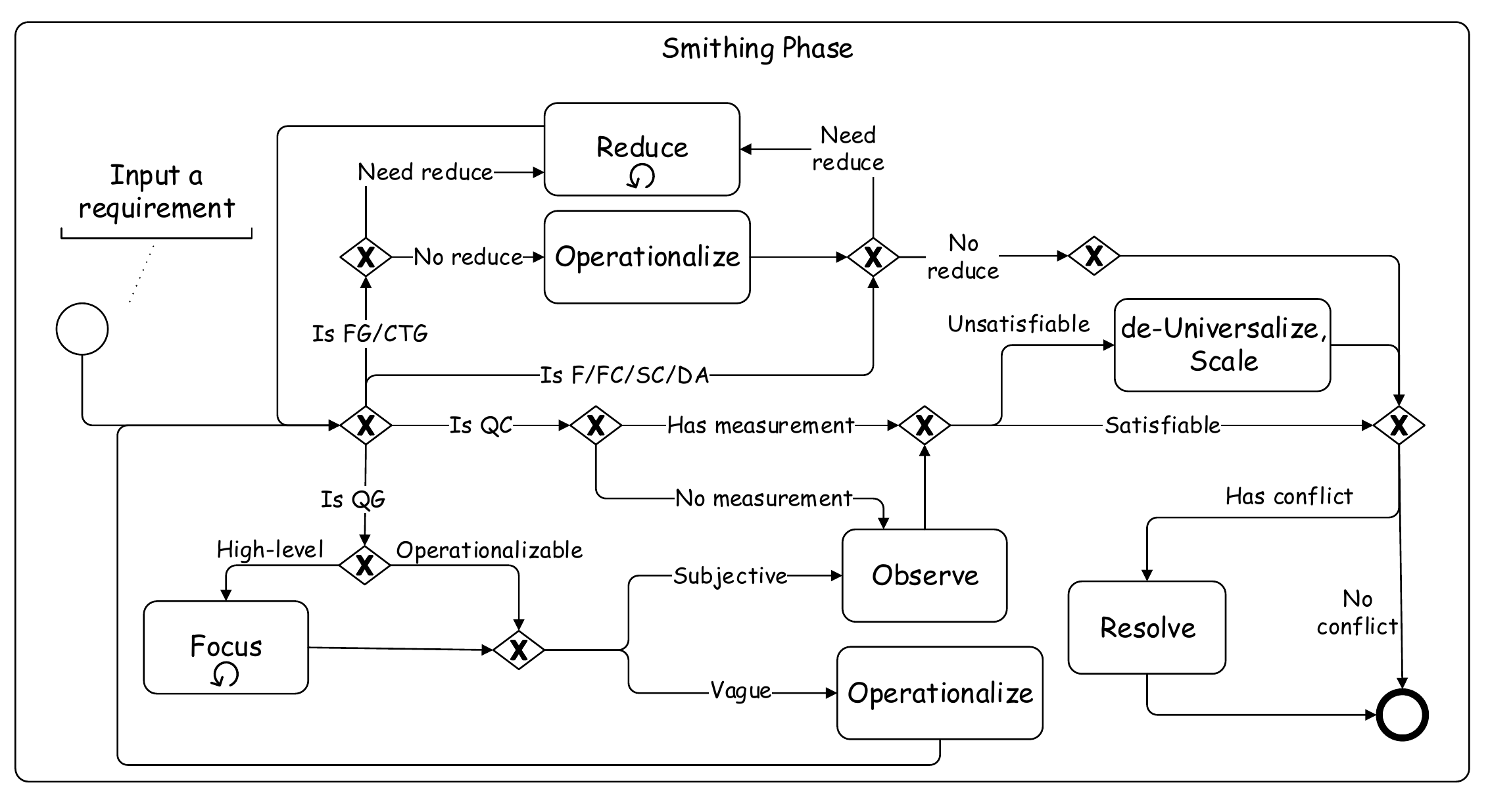}\\
  \vspace {-0.3 cm}
  \caption{The smithing phase}\label{fig:desiree_method_phase3}
\end{figure}

In the rest of this sub-section, we discuss in detail these operational guidelines according to requirements kinds.

\vspace{12pt}
\noindent {\textbf{Functional goals, functions and functional constraints.}} FGs can be refined by using ``$\bm{R_d}$'', and finally operationalized as Fs, FCs, DAs, or combinations thereof with ``$\bm{O_p}$''. Our understanding of manifesting events of functions as polygenic enables us to systematically operationalize FGs. Take the example of ``\emph{the system shall be able to notify realtors with a notification message}'', which implies an functional goal ``$FG_0$ := Realtor $:<$ Notified'' (i.e., realtors be notified). What kind of effect is required to satisfy $FG_0$? Is it the case that $FG_0$ is satisfied by merely a message being sent by the system? Or, alternatively, does this goal also require the message to be properly received by realtors? In the former case, we only need to design a function ``$F_1$ := Send $<$subject: \{the\_system\}$>$ $<$object: Notification\_message$>$ $<$target: Realtor$>$'' and simply assume certain capacities on receiving by adding an assumption ``$DA_2$ := Participant $:<$ $<$has\_capability: Receive $<$ object: Notification\_message$>>$'' (denoted as ``$\bm{O_p}$ ($FG_0$) = \{$F_1$, $DA_2$\}''). In the latter case, we should design both the sending and receiving functions.

%Note that if we treat this example requirement as a function ``$F_0$ := Send $<$actor: \{the\_system\}$>$ $<$object: Message$>$ $<$target: Realtor$>$'', the refinement is similar. The only difference is that we need to reduce rather than operationalize this function $F_0$.

When functional goals are operationalized as functions and/or functional constraints, the derived specification elements can be further refined by using ``$\bm{R_d}$''. In general, a function can be reduced by refining a slot, the description of a slot, or adding/removing a ``$<s:D>$'' pair; a functional constraint can be reduced according to domain knowledge. Note that an application of $\bm{R_d}$ can be either an AND-refinement or an OR-refinement. For example, the function $F_1$ ``Book $<$object: Ticket$>$'' can be AND-reduced to $F_2$ ``$<$object: Airline\_ticket$>$'' and $F_3$ ``$<$object: Train \_ticket$>$'' (denoted as ``${R_d}$ ($F_1$) = \{$F_2$, $F_3$\}''); alternatively, $F_1$ can be reduced to $F_2$ or be reduced to $F_3$ (denoted as ${R_d}$ ($F_1$) = \{$F_2$\} and ${R_d}$ ($F_1$) = \{$F_3$\}). These two refinements are interpreted differently: the former means that both airline and train ticket are needed (in a trip) while the later indicates that only one kind of ticket is needed in a trip.

\vspace{12pt}
\noindent {\textbf{Quality goals and quality constraints.}} QGs can be refined using ``$\bm{F_k}$'', operationalized as QCs using ``$\bm{O_p}$''/``$\bm{O_b}$'', and as Fs and/or FCs using ``$\bm{O_p}$'', and relaxed using ``$\bm{U}$'' and ``$\bm{G_d}$'' (or strengthened with ``$\bm{G_u}$''). In theory, the operationalization and the relaxation/strengthen of a QG can be intertwined. For convenience, we suggest first operationalize a QG as QC(s) and then relax/strengthen the derived QC(s) if needed.

%According to our experience in dealing with requirements in practice

We focus QGs in two ways: via the quality type \emph{Q} or via the subject type \emph{SubjT}. For example, the QG ``Usability (\{the\_product\}) :: Good'' can be focused into ``Learnability (\{the\_product\}) :: Good'' and ``Operability (\{the\_product\}) :: Good'' by following the quality hierarchy in ISO/IEC 25010~\cite{iso/iec_25010_systems_2011}. These quality goals can be further refined along subject hierarchy. For example, a meeting scheduler often has functions like ``set up meeting'' and ``book meeting room'', so the quality ``Learnability'' can be further applied to these functions, obtaining QGs ``Learnability (\{Set\_up\_meeting\}) :: Easy'' and ``Learnability (\{Book\_conference\_room\}) :: Easy'' (the use of curly brackets means a function individual, not its executions).

%\cite{guizzardi_ontological_2014}
%We can then use mathematical techniques such as probability distribution or Collated Voronoi diagram~\cite{decock_what_2014}, using the obtained prototype values to derive corresponding regions~\cite{guizzardi_ontological_2014}.

When operationalizing QGs, vague by nature, to measurable QCs, we suggest using ``prototype values'' to help define quality regions. For example, to operationalize the QG ``\emph{the learning time of meeting scheduler shall be short}'', we first ask stakeholders ``\emph{how long is short}?'' Their answers provide prototype values. We can then use the ``graded membership'' theory discussed in~\ref{sec:interpret_satisficing_nfrs} and the obtained prototype values to derive corresponding regions (defined by graded membership functions).

In our \emph{Desiree} framework, a QG can also be operationalized as functions or functional constraints, through which it becomes implementable. For example, the QG ``Look\_and\_feel (\{the\_interface\}) :: Professional'' (the system shall have a professional interface) can be operationalized as function constraints such as ``Navigation $:<$ $<$means: Standard\_button$>$'' (the system shall use standard button for navigation) and ``Term $:<$ $<$defined\_in: Realtor\_community$>$'' (the system shall use terms that are defined in the realtor community).

%ask observers for judgements, and
We use the ``Observe'' operator to operationalize a subjective QG as QC(s). Using this operator, we delegate the quantifying and measuring process to observers. For example, the subjective QG ``$QG_1$ := Style (\{the\_interface\}) :: Simple'' will be operationalized as a QC ``$QC_2$ := Style (\{the\_interface\}) :: Simple $<$observed\_by: Surveyed\_user$>$'' by employing surveyed users as observers: ``$\bm{O_b}$ ($QG_1$, Surveyed\_user)''.

Note that when the vague region of a QG is made measurable by specifying a QC, the QC is still not operational until the measuring instrument is specified. For example, if we have ``$QC_1$ := Processing\_time(File\_search) :: [0, 30 (\emph{Sec.})]'' we will need to specify a measuring device by applying the ``$\bm{O_b}$'' operator: $QC_2$ := $\bm{O_b}$ ($QC_1$, Stopwatch) = Processing\_time(File\_search) :: [0, 30 (\emph{Sec.})] $<$observed\_by: Stopwatch$>$.

%derived QC operational, we need to specify the measurement instruments for measuring its satisfaction. For example, suppose that we have derived ``$QC_1$ := Processing\_time(File\_search) :: [0, 30 (sec.)]'', we can assign an instrument by applying the ``$\bm{O_b}$'' operator: $QC_2$ := $\bm{O_b}$ ($QC_1$, Stopwatch) = Processing\_time(File\_search) :: [0, 30 (sec.)] $<$observed\_by: Stopwatch$>$.

%In the case that a QG/QC is practically un-satisfiable, we use the operators ``$\bm{U}$'', ``$\bm{G}$'', or a composition thereof to relax it to an acceptable degree. For instance, the requirement ``all the tasks shall be finished within 5 sec.'', captured as ``$QC_1$ := Processing\_time (Tasks) :: $\leq$ 5 (sec.)'', can be relaxed by using the ``$\bm{G_d}$'' operator: ``$QG_2$ := $\bm{G_d}$ ($QC_1$) :: Nearly'' (all the task shall be nearly within 5 sec.), or $\bm{U}$: ``$QC_3$ := $\bm{U}$ (?X, $QC_1$, $<$inheres\_in: ?X$>$, 90\%)'' (90\% of the tasks shall be within 5 sec.),  or even both ``$QG_4$ := $\bm{U}$ (?X, $QG_2$, $<$inheres\_in: ?X$>$, 90\%) :: 90\%'' (90\% of the tasks shall be nearly within 5 sec.).

In the case that a QG/QC is practically un-satisfiable, we use the operators ``$\bm{U}$'', ``$\bm{G}$'', or a composition thereof to relax it to an acceptable degree. For instance, the requirement ``\emph{all the tasks shall be finished within 5 seconds}'', captured as ``$QC_1$ := Processing\_time (Tasks) :: $\leq$ 5 (\emph{Sec.})'', can be relaxed by using the ``$\bm{G_d}$'' operator: ``$QC_2$ := $\bm{G_d}$ ($QC_1$, (1, 1.2))'' (all the task shall be in the region $[0, \; 5 \times 1.2 \; (Sec.)]$ = $[0, \; 6\;(Sec.)]$), or $\bm{U}$: ``$QC_3$ := $\bm{U}$ (?X, $QC_1$, $<$inheres\_in: ?X$>$, 90\%)'' (90\% of the tasks shall be within 5 seconds),  or even both ``$QC_4$ := $\bm{U}$ (?X, $QC_2$, $<$inheres\_in: ?X$>$, 90\%) :: 90\%'' (90\% of the tasks shall in the region $[0, \; 6\;(Sec.)]$). When relaxing a quality requirement, we suggest first operationalize it as a QC if it does not have a measurable region, and then apply the $\bm{U}$ operator to that QC.

%Note that $QG_4$ is still vague since the $\bm{U}$ operator is applied to a quality goal $QG_2$. To make it measurable, we suggest first operationalize

\vspace{12pt}
\noindent {\textbf{Content goals and state constraints.}} CTGs can be operationalized as SC(s) through the ``$\bm{O_p}$'' operator. When operationalizing CTGs, properties of real-world entities being characterized will be mapped to corresponding machine states, often data base schemas. For example, the CTG ``Student $:<$ $<$has\_id: String$>$ $<$has\_name: String$>$$<$has\_GPA: Float$>$'' will be operationalized as a SC ``Student\_record: $<$id: String$>$ $<$name: Varchar$>$ $<$GPA: Float$>$''. State constraints can be further reduced by using the ``$\bm{R_d}$'' operator, if needed.

\vspace{12pt}
\noindent {\textbf{Conflicts.}} Conflicts can be resolved by using the ``$\bm{R_s}$'' operator. We currently provide two mechanisms for resolving a conflict: (1) drop some of the conflicting requirements; (2) weaken some of the conflicting requirements, e.g., applying the ``$\bm{U}$'' or ``$\bm{G_d}$'' operator to QGs/QCs, or applying the ``$\bm{R_d}$'' operator to weaken a function. For example, as in Fig.~\ref{fig:desiree_reslove_conflict}, to resolve the conflict between $F_1$ and $QG_2$, we can either drop $F_1$ or weaken $QG_2$ to $QG_3$ (these two resolutions are represented by the two black thick bars).

\begin{figure}[!htbp]
  \centering
  \vspace {-0.3 cm}
  \includegraphics[width=0.8\textwidth]{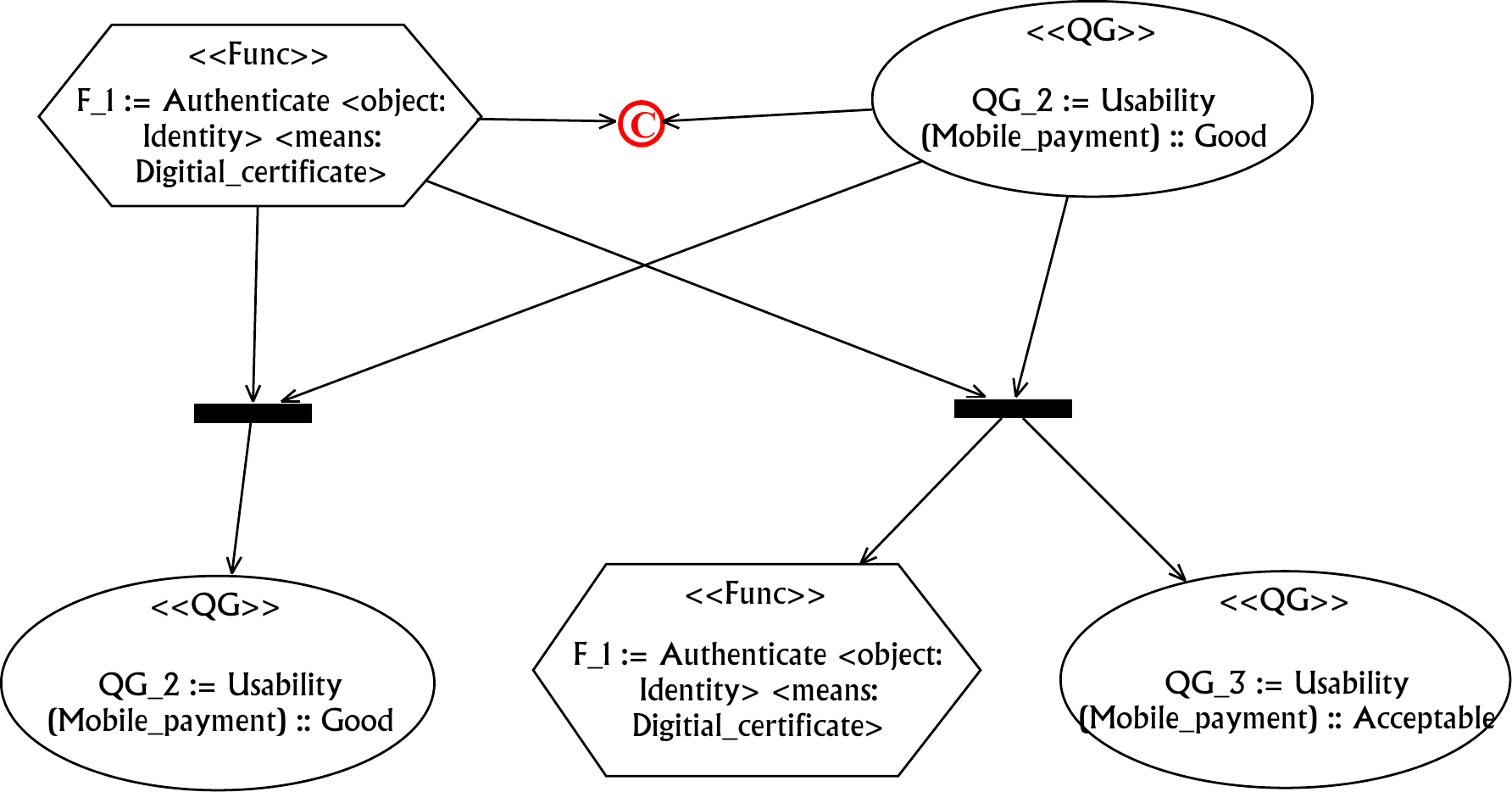}\\
  %\vspace {-0.5 cm}
  \caption{An example resolution of conflict}\label{fig:desiree_reslove_conflict}
\end{figure}

\section{Chapter Summary}
\label{sec:desiree_summary}
In this chapter, we have presented \emph{Desiree}, which includes a rich collection of requirements concepts, operators, a description-based syntax for representing these items, and a methodology for applying these concepts and operators, in order to transform stakeholder requirements into a an eligible (formal, valid, complete enough, unambiguous, consistent, measurable, and practically satisfiable, modifiable, traceable) requirement specification.

Our framework has introduced several new concepts, namely ``(composite) goal'', ``functional constraints'', ``content goal'' and ``state constraint'', that are not captured in traditional goal modeling techniques. Our framework also offers several useful operators, ``Observe'', ``de-Universalize'', ``Scale'', for operationalizing and weakening quality requirements. Our description-based syntax facilitates the identification of requirements issues such as ``ambiguity'', ``incompleteness'', ``un-satisfiable''. The framework will be evaluated through a series of experiments, as described in Chapter~\ref{cha:evaluation}.

%% file: semantics.tex
%\vspace{4cm}
\chapter{The Semantics of \emph{Desiree}}
\label{cha:desiree_semantics}
%We characterize the formal semantics of our language in this section.
In this chapter, we provide the formal semantics of our language and operators. The former will be done using set theory, while the later will transform or relate formulas. In addition, we capture part of the meaning of our language and operators by translation into Description Logic (DL)~\cite{baader_description_2003}, specifically, OWL 2 Web Ontology Language (OWL2)~\cite{group_owl_2009}, in order to obtain a decidable reasoning with requirements.

DL is a family of knowledge representation (KR) formalisms that first defines relevant concepts and relations (captured as roles, in our case, slots) of a domain, then use these concepts and relations to specify properties of individuals occurring in the domain~\cite{baader_description_2003}. In DL, both concepts and roles (properties in OWL2~\cite{group_owl_2009}) are first-class modeling elements, which can be either primitive or composite, and can be automatically classified in a subsumption hierarchy, checked for consistency, etc.

Choosing DL as our computational logic enables us to systematically manage the interrelations between requirements (especially those between FRs and NFRs), detect inconsistences in requirements, perform ``what-if'' analysis on requirements, etc. Moreover, we can make use of existing reasoners to perform these tasks, and do not need to code our own reasoning tool.

% ambiguities
% elicit implicit requirements,
%In this chapter, we provide the formal semantics of our description language and operators. The former will be done using set theory, while the later will transform or relate formulas.

\section{The Semantics of the \emph{Desiree} Language}
\label{sec:semantics_language}

%\noindent \textbf{Translation of Concepts.} The syntax of concepts is given in the second column of Table~\ref{tab:semantics_concepts}. As in Z~\cite{spivey_z_1992}, we start from \emph{atomic} sets/types, for which elements we do not know the structure. In our syntax, these correspond to \emph{ConceptNames}; slots are binary relations, which in this paper, for the sake of brevity, will be assumed to be (mapping) functions; \emph{ElementIdentifiers} are constants; $\bm{T}$ is a recursive function .
%When formalizing the semantics, as in Z~\cite{spivey_z_1992}, we start from \emph{atomic} sets/types, for which elements we do not know the structure.
%\emph{QualityNames},
\noindent \textbf{Translation of \emph{Desiree} descriptions.} We show the syntax of \emph{Desiree} descriptions in the third column of Table~\ref{tab:semantics_descriptions} (see Table~\ref{tab:appendix_desiree_syntax} in Appendix~\ref{cha:appendixes_syntax} for the full syntax). In the forth and fifth column, we give the the translation of \emph{Desiree} descriptions to set-theoretic and DL expressions, using the recursive function $\bm{T}$ and $\bm{L}$, respectively. When formalizing the semantics, as in Z~\cite{spivey_z_1992}, we start from \emph{atomic} sets/types, for which elements we may do not know the structure. In our syntax, these correspond to \emph{\_Names} (e.g., \emph{ConceptNames}, \emph{RegionNames}) and \emph{RegionExpressions} (i.e., intervals and enumeration values); \emph{slots} are binary relations (the inverse of a slot $s$, is denoted as $s^{-1}$); \emph{ElementIdentifiers} are constants. When translating into DL: \emph{Thing} is the universal concept, \emph{Nothing} is the bottom concept; elements of \emph{slots} are roles (properties); \emph{ElementIdentifiers} are individuals, \emph{RegionExpressions} (i.e., intervals and enumeration values) are DL data ranges, and all \emph{\_Names} are DL concepts.

\begin{table}[!htbp]
  \centering
  %\vspace {-0.3 cm}
  \begin{threeparttable}
  \caption {The semantics of \emph{Desiree} descriptions}
  \label{tab:semantics_descriptions}
  \vspace {0.3 cm}
  %\scriptsize
  \scriptsize
  \setlength\tabcolsep{2pt}
  %\begin{tabular}{|c|p{0.8\textwidth}|}
  \begin{tabular}{|c|c|c|c|c|}
  \hline
  \textbf{Id} & & \textbf{Description} \emph{D}  & $\bm{T}$(\emph{D}) & $\bm{L}$(\emph{D})\\ \hline

  1 & \multirow{8}{*}{{\rotatebox[origin=c]{90}{\emph{\textbf{SlotD}}}}}
  & $<s: D>$
  %& $\{x | \forall y. s(x, y) \rightarrow y \in \bm{T}(D) \land |s(x, \bm{T}(D))| = 1 \}$
  %& $ \forall s.\bm{L} (D) \; \sqcap = 1 \; s. \bm{L} (D)$
  & $\{x | \;  |s(x, \bm{T}(D))| = 1 \}$
  & $ 1 \; s. \bm{L} (D)$

  %&
  %\begin{tabular}{@{}c@{}}
  %$ \forall s.\bm{L} (D) \; \sqcap \exists \; s. \bm{L} (D)$ \\
  %\emph{s is a funcitonal property}
  %\end{tabular}
  \\ \cline{1-1}\cline{3-5}

  2 &
  & $<s: \; \leq n \; D>$
  %& $\{x | \forall y. s(x, y) \rightarrow y \in \bm{T}(D)  \land |s(x, \bm{T}(D))| \leq n \}$
  %& $\forall s.\bm{L} (D) \; \sqcap \leq n \; s. \bm{L} (D)$
  & $\{x | \; |s(x, \bm{T}(D))| \leq n \}$
  & $\leq n \; s. \bm{L} (D)$
  \\ \cline{1-1}\cline{3-5}

  3 &
  & $<s: \; \geq n \; D>$
  %& $\{x | \forall y. s(x, y) \rightarrow y \in \bm{T}(D)  \land |s(x, \bm{T}(D))| \geq n \}$
  %& $\forall s.\bm{L} (D) \; \sqcap \geq n \; s. \bm{L} (D)$
  & $\{x |\; |s(x, \bm{T}(D))| \geq n \}$
  & $\geq n \; s. \bm{L} (D)$
  \\ \cline{1-1}\cline{3-5}

  %4 &
  %& $<s: \; =n \;D>$
  %& $\{x | \forall y. s(x, y) \rightarrow y \in \bm{T}(D) \land |s(x, \bm{T}(D))| = n \}$
  %& $\forall s.\bm{L} (D) \; \sqcap = n \; s. \bm{L} (D)$
  %\\ \cline{1-1}\cline{3-5}

  4 &
  & $<s: n \; D>$
  %& $\{x | \forall y. s(x, y) \rightarrow y \in \bm{T}(D)  \land |s(x, \bm{T}(D))| = n \}$
  %& $ \forall s.\bm{L} (D) \; \sqcap = n \; s. \bm{L} (D)$
  & $\{x | \;  |s(x, \bm{T}(D))| = n \}$
  & $ n \; s. \bm{L} (D)$
  \\ \cline{1-1}\cline{3-5}

  5 &
  & $<s: {SOME} \; D>$
  %& $\{x | \forall y. s(x, y) \rightarrow y \in \bm{T}(D)  \land \exists y . s(x,y)\}$
  %& $ \forall s.\bm{L} (D) \; \sqcap \exists \; s. \bm{L} (D)$
  & $\{x | \exists y . s(x,y) \land y \in \bm{T}(D)\}$
  & $ \exists \; s. \bm{L} (D)$
  \\ \cline{1-1}\cline{3-5}

  %5 &
  %& $<s: {EXIST} \; D>$
  %& $\{x | \exists y . s(x,y) \land y \in \bm{T}(D)\}$
  %& $ \exists \; s. \bm{L} (D)$
  %\\ \cline{1-1}\cline{3-5}

  6 &
  & $<s: {ONLY} \; D>$
  & $\{x | \forall y. s(x, y) \rightarrow y \in \bm{T}(D)\}$
  & $ \forall s.\bm{L} (D) $
  \\ \cline{1-1}\cline{3-5}

  7 & &
  $SlotD_1 \; SlotD_2$
  & $\bm{T}(SlotD_1) \cap \bm{T} (SlotD_2)$
  & $\bm{L}(SlotD_1) \sqcap \bm{L} (SlotD_2)$
  \\ \cline{1-5} %\cline{3-5}

  8 & \multirow{8}{*}{{\rotatebox[origin=c]{90}{\emph{\textbf{Concept}}}}}
  & \emph{ConceptName}
  & \emph{ConceptName}
  & \emph{ConceptName}
  \\ \cline{1-1}\cline{3-5}

  %9 &
  %& $AtomicVal$
  %& $AtomicVal$
  %& $AtomicVal$
  %\\ \cline{1-1}\cline{3-5}

  9 &
  & $SlotD$
  & $\bm{T}(SlotD)$
  & $\bm{L}(SlotD)$
  \\ \cline{1-1}\cline{3-5}

  10 &
  & \{$ElemId_1$ ...\}
  & \{$ElemId_1$ ...\}
  & \{$ElemId_1\} \; \sqcup $ ...
  \\ \cline{1-1}\cline{3-5}

  11 &
  & $D.s$
  & $\{x | \exists y. s^{-1}(x, y) \land y \in \bm{T}(D) \}$
  & $\exists s^{-1}. \bm{L} (D)$
  \\ \cline{1-1}\cline{3-5}

  12 &
  & $D_1 \; D_2$
  & $\bm{T}(D_1) \cap \bm{T} (D_2)$
  & $\bm{L}(D_1) \sqcap \bm{L} (D_2)$
  \\ \cline{1-1}\cline{3-5}

  13 &
  & $D_1 \vee D_2$
  & $\bm{T} (D_1) \cup \bm{T} (D_2)$
  & $\bm{L} (D_1) \sqcup \bm{L} (D_2)$
  \\ \cline{1-1}\cline{3-5}

  14 &
  & $D_1 - D_2$
  & $\{x | x \in \bm{T} (D_1) \land x \notin \bm{T} (D_2)\}$
  & $\bm{L} (D_1) \sqcap \bm{L} (\neg D_2)$
  %\\ \cline{1-1}\cline{3-5}

  %15 &
  %& $D_1 :< D_2$
  %& $\forall x. x \in \bm{T} (D_1) \rightarrow x \in \bm{T} (D_2)$
  %& $\bm{L} (D_1) \sqsubseteq \bm{L} (D_2)$
  %\\ \cline{1-1}\cline{3-5}
  \\ \hline

  %16 &
  %& $RgExpr$
  %& $\bm{T}(RgExpr)$
  %& $\bm{L}(RgExpr)$
  %\\ \cline{1-1}\cline{3-5}
  %\\ \hline

  15 & \multirow{3}{*}{{\rotatebox[origin=c]{90}{\emph{\textbf{RgExpr}}}}}
  & $RegionName$
  & $RegionName$
  & $RegionName$
  \\ \cline{1-1}\cline{3-5}

  16 &
  & $[AtomicVal_1, AtomicVal_2]$
  & $\{x|AtomicVal_1 \leq x \leq AtomicVal_2\}$
  & $\geq AtomicVal_1 \; \sqcap \; \leq AtomicVal_2$
  \\ \cline{1-1}\cline{3-5}

  17 &
  & \{$AtomicVal_1$ ...\}
  & \{$AtomicVal_1$ ...\}
  & \{$AtomicVal_1\} \; \sqcup $ ...
  \\ \hline
  \end{tabular}
  \begin{tablenotes}
      \scriptsize
      \item $s$: Slot; $D$: Description; Concept: \emph{Desiree} concept; RgExpr: region expression; $|M|$ denotes the cardinality of the set M, $s(x, M) := \{y | (x, y) \in s \land y \in M\}$, and the inverse of $s$, $s^{-1} := \{(x, y)| (y, x) \in s\}$.
    \end{tablenotes}
  \end{threeparttable}
\end{table}

% in Table~\ref{tab:semantics_descriptions}
%That is, if not explicitly specified, a slot is defaulted as a mapping function (i.e., has only one instance of its description).
%``$=n$'', ``SOME'',
%That is, if not explicitly specified, a slot is defaulted as a total (contra partial) mapping function.
We start with 7 basic rules for translating slot-description pairs (\emph{\textbf{SlotDs}}), the key constructor of our language (rule $1 \sim 7$). By default, a slot relates an individual to one instance that is of and only of type \emph{\textbf{D}}, a description that will be further defined (rule 1). Also, a \emph{SlotD} could have modifiers such as ``$\leq n$'',``$\geq n$'', ``$n$'' (``$n$'' is a positive integer), ``SOME'', or ``ONLY'' constraining its description (rule $2 \sim 6$). For instance, ``$<$register\_for:  $\geq$3 Class$>$'' represents a set of individuals that have registered for at least three classes, and will be translated into the DL expression ``$\geq$3 register\_for.Class''. Two adjacent \emph{SlotDs} will be translated into their intersection (rule 7).

%For instance, ``$<$register\_for:  $\geq$3 Class$>$'' represents a set of individuals that register for only classes and have registered for at least three classes, and will be translated into the DL expression ``($\forall$ register\_for.Class) $\sqcap$ ($\geq$3 register\_for.Class)''.

%requiring its value to belong to a nested concept
%(rule $8 \sim 10$)
As shown in Table~\ref{tab:semantics_descriptions}, a description \emph{\textbf{D}} can be defined in various kinds of ways, such as an atomic concept name (denoting a set, e.g., ``Student\_record''), a slot-description (requiring its value to belong to a nested description, e.g., ``$<register\_for:  \; \geq3 \; Class>$'', as we have shown before), or a set of individuals (e.g., the description ``Interoperable\_DBMS'' can be represented by a set of individuals ``\{MySQL, Oracle, MsSQL\}''); it can also be built using set constructors: intersection (represented by sequencing, e.g., ``Student $<$gender: Male$>$'', which captures ``male students''), union, and set difference (rule $12 \sim 14$). The expression \emph{D.s} (rule 11) is useful for describing the set of individuals related to elements of \emph{D} by \emph{s}. For instance, to capture ``\emph{the collected traffic info shall be in real time}'', we at first define a function ``$F_1$ := Collect $<$object: Traffic\_info$>$'', and then use ``$F_1$.object'', which is translated in to ``$\exists$ $object^{-1}$. $F_1$'' ($object^{-1}$ is the inverse of $object$), to refer to collected, instead of all, traffic info.

%the inverse of

In addition, a description \emph{\textbf{D}} can be a region expression (rule $15 \sim 17$), which can be a region name (e.g., ``\emph{low}'', ``\emph{fast}''), a mathematical region expression (e.g., ``$\geq$20'', as in ``Student $<$gender: Male$>$ $<$age: $\geq$20$>$'', which captures ``\emph{male students older than 20}''), or a set of enumeration values (e.g., ``Student $<$gender: Male$>$$<$age: \{20\}$>$'', which captures ``\emph{male students who are 20-years-old}'').

%``Color (Button) :: \{Light\_gray, Light\_black\}'', which requires the color of buttons to be light gray or light black

%``$C :< D$''
%``$\forall$\emph{x}. \emph{x} $\in$ \emph{C} $\rightarrow$ \emph{x} $\in$ \emph{D}''
%The translation of such formulae to DL axioms can be done according to rule 15 in Table~\ref{tab:semantics_descriptions}.

%\noindent \emph{\textbf{Translation of requirements concepts.}} In general, a goal G will denote a formula with a truth value. Elements of G that are not in any subclass correspond to nullary predicates or propositions. Two subclasses of G, namely FG and CTG, are formulae of the form as $Desc$ in Eq.~\ref{eq:eq_semantics_subsumption}, where \emph{C} and \emph{D} are concepts defined using our language. The same kinds of formulae are used to express FCs, SCs, and DAs. The semantics of such formulae are closed formulae as $AX_{Desc}$ in Eq.~\ref{eq:eq_semantics_subsumption}, and the translation of such formulae to DL axioms can be done according to $DLAxiom_{Desc}$ in Eq.~\ref{eq:eq_semantics_subsumption}.

\vspace{12pt}
\noindent \emph{\textbf{Translation of requirements concepts.}} In general, stakeholder requirements are initially stated using natural language (NL). These early requirements are mere atomic descriptions~\footnote{In our framework, each kind of the requirements concepts, Goal, FG, CTG, QG, F, FC, QC, SC, or DA, can be initially stated in natural language and then be structured using the description-based syntax with the ``Interpret'' operator.}. The instances of an early requirement are all the corresponding problem situations (i.e., all the cases when the corresponding intention is instantiated). For example, for the requirement ``schedule a meeting'', the instances are all the specific requests for a meeting. Further, such requirements can be organized into a subsumption hierarchy by using ``$E_1 :< E_2$'' statements. For example, given $G_1$ ``download contact information for clients'' and $G_2$ ``download clients' contact information'', we can organize them by simply denoting ``$G_2 :< G_1$'' (here $G_1$ is ambiguous since it has another interpretation ``download contact information in the sake of clients'', hence we have only ``$G_2 :< G_1$'' but not the reverse).

%These early requirements are refined into elements with a corresponding structured syntax in our \emph{Desiree} framework (we call them ``early requirements'' or ``stakeholder requirements''  because they are what we start with).
%while later ones, derived through refinements, are structured ones having a more elaborate subsumption relation.
%for the two NL functions

%Two subclasses of G, namely FG and CTG, are formulae of the form as $Desc$ in Eq.~\ref{eq:eq_semantics_subsumption}, where \emph{C} and \emph{D} are concepts defined using our language. The same kinds of formulae are used to express FCs, SCs, and DAs. The semantics of such formulae are closed formulae as $AX_{Desc}$ in Eq.~\ref{eq:eq_semantics_subsumption}, and the translation of such formulae to DL axioms can be done according to $DLAxiom_{Desc}$ in Eq.~\ref{eq:eq_semantics_subsumption}.
%The same kinds of formulae are used to express FCs, SCs, and DAs.
%\emph{Desiree} elements, derived through refinements, are structurally specified.
In our \emph{Desiree} framework, stakeholder requirements are refined into \emph{Desiree} elements with a corresponding structured syntax. Especially, FGs, CTGs, FCs, SCs, and DAs are formulae of the form as $Desc$ in Eq.~\ref{eq:eq_semantics_subsumption}, where \emph{C} and \emph{D} are concepts defined using our language. The semantics of such formulae are closed formulae as $AX_{Desc}$ in Eq.~\ref{eq:eq_semantics_subsumption}, and the translation of such formulae to DL axioms can be done according to $DLAxiom_{Desc}$ in Eq.~\ref{eq:eq_semantics_subsumption}. For example, the semantics of the FC ``$FC_1$ := Data\_table $:<$ $<$accessed\_by: ONLY Manager$>$'' will then emerge as ``$\forall$\emph{x}/Data\_table $\forall$\emph{y} accessed\_by (\emph{x}, \emph{y}) $\rightarrow$ \emph{y} $\in$ Manager'', where ``$\forall x/C$'' is a shorthand for ``$\forall x.x \in C$''. Further, this FC will be translated to the DL axiom ``Data\_table $\sqsubseteq$ $\forall$ accessed\_by.Manager''.\\
\begin{equation}\label{eq:eq_semantics_subsumption}
    \begin{aligned}
        Desc := \; & C :< D \\
        AX_{Desc} \equiv \; & \forall x.  x \in  \bm{T}(C) \rightarrow  x \in  \bm{T}(D) \\
        DLAxiom_{Desc} \equiv \; & \bm{L} (C) \sqsubseteq \bm{L} (D) \\
    \end{aligned}
\end{equation}
%DLConcept_{Desc} \equiv \; & \exists subsumee.\bm{L} (C) \sqcap \exists subsumer.\bm{L} (D)

In addition, we translate formulae of the syntactic form ``C $:<$ D'' in to DL concepts as $DLConcept_{Desc}$ in Eq.~\ref{eq:eq_subsumption_concept}, where ``subsumee'' and ``subsumer'' are reserved object properties that have ``$\bm{L} (C)$'' and ``$\bm{L} (D)$'' as ranges (fillers), respectively. For example, we translate the above $FC_1$ into a DL concept ``$\exists$ subsumee.Data\_table $\sqcap$ $\exists$ subsumer.($\forall$ accessed\_by.Manager)''.
\begin{equation}\label{eq:eq_subsumption_concept}
    \begin{aligned}
        DLConcept_{Desc} \equiv \; & \exists \; subsumee.\bm{L} (C) \sqcap \exists \; subsumer.\bm{L} (D)
    \end{aligned}
\end{equation}

%At the same time, we associate with this subsumption axiom a DL concept ``$\exists$ subsumee.Data\_table $\sqcap$ subsumer.($\forall$ accessed\_by.Manager)''. In the rest of this section,  we focus on the semantics of the other three kinds of requirement concepts, namely Functions, QGs and QCs.
This is mainly for representing subsumption relations between elements in the form of ``$C :< D$''. For example, ``$FC_1 := A :< B$'' and ``$FC_2 := C :< D$'' can be organized in a subsumption hierarchy if they can be translated into DL concepts as in Eq.~\ref{eq:eq_subsumption_concept}, e.g., $FC_2 :< FC_1$, $FC_1 :< FC_2$, or neither. If $FC_1$ and $FC_2$ are in their original forms, which represent assertions, then they can not be organized in such a hierarchy.

In the rest of this section,  we focus on the semantics of the other three kinds of requirement concepts, namely Functions, QGs and QCs.

\vspace{12pt}
\noindent \emph{\textbf{Translation of Functions.}} In \emph{Desiree}, a function description consists of a function name and a list of optional slot-description pairs (\emph{SlotDs}). Ontologically, it represents a set of manifestations (i.e., runs) of a capability. For example, the expression ``$F_1$ := Activate $<$actor: Manager$><$object: Debit\_card$>$'' says that each run of $F_1$ is an activation with a manager as its actor and a debit card as its object. Our description translation gives the semantics of $F_1$ as $AX_{F_1}$ in Eq.~\ref{eq:eq_semantics_f_example_true}, where ``$||$'' denotes the cardinality of a set, and ``$s(x, C)$'' represents a set ``$\{y | (x, y) \in s \land y \in C\}$''.
\begin{comment}
\begin{equation}\label{eq:eq_semantics_f_example_true}
    %\small
    \begin{aligned}
    AX_{F_1} \equiv \:  & \forall x.F_1 (x) \rightarrow Activate (x)   \\
                    & \land (\forall y. actor (x, y) \rightarrow y \in Manager) \land |actor(x, Manager)| = 1 \\
                    & \land (\forall y . object (x, y) \rightarrow y \in Debit\_card) \land |object(x, Debit\_card)|=1\\
    \end{aligned}
\end{equation}
\end{comment}
\begin{equation}\label{eq:eq_semantics_f_example_true}
    %\small
    \begin{aligned}
    AX_{F_1} \equiv \:  & \forall x.F_1 (x) \rightarrow Activate (x)   \\
                    & \land |actor(x, Manager)| = 1 \land |object(x, Debit\_card)|=1\\
    \end{aligned}
\end{equation}

%``$F_{2}$ := Send $<$object: $\geq 1$ Notification$>$'', the ``object'' slot relates an execution of $F_{2}$ to a set of notifications ($\geq 1$).
Note that we can not define the semantics of $F_1$ as $AX_{F_1}'$ in Eq.~\ref{eq:eq_semantics_f_example_false}, and restrict ``actor'' and ``object'' to be global functional slots (i.e., having only one instance of its description). This is because these slots may have multiple instances of their descriptions in other requirements in the same specification. For example, in the same specification, there could be another function ``$F_2$ := Activate $<$actor: Manager$>$$<$object: $\geq 1$ Debit\_card$>$'', where the ``object'' slot relates an execution of $F_2$ to a set of ($\geq 1$) debit cards. That is, $F_2$ allows a manager to activate debit cards in a batch mode.\\
\begin{comment}
\begin{equation}\label{eq:eq_semantics_f_example_false}
    %\small
    \begin{aligned}
    AX_{F_1}' \equiv \:  & \forall x.F_1 (x) \rightarrow Activate (x)   \\
                    & \land (\forall y. actor (x, y) \rightarrow y \in Manager) \land \exists y.actor(x,y) \\
                    & \land (\forall y . object (x, y) \rightarrow y \in Debit\_card) \land \exists y.object(x,y)\\
    \end{aligned}
\end{equation}
\end{comment}
\begin{equation}\label{eq:eq_semantics_f_example_false}
    %\small
    \begin{aligned}
    AX_{F_1}' \equiv \:  & \forall x.F_1 (x) \rightarrow Activate (x)   \\
                    & \land \exists y.actor(x,y) \land \exists y.object(x,y)\\
    \end{aligned}
\end{equation}

%(``$s$'' is a functional slot)
In general, the semantics of a function specified as ``$F_e$ := \emph{FName} $<s: D>$'' can be generalized as in Eq.~\ref{eq:eq_semantics_f_general}.
\begin{comment}
\begin{equation}\label{eq:eq_semantics_f_general}
    %\small
    \begin{aligned}
    AX_{F_e} \equiv \;  & \forall x/F_e \rightarrow FName(x)\\
                        & \land (\forall y.s(x,y) \rightarrow y \in \bm{T}(D)) \land |s(x,\bm{T}(D))|=1
    \end{aligned}
\end{equation}
\end{comment}
\begin{equation}\label{eq:eq_semantics_f_general}
    %\small
    \begin{aligned}
    AX_{F_e} \equiv \;  & \forall x/F_e \rightarrow FName(x) \land |s(x,\bm{T}(D))|=1
    \end{aligned}
\end{equation}
%\land |s(x,\bm{T}(D))|=1
%\exists y.s(x,y)
%which has a necessary condition ``$\forall$ actor.Thing $\sqcap$ $\forall$ object.Thing'' (i.e., ``Function $\sqsubseteq$ $\forall$ actor.Thing $\sqcap$ $\forall$ object.Thing'') that constrains the basic slots (``actor'' and ``object''),
% which has a necessary condition ``$\forall$ inheres\_in.SubjT $\sqcap$ $\exists$ inheres\_in.SubjT'' (i.e., ``Function $\sqsubseteq$ $\forall$ inheres\_in.SubjT $\sqcap$ $\exists$ inheres\_in.SubjT''),
%When translating a function description to DL, the function name will be a sub-class of ``{\textbf{Function}}'', which has a necessary condition ``$\forall$ exhibited\_by.SubjT $\sqcap$ $\exists$ exhibited\_by.SubjT'' (i.e., Function $\sqsubseteq$ $\forall$ exhibited\_by.SubjT $\sqcap$ $\exists$ exhibited\_by.SubjT)
% exhibited\_by.SubjT'' (i.e., Function $\sqsubseteq$ $\exists$ exhibited\_by.SubjT)~\footnote{Alternatively, we can say ``\{Function\} $\sqsubseteq$ $\forall$ inheres\_in.SubjT $\sqcap$ $\exists$ inheres\_in.SubjT''

When translating a function description to DL, the function name will be a sub-class of ``{\textbf{Function}}'', which has a necessary condition ``$\exists$ exhibited\_by.SubjT'' (i.e., Function $\sqsubseteq$ $\exists$ exhibited\_by.SubjT)~\footnote{Alternatively, we can say ``\{Function\} $\sqsubseteq$ $\exists$ inheres\_in.SubjT'', where ``inheres\_in'' is also a functional slot. The difference is that ``Function'' indicates a set of manifestations (executions), while ``\{Function\}'' represents a function capability (an individual). We suggest use ``Function'' here for convenience.}, where ``\emph{exhibited\_by}'' is a functional slot, ``\emph{SubjT}'' is a concept defined using our language; that is, an execution will be manifested by a function individual, and accordingly exhibited by the subject that the function individual inheres in; the slot-description pairs will be translated according to rule $1 \sim 7$ in Table~\ref{tab:semantics_descriptions}. For example, the function description $F_1$ above will result in the declaration of concept $F_1$ as in Eq.~\ref{eq:eq_translation_f_example}.
\begin{comment}
\begin{equation}\label{eq:eq_translation_f_example}
    \begin{aligned}
        F_1 \; \equiv \;& \textbf{Function} \sqcap Activate \\
                        &\sqcap \forall actor.Manager \; \sqcap 1 \; actor.Manager \\
                        &\sqcap \forall object.Debit\_card \; \sqcap 1 \; object.Debit\_card \\
    \end{aligned}
\end{equation}
\end{comment}
\begin{equation}\label{eq:eq_translation_f_example}
    \begin{aligned}
        F_1 \; \equiv \;& \textbf{Function} \sqcap Activate \\
                        &\sqcap 1 \; actor.Manager \sqcap 1 \; object.Debit\_card \\
    \end{aligned}
\end{equation}

 %(``$s$'' is a functional property).
Generally, a function description $F_e$ as above will be translated into a DL concept $F_e$ as in Eq.~\ref{eq:eq_translation_f_general}.
\begin{comment}
\begin{equation}\label{eq:eq_translation_f_general}
    \begin{aligned}
       {F_e} \; \equiv \;& \textbf{Function} \sqcap FName \sqcap \forall s.\bm{L}(D) \; \sqcap 1 s.\bm{L}(D)
    \end{aligned}
\end{equation}
\end{comment}
\begin{equation}\label{eq:eq_translation_f_general}
    \begin{aligned}
       {F_e} \; \equiv \;& \textbf{Function} \sqcap FName \sqcap 1 \; s.\bm{L}(D)
    \end{aligned}
\end{equation}

We distinguish between a function individual $\{F\}$ and its manifestations $F$ since a function (capability) could have been implemented but not manifested if its activating situation does not hold. For example, a web site may have a keyword search capability, but will not be manifested if nobody use it. Further, the distinction between function (capability) and its manifestations allows us to specify requirements on the function (capability) itself. For example, we can not only require the ``schedule meetings'' function of a meeting scheduler (its manifestations, i.e., executions) to be fast, but also require the function (capability) itself to be easy to learn. These two requirements can be captured as $QG_{2-2}$ and $QG_{2-3}$ in Eq.~\ref{eq:eq_findividual_fruns}, respectively. Their semantics will be discussed later.\\
\begin{equation}\label{eq:eq_findividual_fruns}
    \begin{aligned}
    F_{2-1} :=  & \; Schedule <actor: Organizer> \\
                & <object: Meeting>\\
    QG_{2-2} := & \; Processing\_time \; (F_{2-1}) :: Fast \\
    QG_{2-3} := & \; Learnability \; (\{F_{2-1}\}) :: Good
    \end{aligned}
\end{equation}

\vspace{6pt}
\noindent \emph{\textbf{Translation of QGs and QCs.}} A QGC (QG/QC) is a requirement that requires a quality to take its value in a desired region. For example, the quality constraint ``$QC_3$ := Processing\_time (File\_search) :: [0, 30 (Sec.)]'' requires each run of file search to take less than 30 seconds. Its semantics will be expressed by the formula $AX_{QC_3}$ in Eq.~\ref{eq:eq_semantics_qgc_example}, where ``\emph{inheres\_in}'' and ``\emph{has\_value\_in}'' are reserved binary predicates used to express the semantics of QGC.
\begin{equation}\label{eq:eq_semantics_qgc_example}
    \begin{aligned}
     AX_{QC_3} \equiv \;& \forall s/File\_search \; \forall q/Processing\_time \; inheres\_in( q, s) \\
                        & \rightarrow  has\_value\_in( q, time\_region ( x, 0, 30, Sec.))
    \end{aligned}
\end{equation}

%(here ``SubjT'' is a concept defined in our language)
This can be generalized to QGCs, which have the syntax ``Q (SubjT) :: QRG'', by Eq.~\ref{eq:eq_semantics_qgc_general}.
\begin{equation}\label{eq:eq_semantics_qgc_general}
    \begin{aligned}
        AX_{QGC_e} \equiv \;  &\forall s/\bm{T}(SubjT) \; \forall q/\bm{T}(Q) \; inheres\_in (q, s) \\
                            &\rightarrow  has\_value\_in(q, \bm{T}(QRG))
    \end{aligned}
\end{equation}

%Note that the formal treatment of quantitative regions that are intervals is obvious; and qualitative regions can be treated as we have discussed in Section~\ref{sec:interpret_satisficing_nfrs}.
A QG has a qualitative region, e.g., ``\emph{fast}'', which is imprecise and is translated to a primitive concept, e.g., ``\emph{Fast}''. A QC has a quantitative region, which is a mathematically specified precise region. Since OWL2~\cite{group_owl_2009} supports one-dimensional primitive types only (i.e., it cannot represent points in $\mathbb{R}^3$), we restrict ourselves to single-dimensional regions, which are intervals of the form $[Bound_{low}, Bound_{high}]$ on the integer or decimal line.

%(has only one instance of its description)
%As such, the quality constraint $QC_1$ will result in the declaration of concept $QC_1$ as in Eq.
%Note that as with function descriptions, QG and QC are not assertions (true or false), but requirements to be fulfilled. Therefore, QG and QC will be translated to DL concepts.
%Note that as with function descriptions, QG and QC are not assertions (true or false), and will be translated to DL concepts. In general, QGs (resp. QCs) will be subclasses of the concept \textbf{QG} (resp. \textbf{QC}) that has necessary conditions ``Q $\sqcap$ $\forall$ inheres\_in.SubjT $\sqcap$ $\exists$ inheres\_in.SubjT $\sqcap$ $\forall$ has\_value\_in.QRG $\sqcap$ $\exists$ has\_value\_in.QRG'',
As with function descriptions, QG and QC are translated to DL concepts. In general, QGs (resp. QCs) will be subclasses of the concept \textbf{QG} (resp. \textbf{QC}) that has necessary conditions ``Q $\sqcap$ $\exists$ inheres\_in.SubjT $\sqcap$ $\exists$ has\_value\_in.QRG'', where ``\emph{QRG}'' can be a region name, an interval, or a set of enumeration values, ``\emph{inheres\_in}'' and ``\emph{has\_value\_in}'' are functional slots. Here we overload the notation ``\emph{has\_value\_in}'': it is an object property in the case of QGs but a data property in the case of QCs.

As such, the above quality constraint $QC_3$ be translated into the DL concept $QC_3$ as in Eq.~\ref{eq:eq_translation_qgc_example}.
\begin{comment}
\begin{equation}\label{eq:eq_translation_qgc_example}
    \begin{aligned}
     QC_3 \equiv \; & \textbf{QC} \sqcap Processing\_time \\
                    & \sqcap \forall inheres\_in.File\_search \sqcap \exists \; inheres\_in.File\_search \\
                    & \sqcap \forall has\_value\_in.((\geq 0) \sqcap (\leq 30)) \sqcap \exists \; has\_value\_in.((\geq 0) \sqcap (\leq 30))	
    \end{aligned}
\end{equation}
\end{comment}
\begin{equation}\label{eq:eq_translation_qgc_example}
    \begin{aligned}
     QC_3 \equiv \; & \textbf{QC} \sqcap Processing\_time \\
                    & \sqcap \exists \; inheres\_in.File\_search \\
                    & \sqcap \exists \; has\_value\_in.((\geq 0) \sqcap (\leq 30))	
    \end{aligned}
\end{equation}

%Generally, a QGC in the form of ``$QGC_e$ := Q(SubjT) :: QRG'' is semantically equivalent to ``Q $<$inheres\_in: SubjT$>$$<$has\_value\_in: QRG$>$'', and will be translated to the DL concept as in
Generally, a QGC in the form of ``$QGC_e$ := Q(SubjT) :: QRG'' will be translated to the DL concept as in Eq.~\ref{eq:eq_translation_qgc_general}.\\
\begin{comment}
\begin{equation}\label{eq:eq_translation_qgc_general}
    \begin{aligned}
     QGC_e \equiv \;& \textbf{QGC} \sqcap \bm{L}(Q) \\
                    &\sqcap \forall inheres\_in. \bm{L}(SubjT) \sqcap \exists \; inheres\_in. \bm{L}(SubjT) \\
                    &\sqcap \forall has\_value\_in. \bm{L}(QRG) \sqcap \exists \; has\_value\_in. \bm{L}(QRG)
    \end{aligned}
\end{equation}
\end{comment}
\begin{equation}\label{eq:eq_translation_qgc_general}
    \begin{aligned}
     QGC_e \equiv \;& \textbf{QGC} \sqcap \bm{L}(Q) \\
                    & \sqcap \exists \; inheres\_in. \bm{L}(SubjT) \\
                    & \sqcap \exists \;  has\_value\_in. \bm{L}(QRG)
    \end{aligned}
\end{equation}

\section{The Semantics of the Requirement Operators}
\label{sec:semantics_operator}

In our \emph{Desiree} framework, there are two kinds of semantics, namely \emph{entailment} and \emph{fulfillment}, for the operators.

The first kind is ``\emph{entailment}'' semantics. When refining a goal $G_1$ to $G_2$,
%In this section, we provide the \emph{entailment} semantics for the requirement operators:
\begin{enumerate}
  \item If each solution for $G_2$ is also a solution for $G_1$, then $G_2$ entails $G_1$ (denoted as $G_2$ $\models$ $G_1$), and this refinement is a \emph{strengthening};
  \item If each solution for $G_1$ is also a solution for $G_2$, then $G_1$ entails $G_2$ (denoted as $G_1$ $\models$ $G_2$), and this refinement is a \emph{weakening};
  \item In case $G_1$ and $G_2$ mutually entail each other (denoted as $G_1$ $\doteq$ $G_2$), we say that the two goals are equivalent and we term this refinement \emph{equating}.
\end{enumerate}

For example, the refinement from $G_1$ ``\emph{the system be secure}'' to $G_2$ ``\emph{the data storage be secure}'' is a weakening, because any solution \emph{sol} that can fulfill $G_1$, can also fulfill $G_2$, but not vice versa. This also means that there are potentially fewer solutions for $G_1$ than for $G_2$. This entailment semantics is of importance to requirements refinement: we need to weaken a requirement if it is too strong (e.g., practically unsatisfiable, and conflicting) and constrain it if it is arbitrary (e.g., ambiguous, incomplete, and vague).

The second kind is ``\emph{fulfillment}'' semantics: for each operator, we consider the propagation of fulfillment from its output elements to its input one(s). For example, in the case a goal $G_1$ ``\emph{the system be secure}'' is refined to (only) $G_2$ ``\emph{the data storage be secure}'', we say that if $G_2$ is fulfilled, then $G_1$ is also fulfilled. Note that fulfillment is different from entailment: in this example, $G_2$ does not entail $G_1$, but the fulfillment of $G_2$ indicates the fulfillment of $G_1$. This fulfillment semantics is of importance to the ``what-if'' analysis in goal-oriented requirements models, e.g., what kinds of elements in your model will be affected if some Function, FCs, or QCs are fulfilled while others are not?

\subsection{Entailment Semantics}
\label{sec:semantics_entailment}

In the \emph{Desiree} framework, an application of any requirements operator will be one of the three kinds of refinements: strengthening, weakening or equating. An overview of the entailment semantics of each operator is shown in Table~\ref{tab:semantics_operators}.
%A requirement operator that is ambiguous as to its strength status needs an additional argument to make the appropriate choice. For example, because an application of ``Reduce'' can be either a strengthening or an equating, we could denote the reduce from $G_1$ to $G_2$ as $R_d$ ($G_1$, $\Dashv$, \{$G_2$\}) if the refinement is a strengthening, and $R_d$ ($G_1$, $\doteq$, \{$G_2$\}) if it is an equating~\footnote{This additional argument will be filled in by our supporting tool based on users' selection of requirements operators. For example, the tool will add `$\Dashv$'' as a strengthening indicator to a refinement if a user choose ``Reduce (AND/OR refinement)'', or ``$\doteq$'' if a user choose ``Reduce (Separate Concerns)''.}.

\begin{table}[!htbp]
  \caption {The semantics of requirement operators}
  \label{tab:semantics_operators}
  \vspace {0.3 cm}
  \centering
  \small
  \setlength\tabcolsep{2pt}
  %\begin{tabular}{|c|p{0.8\textwidth}|}
  \begin{tabular}{|c|c|}
  \hline
  \textbf{Refinement} & \textbf{Operators} \\ \hline
  \multirow{5}{*}
  {Strengthening} & Interpret (Disambiguation) \\
	& Reduce (AND- and OR-refinement) \\
	& Reduce (Adding a SlotD, specializing the description of a slot) \\
	& Scale up (Shrinking QRG) \\
	& Operationalize (Goal as F, FC, QC, SC with optional DA) \\
	& Observe \\ \hline
  \multirow{5}{*}
  {Weakening} & de-Universalize \\
	& Scale down (Enlarging QRG) \\
	& Focus (Partial set of sub-elements) \\
    & Reduce (Removing a SlotD, generalizing the description of a slot) \\
	& Resolve \\
	& Operationalize (Goal as only DAs) \\ \hline
  \multirow{3}{*}
  {Equating} & Interpret (Encoding) \\
    & Reduce (Separating concern) \\
    & Focus (Full set of sub-elements) \\
    \hline
  \end{tabular}
\end{table}

%$R_d$ ($G_1$, $\vDash$, \{$G_2$\}) if it is a weakening,
There are two points to be noted. First, a requirement operator that is ambiguous as to its strength status needs an additional argument to make the appropriate choice. We use `$\Dashv$' for a strengthening, `$\vDash$' for a weakening, and `$\doteq$' for an equating. For example, because an application of ``Reduce'' can be a strengthening, a weakening, or an equating, we could denote the reduce from $G_1$ to $G_2$ as $R_d$ ($G_1$, $\Dashv$) = \{$G_2$\} if the refinement is a strengthening, $R_d$ ($G_1$, $\vDash$) = \{$G_2$\} if it is a weakening, and $R_d$ ($G_1$, $\doteq$) = \{$G_2$\} if it is an equating~\footnote{This additional argument will be filled in by our supporting tool based on analysts'/engineers' selection of requirements operators. For example, the tool will add ``$\Dashv$'' as a strengthening indicator to a refinement if an analyst/engineer choose ``Reduce (AND/OR refinement)'', or ``$\doteq$'' if an analyst/engineer choose ``Reduce (Separate concerns)''.}. Second, we distinguish between ``Relators'', operators that assert relationships between existing elements (``Interpret'', ``Reduce'', ``Operationalize'', and ``Resolve''), and ``Constructors'', operators that construct in a precise way new elements from their arguments (``Focus'', ``Scale'', ``Observe'', ``de-Universalize'').
%(``Interpret'', ``Reduce'', ``Operationalize'', and ``Resolve'')
%(``Focus'', ``Scale'', ``Observe'', ``de-Universalize'')
%\footenote{Users do not need to explicitly specify the strength status when applying an operator; instead, they choose an appropriate operation when applying ait is the supporting tool that offers necessary support. For example, when applying the reduce operator, the supporting tool shall provide }.

%\noindent \emph{\textbf{Relators.}} We distinguish between ``Relators'', operators that assert relationships between existing elements, and ``Constructors'', remaining ones that construct in a precise way new elements from their arguments. We will discuss each operator in detail in this subsection.
%(``Interpret'', ``Reduce'', ``Operationalize'', and ``Resolve'')
%(``Focus'', ``Scale'', ``Observe'', ``de-Universalize'')

We discuss each operator in detail in this sub-section. Specifically, for each operator, we first introduce the entailment (strengthening/weakening) semantics, and then discuss about the DL translation such that we can use DL subsumption to simulate entailment. For relators, we simply constrain the output elements to be subsumed by the input element(s) in case of a strengthening, and add a subsumption axiom in reverse in case of an equating. For constructors, output elements are constructed from input elements by following specific ways; once translated in to DL concepts using the rules introduced in Section~\ref{sec:semantics_language}, a DL reasoner (e.g., Hermit~\cite{shearer_hermit:_2008}) is able to infer the subsumption relations between input and output elements. However, for some constructors, the entailment and DL subsumption are not conformative. For example, when focusing $QG_1$ ``Security (\{the\_system\}) :: Good'' to $QG_2$ ``Security (\{the\_data\_module\}) :: Good'', we have $QG_1 \models QG_2$ (i.e., if the system is secure, then its date module is also secure); however, when translated into DL expressions, we will have $QG_2 \sqsubseteq QG_1$ as ``the\_data\_module'' is part of ``the\_system'' (if we capture this as $\{the\_data\_module\} \sqsubseteq \{the\_system\}$). Therefore, we will use some tricks when applying such constructors in order to obtain conformative DL subsumptions with regarding to their entailment semantics.

\vspace{12pt}
\noindent \emph{\textbf{Relators.}} Four out of the eight operators, namely ``Interpret'', ``Reduce'', ``Operationalize'', and ``Resolve'', are relators. These operators need to have their arguments ready, and then relate an input element to the corresponding output element(s) that are of concern. For example, given an ambiguous goal $G$, what the ``Interpret'' operator does is to choose one intended meaning from its multiple possible interpretations (i.e., the possible interpretations are already there, what we need to do is to discover them and choose the intended one from them). This is similar for ``Reduce'' or ``Operationalize'': choosing one kind of refinement/operationalization from multiple possible refinements/operationalizations~\footnote{``Reduce'' and ``Operationalize'' (also other operators) can be applied to the same input element more than one time. In an application, an input element can be refined/operationalized to multiple sub-elements. This captures traditional AND. The multiple applications captures traditional OR.}.
%-refinement/operationalization
%(i.e., those chosen from existing elements)

\vspace{6pt}
\noindent \emph{\textbf{Reduce}} ($\bm{R_d}$). In general, an application of ``$\bm{R_d}$'' is a strengthening, but can also be an equating or a weakening. In general, traditional ``AND'' and ``OR'' refinements are strengthening. For example, when reducing a goal $G_1$ ``\emph{trip be scheduled}'' to $G_2$ ``\emph{hotel be booked}'' and $G_3$ ``\emph{airline ticket be booked}'', a solution that can satisfy both $G_2$ and $G_3$ can also satisfy $G_1$, but not the opposite since $G_1$ can be satisfied by other solutions (e.g., ``hostel be booked'' and ``train ticket be booked''). In this situation, we have $\bm{R_d}$ ($G_1$, `$\Dashv$') = \{$G_2$, $G_3$\}, which asserts $G_2, G_3 \models G_1$.

When reducing a complex requirement (a requirement with multiple concerns) to atomic ones (a requirement with a single concern), the refinement is an equating. For example, the reduction of $G_1$ ``\emph{the system shall collect real time traffic info}'' to $G_2$ ``\emph{traffic info be collected}'' and $G_3$ ``\emph{collected traffic info shall be in real time}''. In this case, having $\bm{R_d}$ ($G_1$, `$\doteq$') = \{$G_2$,$G_3$\} asserts $G_2$, $G_3$ $\doteq$ $G_1$.

%Also, the adding of a slot-description (\emph{SlotD}) or a specialization of the description of a slot is a strengthening~\footnote{In theory, it is also possible to remove a \emph{SlotD} or generalize the description of a slot. However, this is not the case in RE because that means pushing for higher abstraction and hiding details in a requirements specification.}. For example, the refinement from ``$F_1$ := Book $<$object: Ticket$>$'' to ``$F_1'$ := Book $<$object: Airline\_ticket$>$'' is a strengthening since $F_1$ can be fulfilled by any solution that is able to book a ticket, but $F_1'$ can only be fulfilled if a solution is able to book an airline ticket. That is, $F_1'$ has fewer solutions. This refinement can be captured as $\bm{R_d}$ ($F_1$, `$\Dashv$') = \{$F_1'$\}, and asserts $F_1' \models F_1$.

When applying ``Reduce'' to a function description, the adding of a slot-description pair (\emph{SlotD}) or a specialization of the description of a slot is a strengthening, while the removing of a \emph{SlotD} or the generalization of the description of a slot is a weakening~\footnote{We do not allow strengthening a slot while weakening another in a single refinement.}. For example, the refinement from ``$F_1$ := Book $<$object: Ticket$>$'' to ``$F_1'$ := Book $<$object: Airline\_ticket$>$'' is a strengthening since $F_1$ can be fulfilled by any solution that is able to book a ticket (e.g., airline ticket, bus ticket, train ticket, etc.), but $F_1'$ can only be fulfilled if a solution is able to book an airline ticket. That is, $F_1'$ has fewer solutions. This refinement can be captured as $\bm{R_d}$ ($F_1$, `$\Dashv$') = \{$F_1'$\}, which asserts $F_1' \models F_1$.

We use DL subsumption to simulate the entailment semantics of ``Reduce'': if it is used as a strengthening, we constrain the intersection of the output elements to be subsumed by the input element; if it is used as an equating, we add an extra subsumption axiom, restricting the input element as a sub-class of the intersection of the output elements. For example, for ``\emph{trip be scheduled}'', we specify $G_2 \sqcap G_3 \sqsubseteq G_1$; for ``\emph{collect real time traffic info}'', we use both $G_2 \sqcap G_3 \sqsubseteq G_1$ and $G_1 \sqsubseteq G_2 \sqcap G_3$.

In the case of refining a function description, the subsumption between input and output elements can also be inferred by a DL reasoner: (1) when adding or removing a \emph{SlotD} to the structured description (e.g., the corresponding DL concept of ``Book $<$object: Airline\_ticket$>$$<$means: Credit\_card$>$'' will be subsumed by that of ``Book $<$object: Airline\_ticket$>$''); (2) when specializing or generalizing the description of a slot, and the relation between descriptions are given (e.g., if we have \emph{Airline\_ticket} $\sqsubseteq$ \emph{Ticket}, a DL reasoner is able to infer that $F_1' \sqsubseteq F_1$).

%The DL translation for the ``Reduce'' operator can be also applied to the other three relators, namely ``Interpret'',``Operationalize'', and ``Resolve''. Therefore we will not discuss the DL translation for these operators.

\vspace{6pt}
\noindent \emph{\textbf{Interpret}} ($\bm{I}$). In \emph{Desiree}, an interpretation of an ambiguous or under-specified requirement is a strengthening, and an encoding of a natural language requirement is an equating. This is because an ambiguous/under-specified requirement has more than one interpretation, hence possessing more solutions. For example, when an ambiguous goal $G_1$ ``\emph{notify users with email}'' is interpreted into $G_2$ ``\emph{notify users through email}'', each solution for $G_2$ could also be a solution for $G_1$, but not vice versa: $G_1$ has another interpretation $G_3$ ``\emph{notify users who have email}'', a solution for $G_3$ can fulfill $G_1$, but not $G_2$. In the case we simply encode a (natural language) requirement using our syntax, the encoding is an equating because the solution space does not change.

Therefore, for two \emph{Desiree} elements $E_1$ and $E_2$, we have $E_2$ $\models$ $E_1$ if $\bm{I}$ ($E_1$, `$\Dashv$') = $E_2$, or $E_2$ $\doteq$ $E_1$ if $\bm{I}$ ($E_1$, `$\doteq$') = $E_2$. In the former case, we have only $E_2 \sqsubseteq E_1$; in the latter case, we have both $E_2 \sqsubseteq E_1$ and $E_1 \sqsubseteq E_2$.

\vspace{6pt}
\noindent \emph{\textbf{Operationalize}} ($\bm{O_p}$). Operationalization is similar to reduction ``$R_d$''. In \emph{Desiree}, the ``${O_p}$'' operator is overloaded in several ways. When operationalizing an FG to Function, FCs, DAs, or combinations thereof, or operationalizing a QG to QC(s) by making clear its vague region, to F and/or FCs to make it implementable, or operationalizing a CTG to SC(s), it is a strengthening. In case a goal is merely operationalized as DA(s), it is a weakening (a DA can be simply treated as true and can be fulfilled by any solution; that is, the solution space is enlarged to infinite when operationalizing a goal merely as DAs).

The $O_p$ operator has a similar semantics as ``Reduce''. However, if instances of output elements and instances of the input elements are not of the same type (e.g., operationalizing an FG as functions), its semantics is subtly different. For example, when an FG ``$FG_1$ := Meeting\_notification $:<$ Sent'' is operationalized as a function ``$F_2$ := Send $<$subject: \{the\_system\}$>$ $<$object: Meeting\_notification$>$'', we will have ``$F_2$.effect $\models$ $FG_1$'', where ``$F_2$.effect'' represent the situation brought about by the execution of $F_2$ (i.e., a message is sent), and accordingly ``$F_2$.effect $\sqsubseteq$ $FG_1$''. Note that here we can not use ``$F_2 \models FG_1$'', which leads to ``$F_2$ $\sqsubseteq$ $FG_1$'', because the set of instances of $F_2$ and that of $FG_1$ can not be subsumed by each other: instances of $F_2$ are a set of its executions while instances of $FG_1$ are a set of problem situations, or, alternatively, instantiations of the corresponding intention (the intention of a message being sent, in this case). In the case $FG_1$ is operationalized to a set of elements, say $F_2$ and $F_3$, we will have $F_2$.effect, $F_3$.effect $\models$ $FG_1$, and accordingly $F_2$.effect $\sqcap$ $F_3$.effect $\sqsubseteq$ $FG_1$.

%In the case $FG_1$ is operationalized to a set of elements, say $F_2$ and $F_3$, we will have $F_2$.effect, $F_3$.effect $\models$ $FG_1$, and accordingly $F_2$.effect $\sqcap$ $F_3$.effect $\sqsubseteq$ $FG_1$.

\vspace{6pt}
\noindent \emph{\textbf{Resolve}} ($\bm{R_s}$). An application of the ``Resolve'' operator to a set of conflicting requirements $S_i$ will produce a set of conflict-free requirements $S_o$. Being a weakening, such a refinement will be denoted as ``$\bm{R_s}$($S_i$, `$\vDash$') = $S_o$''. A conflict among a set of requirements means that they cannot be satisfied simultaneously; that is, there is an empty set of solutions for $S_i$. Once resolved, then there should be some possible solutions; that is, there is a non-empty set of solutions for $S_o$. Hence this adds the meta-statement $S_i \models S_o$.

We do not add any DL subsumption axioms for an application of $R_s$. This is because when we constraining the intersection of the input elements $S_i$ to be subsumed by the intersection of the output elements $S_o$, we are actually specifying a trivial subsumption $\emptyset \sqsubseteq E_{o1} \sqcap ... \sqcap E_{on}$, where ``$E_{o1}$ ... $E_{on}$'' are elements of $S_o$. This is needless.
%(in fact, this intersection corresponds an empty set of solutions)

\vspace{12pt}
\noindent \emph{\textbf{Constructors.}} The remaining four operators, namely ``Focus'', ``Scale'', ``Observe'', ``de-Universalize'', are constructors. These operators take as input \emph{Desiree} elements (specifically, QGs or QCs) and necessary arguments, and create/construct new elements, for which entailment can be computed by logic.

\vspace{6pt}
\noindent \emph{\textbf{Focus}} ($\bm{F_k}$). The ${F_k}$ operator narrows the scope of a quality or subject by following certain hierarchies , e.g., ``dimension-of'' ``part-of'', and makes a QGC easier to fulfill. Its application leads to a weakening (e.g., focusing ``security'' to a sub-set of its dimensions, say ``confidentiality'', ``the system'' to some of its sub-parts ``interface''), but sometimes an equating (e.g., focusing ``security'' to the full set of its sub-dimensions, ``confidentiality'', ``integrity'' and ``availability''). In the former case, where $\bm{F_k}$ ($QGC_1$, $Qs$/$SubjTs$, `$\vDash$') = $QGC_{partial}$ ($QGC_{partial}$ is a sub-set of the potential resultant QGCs), we have $QGC_1$ $\models$ $QGC_{partial}$; in the latter case, where $\bm{F_k}$ ($QGC_1$, $Qs$/$SubjTs$, `$\doteq$') = $QGC_{full}$ ($QGC_{full}$ is the full-set of the the potential resultant QGCs), we have $QGC_1$ $\doteq$ $QGC_{full}$, i.e., $QGC_1$ $\models$ $QGC_{full}$ and $QGC_{full}$ $\models$ $QGC_1$.

%In the former case, where $\bm{F_k}$ ($QGC_1$, $\bm{\wp}(Q)$/$\bm{\wp}(SubjT)$, $\vDash$) = $S_{partial}$, we have $QGC_1$ $\models$ $S_{partial}$; in the latter case, where $\bm{F_k}$ ($QGC_1$, $\bm{\wp}(Q)$/$\bm{\wp}(SubjT)$, $\doteq$) = $S_{full}$), we have $QGC_1$ $\doteq$ $S_{full}$.

To have conformative DL subsumptions, we use a trick when applying the $F_k$ operator: as shown in Eq.~\ref{eq:eq_semantics_Fk_example}, when focusing $QG_{4-1}$ via subject type, its subject will be expanded to ``\{the\_system\} $\lor$ \{the\_data\_module\}'' rather than be replaced by ``\{the\_data\_module\}''. As such, when $QG_{4-1}$ and $QG_{4-2}$ are translated to DL concepts, we will have the DL subsumption $QG_{4-1} \sqsubseteq QG_{4-2}$, which conforms to the entailment $QG_{4-1} \models QG_{4-2}$. This trick is also used when focusing a QGC via its quality type.\\
\begin{equation}\label{eq:eq_semantics_Fk_example}
    \begin{aligned}
    QG_{4-1} := \;& Security \; (\{the\_system\}) :: Good \\
    QG_{4-2} := \;& \bm{F_k} \; (QG_{4-1}, \{the\_data\_module\})\\
            =   \;& Security \; (\{the\_system\} \lor \{the\_data\_module\}) :: Good\\
    %QG_{4-3} := \;& \bm{F_k} \; (QG_{4-1}, Availability)\\
    %        =   \;& Security \lor Availability \; (\{the\_system\}) :: Good\\
    \end{aligned}
\end{equation}

\vspace{6pt}
\noindent \emph{\textbf{Scale}} (\emph{\textbf{G}}). The \emph{G} operator is used to enlarge or shrink quality regions. When enlarging a region of $QGC_1$ to $QGC_2$, e.g., scale ``\emph{fast}'' to ``\emph{nearly fast}'' or ``[0, 30 (\emph{Sec.})]'' to ``[0, 40 (\emph{Sec.})]'', logic gives $QGC_1 \models QGC_2$ (i.e., it is a weakening). When shrinking a region of $QGC_1$ to $QGC_2$, e.g., strengthen ``\emph{fast}'' to ``\emph{very fast}'' or ``[0, 30 (\emph{Sec.})]'' to ``[0, 20 (\emph{Sec.})]'', logic implies $QGC_2 \models QGC_1$ (i.e., it is a strengthening).

When scaling qualitative regions of QGs, analysts/engineers may declare as DA axioms concerning relations between them. For example, the region ``\emph{fast}'' can be relaxed to ``\emph{nearly\_fast}'' or strengthened to ``\emph{very\_fast}'', the relations between them would be \emph{very\_fast} $\sqsubseteq$ \emph{fast} $\sqsubseteq$ \emph{nearly\_fast}. When regions to be scaled are mathematically specified, a DL reasoner is able to automatically infer the subsumption relations. For example, ``[0, 30]'' is subsumed by ``[0, 40]''. Therefore, by translating the resulting QGCs to DL concepts as we have done for QGs/QCs in Section~\ref{sec:semantics_language}, we are able to obtain conformative DL subsumptions, i.e., $QGC_1 \sqsubseteq QGC_2$ when $QGC_1 \models QGC_2$ and $QGC_2 \sqsubseteq QGC_1$ when $QGC_2 \models QGC_1$.

%When applying the ``$\bm{G}$'' operator to a QGC, the resulting QGC is still in the syntactic form of ``Q (SubjT) :: QRG'', and can be translated to DL expressions as we have done for QGs/QCs in Section~\ref{sec:semantics_language}. Moreover, this translation is able to give conformative DL subsumptions, i.e., $QGC_1 \sqsubseteq QGC_2$ when $QGC_1 \models QGC_2$ and $QGC_2 \sqsubseteq QGC_1$ when $QGC_2 \models QGC_1$.

\vspace{6pt}
%\noindent \emph{\textbf{de-Universalize}} ($\bm{U}$). The $U$ operator is used for weakening. For example, instead of requiring all the runs of file search to take less than 30 seconds, we can relax it to 80\% of the runs to be so by applying $U$. Formally, the semantics of ``$QG_{1-2}$ := $\bm{U}$ (?X, $QG_{1-1}$, $<$inheres\_in: ?X$>$, 80\%)'', derived from ``$QG_{1-1}$ := Processing\_time (File\_search) :: Fast'', is expressed as $AX_{QG_{1-2}}$ in Eq.~\ref{eq:eq_semantics_U_example}.\\

\noindent \emph{\textbf{de-Universalize}} ($\bm{U}$). The $U$ operator is used for weakening. For example, instead of requiring all the runs of file search to take less than 30 seconds, we can relax it to 80\% of the runs to be so by applying $U$. Formally, the semantics of $QG_{5-2}$, derived from $QG_{5-1}$, is expressed as $AX_{QG_{5-2}}$ in Eq.~\ref{eq:eq_semantics_U_example}.\\
\begin{equation}\label{eq:eq_semantics_U_example}
    \begin{aligned}
    QG_{5-1} := \;& Processing\_time (File\_search) :: Fast \\
    QG_{5-2} := \;& \bm{U} (?X, QG_{5-1}, <inheres\_in: ?X>, 80\%)\\
            =   \;& Processing\_time (File\_search \; \lor <pct: [80\%, 100\%]) :: Fast \\
    AX_{QG_{5-2}} \equiv \; & \exists ?X/\bm{\wp}(File\_search) (|?X|/|File\_search| > 0.8 \\
                            & \land \forall s/?X \; \forall q/Processing\_time \; inheres\_in (q, s) \\
                            &  \rightarrow has\_value\_in(q,  Fast))
    \end{aligned}
\end{equation}

%Semantically, $QG_{5-1}$ is equivalent to the description ``Processing\_time $<$inheres\_in: File\_search$>$ $<$has\_value\_in: Fast$>$''.
Recall that $QG_{5-1}$ is translated to the description ``Processing\_time $<$inheres\_in: File\_ search$>$ $<$has\_value\_in: Fast$>$''. Through pattern matching, we are able to identify that ``?X'' represents a sub set of ``File\_search'', with which the required percentage ``[80\%, 100\%]'' will be associated. When expanding the description, we use a trick: the subject of $QG_{5-2}$ will be ``File\_search $\lor$ $<$pct: [80\%, 100\%]$>$'' rather than ``File\_search $<$pct: [80\%, 100\%]$>$''.
%The reason will be shown later.
%which can be translated to DL expressions according to the rules in Table~\ref{tab:semantics_descriptions}.

In general, the semantics of a QGC that is applied with $U$ (UQGC for short), of the form ``$\bm{U}$ (?X, Q (SubjT) :: QRG, $<$inheres\_in: ?X$>$, Pct)'', is given in Eq.~\ref{eq:eq_semantics_U}, where a percentage ``Pct'' indicates a region [Pct, 100\%]. We can see that if $QGC_1$ is relaxed to $QGC_2$ by $U$, logic gives $QGC_1$ $\models$ $QGC_2$. \\
\begin{equation}\label{eq:eq_semantics_U}
    \begin{aligned}
    AX_{UQGC} \equiv \;  & \exists ?X/\bm{\wp}(\bm{T}(SubjT)) \; (|?X|/|\bm{T}(SubjT)| > Pct \\
                        & \land \forall s/?X \; \forall q/\bm{T}(Q) \; inheres\_in (q, s) \\
                        & \rightarrow has\_value\_in (q,  \bm{T}(QRG)))	
    \end{aligned}
\end{equation}

%Recall that $QG_{1-1}$ is semantically equivalent to the description ``Processing\_time $<$inheres\_in: File\_search$>$ $<$has\_value\_in: Fast$>$''. Through pattern matching, we are able to identify ``?X'' as a sub set of ``File\_search'', with which the required percentage ``[80\%, 100\%]'' will be associated. That is, the subject of $QG_{1-2}$ will be updated to ``File\_search $\lor$ $<$pct: [80\%, 100\%]$>$'', which can be translated to DL expressions according to the rules in Table~\ref{tab:semantics_descriptions}.
We translate a UQGC to the DL concept as in Eq.~\ref{eq_translation_uqgc_general}.\\
\begin{comment}
\begin{equation}\label{eq_translation_uqgc_general}
    \begin{aligned}
    UQGC_e \equiv \;& UQGC \sqcap \bm{L}(Q) \\
                    & \sqcap \forall inheres\_in.(\bm{L}(SubjT) \sqcup (\forall pct.\bm{L}(Pct) \sqcap \exists pct. \bm{L}(Pct))) \\
                    & \sqcap \exists inheres\_in.(\bm{L}(SubjT) \sqcup ( \forall pct.\bm{L}(Pct) \sqcap \exists pct. \bm{L}(Pct))) \\
                    & \sqcap \forall has\_value\_in. \bm{L}(QRG) \\
                    & \sqcap \exists has\_value\_in. \bm{L}(QRG)	
    \end{aligned}
\end{equation}
\end{comment}
\begin{equation}\label{eq_translation_uqgc_general}
    \begin{aligned}
    UQGC_e \equiv \;& UQGC \sqcap \bm{L}(Q) \\
                    & \sqcap \exists inheres\_in.(\bm{L}(SubjT) \sqcup \exists \; pct.\bm{L}(Pct)))\\
                    & \sqcap \exists has\_value\_in. \bm{L}(QRG)	
    \end{aligned}
\end{equation}

%Therefore, if a QGC $QGC_1$ is relaxed to $QGC_2$ by U, we will have $QGC_1$ $\sqsubseteq$ $QGC_2$, which means the satisfaction of $QGC_2$ would infer that of $QGC_1$.
In this way, the $\bm{U}$ operator would have the property as shown in Eq.~\ref{eq_translation_uqgc_property}, where $0\% \leq Pct_2 < Pct_1 \leq 100\%$. That is, if $QGC_1$ is relaxed to $QGC_2$ by U, we will have $QGC_1$ $\sqsubseteq$ $QGC_2$. We can see now that the trick of using ``File\_search $\lor$ $<$pct: [80\%, 100\%]$>$'' instead of ``File\_search $<$pct: [80\%, 100\%]$>$'' helps us to derive this DL subsumption. In fact, if we use the latter expansion (i.e., ``File\_search $<$pct: [80\%, 100\%]$>$''), we will get the opposite subsumption (i.e., $QGC_2$ $\sqsubseteq$ $QGC_1$), which is not intended in our case.\\
\begin{equation}\label{eq_translation_uqgc_property}
    \begin{aligned}
    \bm{U} (?X, QGC, <inheres\_in: \; ?X>, Pct_1) \\ \sqsubseteq \bm{U} (?X, QGC, <inheres\_in: \; ?X>, Pct_2)
    \end{aligned}
\end{equation}

In some cases, the U operator may be nested. For example, ``(\emph{at least}) \emph{80\% of the system functions shall be fast at least 90\% of the time}''. To capture this requirement, we first define a quality goal $QG_{6-1}$ as in Eq.~\ref{eq:eq_example_nested_u}, and apply $\bm{U}$ twice to obtain $QG_{6-3}$. In this example, $QG_{6-3}$ will be expanded as in Eq.~\ref{eq:eq_example_nested_u}, which can be translated to DL expressions according to the rules in Table~\ref{tab:semantics_descriptions}. We refer interested readers to Appendix~\ref{cha:appendixes_semantics_u} for detail about the semantics.
\begin{equation}\label{eq:eq_example_nested_u}
    \begin{aligned}
    QG_{6-1} :=  \; &Processing\_time (Run <run\_of: System\_function>) :: Fast\\
    QG_{6-2} :=  \; &\bm{U} \; (?F, \; QG_{6-1}, <inheres\_in: <run\_of: \; ?F>>, 80\%)\\
              =  \; &Processing\_time (Run <run\_of: \\
                    &System\_function \lor <pct: [80\%, 100\%]>>) :: Fast \\
    QG_{6-3} :=  \; &\bm{U} \; (?Y, \; QG_{6-2}, <inheres\_in: \; ?Y>, 90\%)\\
              =  \; &Processing\_time (Run <run\_of: \\
                    &System\_function \lor <pct: [80\%, 100\%]>> \\
                    &\lor <pct: [90\%, 100\%]>) :: Fast
    \end{aligned}
\end{equation}

Note that our DL translation is not able to capture the entailment semantics of nested $\bm{U}$, which is expressed by using second-order (even higher-order) logic. For example, our DL translation is able to infer that $G_1 \models G_2$, where $G_1$ is ``\emph{the file search function shall take less than 30 seconds} (\emph{at least}) \emph{$80\%$ of the time}'', and $G_2$ is ``\emph{the file search function shall take less than 30 seconds} (\emph{at least}) \emph{$70\%$ of the time}''. However, it can not tell the entailment between $G_3$ ``\emph{$80\%$ of the system functions shall be fast at $90\%$ of the time}'' and $G_4$ ``\emph{$90\%$ of the system functions shall be fast at $80\%$ of the time}''.

\vspace{6pt}
\noindent \emph{\textbf{Observe}} ($\bm{O_b}$). An application of $O_b$ to a QGC will append a \emph{SlotD} ``$<$observed\_by: Observer$>$'' to the QGC, and hence strengthen it. That is, we will have $QGC_2 \models QGC_1$ when applying the $O_b$ operator to $QGC_1$: $\bm{O_b}(QGC_1, Observer)$ = $QGC_2$~\footnote{We do not use ``$\Dashv$'' as an argument because an application of $O_b$ is always a strengthening.}. For example, by applying $O_b$, $QG_{7-1}$ ``Style (\{the\_interface\}) :: Simple'' becomes ``$QC_{7-2}$ := $QG_{7-1}$ $<$observed\_by: Surveyed\_user$>$'', the semantics of which is shown in Eq.~\ref{eq:eq_semantics_Ob_example}.\\
\begin{equation}\label{eq:eq_semantics_Ob_example}
    \begin{aligned}
        AX_{QC_{7-2}} \equiv \; & \forall o/Surveyed\_user \; \forall s/\{the\_interface\} \; \forall q/Style \\
                                & (inheres\_in (q, s) \rightarrow observed\_by (q, o)) \\
                                & \rightarrow has\_value\_in (q, Simple)
    \end{aligned}
\end{equation}

To obtain the general semantics of $O_b$, we only need to replace the specific quality type (e.g., ``Style''), subject type (e.g., ``{the\_interface}''), quality region (e.g., ``Simple''), and observer (e.g., ``surveyed\_user'') with general \emph{Q}, \emph{SubjT}, \emph{QRG} and \emph{Observers}, as in Eq.~\ref{eq:eq_semantics_Ob_general}.\\
\begin{equation}\label{eq:eq_semantics_Ob_general}
    \begin{aligned}
        AX_{QGC_{O_b}} \equiv \; & \forall o/\bm{T}(Observer) \; \forall s/\bm{T}(SubjT) \; \forall q/\bm{T}(Q) \\
                                & (inheres\_in (q, s) \rightarrow observed\_by (q, o)) \\
                                & \rightarrow has\_value\_in (q, \bm{T}(QRG))
    \end{aligned}
\end{equation}

%The generalized DL expression for $\bm{O_b}$ can be obtained by translating ``$\bm{L}$(QGC) $\sqcap$ $\forall$ observed\_by.$\bm{L}$(Observer) $\sqcap$ 1 observed\_by.$\bm{L}$(Observer)'',
The generalized DL expression for $\bm{O_b}$ can be obtained by translating ``$\bm{L}$(QGC) $\sqcap$ 1 observed\_by.$\bm{L}$(Observer)'', where ``QGC'' is the QG/QC to be observed, ``Observer'' is set of users who are observing. Here, ``observed\_by'' is not a functional property since an individual could be observed by more than one observer. As such, for an application of $O_b$, i.e., $\bm{O_b}(QGC_1, Observer)$ = $QGC_2$, we accordingly have the DL concept subsumption axiom $QGC_2 \sqsubseteq QGC_1$.
%(the description of the slot ``observed\_by'').

In the above example, $QC_{7-2}$ is hard to satisfy since it requires all the surveyed users to agree that the interface is simple. In practice, it is often relaxed by using $U$. For instance, we could have ``$QC_{7-3}$ := $\bm{U}$ (?O, $QC_{7-2}$, $<$observed\_by: ?O$>$, 80\%)'', which requires only 80\% of the users to agree. We show the semantics of $QC_{7-3}$ in Eq.~\ref{eq:eq_semantics_Ob_U_example}. \\
\begin{equation}\label{eq:eq_semantics_Ob_U_example}
    \begin{aligned}
        AX_{QC_{7-3}} \equiv \; & \exists ?O/\bm{\wp}(Surveyed\_user).[|?O|/|Surveyed\_user|> 0.8 \\
                                & \land \forall o/?O \; \forall s/\{the\_interface\} \; \forall q/Style \\
                                & (inheres\_in (q, s) \rightarrow observed\_by (q, o)) \\
                                & \rightarrow has\_value\_in (q, Simple) ]
    \end{aligned}
\end{equation}
%and refer interested readers to Appendix~\ref{cha:appendixes_semantics_u} for more detail about the semantics of combined use of $\bm{U}$ and $\bm{O_b}$\\
%In practice, the $\bm{U}$ and $\bm{O_b}$ operator can be used together, e.g., ``80\% of the surveyed users shall agree that 90\% of the functions of the system is fast''. In this case, we first define a QG ``QG := Processing\_time (System\_function) :: Fast'', and then apply $\bm{U}$, $\bm{O_b}$ and $\bm{U}$ one after another. We refer interested readers to Appendix~\ref{cha:appendixes_semantics_u} for more detail about the semantics.

\subsection{Fulfillment Semantics}
\label{sec:semantics_fulfillment}
We briefly discuss the fulfillment semantics according to the mapping relation between an operator's input and output. In general, we have two classes of operators: (1) ``one-to-one''(``$\bm{I}$'', ``$\bm{U}$'', ``$\bm{G}$'', and ``$\bm{O_b}$''); and (2) ``one-to-many'' (``$\bm{R_d}$'', ``$\bm{F_k}$'', and ``$\bm{O_p}$''). In general, a \emph{Desiree} operator can be applied to an input element more than one time; in one of those possible applications, some operators relate an input element to one output element, while the others relate the same input element to a set of ($\geq1$) output elements. Note that the ``$\bm{R_s}$'' operator does not possess fulfillment semantics because its input elements are conflicting and cannot be fulfilled at the same time.

%such an operator relates one input element to one output element in an application
%such an operator relates one input element to a set of output elements ($\geq 1$) in an application

%There are two points to be noted. First, a \emph{Desiree} operator can be applied to the same input element more than one time; in one of those possible applications, some operators relate an input element to one output element, while the others relate an input element to a set. Second, the ````$\bm{R_s}$''`` operator does not possess fulfillment semantics because its input elements are conflicting and cannot be fulfilled at the same time.

When a \emph{Desiree} element $E_1$ is refined/operationalized to another, say $E_2$, the fulfillment of $E_2$ would imply that of $E_1$. Formally, it can be expressed as ``$\forall E_1 \; \exists E_2 \; P_1 (E_1, E_2) \land fulfilled (E_2) \rightarrow fulfilled (E_1)$``, where ``$P_1$`` is one of the four operators: ``$\bm{I}$'', ``$\bm{U}$'', ``$\bm{G}$'', and ``$\bm{O_b}$''. When $E_1$ is refined/operationalized to a set of elements $E_S$, then each element in $E_S$ shall be fulfilled to make $E_1$ fulfilled. This can be formally stated as ``$\forall E_1 \; \exists \; E_S \; P_2 (E_1, E_S) \land fulfilled (E_S) \rightarrow fulfilled (E_1)$``, where ``$P_2$`` is a one of the three operators, ``$\bm{R_d}$'', ``$\bm{F_k}$'', and ``$\bm{O_p}$'', and ``$fulfilled (E_S) =_{\mathrm{def}} \forall e.e \in E_S \rightarrow fulfilled \;(e)$''.

We use ``Interpret'' and ``Reduce'' as the representatives to discuss the DL translation for the fulfillment semantics, which is inspired by Horkoff et al.~\cite{horkoff_making_2012}.
%For instance, the semantics of ``Operationalize'' are similar to those of ``Reduce''.

\vspace{6pt}
\noindent \emph{\textbf{One-to-one.}} ``Interpret'' is a ``one-to-one`` operator. When a goal $G_1$ is interpreted to another, e.g., $G_2$, we say that the fulfillment of $G_2$ would imply that of $G_1$. Formally, it can be denoted as ``$\forall G_1 \; \exists G_2 \; interpret\_to (G_1, G_2) \land fulfilled (G_2) \rightarrow fulfilled (G_1)$''. To have this inference, we add an assertion as in Eq.~\ref{eq:eq_translation_satthing}: a \emph{Fulfilled\_Thing} is a \emph{Thing} that is related (specifically, interpreted, scaled, observed, or de-Universalized) to at least one \emph{Fulfilled\_Thing}. When $G_1$ is interpreted to $G_2$, we write a DL axiom $G_1 \sqsubseteq \exists \; interpret\_to.G_2$. As such, in the case $G_2$ is satisfied (i.e., $G_2 \sqsubseteq Fulfilled\_Thing$), a DL reasoner will infer that $G_1 \sqsubseteq Fulfilled\_Thing$.\\
\begin{equation}\label{eq:eq_translation_satthing}
    \begin{aligned}
        Fulfilled\_Thing \equiv & \; \exists \; relate\_to. Fulfilled\_Thing \\
        relate\_to\_one \sqsubseteq & \; relate\_to \\
        interpret\_to \sqsubseteq & \; relate\_to\_one \\
    \end{aligned}
\end{equation}

%or, shortly, $ (G_2 \sqcup G_3) \sqsubseteq Fulfilled\_Thing$,
\vspace{6pt}
\noindent \emph{\textbf{One-to-many.}} ``Reduce'' is a one-to-many operator. For a reduce refinement, to fulfill the parent goal, all its sub-goals need to be fulfilled. This meaning can be formally stated as ``$\forall G_1 \; \exists G_{set} \; reduce\_to \; (G_1, G_{set}) \land fulfilled(G_{set}) \rightarrow fulfilled (G_1)$''. To capture such inference, we define \emph{ALL\_Fulfilled\_Thing } as a sub-concept of \emph{Fulfilled\_Thing} (Eq.~\ref{eq:eq_translation_allsatthing}). An element is an \emph{ALL\_Fulfilled\_Thing} only if all the elements related (specifically, reduced, focused, or operationalized) to it are fulfilled (Eq.~\ref{eq:eq_translation_allsatthing}). In the case a goal $G_1$ is reduced to $G_2$ and $G_3$, we need the following DL axioms: ``$G_1 \sqsubseteq \exists \; reduce\_to. G_2$'', ``$G_1 \sqsubseteq \exists \; reduce\_to. G_3$'', and ``$G_1 \sqsubseteq 2 \; reduce\_to.Thing$''. Now if $G_2$ and $G_3$ are satisfied, i.e., $G_2 \sqsubseteq Fulfilled\_Thing$ and $G_3 \sqsubseteq Fulfilled\_Thing$, then a DL reasoner will give the inference $G_1 \sqsubseteq ALL\_Fulfilled\_Thing \sqsubseteq Fulfilled\_Thing$. \\
\begin{equation}\label{eq:eq_translation_allsatthing}
    \begin{aligned}
        ALL\_Fulfilled\_Thing \sqsubseteq & \; Fulfilled\_Thing	\\
        ALL\_Fulfilled\_Thing \equiv & \; \forall relate\_to\_many. Fulfilled\_Thing \\
        relate\_to\_many \sqsubseteq & \; relate\_to \\
        reduce\_to \sqsubseteq & \; relate\_to\_many \\
    \end{aligned}
\end{equation}
%Interested readers can refer to Horkoff et al.~\cite{horkoff_making_2012} for more detail about the DL translation of the fulfillment semantics.

The other operators can be handled in a similar way. Take the ``Operationalize'' operator for example, we only need to: (1) add an DL assertion ``$operationalize\_to \sqsubseteq one\_to\_many$'' for the operator; (2) capture each operationalization using DL subsumptions, e.g., we will have ``$G_1 \sqsubseteq \exists \; operationalize\_to.F_2$'', ``$G_1 \sqsubseteq \exists \; operationalize\_to.F_3$'' and ``$G_1 \sqsubseteq 2 \; operationalize\_to. Thing$'' for the operationalization ``$\bm{O_p}(G_1)=\{F_2, F_3\}$''.

%\begin{comment}
As pointed by Horkoff et al.~\cite{horkoff_making_2012}, although this fulfillment reasoning is supported by existing goal-model reasoners~\cite{giorgini_reasoning_2003}\cite{sebastiani_simple_2004}, DL formulation allows exploration of less restrictive prorogation (intermediate form between AND and OR). For example, if a goal $G$ in AND-operationalized to 4 functions, we can relax the fulfillment of $G$ as such: if at least 3 functions are fulfilled, then $G$ is fulfilled. This is akin to setting a threshold for the degree of fulfillment of a ``good-enough'' QR (e.g., if a QR is fulfilled to a degree of 0.75, then it is said to be fulfilled) and can be done by using DL quantified number restriction as in Eq.~\ref{eq:eq_translation_approx_satthing}.\\
\begin{equation}\label{eq:eq_translation_approx_satthing}
    \begin{aligned}
        Approximate\_Fulfilled\_Thing \sqsubseteq & \; Fulfilled\_Thing	\\
        Approximate\_Fulfilled\_Thing \equiv & \; \geq3 \; relate\_to\_many.Fulfilled\_Thing\\
    \end{aligned}
\end{equation}
%\end{comment}

\section{Chapter Summary}
\label{sec:desiree_summary}
In this chapter, we have formalized the semantics of our \emph{Desiree} language and operators. For the language, we have provided the semantics by using set-theoretical expressions, and translated it to DL expressions in order to get a decidable reasoning with requirements. For the operators, we have discussed two kinds of semantics: \emph{entailment semantics}, which is of importance to requirements refinement; and \emph{fulfillment semantics}, which is of importance to the ``what-if'' analysis in goal-oriented requirements models, e.g., what kinds of elements in your model will be affected if some Function, FCs, or QCs are fulfilled while others are not? We have also discussed about how to use DL subsumption to simulate these two kinds of semantics.

%However, our DL translation is not able to capture the entailment semantics of nested $\bm{U}$, the semantics of which is expressed by using second-order (even higher-order) logic. For example, our DL translation is able to infer that $G_1 \models G_2$, where $G_1$ is ``the file search function shall take 30 seconds (at least) 80\% of the time'', and $G_2$ is ``the file search function shall take 30 seconds (at least) 70\% of the time''. However, it can not tell the entailment between $G_3$ ``80\% of the system functions shall be fast at 90\% of the time'' and $G_4$ ``90\% of the system functions shall be fast at 80\% of the time''.

%% file: tool.tex
%\vspace{4cm}
\vspace{-6pt}
\chapter{The \emph{Desiree} Tool}
\label{cha:tool}
\vspace{-6pt}

\setlength\epigraphwidth{0.8\textwidth}
\setlength\epigraphrule{0pt}

In this chapter, we present a prototype tool that is developed in support of our \textit{Desiree} framework. We first give an overview of the tool, and then introduce its key components in detail, and at last show an example model, demonstrating how to capture requirements.

\section{Overview}
\label{sec:tool_overview}

\begin{figure}[!htbp]
  \centering
  \vspace {-0.5 cm}
  \includegraphics[width=0.8\textwidth]{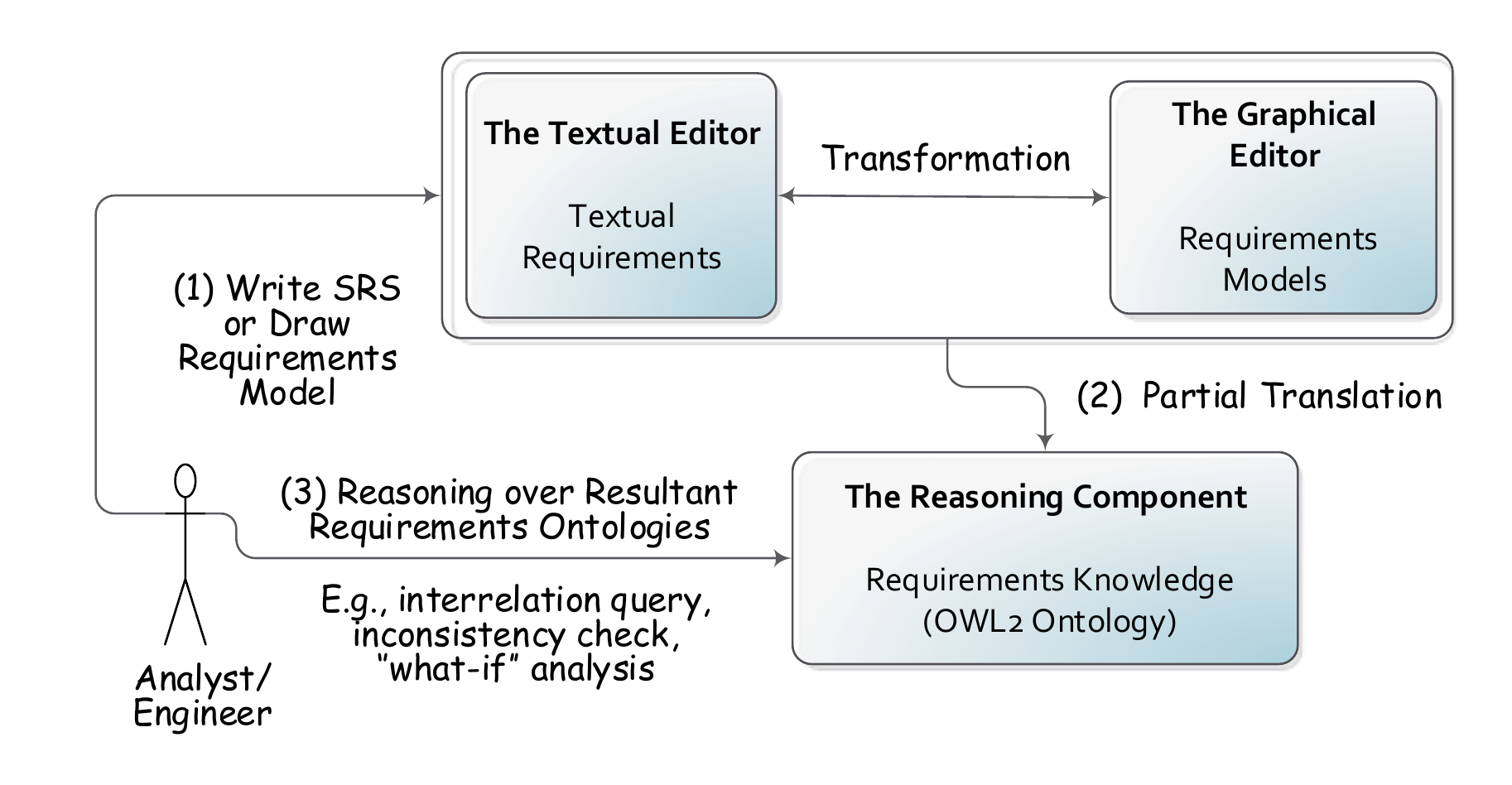}\\
  \vspace {-0.5 cm}
  \caption{The usage of the \emph{Desiree} tool}\label{fig:tool_desiree_usage}
\end{figure}

%We show the usage of our \emph{Desiree} tool in Fig.~\ref{fig:tool_desiree}.
%The usage of our \emph{Desiree} tool is shown in Fig.~\ref{fig:tool_desiree_usage}.
We show the usage of our \emph{Desiree} tool in Fig.~\ref{fig:tool_desiree_usage}. The \emph{Desiree} tool consists of three key components: (1) a textual editor, which allows analysts/engineers to write requirement using our language; (2) a graphical editor, which allows users to draw requirements models through a graphic user interface; (3) a reasoning component, which translates requirements (texts or models) to OWL2 ontologies, and make use of existing reasoners (e.g., Hermit~\cite{shearer_hermit:_2008}) to perform reasoning tasks.

\begin{figure}[!htbp]
  \centering
  \vspace {-0.1 cm}
  \includegraphics[width=0.8\textwidth]{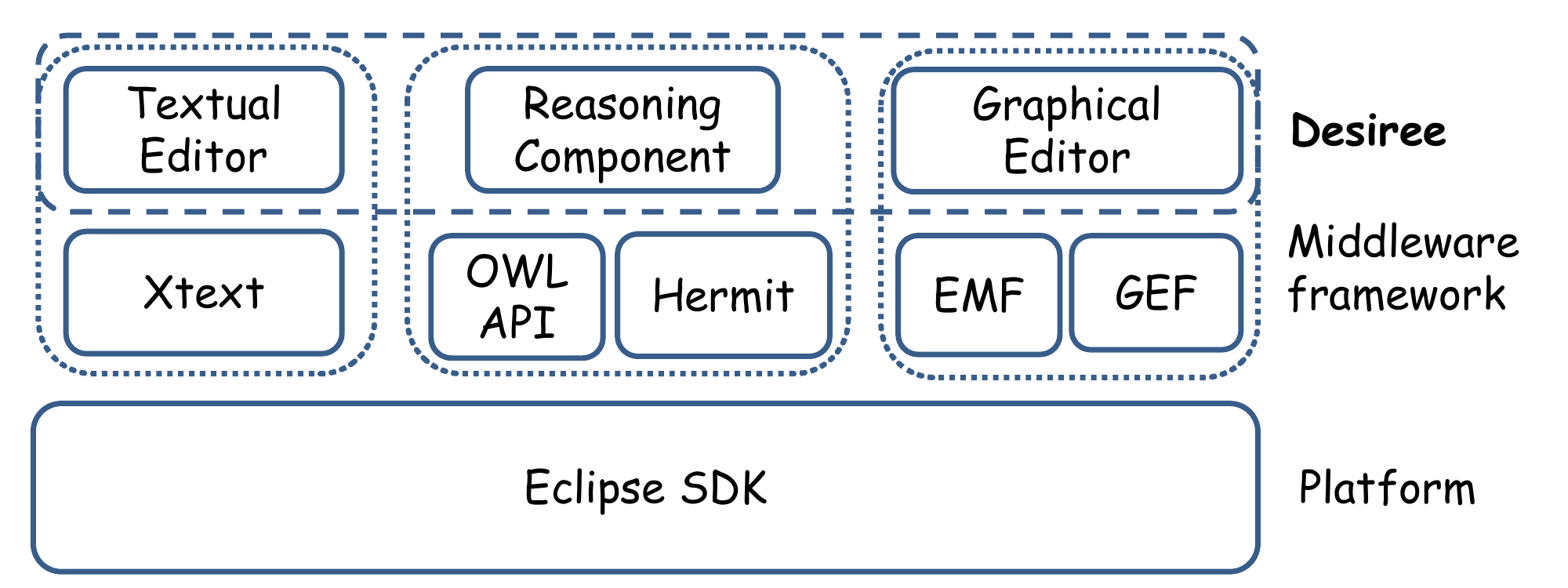}\\
  \vspace {-0.1 cm}
  \caption{The architecture of the \emph{Desiree} tool}\label{fig:tool_desiree_arc}
\end{figure}

%The architecture of our \emph{Desiree} tool is shown in Fig.~\ref{fig:tool_desiree_arc}.
We present the architecture of our \emph{Desiree} tool in Fig.~\ref{fig:tool_desiree_arc}. As a Rich Client Platform (RCP) application, the \emph{Desiree} tool is developed based on the Eclipse SDK 4.4.1~\footnote{http://archive.eclipse.org/eclipse/downloads/drops4/R-4.4.1-201409250400/}. Among the major components, the textual editor is developed based on Xtext 2.7.3~\footnote{https://eclipse.org/Xtext/}, a framework for developing domain specific languages; the graphical editor is developed based on EMF 2.10.2~\footnote{http://eclipse.org/modeling/emf/} (Eclipse Modeling Framework), an Eclipse-based modeling framework for data modeling and code generation, and GEF 3.9.100 ~\footnote{https://eclipse.org/gef/} (Graphical Editing Framework), which provides technology to realize rich graphical editors and views; the reasoning component is developed based on OWL API 3.2.4.1806~\footnote{http://owlapi.sourceforge.net/}, a Java API and reference implementation for creating, manipulating and serialising OWL Ontologies, and Hermit 1.3.8~\footnote{http://hermit-reasoner.com/}, a Java-based reasoner for DL ontologies.

The \emph{Desiree} Tool supports the following functionalities:

\begin{enumerate}
  \item Writing textual requirements specifications using the description-based syntax.
  \item Creating graphical requirements models using graphic notations.
  \item Transforming textual requirements specifications to graphical requirements models, and vice versa.
  \item Supporting automatic layout of modeling elements.
  \item Dynamically creating user-defined views, i.e., analysts/engineers are able to model the requirements of a project in multiple modeling views (canvas pages) as needed, instead of only one.
  \item Laying the modeling elements in a view, i.e., analysts/engineers can choose to show what kinds of elements.
  \item Transforming requirements models to OWL2 ontologies, and allowing analysts/engineers to perform reasoning tasks such as interrelations query, inconsistency check, and ``what-if'' analysis over the resultant requirements  ontologies.
\end{enumerate}

One of the key features of the \emph{Desiree} tool is its strong support for scalability. This support is two-fold: (1) we allow analysts/engineers to write textual requirements specifications, and automatically create graphical models from textual specifications with built-in automatic layout; (2) we allow analysts/engineers to dynamically create user-defined views for large models, with each view showing a sub-part of the model.

In the rest of this chapter, we will discuss each component in detail.

\subsection{The Textual Editor}
\label{sec:tool_textual_editor}
\begin{figure}[!htbp]
  \centering
  \includegraphics[width=\textwidth]{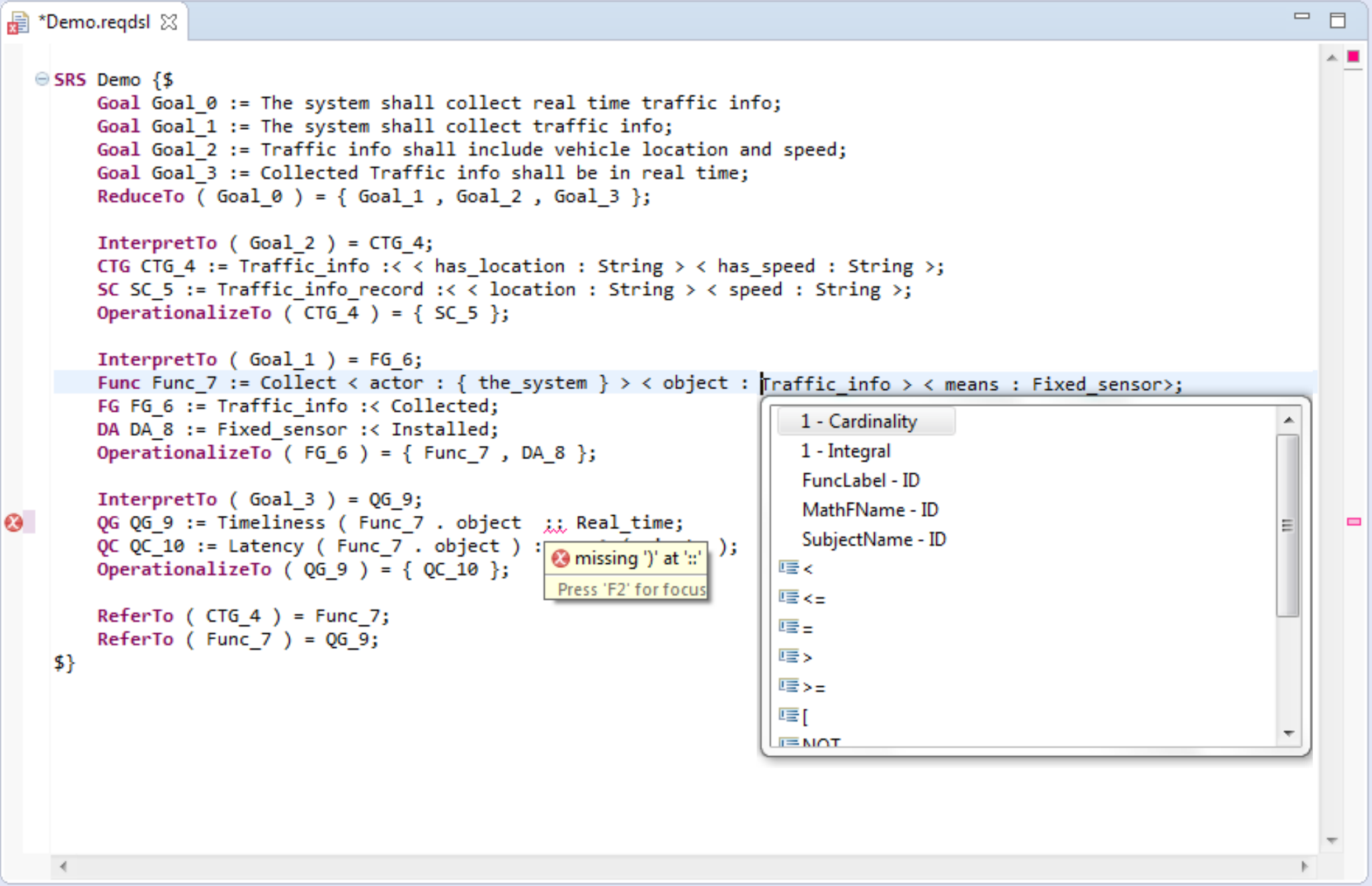}\\
  %\vspace {-0.2 cm}
  \caption{The textual editor}\label{fig:tool_textual_editor}
\end{figure}

%The textual editor is developed based on Xtext~\footnote{https://eclipse.org/Xtext/}, a framework developing domain specific languages.
Our textual editor offers several useful features: (1) syntax coloring, highlighting keywords in our language (e.g., the name of requirement concepts and operators, which are colored as red in in Fig.~\ref{fig:tool_textual_editor}); (2) content assistance, providing users with syntactic hints for our language (e.g., the drop-down box in Fig.~\ref{fig:tool_textual_editor} suggests the syntactic candidates for the description of the slot ``object'' when defining ``Func\_7''); (3) error checking, reporting syntax errors in requirements, e.g., the small pop-up window in Fig.~\ref{fig:tool_textual_editor} says that a ``)'' symbol is missing in the expression of ``QG\_9''.

\subsection{The Graphical Editor}
\label{sec:tool_graphical_editor}
%The graphical editor is developed based on the Eclipse Modeling Framework (EMF)~\footnote{http://eclipse.org/modeling/emf/} and the Graphical Editing Framework (GEF)~\footnote{https://eclipse.org/gef/}.

\begin{figure}[!htbp]
  \centering
  \includegraphics[width=\textwidth]{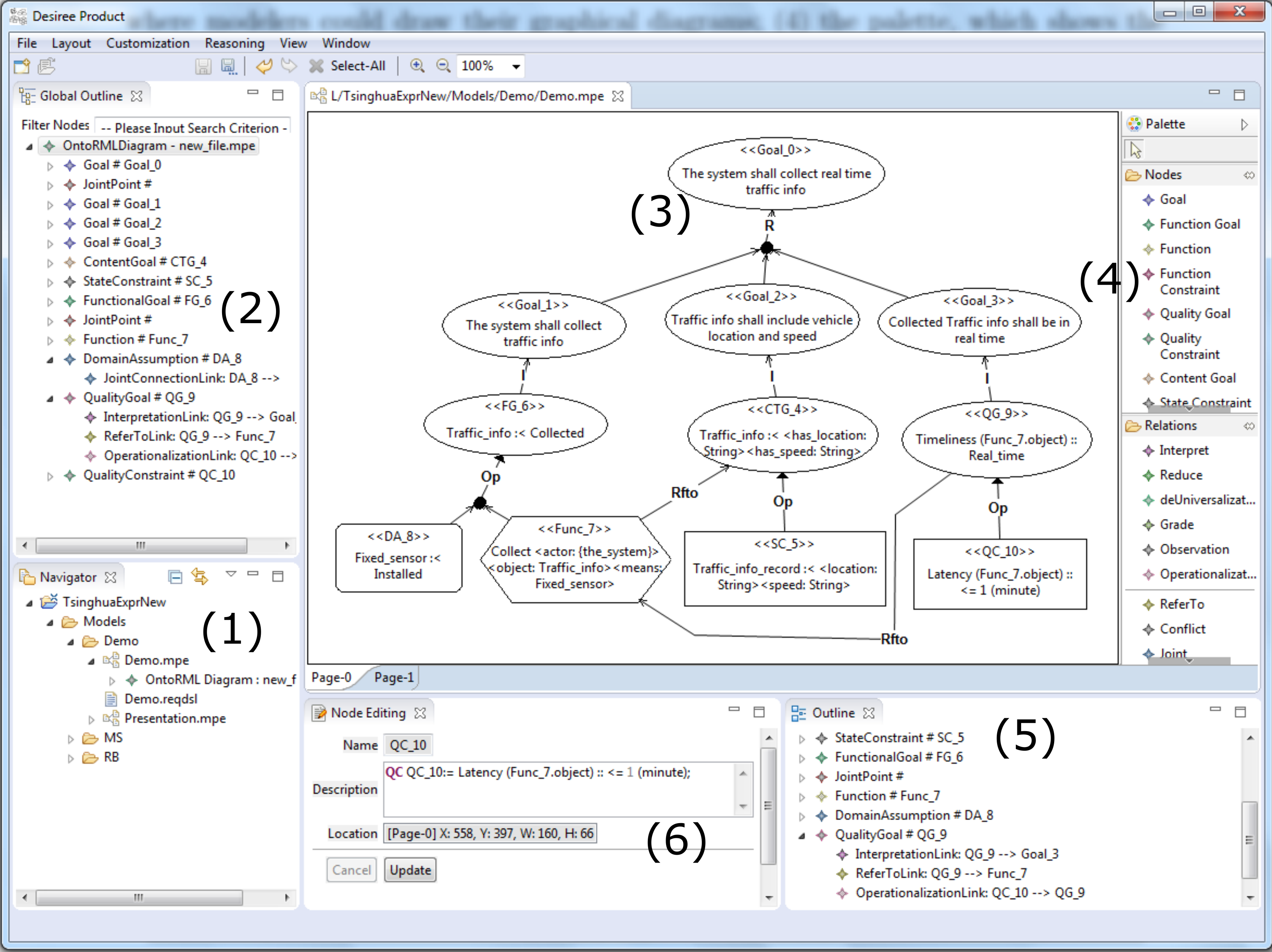}\\
  %\vspace {-0.2 cm}
  \caption{The graphical editor}\label{fig:tool_graphical_editor}
\end{figure}

The graphical editor supports all the 9 requirements concepts and 8 operators of the \emph{Desiree} framework, and allows analysts/engineers to write requirements using our syntax. As shown in Fig.~\ref{fig:tool_graphical_editor}, the graphical editor includes a set of tooling views~\footnote{We distinguish between tooling views and modeling views: a tooling view is a view of the \emph{Desiree} tool while a modeling view is a view of a requirements model that resides in the \emph{Desiree} tool.}: (1) the navigator, which displays the resources (e.g., folders, files) of a \emph{Desiree} project; (2) the global outline, which organizes all the modeling elements (concepts and operators) of a \emph{Desiree} model, which can be distributed in multiple canvas pages (modeling views); (3) the canvas page (modeling view), where analysts/engineers could draw their graphical diagrams; (4) the palette, which shows the graphical notations of the concepts and operators that analysts/engineers can use to build requirements models; (5) the local outline, which organizes the modeling elements (concepts and operators) in one canvas page (modeling view); (6) the editing view, where analysts/engineers are able to edit the concept nodes that are located in a canvas page (modeling view).
%The model in Fig .3 demonstrates the graphical representation in our tool.

Note that a \emph{Desiree} requirements model consists of multiple modeling views (canvas pages), e.g., the model in Fig.~\ref{fig:tool_graphical_editor} consists of two modeling views, ``Page-0" and ``Page-1''. For a requirements model, the global outline differs from a local outline in that the former sketches all the elements in the model (i.e., all the elements in its constituting canvas pages) while the latter outlines only the element in a specific canvas page.

Our graphical modeling tool provides several practical features: (1) it is able to transform textual requirements written in our syntax to graphical models with automatic layout, decreasing the efforts of analysts/engineers on drawing large models; (2) it maintains a central repository for each requirement model, and allows analysts/engineers to dynamically create views as needed (e.g., one can create a view for an important function and its refinement), aiming to address the scalability of large models; (3) it allows analysts/engineers to filter models by choosing specific requirement concepts or refinements (e.g., one can choose to see only the reduce refinements in a user-defined modeling view).

\subsection{The Reasoning Component}
\label{sec:tool_reasoning}
This component includes two parts: (1) a parser that is built on OWL API and translates requirements (texts or models) specified using our language to OWL2 ontologies; (2) a reasoning module that makes use of an existing reasoner, i.e., Hermit~\cite{shearer_hermit:_2008}, to perform reasoning tasks. The translation to OWL2 captures part of the semantics of our language (as its expressiveness is much weaker, e.g., the expressions of nested \textbf{U} cannot be supported), but this translation still allows us to do some interesting reasoning such as interrelations query, inconsistency check, and ``what-if'' analysis.
%For example, since the interrelations between qualities and functions is captured using our syntax for QGs, we can ask what kinds of subjects a quality will refer to, what kinds of qualities will be affected if we change a function, etc. See an example in our previous work~\cite{li_stakeholder_2015}.

\emph{Interrelations query.} Our description-based syntax captures a rich set of interrelations between requirement. For example, the QC ``Processing\_time ($F_1$) :: $\leq$ 10 (\emph{Sec.})'' (the processing time of searching meeting room records shall be less than 10 seconds) specifies a quality ``Processing\_time'' that inheres in a function ``$F_1$ := Search $<$actor: Meeting\_organizer$>$ $<$object: Meeting\_room\_record$>$'' (the system shall allow meeting organizers to search meeting room records), which refers to another content requirement ``Meeting\_room\_record $:<$ $<$has\_id: String$>$ $<$has\_name: String$>$$<$has\_type: String$>$'' (a meeting room record shall includes id, name and type). Using the reasoning power of DL subsumption, we can ask ``what kinds of functions does the quality \emph{processing\_time} inheres in?'', ``what kinds of qualities do users concern about the search function ($F_1$) ?'', ``what kinds of functions operate on meeting\_room\_records?'', so on and so forth.

\emph{Inconsistency check.} Once a requirements specification is translated into an OWL2 ontology, we are able to identify possibly inconsistent issues by utilizing the DL reasoning power. For example, in the \emph{Nursing Scheduler} project chosen from the PROMISE requirements dataset~\cite{menzies_promise_2012}, there are two potentially inconsistent requirements: ``Student\_ personal\_information $:<$ $<$accessed\_by: ONLY Authorized\_user$>$ '' (only authorized users shall have access to students' personal information) and ``Student\_personal\_information $:<$ $<$accessed\_by: ONLY \{Dr\_Susan\_Poslusny, Dr\_Julie\_Donalek\}$>$ '' (Dr Susan Poslusny and Dr Julie Donalek are the only people who shall have access to students' personal information). In this case, the Hermit reasoner would imply that Dr Susan Poslusny and Dr Julie Donalek are authorized users. If they have been explicitly stated as unauthorized user, the Hermit reasoner would report an inconsistency (supposing that analysts/engineers have stated that ``Authorized ($\cap$) Unauthorized $:<$ Nothing'').

\emph{``What-if'' analysis.} We have discussed the fulfillment semantics for each operator in Section \ref{sec:semantics_fulfillment} and how to simulate the fulfillment semantics using DL translation. The DL formulation allows us to reason about ``what if some specification elements (e.g., Fs, FCs, QCs or SCs) are fulfilled, how about the goals that they are derived from?''. To express that a specification element \emph{SE} is fulfilled, we add a DL axiom ``$SE \sqsubseteq Fulfilled\_Thing$'' for each element. For example, if a goal $G_1$ ``\emph{ticket be booked}'' is operationalized into a function $F_2$ ``Book $<$object: Ticket$>$'', which is further AND-reduced to $F_3$ ``Book $<$object: Airline\_ticket$>$ $<$means: Credit\_card$>$'' and $F_4$ ``Book $<$object: Bus\_ticket$>$ $<$means: Cash$>$'', we can ask ``what if $F_3$ and $F_4$ are fulfilled, how about $G_1$ ?'' by adding ``$F_3 \sqsubseteq Fulfilled\_Thing$'' and ``$F_4 \sqsubseteq Fulfilled\_Thing$''. The Hermit reasoner is then able to infer whether $F_2$ and $G_1$ are subclasses of ``\emph{Fulfilled\_Thing}'' or not.

\section{An Illustrative Example}
\label{cha:tool_illustration}
%Our \emph{Desiree} framework takes as input informal stakeholder requirements, and outputs an eligible specification through incremental applications of requirements operators.

We use a simple requirement ``\emph{The system shall collect real time traffic info}'' to illustrate the use of our three-staged \emph{Desiree} method and tool.

\begin{figure}[!htbp]
  \centering
  %\vspace {-0.1 cm}
  \includegraphics[width=1.0\textwidth]{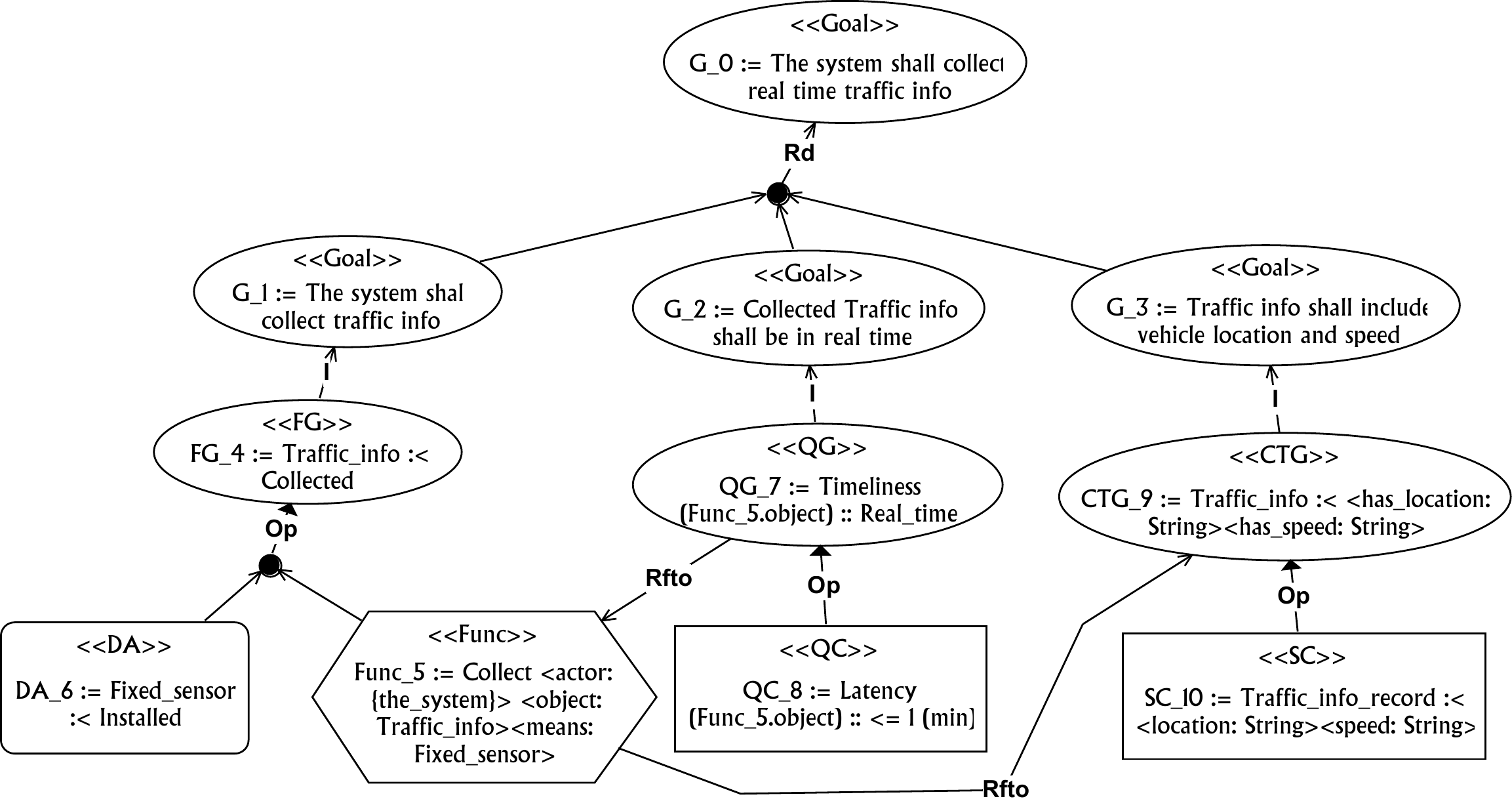}\\
  %\vspace {-0.2 cm}
  \caption{An illustrative example for the \emph{Desiree} method (with stereotypes on nodes)}\label{fig:tool_demo}
  %\vspace {-0.6 cm}
\end{figure}

\emph{The informal stage}. We first capture this requirement as a goal $G_0$. We then identify its concerns by asking ``\emph{what does it concern}?'': a function ``collect'', a quality ``timeliness'' of collected traffic info, and a content concern ``traffic info''. We then accordingly reduce $G_0$ to $G_1$, $G_2$ and $G_3$ by using the ``Reduce'' operator, as shown in in Fig. \ref{fig:tool_demo}.

%Consider it more carefully, we can find a missing content requirement: what does ``traffic info'' contain? We then accordingly reduce $G_0$ to three subgoals $G_1$, $G_2$ and $G_3$, as shown in Fig. \ref{fig:tool_demo}.

%\emph{The interpretation stage}. At this stage, we interpret $G_1$ to a functional goal $FG_6$, $G_2$ to a quality goal $QG_9$, $G_3$ to a content goal $CTG_4$, and encode the derived goals using our \emph{Desiree} syntax.

\emph{The interpretation stage}. At this stage, we interpret $G_1$ to a functional goal $FG_4$, $G_2$ to a quality goal $QG_7$, $G_3$ to a content goal $CTG_9$, and encode the derived goals using our \emph{Desiree} syntax. This step is conducted by employing the ``Interpret'' operator.

%This step is conducted by employing the ``Interpret'' operator.

%\emph{The smithing stage}. At this stage, we operationalize the structurally specified goals into specification elements. For example, we operationalize $FG_6$ ``Traffic\_info $:<$ Collected'' into a function ``$Func_7$ := Collect $<$actor: \{the\_system\}$>$ $<$object: Traffic\_info$>$ $<$means: Fixed\_sensor$>$'' and a domain assumption ``$DA_8$ := Fixed\_sensor $:<$ Installed''.

\emph{The smithing stage}. At this stage, we operationalize the structurally specified goals into specification elements. For example, we operationalize $FG_4$ ``Traffic\_info $:<$ Collected'' into a function ``$Func_5$ := Collect $<$actor: \{the\_system\}$>$ $<$object: Traffic\_info$>$ $<$means: Fixed\_sensor$>$'' and a domain assumption ``$DA_6$ := Fixed\_sensor $:<$ Installed''.

There are two ``\emph{refer\_to}'' relations in this model. The fist link means that the function ``$Func_5$'' operates on ``Traffic\_info'', which is defined in ``$CTG_9$''; the second one indicates that the quality goal ``$QG_7$'' takes the object of ``$Func_5$'' (i.e., traffic info collected by ``$Func_7$'', not traffic info in general) as subject, constraining it to be in real time.

\begin{figure}[!htbp]
  %\centering
  \hspace*{-0.8cm}
  \includegraphics[width=1.1\textwidth]{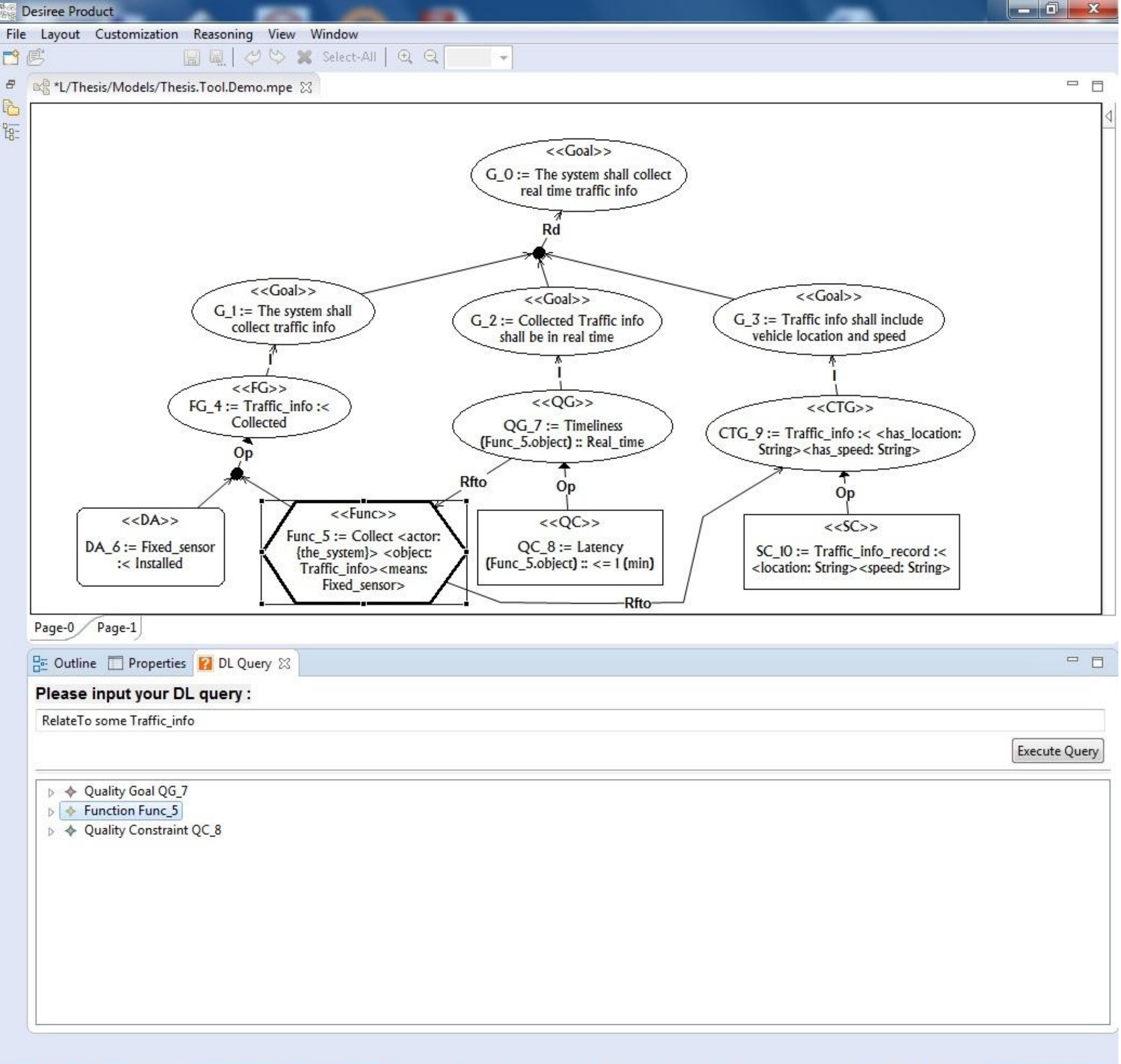}\\
  %\vspace {-0.3 cm}
  \caption{An example DL query}\label{fig:tool_query}
\end{figure}

Once we have derived the specification from requirement(s), our tool can automatically translate it into an OWL2 ontology, based on which we are able to perform some interesting reasoning tasks. For example, we can check which requirements are related to ``Traffic\_info'' as shown in Fig.~\ref{fig:tool_query}. We are also able to perform inconsistency check and ``what-if'' analysis over the resultant ontology piece, but in Prot\'eg\'e~\footnote{http://protege.stanford.edu/}, an open-source ontology editor and framework. The graphical representation for these two reasoning tasks will be developed in our next step work.

\section{Chapter summary}
\label{cha:too_chapter_summary}
In this chapter, we have introduced a prototype tool developed in support of the \textit{Desiree} framework, and demonstrated how to capture requirements by using the \emph{Desiree} method within the tool. We will use this tool for our experimental evaluations in the Section \ref{cha:evaluation}.

%% file: evaluation.tex
%\vspace{4cm}

\chapter{Evaluation}
\label{cha:evaluation}

\setlength\epigraphwidth{0.8\textwidth}
\setlength\epigraphrule{0pt}

In this chapter, we present a set of empirical evaluations conducted to evaluate our proposal, including (1) the coverage of our requirements ontology; (2) the expressiveness
of our description-based language; (3) the applicability of our methodology; and (4) the effectiveness of the entire \emph{Desiree} framework.

\section{The PROMISE Requirements Dataset}
\label{sec:eval_promise}

The PROMISE (PRedictOr Models in Software Engineering)~\cite{menzies_promise_2012} dataset includes 625 requirements collected from 15 software development projects~\footnote{http://openscience.us/repo/requirements/requirements-other/nfr.html}. Among the 625 requirements, 255 items are marked as functional requirements (FRs) and the remaining 370 non-functional requirements (NFRs) items are classified into 11 sub-categories, such as Security, Performance and Usability. Classification counts are shown in Table~\ref{tab:promise}.

\begin{table}[!htbp]
  \caption {Statistics of the PROMISE requirements data set}
  \label{tab:promise}
  \vspace {0.3 cm}
  \centering
  \begin{tabular}{|c|c|c|}
  \hline
  \textbf{} & \textbf{Sub-kind} & \textbf{Counts} \\ \hline
  \textbf{Functional} & Functional (F) & 255 \\ \hline
  \multirow{11}{*}{\textbf{Non-functional}} & Usability (US) & 67 \\
   & Security (SE) & 66 \\
   & Operational (O) & 62 \\
   & Performance (PE) & 54 \\
   & Look and Feel (LF) & 38 \\
   & Availability (A) & 21 \\
   & Scalability (SC) & 21 \\
   & Maintainability (MN) & 17 \\
   & Legal (L) & 13 \\
   & Fault tolerance (FT) & 10 \\
   & Portability (PO) & 1 \\ \hline
  Total & - & 625 \\ \hline
  \end{tabular}
\end{table}

These requirements were developed by DePaul University Master's students as part of a 10-week RE course project (in 2007, probably). The students were told to only write functional requirements for one feature, but to write a more general set of extra-functional requirements. They are also required to use the Volere template~\cite{robertson_mastering_2012}, which requires the categorization of requirements by quality type, to write NFRs. Approximately 80\% of the students in this class have two or more years of experience in the IT industry.

We use this requirements set for our evaluation. We classify all the 625 requirements to assess the coverage of our requirements ontology, rewrite all of them to evaluate the expressiveness of our description-based language. We conduct a realistic \emph{Meeting Scheduler} case study, which is selected from this set, to demonstrate our methodology. We use two of the projects, \emph{Meeting Scheduler} and \emph{Realtor Buddy}, for our controlled experiments, to evaluate the entire \emph{Desiree} framework.

\section{Evaluating the Requirements Ontology}
\label{sec:eval_ontology}

%We went over the full dataset, identified the key concern(s) of each requirement, and classified them by following our proposed classification guidelines. For the classification, we treated a functional requirement as a function (F) rather than a functional goal (FG), which is more commonly accepted in practice. For example, the requirement ``the system shall be able to send meeting notification'', which states an implicit functional goal ``meeting notification be sent'', was classified as a function. We did not distinguish between quality goals (QGs, vague quality requirements) and quality constraints (QCs, measurable quality requirements), and treat them as quality requirements in general (QGC). We show our classification counts in Table~\ref{tab:eval_ontology}, where ``+'' indicates a combination of concerns within a requirement (e.g., F+QGC means a mix of F and QGC).

We went over the full dataset, identified the key concern(s) of each requirement and classified them using our proposed classification guidelines. Our classification includes five basic categories: ``functional requirements (FR$'$)'', ``functional constraints (FC)'', ``quality requirement (QR)'', ``content requirements (CTR)'' and ``domain assumptions (DA)'', which will be accordingly modeled using the modeling elements shown in Table~\ref{tab:promise_classification_scheme}.

\begin{table}[!htbp]
    \caption {The classification scheme for the ontology evaluation}
    \label{tab:promise_classification_scheme}
    \vspace {0.3 cm}
    \centering
    \setlength\tabcolsep{2pt}
    \begin{tabular}{|c|c|}
    \hline
    \textbf{Classification Type} & \textbf{Modeling Elements} \\ \hline
    Functional requirement (FR$'$) & Functional goal (FG) $\lor$ Function (F) \\ \hline
    Functional constraint (FC) & Functional constraint (FC) \\ \hline
    Quality requirement (QR) & Quality goal (QG) $\lor$ Quality constraint (QC) \\ \hline
    Content requirement (CTR) & Content goal (CTG) $\lor$ State constraint (SC) \\ \hline
    Domain assumption (DA) & Domain assumption (DA) \\ \hline
    \end{tabular}
\end{table}

At this moment, we did not distinguish between functional goals (FGs, e.g., ``\emph{meeting notification be sent}'') and functions (Fs, e.g., ``\emph{the system shall be able to send meeting notification}''), ``quality goals (QGs)'' and ``quality constraints (QCs)'', and ``content goals (CTGs)'' and ``state constraints (SCs)'', and delay these distinctions to the follow-up interpretation (modeling) process/stage.

Note that ``functional constraints (FCs)'' are function related requirements. We separated them from functional requirements since they need more attention in practice (see the analysis of the classification result). As such, we use the symbol ``FR$'$'' to indicate functional requirements (FR) without functional constraints. We show our classification counts in Table~\ref{tab:eval_ontology}, where ``+'' indicates a combination of concerns within a requirement (e.g., FR$'$+QR means a mix of FR$'$ and QR).

\begin{table}[!htbp]
  \begin{threeparttable}
  \caption {The ontological classification of the 625 requirements}
  \label{tab:eval_ontology}
  \vspace {0.3 cm}
  \centering
  \small
  \setlength\tabcolsep{2pt}
  %\begin{tabularx}{\textwidth}{|c|c|c|c|c|c|c|c|c|c|}
  \begin{tabular}{|c|c|c|c|c|c|c|c|c|c|c|}
  \hline
  \textbf{Kind} & \textbf{Org.} & \textbf{FR$'$} & \textbf{QR} & \textbf{FC} & \textbf{CTR} & \textbf{FR$'$+QR} & \textbf{FC+QR} & \textbf{FR$'$+FC} & \textbf{FR$'$+CTR} & \textbf{DA}\\ \hline
    Functional & 255 & 183 & 6 & 9 & 21 & 1 & 0 & 6 & 29 & 0\\ \hline
    Usability & 67 & 7 & \textbf{46} & 2 & 0 & 11 & 1 & 0 & 0 & 0\\ \hline
    Security & 66 & 11 & 2 & 39 & 0 & 9 & 2 & 3 & 0 & 0\\ \hline
    Operational & 62 & 14 & 10 & 13 & 0 & 10 & 2 & 6 & 0 & 7\\ \hline
    Performance & 54 & 3 & \textbf{43} & 1 & 0 & 4 & 1 & 1 & 0 & 0\\ \hline
    Look and Feel & 38 & 9 & \textbf{20} & 0 & 1 & 6 & 2 & 0 & 0 & 0\\ \hline
    Availability & 21 & 0 & \textbf{20} & 1 & 0 & 0 & 0 & 0 & 0 & 0\\ \hline
    Scalability & 21 & 1 & \textbf{19} & 0 & 0 & 0 & 0 & 1 & 0 & 0\\ \hline
    Maintainability & 17 & 1 & 10 & 3 & 0 & 2 & 1 & 0 & 0 & 0\\ \hline
    Legal & 13 & 1 & 11 & 1 & 0 & 0 & 0 & 0 & 0 & 0\\ \hline
    Fault tolerance & 10 & 4 & 4 & 0 & 0 & 2 & 0 & 0 & 0 & 0\\ \hline
    Portability & 1 & 0 & 0 & 0 & 0 & 0 & 0 & 0 & 0 & 1\\ \hline
    Total & 625 & 234 & 191 & 69 & 22 & 46 & 9 & 17 & 29 & 8\\ \hline
  \end{tabular}
  \begin{tablenotes}
      \small
      \item Org.: original classification; FR$'$: functional requirement (except FC); FC: functional constraint; QR: quality requirement; CTR: content requirement; DA: domain assumption
    \end{tablenotes}
  \end{threeparttable}
\end{table}

From each row of table~\ref{tab:eval_ontology}, we can see how the original categorization of requirements is distributed across our ontological classification. For example, from the original 255 FRs, we identified 183 FR$'$s, 6 QRs, 9 FCs, 21 CTRs, 1 FR$'$+QR, 6 FR$'$+FCs, and 29 FR$'$+CTRs. Here 51 out of the 255 (20\%) FRs concern content. We found that most of the original security NFRs are often functional (FR$'$) or functional constraint (FC) related (the third row): 97\% of them (66 in total) are identified as FR$'$s, FCs, or combination with other concerns (11 FR$'$s, 39 FCs, 9 FR$'$+QRs, 2 FC+QRs, and 3 FR$'$+FCs). For example, ``\emph{only managers are able to deactivate user accounts}'', originally classified as a security NFR, should actually be a functional constraint (FC). This is because the system needs to check whether the actor is a manager or not when the deactivation function is accessed. In the PROMISE requirements set, there are many requirements of the form ``only users with $<$role$>$ are allowed to perform $<$action$>$ or access $<$asset$>$''. In addition, we discovered that 16.16\% of the requirements (101/625) mix multiple concerns (i.e., with ``+'' in their labels).

Examining the data more closely, we found that 148 out of the 191 QRs (148/191, 77.49\%) are classified under usability, performance, availability, look and feel, and scalability. We further analyzed the 246 quality-related NFRs (191 QRs, 46 FR$'$+QRs and 9 FC+QRs), and identified 67 unique qualities with 327 occurrences. The most frequent ones are operability, availability, processing/response time, and scalability. That is, these kinds of quality requirements may appear more frequently in a requirements specification.

Our evaluation shows that functional constraints (FCs), content requirements (CTRs), and the mix of concerns such as FR$'$+FC, FR$'$+QR, and FR$'$+CTR are not trivial and need more attention in practice. The results also provide evidence that our requirements ontology is adequate for covering requirements in practice.

%\begin{figure}[!htbp]
  %\centering
  %\vspace {-0.5 cm}
  %\includegraphics[width=0.8\textwidth]{./pics/Tool_framework.pdf}\\
  %\vspace {-0.5 cm}
  %\caption{The Desiree tool}\label{fig:tool_desiree}
%\end{figure}

\section{Evaluating the Requirements Language}
\label{sec:eval_language}

\subsection{Evaluating the Language Expressiveness}
\label{sec:eval_language_expressiveness}

After classification, we rewrote the set of all 625 requirements using our language to evaluate its expressiveness. In this step, we separated the concerns of a requirement if it was composite, and encoded it by using the description-based syntax. Our syntax was able to capture all 625 requirements, resulting in 1276 statements (nearly double the amount of original requirements), including 419 FGs/Fs, 313 FCs, 375 QGs/QCs, 90 CTGs, and 79 DAs. Note that there are 7 instance-level constraints (7/625, 1.12\%) identified in the evaluation. We were able to express these constraints by using the ``\emph{same\_as}'' DL constructor~\cite{cohen_learning_1994}; however, the use of ``\emph{same\_as}'' imposes severe limitations on reasoning.

The counts of each type of statement in our language does not strictly correspond to the classification counts in Table~\ref{tab:eval_ontology}. For example, we have 22 CTRs and 29 FR$'$+CTRs in Table~\ref{tab:eval_ontology}, but ultimately 90 rather than 51 CTGs in this run. This is because the original dataset includes many composite and nested requirements, such as sibling functions, nested qualities and content. For example, there are two CTGs, namely ``\emph{billing information}'' and ``\emph{contact information}'', in the requirement ``\emph{users shall be able to update their billing and contact information}''. We broke these up into separate requirements when encoding them. In addition, we treated domain knowledge as domain assumption(s). For instance, ``\emph{Open source examples include Apache web server Tomcat}'' was captured as ``DA := Tomcat $:<$ Web\_server $\land$ Open\_source''.

Our language and guidelines facilitate the identification of ambiguity. In the interpretation process, we identified 24 ambiguous requirements (24/625, 3.84\%), and eliminated the ambiguity by choosing the most likely interpretation. For example, ``\emph{notify users with email}'' will be encoded as ``notify $<$object: user$>$ $<$means: email$>$''. Note that although we could have found some ambiguities by reading natural language requirements text, using a more ad-hoc, less systematic approach, such an approach would likely cause us to miss many ambiguities; as such naive approaches do not force the user to carefully analyze and classify the requirement text. This is evidenced in our controlled experiments in Section~\ref{sec:eval_framework}. Furthermore, once ambiguities are found, an ad-hoc approach would not tell us what to do when an ambiguous requirement is found. Our approach provides a systematic way for not only identifying but also dealing with ambiguities in requirements.

Our language and guidelines also contribute to making requirements more accurate and complete. For example, for a rather informal statement ``\emph{the product shall make the users want to use it}'', we can identify its focus by asking ``\emph{what does it concern}?'', and restate it as a quality goal ``$QG_1$ := Attractiveness (\{the\_product\}) :: Good'', which can be further refined/operationalized, e.g., ``$QC_2$ := Number\_of\_users (\{the\_product\} $<$period: One\_week\_after\_its\_launch$>$) :: $\geq$ 1000''. Moreover, to satisfice $QG_1$, we may need to design in the system certain functions and/or functional constraints, e.g., ``\emph{the product shall use site maps for navigation}'', which can be captured as ``$F_3$ := Navigate $<$actor: \{the\_product\}$>$$<$object: User$>$'' and ``$FC_4$ := $F_3$ $:<$ $<$means: Site\_map$>$''.

\subsection{Evaluating the Need of the U, G, O Operators}
\label{sec:eval_language_need}
We focused on analyzing the 370 NFRs in the PROMISE requirements set, in order to evaluate the need of the three operators, namely ``de-Universalize''(\textbf{U}), ``Scale''(\textbf{G}) and ``Observe''(\textbf{O}), which are introduced for relaxing quality requirements (QGs and QCs).

Our analysis of the 370 NFRs resulted in 481 requirements statements, which were further classified as shown in Table~\ref{tab:nfr_satisfication_type}. Note that a statement can be tagged with more than one type, e.g., ``\emph{all users shall be authenticated}'' is practically unsatisfiable, but also measurable, thus the sum of this classification is greater than 481 (in fact, 567). In this classification, we found 15.17\% (86/567) of the statements to be practically unsatisfiable, 25.22\% (143/567) vague and only 58.73\% (333/567) measurable.

\begin{table}[!htbp]
    \caption {The statistics of satisfaction type}
    \label{tab:nfr_satisfication_type}
    \vspace {0.3 cm}
    \centering
    \begin{tabular}{|c|c|}
    \hline
    \textbf{Satisfaction Type} & \textbf{\#NFRs} \\ \hline
    Ambiguous & 5 \\ \hline
    Unsatisfiable & 86 \\ \hline
    Vague & 143 \\ \hline
    Measurable & 333 \\ \hline
    Total & 567 \\ \hline
    \end{tabular}
\end{table}

We analyzed the implicit application of \textbf{U}, \textbf{G}, and \textbf{O} to the 481 requirements statements. For example, ``\emph{$80\%$ of the users shall report the user interface is simple}'' captures an observation (i.e., an application of \textbf{O}) and a de-Universalization (\textbf{U}). We show these counts in Table~\ref{tab:nfr_implicit_application}: 50 \textbf{U} (i.e., stating percentages), 10 \textbf{G}, and 16 \textbf{O}.
%Our analysis shows that few of the statements have (implicitly) used U and A. Particularly, G is rarely used. Meanwhile, we found that many NFRs, which need to be relaxed to become satisfiable and measurable, have not been adequately dealt with.

\begin{table}[!htbp]
    \caption {Implicit presence of the U, G, O operators on the 370 NFRs}
    \label{tab:nfr_implicit_application}
    \vspace {0.3 cm}
    \centering
    \begin{tabular}{|c|c|}
    \hline
    \textbf{Weakening operators} & \textbf{\#NFRs} \\ \hline
    de-Universalize (U) & 50 \\ \hline
    Scale (G) & 10 \\ \hline
    Observe (O) & 16 \\ \hline
    Total & 76 \\ \hline
    \end{tabular}
\end{table}

%to be properly treated
Among the 481 statements, 86 of them (in addition to the 50 implicit presence of \textbf{U}) implicitly or explicitly use universal quantifiers, e.g., ``all'', ``any'' and ``each''. They are counted as (potentially) unsatisfiable in Table~\ref{tab:nfr_satisfication_type}, and likely need to be relaxed by using \textbf{U}. Also, 36 subjective statements are identified, e.g., ``look'', ``readability'', ``usefulness'', etc., indicating that at least another 20 requirements (except the 16 presence of \textbf{O}) should be observed by using \textbf{O}. Lastly, \textbf{G} could be applied to unsatisfiable, vague or measurable requirements, thus all the 476 items (except the 5 ambiguous ones) are candidates for the \textbf{G} operator, but only 10 actually (implicitly) used it (e.g., ``almost all'', ``fast enough''). For example, ``\emph{the interface shall be appealing}'' as found in the dataset, is clearly gradable.

Our analysis shows that many NFRs are (practically) unsatisfiable, vague, and subjective, demonstrating the need for the operators as introduced in our language.

\subsection{Lessons Learned}
\label{sec:eval_language_lessons}

When rewriting the set of all 625 natural language requirements, we observed several important phenomena, including instance-level constraints, contextualized requirements, temporal relations, the use of ``AND'', ``OR'', and universal quantifiers. We discuss them in detail below.

\vspace {6pt}
\noindent\textbf{Instance-level Constraints.} In the interpretation process, we have found 20 (20/625, 3.20\%) instance-level constraints. For 13 out of the 20 items, we are able to capture them by using \emph{role chain}, which is composed of DL roles (i.e., slots) through the composition operator ($\circ$)~\cite{baader_description_2003}. For example, the requirement ``\emph{administrators shall be able to add a student who has registered for a clinical class to a clinical lab section for that class}'' can be captured as in Example 1 below. We first define $F_1$, which includes three established relations (slots): ``\emph{registered\_for}'' (domain: \emph{Students}; co-domain: \emph{Clinical\_classes}), ``\emph{associated\_with}'' (domain: \emph{Lab\_sections}; co-domain: \emph{Clinical\_classes}) and ``\emph{added\_to}'' (domain: \emph{Students}; co-domain: \emph{Lab\_sections}), and then use $FC_1$ to reflect the restriction: the set of student who have been added to a lab section that is associated with a class, shall be subsumed by the set of students who have registered for that class.

%, and supported by Web Ontology Language (OWL)~\cite{mcguinness_owl_2004}.
\begin{table}[!htbp]
    \centering
    \small
    \begin{tabular}{|rp{0.75\textwidth}|}
        \hline
        \multicolumn{2}{l}{|\textbf{Example 1}|} \\ \hline
        $F_1$ := & Add $<$actor: Administrator$>$ $<$object: Student $<$registered\_for: Clinical\_class$>>$
        $<$target: Lab\_section $<$associated\_with: Clinical\_class$>>$  \\
        $FC_1$ := & added\_to $\circ$ associated\_with $:<$ registered\_for \\
        \hline
    \end{tabular}
\end{table}

To express the other 7 examples (7/625, 1.12\%), we use the ``\emph{same\_as}'' DL constructor~\cite{borgida_adding_1999}, which is more generally known as ``\emph{role value map}''~\cite{schmidt-schaus_s_subsumption_1988}. For example, ``\emph{Managers shall be able to move a student from one clinical lab section to another clinical lab section corresponding to the same clinical class}'' can be captured as in Example 2, where we use ``\emph{same\_as}'' to reflect ``{the same clinical class}'' restriction. Note that the ``\emph{same\_as}'' constructor would cause the problem of undecidability on its general form~\cite{schmidt-schaus_s_subsumption_1988}, and thus imposing limitations on reasoning.

\begin{table}[!htbp]
    \centering
    \small
    \begin{tabular}{|rp{0.75\textwidth}|}
        \hline
        \multicolumn{2}{l}{|\textbf{Example 2}|} \\ \hline
        $F_2$ := & Move $<$actor: Manager$>$ $<$object: Student$>$
                      $<$source: Lab\_section $<$associated\_with: Clinical\_class$>>$
                      $<$target: Lab\_section $<$associated \_with: Clinical\_class$>>$  \\
                % & \emph{\textbf{same\_as}} (source $\circ$ associated\_with, target $\circ$ associated\_with) \\
        \multicolumn{2}{|l|}{\emph{\textbf{same\_as}} (source $\circ$ associated\_with, target $\circ$ associated\_with)} \\
        \hline
    \end{tabular}
\end{table}

One alternative way is to use SWRL rules~\cite{horrocks_swrl:_2004}. For instance, we can write the rule as shown in Example 3: if the lab sections \emph{y} and \emph{z} are different and belong to the same clinical class \emph{c}, then a student \emph{x} can be moved from \emph{y} to \emph{z}. Each time when moving a student \emph{x} from lab section \emph{y} to \emph{z}, the system needs to check whether both \emph{in\_lab\_section} (?\emph{x}, ?\emph{y}) and \emph{moveable} (?\emph{x}, ?\emph{z}) hold.

\begin{table}[!htbp]
    \centering
    \small
    \begin{tabular}{|rp{0.72\textwidth}|}
        \hline
       \multicolumn{2}{l}{|\textbf{Example 3}|} \\ \hline
        $SWRL_3$ := & Student(?\emph{x}) $\land$ Clinical\_class(?\emph{c}) $\land$ Clinical\_section (?\emph{y}) $\land$ Clinical\_section (?\emph{z}) $\land$ differ\_from (?\emph{y}, ?\emph{z}) $\land$ belong\_to\_class (?\emph{y}, ?\emph{c}) $\land$ belong\_to\_class (?\emph{z}, ?\emph{c}) $\land$ in\_lab\_section (?\emph{x}, ?\emph{y}) $\Rightarrow$ moveable (?\emph{x}, ?\emph{z})
        \\ \hline
    \end{tabular}
\end{table}

\vspace {6pt}
\noindent\textbf{Contextualized Requirements.} In the evaluation, we have found 40 contextualized requirements (40/625, 6.40\%), which specify pre-conditions (15/40) or triggers (25/40). In general, we can treat a pre-condition/trigger as a property of a corresponding function, and denote it as a slot-description pair of that function. For example, ``\emph{the system shall notify the realtor in a timely fashion when a seller or buyer responds to an appointment request}'' can be captured as ``Notify $<$subject: \{the\_system\}$>$ $<$target: Realtor$>$$<$trigger: Respond $<$subject: Seller $\lor$ Buyer$>$ $<$target: Appointment\_request$>>$''). If we plan to accommodate evolutionary changes (e.g., adding a pre-condition/trigger to a function) in the future, it is better to specify contextualized
requirements by using functional constraints as shown in Example 4.
%We treat a pre-condition/trigger as a property of a corresponding function. For example, ``the system shall notify the realtor in a timely fashion when a seller or buyer responds to an appointment request'' can be captured as FC4-2 in Example 4.

\begin{table}[!htbp]
    \centering
    \small
    \begin{tabular}{|rp{0.7\textwidth}|}
        \hline
        \multicolumn{2}{l}{|\textbf{Example 4}|} \\ \hline
        $F_{4-1}$ := & Notify $<$subject: {the\_system}$>$ $<$target: Realtor$>$ \\
        $FC_{4-2}$ := & $F_{4-1}$ $:<$ $<$trigger: Respond $<$subject: Seller $\lor$ Buyer$>$ $<$target: Appointment\_request$>>$ \\
        %QG4-3 := & Timeliness (F4-1) :: Timely \\
        \hline
    \end{tabular}
\end{table}

%an attributive component
%Note that we could also denote a pre-condition/trigger as a slot-restriction of a function (e.g., ``notify $<$subject: {the\_system}$>$ $<$target: realtor$>$$<$trigger: respond $<$subject: seller $\lor$ buyer$>$ $<$target: appointment\_request$>>$''). We suggest specify contextualized requirements by using FCs in order to support possible evolutionary changes (e.g., adding a pre-condition/trigger to a function) in the future.

\vspace {6pt}
\noindent\textbf{Temporal Relations.} In the evaluation, we have identified 6 requirements (6/625, 0.96\%) that involve temporal relations, such as ``after'', ``before'', ``prior to'', ``since'' and ``until''. We capture these temporal relations as axioms or slots.
%Before ($F_{5-1}$, $F_{5-2}$)} \\

\begin{enumerate}
  \item If a temporal constraints is imposed on two functions, we capture each FR as a function and model the temporal relation as a link between them. For example, ``\emph{the website shall authorize credit card payment before allowing a user to stream a movie}'' can be captured as in Example 5.
  \begin{table}[!htbp]
    \centering
    \small
    %\vspace{-0.2cm}
    \begin{tabular}{|rp{0.72\textwidth}|}
        \hline
        \multicolumn{2}{l}{|\textbf{Example 5}|} \\ \hline
        $F_{5-1}$ := & Authorize $<$actor: \{the\_product\}$>$$<$object: Credit\_info $<$associated \_with: User $>$$>$ \\
        $F_{5-2}$ := & Stream $<$subject: \{the\_product\}$>$ $<$actor: User$>$$<$object: Movie$>$ \\
        %             & Before $<F_{5-1}, \; F_{5-2}>$ \\
        $FC_{5-3}$ := & $F_{5-1}$ $:<$ $<$before: $F_{5-2}$$>$ \\
        %\multicolumn{2}{|l|}{\,\,\, \, $F_{5-1}$ $:<$ \; $<$before: $F_{5-2}$$>$} \\
        \hline
    \end{tabular}
  \end{table}
  \vspace{-0.2cm}

  \item If a temporal constraint is imposed on a function and a timepoint/state, we capture such temporal relation as a slot of the corresponding function. For instance, ``\emph{the system shall allow on demand generation of IQA} (\emph{Inventory Quantity Adjustment}) \emph{documents since certain point of time}'' can be captured as in Example 6, where ``Certain\_time\_point'' is vague and needs to be further refined.
  \begin{table}[!htbp]
    \centering
    \small
    %\vspace{-0.2cm}
    \begin{tabular}{|rp{0.75\textwidth}|}
        \hline
        \multicolumn{2}{l}{|\textbf{Example 6}|} \\ \hline
        $F_{6-1}$ := & Generate $<$subject: \{the\_product\}$>$$<$object: IQA\_document$>$ $<$since: Certain\_time\_point$>$ \\
        $FC_{6-2}$ := & $F_{6-1}$ $:<$ $<$trigger: Requested$>$ \\
        \hline
    \end{tabular}
\end{table}

\end{enumerate}

\vspace {6pt}
\noindent\textbf{The Use of ``AND'' in NL Requirements.} We found in our evaluation the use of ``AND'' in 206 requirements (206/625, 32.96\%). This simple but common conjunction word has many different kinds of interpretations.

%  \item Sibling functions or subjects of QGs/QCs (62/206, 30.10\%): we define them as separate functions or QGs/QCs. For instance, ``The product shall use symbols and words that are naturally understandable by the realtor community'' can be captured as two QGs, as shown in Example 7.

\begin{enumerate}
  \item Sibling functions, qualities or subjects (62/206, 30.10\%): we capture them as separate functions, QGs or QCs. For instance, the requirement ``\emph{The system shall allow managers to add, update and delete students}'' can be captured as three functions, which are shown in Example 7.

      \begin{table}[!htbp]
        \centering
        \small
        %\vspace{-36pt}
        \setlength\tabcolsep{2pt}
        \begin{tabular}{|rp{0.75\textwidth}|}
            \hline
            \multicolumn{2}{l}{|\textbf{Example 7}|} \\ \hline
            %QG7-1 := & Understandability (symbol $<$ref\_actor: Realtor\_community$>$) :: natural \\
            %QG7-2 := & Understandability (word $<$ref\_actor: Realtor\_community$>$) :: natural\\
            $F_{7-1}$ := & Add $<$subject: \{the\_system\}$>$$<$actor: Manager$>$$<$oject: Student$>$\\
            $F_{7-2}$ := & Update $<$subject: \{the\_system\}$>$$<$actor: Manager$>$$<$oject: Student$>$\\
            $F_{7-3}$ := & Delete $<$subject: \{the\_system\}$>$$<$actor: Manager$>$$<$oject: Student$>$\\
            \hline
        \end{tabular}
      \end{table}
      \vspace{-0.2cm}

  \item Conjunct information that needs to be integrated (59/206, 28.64\%): we capture them as content goals (CTGs, akin to UML classes). For example, to capture the requirement ``\emph{the system shall display movie title, actor, and director}'', we first define a CTG and then specify a function as in Example 8.

      \begin{table}[!htbp]
        \centering
        \small
        %\vspace{-36pt}
        \setlength\tabcolsep{2pt}
        \begin{tabular}{|rp{0.75\textwidth}|}
            \hline
            \multicolumn{2}{l}{|\textbf{Example 8}|} \\ \hline
            $CTG_{8-1}$ := & Movie\_detail\_info $:<$$<$has\_title: String$>$ $<$has\_actor: Person$>$ $<$has\_ director: Person$>$ \\
            $F_{8-2}$ := & Display $<$actor: \{the\_system\}$>$ $<$object: Movie\_detail\_info$>$ \\
            \hline
        \end{tabular}
      \end{table}
        \vspace{-0.2cm}

  \item Conjunct information that needs to be separated (33/206, 16.02\%): we use ``$\lor$'' and reduce it to separate functions. For example, ``\emph{users shall be able to update their billing and contact information}'' can be captured as in Example 9, where $F_9$ is OR-reduced into $F_{9-1}$ and $F_{9-2}$. Note that here ``OR ($\lor$) '' means alternatively at run-time, and $F_{9-1}$ and $F_{9-2}$ need to be implemented at design time.

      \begin{table}[!htbp]
        \centering
        \small
        %\vspace{-36pt}
        \setlength\tabcolsep{2pt}
        \begin{tabular}{|rp{0.75\textwidth}|}
            \hline
            \multicolumn{2}{l}{|\textbf{Example 9}|} \\ \hline
            $F_9$ := & Update $<$actor: User$>$$<$object: Billing\_info $\lor$ Contact\_info $>$ \\
            $F_{9-1}$ := & Update $<$actor: User$>$$<$object: Billing\_info$>$ \\
            $F_{9-2}$ := & Update $<$actor: User$>$$<$object: Contact\_info$>$ \\
            %FC9-3 := & \{F9-1\} $:<$ $<$inhere\_in: \{the\_system\}$>$ \\
            %FC9-4 := & \{F9-2\} $:<$ $<$inhere\_in: \{the\_system\}$>$ \\
            \hline
        \end{tabular}
      \end{table}
      \vspace{-0.2cm}

  \item Set of individuals (12/206, 5.83\%): we define them as new types of entities based on set. For example, ``\emph{the system must be able to interface with the following browsers: IE 5.x, 6.0, Netscape 6.x, 7.x, 8.x, and Firefox 1.0} (6 kinds of browsers in total)'' can be captured as in Example 10.

      \begin{table}[!htbp]
        \centering
        \small
        %\vspace{-36pt}
        \setlength\tabcolsep{2pt}
        \begin{tabular}{|rp{0.75\textwidth}|}
            \hline
            \multicolumn{2}{l}{|\textbf{Example 10}|} \\ \hline
            $DA_{10-1}$ := & Required\_Browser $\equiv$ \{IE 5.x, 6.0, Netscape 6.x, 7.x, 8.x, Firefox 1.0\}\\
            $F_{10-2}$  := & Interface $<$actor: \{the\_system\}$>$ $<$object: Required\_Browser$>$ \\
            $QC_{10-3}$ := & Number\_of\_browser\_kind ($F_{10-2}$.object) :: \{6\} \\
            \hline
        \end{tabular}
      \end{table}
      \vspace{-0.2cm}

  \item Logic ``OR'' (13/206, 6.31\%): we use ``$\lor$''. For instance, for ``\emph{the system will notify affected parties when changes occur affecting clinical, including but not limited to clinical section capacity changes, and clinical section cancellations}'', either kind of changes would trigger the notification (see Example 11).

      \begin{table}[!htbp]
        \centering
        \small
        %\vspace{-36pt}
        \setlength\tabcolsep{2pt}
        \begin{tabular}{|rp{0.75\textwidth}|}
            \hline
            \multicolumn{2}{l}{|\textbf{Example 11}|} \\ \hline
            $F_{11-1}$ := & Notify $<$actor: \{the\_system\}$>$$<$object: Affected\_parts$>$ \\
            $FC_{11-2}$ := & $F_{11-1}$ $:<$ $<$trigger: Occur $<$subject: Capacity\_change $\lor$ Cancellation$>>$ \\
            \hline
        \end{tabular}
      \end{table}
      \vspace{-0.2cm}

  \item Logic ``AND'' (15/206, 7.28\%): we use the ``$\land$''. For example, ``\emph{the product shall be able to be operated in a repair facility during dirty and noisy conditions}'' can be captured in Example 12.

      \begin{table}[!htbp]
        \centering
        \small
        %\vspace{-36pt}
        \setlength\tabcolsep{2pt}
        \begin{tabular}{|rp{0.72\textwidth}|}
            \hline
            \multicolumn{2}{l}{|\textbf{Example 12}|} \\ \hline
            $QG_{12}$ := & Availability (\{the\_product\} $<$location: Repair\_facility$>$ $<$situation: Dirty  $\land$  Noisy$>$): 100\% \\
            \hline
        \end{tabular}
      \end{table}
      \vspace{-0.2cm}

  \item Compound words: we define them as concepts (12/206, 5.83\%). For example, ``\emph{look and feel}'' indicates an interface, ``\emph{between 8 AM and 6 PM}'' means a period, etc.
\end{enumerate}

\vspace{6pt}
\noindent\textbf{ The Use of ``OR'' in NL Requirements.} We found in our evaluation the use of ``OR'' in 52 requirements (52/625, 8.32\%). Being similar with ``AND'', the term ``OR'' in NL requirements also has different kinds of interpretations.

\begin{enumerate}
  \item Sibling functions, qualities or subjects (6/52, 11.54\%): we define separate functions, QGs or QCs. For example, ``\emph{the system shall update or create new property listings}'' is captured in Example 13, akin to the first interpretation of ``AND'' (Example 7).
      %, except that we need two extra FCs, which constrain both functions to be implemented in the system.

      \begin{table}[!htbp]
        \centering
        \small
        %\vspace{-0.3cm}
        \begin{tabular}{|rp{0.75\textwidth}|}
            \hline
            \multicolumn{2}{l}{|\textbf{Example 13}|} \\ \hline
            %F13-1 := & Update $<$actor: \{the\_system\}$>$ $<$object: Property\_listing $<$where: \{the\_MLS\}$>>>$ \\
            %F13-2 := & Create $<$actor: \{the\_system\}$>$ $<$object: New\_property\_listing $<$where: \{the\_MLS\}$>>>$ \\
            $F_{13-1}$ := & Update $<$actor: \{the\_system\}$>$ $<$object: property\_listing$>$ \\
            $F_{13-2}$ := & Create $<$actor: \{the\_system\}$>$ $<$object: New\_property\_listing$>$ \\
            %FC13-3 := & \{F13-1\} $:<$ $<$inhere\_in: \{the\_system\}$>$ \\
            %FC13-4 := & \{F13-2\} $:<$ $<$inhere\_in: \{the\_system\}$>$ \\
            \hline
        \end{tabular}
      \end{table}
      \vspace{-0.2cm}

  \item Conjunct content that needs to be separated (17/52, 32.70\%): we use ``$\lor$''. For example, ``\emph{the system shall allow user to search for movies by title, actor or director}'' can captured as in Example 14 ($F_{14}$ is OR-reduced to $F_{14-1}$, $F_{14-2}$ and $F_{14-3}$), which is similar to the third interpretation of ``AND'' (Example 9).
      %. where F14 can be reduced to F14-1 and F14-2. Similarly, if we want both functions to be implemented, we need FC14-3 and FC14-4.

      \begin{table}[!htbp]
        \centering
        \small
        %\vspace{-0.3cm}
        \begin{tabular}{|rp{0.78\textwidth}|}
            \hline
            \multicolumn{2}{l}{|\textbf{Example 14}|} \\ \hline
            $F_{14}$ := & Search $<$subject: \{the\_product\}$>$ $<$object: Movie$>$ $<$param: title $\lor$ actor $\lor$ director $>$ \\
            $F_{14-1}$ := & Search $<$subject: \{the\_product\}$>$ $<$object: Movie$>$ $<$param: Title $>$ \\
            $F_{14-2}$ := & Search $<$subject: \{the\_product\}$>$ $<$object: Movie$>$ $<$param: Actor $>$ \\
            $F_{14-3}$ := & Search $<$subject: \{the\_product\}$>$ $<$object: Movie$>$ $<$param: Director $>$ \\
            %FC14-3 := & \{F14-1\} $:<$ $<$inhere\_in: \{the\_system\}$>$ \\
            %FC14-4 := & \{F14-2\} $:<$ $<$inhere\_in: \{the\_system\}$>$ \\
            \hline
        \end{tabular}
     \end{table}
     \vspace{-0.2cm}

  \item Sets of individuals (2/52, 3.84\%): we define them as new types of entities based on set. For example, ``\emph{the product shall be able to interface with a database management system such as Oracle DB2 MySql or HSQL}''. We can capture such requirements as in Example 10.
  \item Logic ``OR'' (22/52, 42.31\%): we use ``$\lor$''. For instance, ``\emph{the product shall prevent a player from viewing the offensive or defensive grids of the other player}''. In this example, the opponent's status can be either offensive or defensive, but not both (see Example 15).

      \begin{table}[!htbp]
        \centering
        \small
        %\vspace{-0.3cm}
        \begin{tabular}{|rp{0.78\textwidth}|}
            \hline
            \multicolumn{2}{l}{|\textbf{Example 15}|} \\ \hline
            $F_{15}$ := & Prevent $<$subject: \{the\_product\}$>$ $<$object: Player$>$ $<$target: View $<$ object: (Offensive\_grid $\lor$ Defensive\_grid) $<$associated\_with: Opponent$>$$>$$>$ \\
            \hline
        \end{tabular}
      \end{table}
      \vspace{-0.2cm}

  \item Compound word (5/52, 9.61\%): we define them as concepts. For example, ``\emph{himself or herself}'', ``\emph{his or her status}'', etc.
\end{enumerate}

\vspace {6pt}
\noindent\textbf{The Use of Universal Quantifier in NL Requirements.} In our evaluation, we found that 115 requirements (115/625, 18.40\%) have used universal quantifiers such as ``all'', ``any'', ``100\%'', ``every'' and ``each'' (86 of them are in the 370 NFRs, see Table~\ref{tab:nfr_satisfication_type}). The use of universal in a requirement is dangerous, meaning, fulfillment of the requirement is often practically unrealizable or at least expensive. Accordingly, we need to use the \textbf{U} operator to relax the universality (i.e., we require only a certain percentage of the set of subjects to satisfy a requirement). For example, the quality requirement ``\emph{the processing time of all tasks shall be less than 5 seconds}'', captured as ``$QC_1$ := Processing\_time (Tasks) :: $\leq$ 5 (\emph{Sec.})'', can be relaxed by \textbf{U}: ``$QC_2$ := $\bm{U}$ (?X, $QC_1$, $<$inheres\_in: ?X$>$, 90\%)'', requiring 90\% of the tasks shall take less than 5 seconds.

On the other hand, sometimes stakeholders may need to impose constraints on a whole set rather than part of them. For example, ``\emph{the owner shall have free access to all of the streaming movies}''. It is a strict ``all'' here because it is unreasonable if an owner cannot access some of his/her streaming movies. At such situation, we use FCs to specify the universality requirement, as shown in Example 16.

\begin{table}[!htbp]
    \centering
    \small
    \begin{tabular}{|rp{0.75\textwidth}|}
        \hline
        \multicolumn{2}{l}{|\textbf{Example 16}|} \\ \hline
        $F_{16-1}$ := & Access $<$actor: Owner$>$$<$object: Streaming\_movie$>$ $<$mode: Free$>$ \\
        $FC_{16-2}$ := & $F_{16-1}$.object $:<$ $<$accessedBy: Owner$>$ \\
        \hline
    \end{tabular}
\end{table}

\section{Evaluating the Methodology}
\label{sec:eval_method}
We performed a case study on the \emph{Meeting Scheduler}(MS) project, adopted from the PROMISE dataset, to illustrate how our method can be applied to realistic requirements.
%to illustrate how our \emph{Desiree} language and methodology can be applied to realistic case studies.
%to demonstrate our methodology.

\subsection{Meeting Scheduler: Modeling}
\label{sec:eval_ms_modeling}

The \emph{Meeting Scheduler} (MS) project has 74 requirements, including 27 FRs and 47 NFRs. Functionally, the meeting scheduler is required to create meetings, send meeting invitations, book meeting rooms, book room equipment, and so on. The non-functional requirements cover different aspects of the system, such as ``Usability'', ``Configurability'', ``Look and feel'', ``Inter-operability'', ``Security'', and ``Maintainability''.

We classified the 74 requirements according to our requirement ontology, obtaining 34 FR$'$s, 5 FCs, 18 QRs, 3 FR$'$+FCs, 9 FR$'$+QRs, 4 FC+QRs, and 1 DA (74 in total). Following the proposed methodology in Section~\ref{sec:transform_process}, we captured each requirement as a goal, separated the concerns of a goal if needed, and then encoded them by using our description-based syntax. In this interpreting process, we identified several kinds of requirements issues, such as ambiguous, incomplete, unverifiable, and unsatisfiable~\footnote{These issues can be defined by identifying the inverse of each characteristic in Table~\ref{tab:req_desired_properties}, see Table~\ref{tab:req_issues_kinds} for the detailed definitions.}. We resolved these kinds of issues by using the set of our provided operators (e.g., ``Interpret'', {``Reduce''}, {``Operationalize/Observe''}, and {``de-Universalize''}) to make them unambiguous, complete (enough), verifiable, and practically satisfiable.

%and refined the structurally specified goals; unverifiability
%redundant (leading to un-modifiable) and term inconsistent

\begin{comment}
\begin{table}[!htbp]
  \centering
  \begin{threeparttable}
  \caption {Statistics of the \emph{Meeting Scheduler} case study}
  \label{tab:ms_stats}
  \vspace {0.3 cm}
  \small
  \setlength\tabcolsep{2pt}
  \begin{tabular}{|c|c|c|c|c|c|c|c|c|c|c|c|c|c|}
  \hline
  & \textbf{FG} & \textbf{F}  & \textbf{FC} & \textbf{QG} & \textbf{QC} & \textbf{CTG} & \textbf{SC} & \textbf{FG+QG} & \textbf{F+FC} & \textbf{F+QG} & \textbf{FC+QG} & \textbf{DA} & \textbf{Total} \\ \hline
   \emph{Classification} & 5 & 27  & 4 & 17 & 0 & 0 & 0& 2 & 6 & 8 & 4 & 1 & 74\\ \hline
   \emph{Specification} & 0 & 67  & 8 & 0 & 67 & 0 & 3 & 0 & 0 & 0 & 0 & 10 & 155\\ \hline
  \end{tabular}
  \begin{tablenotes}
      \small
      \item F: function; FC: functional constraint; QGC: quality goal (QG) or quality constraint (QC); CTG: content goal; DA: domain assumption; ``+'': mixture of concerns
    \end{tablenotes}
  \end{threeparttable}
\end{table}
\end{comment}

%In the modeling process, we found several kinds of requirements issues as follows.

%Meeting notification vs. meeting invitation. Req. 24 vs. Req. 2
%Add Goal_13_1 and Goal_14_1, reserve room and reserve equipment

%Meeting Type => CTG
%Transportation Status ==> CTG
%``Send $<$means: Email $\lor$ SMS$>$''
\begin{itemize}
  \item \emph{Incomplete}. We found 22 incomplete requirements~\footnote{In this case study, we counted the number of problematic requirements instead of the number of issues in each requirement, which is the case in our follow-up controlled experiment.} (22/74, 29.73\%). A requirement is incomplete if necessary information is missing for its implementation. For example, the requirement $R_1$ ``\emph{The product shall be able to send meeting notifications via different kinds of end-user specified methods}'', captured as ``$F_2$ := Send $<$actor: \{the\_product\}$>$ $<$object: Meeting\_notification$>$ $<$means: User\_defined\_method$>$'', is incomplete since several pieces of information is missing: (1) Who will send? (2) Send to whom? and (3) Send how many notifications at a time? With \emph{Desiree}, we are able to add the missing information through updating slot-description pairs by applying the ``Reduce'' operator: ``Send $<$actor: Organizer$>$ $<$object: $\geq$ 1 Meeting\_notification$>$ $<$target: Participant$>$ $<$means: User\_defined\_method$>$''. Here ``$\geq$ 1'' indicates a set, and ``User\_defined\_method'' needs to be further refined.
  \item \emph{Ambiguous}. We identified 4 ambiguous requirements (4/74, 5.41\%). For example, the above requirement $R_1$ is ambiguous because it is unclear whether a notification message will be sent by all the specified methods or only one of them at run-time. In our framework, this requirement will be captured as $F_2$ (as above), which can be AND-reduced (resp. OR-reduced) into different functions, e.g., $\bm{R_d}(F_2)=\{F_3, F_4\}$, where $F_3$ is ``Send $<$means: Email$>$'' and $F_4$  is ``Send $<$means: SMS$>$'' (resp. $\bm{R_d}(F_2)=\{F_3\}$ and $\bm{R_d}(F_2)=\{F_4\}$), which corresponds to the first (resp. the second) interpretation.
  \item \emph{Un-verifiable}. We identified 19 unverifiable requirements that use vague or subjective words such as ``multiple'', ``various'' and ``intuitive'' (19/74, 25.69\%). We make these kinds of requirements measurable by applying the ``Operationalize'' or the ``Observe'' operator. If a requirement is objective, we use ``Operationalize''. For example, the QR ``\emph{the product shall be able to interface with various kinds of DBMSs}'' will be operationalized as a QC ``Kinds\_of\_interoperable\_DBMS (\{the\_product\}) ::  \{MySQL, SQLServer, HSQL\}''. If a requirement is subjective, we use ``Observe''. For example, the QR ``\emph{the product shall have an intuitive interface}'' will be operationalized as a QC ``Understandability (\{the\_product\}) :: Intuitive $<$observed\_by: Surveyed\_user$>$'' by observing, which can be further relaxed to make it practically satisfiable.
  \item \emph{Un-satisfiable}.  We found 8 practically un-satisfiable requirements (8/74, 10.81\%). A requirement is (practically) un-satisfiable if it is too costly or technically unfeasible to implement  it. This issue is often caused by the un-restricted use of universals such as ``all'', ``any'', ``every'' and ``100\%''. In our framework, we make such requirements practically satisfiable by using the ``de-Universalize'' operator. For example, the QR ``\emph{the search function shall take less 5 seconds} (\emph{every time})'', captured as a QC ``Processing\_time (Search) :: $\leq$ 5 (\emph{Sec.})'', will be relaxed as ``\textbf{U} (?X, Processing\_time (Search) :: $\leq$ 5 (\emph{Sec.}), $<$inhere\_in: ?X$>$, 95\%)'', which requires the search function to take less than 5 seconds at 95\% of the time. Here the variable ``?X'' indicates a sub-set of the search runs.
  \item \emph{Redundant}. We identified 6 redundant requirements (6/74, 8.11\%). For example, the requirement ``\emph{The product shall allow an organizer to invite other employees to meetings}'' repeats ``\emph{The product will notify employees of meeting invitations}''. We choose to model the more complete requirement (e.g., the formmer) and abandon the less complete one (e.g., the latter).
  \item \emph{Inconsistent Terms}. We found term inconsistencies in 5 requirements with regarding to the rest ones (5/74, 6.76\%). A term inconsistency arises when different terms are used to refer to the same (type of) thing, e.g., ``\emph{meeting confirmation}'' vs. ``\emph{meeting acknowledgement}'' (1 occurrence), ``\emph{the system}'' vs. ``\emph{the product}'' (2 occurrences) and ``\emph{meeting agenda}'' vs. ``\emph{meeting schedule}'' (1 occurrence), or the same term is used for referring to different kinds of things, e.g., ``\emph{meeting room}'' as a real-world object vs. ``\emph{meeting room}'' as an information object (1 occurrence). We address the first kind of issue by adding a DA axiom that equating two concepts (e.g., ``$DA_1$ := Meeting\_agenda $\equiv$ Meeting\_schedule''), and address the second kind of issue by defining different concepts (e.g., ``\emph{meeting room}'' for real-world objects, and ``\emph{meeting room record}'' for information objects).
\end{itemize}

% ``the product'' vs. ``the system'' (2): req-60, req-70
% ``meeting room'' vs. ``conference room''
% req-16, 29, 43, 44, 48
%``Concept\_A $\equiv$ Concept\_B''

\begin{figure}[!htbp]
  \centering
  %\vspace {-2cm}
  %\hspace*{-0.8cm}
  \includegraphics[width=\textwidth]{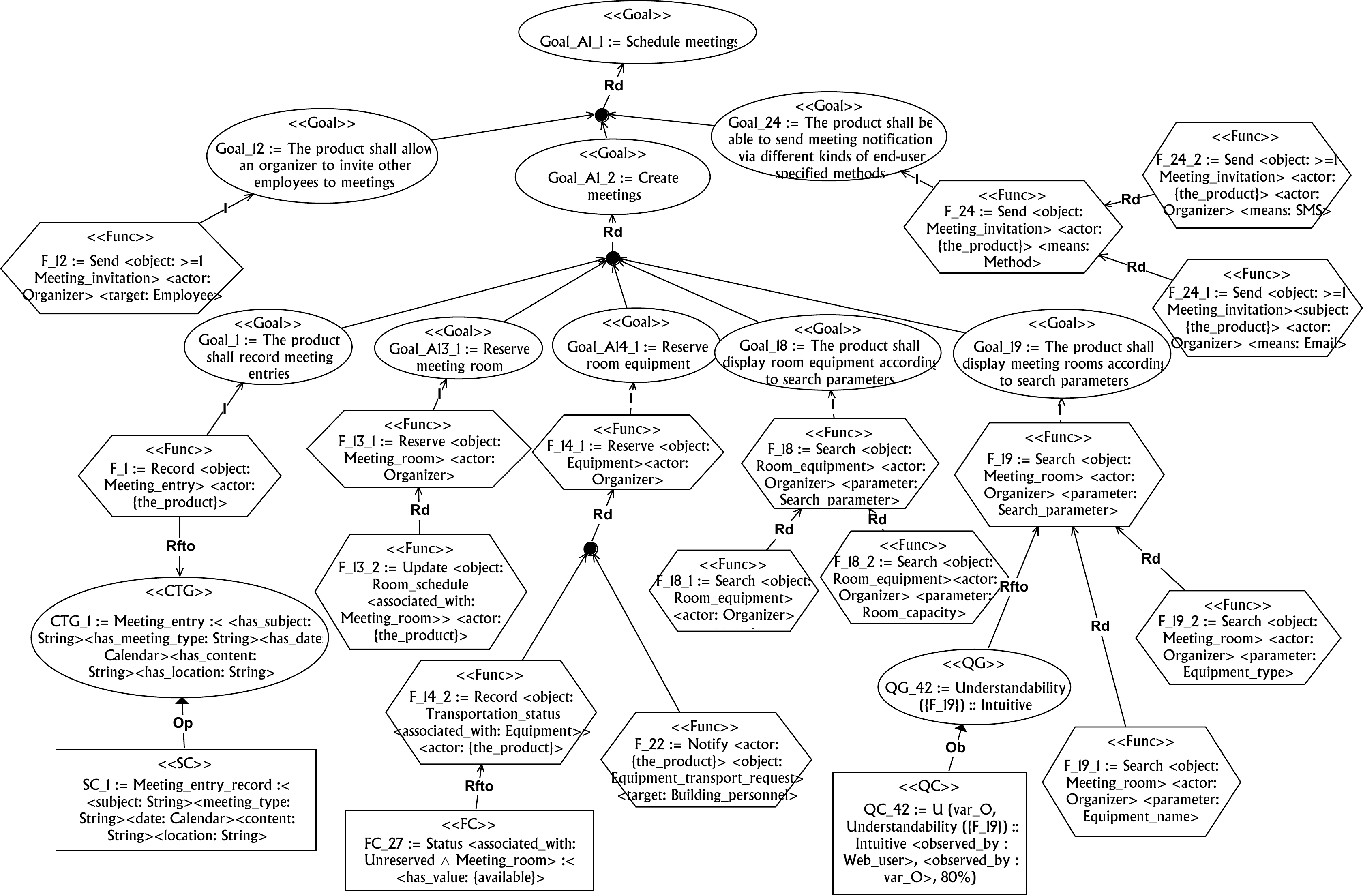}\\
  \caption{The overall functionality of the \emph{Meeting Scheduler} project}
  \label{fig:ms_fr}
\end{figure}

\begin{figure}[!htbp]
  \centering
  %\vspace {-2cm}
  \includegraphics[width=\textwidth]{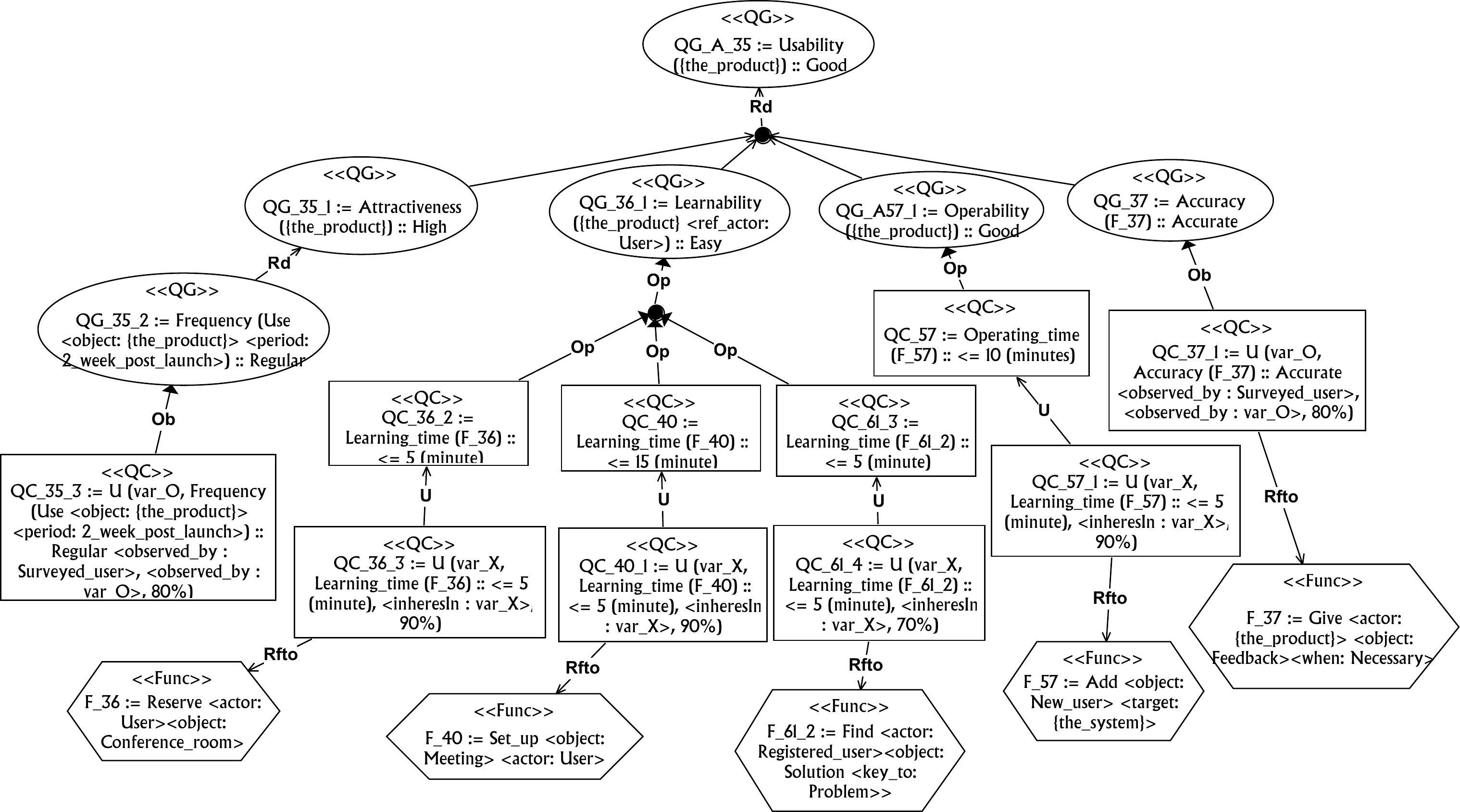}\\
  %\caption{An example category of NFRs (``Usability'') of the \emph{Meeting Scheduler} project}
  \caption{Example \emph{Usability} NFRs of the \emph{Meeting Scheduler} project}
  \label{fig:ms_nfr}
  \vspace{-0.5cm}
\end{figure}

We finally obtained a specification, which consists of 58 functions, 54 QCs, 10 FCs, 8 SCs and 13 DAs (143 elements in total). We show the overall functionality and an example category of NFRs (i.e., usability requirements) of the meeting scheduler project in Fig.~\ref{fig:ms_fr} and Fig.~\ref{fig:ms_nfr}, respectively. The full case study can be found at https://goo.gl/oeJ9Fi.

Here, each modeling element has an associated stereo type (e.g., ``$<<$QG$>>$'' indicates a quality goal), the ``\textbf{I}'', ``$\bm{R_d}$'', ``$\bm{O_p}$'', ``$\bm{O_b}$'', ``\textbf{U}''  links represent the applications of ``Interpret'', ``Reduce'', ``Operationalize'', ``Observe'', and ``de-Universalize'', respectively. A ``\textbf{Rfto}'' link means that a modeling element refer to another, e.g., the function ``$F_1$ := Record $<$object: Meeting\_entry$>$$<$actor: \{the\_product\}$>$'' refers to a content goal ``$CTG_1$ := Meeting\_entry $:<$ $<$has\_subject: String$>$$<$has\_meeting\_type: String$>$$<$has\_date: Calendar$>$$<$has\_content: String$>$$<$has\_location: String$>$''.

The use of a small black circle with a ``$\bm{R_d}$'' (resp. ``$\bm{O_p}$'') link represents an ``AND-Reduce'' (resp. ``AND-Operationalize''). For example, the function ``$F_{14-1}$ := Reserve $<$object: Equipment$>$$<$actor: Organizer$>$'' is AND-reduced to ``$F_{14-2}$ := Record $<$object: Transportation\_status $<$associated\_with: Equipment$>>$ $<$actor: \{the\_product\}$>$'' and ``$F_{22}$ := Notify $<$actor: \{the\_product\}$>$ $<$object: Equipment\_transport\_request$>$ $<$target: Building\_personnel$>$''. This means both $F_{14-2}$ and $F_{22}$ need to be fulfilled in order to make $F_{14-1}$ fulfilled.

To represent ``OR-Reduce'' (resp. ``OR-Operationalize''), we applied the ``Reduce'' (resp. ``Operationalize'') operator more than one time (can be just one). For example, ``$F_{19}$ := Search $<$object: Meeting\_room$>$ $<$actor: Organizer$>$ $<$parameter: Search\_parameter$>$'' is reduced to `` $F_{19-1}$ := Search $<$object: Meeting\_room$>$ $<$actor: Organizer$>$ $<$parameter: Equipment\_name$>$'' and ``$F_{19-2}$ := Search $<$object: Meeting\_room$>$ $<$actor: Organizer$>$ $<$parameter: Equipment \_type$>$'', respectively. This means that a search can be performed according to either an equipment name or an equipment type, but not both. In the case both name and type are needed for a combined search, we can define a CTG that with the two attributes as its slots, and then use the newly defined CTG as the description of the ``parameter'' slot.

\subsection{Meeting Scheduler: Reasoning}
\label{sec:eval_ms_reasoning}

We kept the requirements (including 54 Goals, 3 FGs, 35 QGs and 8 CTGs), specifications (including 58 functions, 54 QCs, 10 FCs, 8 SCs and 13 DAs), and the derivation process (refinements and operationalizations) in a \emph{Desiree} model, and translated the entire model to OWL2 (OWL2 Web Ontology Language)~\cite{mcguinness_owl_2004}.

To support this process, we have developed a translation component based on OWL API~\footnote{http://owlapi.sourceforge.net/} (as part of our \emph{Desiree} tool) for systematically and automatically translating a description-based specification into an OWL2-ontology. Interested readers can download the resultant ontology, and import it to Prot\'eg\'e~\footnote{http://protege.stanford.edu/} to try some reasoning tasks as discussed below~\footnote{The resultant ontology and our \emph{Desiree} tool is available at https://goo.gl/oeJ9Fi}. Alternatively, one can also try to do the reasoning tasks with our \emph{Desiree} tool, which has integrated the Hermit reasoner~\footnote{http://hermit-reasoner.com/}~\cite{shearer_hermit:_2008}.
%We have tested the ontology on Protege 5.0 and Hermit 1.3.8. Note that to get correct results, one needs to first perform reasoning and then execute corresponding queries.

By translating a requirements specification into an OWL2-ontology, we are able to: (1) query the interrelations between functions, qualities, and content (i.e., it contributes to improving the traceability of a requirements specification, see the characteristics of an eligible specification in Table~\ref{tab:req_desired_properties}); (2) identify (some) inconsistency issues such as term inconsistencies and normative conflicts (i.e., it helps to reducing inconsistencies in a specification); (3) reason about the satisfaction of goals. In the rest of this section, we discuss each of them in detail.

The major benefit of such a translation is the convenience of obtaining an overview of concerns (e.g., functions, qualities and entities) and interrelations query: we are able to ask a list of questions as shown in Table~\ref{tab:ms_query} (technically, these questions will be translated into DL queries). For instance, we can ask ``$<$inheres\_in: \{the\_product\}$>$'' (an instantiation of Q2) to retrieve the set of qualities that inhere in ``the product''. Note that these questions are not exhaustive. If desired, we can ask more complex questions like ``what functions are required to finish within 5 sec.?'' in the form of ``$<$has\_quality: Processing\_time $<$has\_value\_in: $\leq$ 5 (sec.)$>>$''.

%\footnote{Interested readers can download the ontology from https://goo.gl/oeJ9Fi and import it to Protege (http://protege.stanford.edu/) to have a try. To get correct results, one needs to first perform reasoning using some reasoner (e.g., Hermit) and then execute corresponding queries.}
\begin{table}[!htbp]
  %\vspace{-0.5cm}
  \caption {Example queries over the \emph{Meeting Scheduler} requirements specification }
  \label{tab:ms_query}
  \vspace {0.3 cm}
  \small
  \centering
  \begin{tabular}{|c|c|c|}
  \hline
  \textbf{ID} & \textbf{Concerned Questions } & \textbf{Our Syntax} \\ \hline
  Q1 & What kinds of subjects does a quality refer to?	& $<$has\_quality: QualityName$>$ \\ \hline
  Q2 & What qualities are of concern for a subject?	& $<$inheres\_in: SubjT$>$ \\ \hline
  Q3 & Who performs the function? &	$<$is\_actor\_of: F$>$ \\ \hline
  Q4 & What is the function operating on? & $<$is\_object\_of: F$>$ \\ \hline
  Q5 & What are the functions that an object is involved in? & $<$object: SubjT$>$ \\ \hline
  \end{tabular}
\end{table}

A second benefit of such a translation is the identification of (some) inconsistencies. For example, the term ``user'' in ``\emph{users shall be able to register within 2 minutes}'' refers to a person in the real world, but a symbolic or representational entity in the system in ``\emph{managers shall be able add users into the system}''. Here, the former requirement refers to a usability requirement (with regarding to a set of users), and the latter is a mix of function and content concerns (what kinds of slots/properties shall be used to describe a user profile?). To detect such inconsistencies, we add two axioms, ``System\_function $:<$ $<$object: ONLY Information\_entity$>$'' and ``Information\_entity ($\land$) Real\_world\_entity $:<$ Nothing'', constraining the object of a system function to be only information entities, which are disjoint with real-world entities~\footnote{Contra system functions, interface functions could act on real-world entities, e.g., ``\emph{navigate users}''.}. When specifying the function ``Register $<$actor: User$>$ $<$target: \{the\_system\}$>$'', we will add a DA axiom ``User $:<$ Real\_world\_entity''. If we have another function ``Add $<$actor: Manager$>$$<$object: User$>$$<$target: \{the\_system\}$>$'', and all these descriptions are translated into DL formulae, a DL reasoner is able detect an inconsistency therein: the term ``User'' cannot be used to refer to both a real-world entity and an information entity, which are disjoint classes, at the same time. Moreover, the identification of such inconsistencies helps us to discover implicit or missing requirements, e.g., there is an implicit content requirement about user profile in this example. In the \emph{Meeting Scheduler} project, we identified 3 such inconsistencies, the other two are ``\emph{meeting room}'' and ``\emph{room equipment}'' (being a real-world entity vs. being an information entity).

%Moreover, the identification of such inconsistencies helps us to discover implicit or missing requirement, e.g., there are three implicit content requirements in these examples: what information shall be included in a user profile, a meeting room record and a room equipment record?

Since we have captured each requirement/operationalization using DL formulae, we are also able to perform ``what if'' analysis on the resultant requirements ontology. To do so, we add some extra DL axioms that assert certain specification elements to be fulfilled, and then check whether the goal of concern is fulfilled. For example, as in Fig.~\ref{fig:ms_fr}, the goal $G_{19}$ ``\emph{search meeting room}'' is interpreted as $F_{19}$ ``Search $<$object: Meeting\_room$>$ $<$actor: Organizer$>$ $<$parameter: Search\_parameter$>$'', which is OR-deuced to $F_{19-1}$ ``Search ... $<$parameter: Equipment\_name$>$'' and $F_{19-2}$ ``Search ... $<$parameter: Equipment\_type$>$''. We assume $F_{19-1}$ to be fulfilled by adding an axiom ``$F_{19-1} \sqsubseteq Sulfilled\_Thing$'', and then check if $G_{19}$ is fulfilled (i.e., whether ``$G_{19} \sqsubseteq Sulfilled\_Thing$'' holds) by running the DL subsumption reasoning task.

%AS we have mentioned in Section~\ref{sec:semantics_operator} and~\ref{sec:tool_reasoning},
\begin{comment}
Since we have captured each requirement and operationalization using DL formulation, we are also able to perform ``what if'' analysis. We can add some extra DL axioms to express that which specification elements are fulfilled (e.g., $Func_{75} \sqsubseteq Sulfilled\_Thing$ means that we assume $Func_{75}$ to be fulfilled), and then check whether the goal of concern is fulfilled. As pointed by Horkoff et al.~\cite{horkoff_making_2012}, although ``what-if'' analysis is supported by existing goal-model reasoners~\cite{giorgini_reasoning_2003}\cite{sebastiani_simple_2004}, DL formulation allows exploration of less restrictive prorogation. For example, if a goal $G$ in AND-operationalized to 4 functions, we can relax the fulfillment of $G$ as such: if at least 3 functions are fulfilled, then $G$ is fulfilled. This is akin to setting a threshold for the degree of fulfillment of a ``good-enough'' QR (e.g., if a QR is fulfilled to a degree of 0.75, then it is said to be fulfilled) and can be implemented by using DL quantified number restriction as in Eq.~\ref{eq:eq_translation_approx_satthing}.\\
\begin{equation}\label{eq:eq_translation_approx_satthing}
    \begin{aligned}
        Approximate\_Fulfilled\_Thing \sqsubseteq & \; Fulfilled\_Thing	\\
        Approximate\_Fulfilled\_Thing \equiv & \; \geq 3 \; relate\_to\_many.Fulfilled\_Thing\\
    \end{aligned}
\end{equation}
\end{comment}

Another benefit, which is closely related to the interrelation management, is the support for impact analysis when changes occur. For instance, for security reasons, a stakeholder may require their email to be invisible to others. Given this new requirement, we can at first find out the functions that are related to emails through query (suppose that we have ``$F_x$ := Send $<$object: Meeting\_invitation$>$ $<$means: Email$>$'' through the query ``$<$means: Email$>$''), and then add the new requirement according to some mechanisms (e.g., reduce $F_x$  to $F_x'$ := Send $<$object: Meeting\_invitation$>$ $<$means: Email $<$kind: BCC$>>$'', making ``\emph{blind carbon copy} (\emph{bcc})'' a nested SlotD that modifies ``Email'', the means of $F_x$). Next, we need to evaluate how this change would impact other elements (e.g., the influence on the inhering performance qualities of $F_x$). This interesting topic will be explored in the next steps of our work.

\section{Evaluating the \emph{Desiree} Framework}
\label{sec:eval_framework}
%(the first at University of Trento, Italy, and the second and the third at Tsinghua University, China)
In this section, we describe a set of three controlled experiments, conducted to assess whether \emph{Desiree} can indeed help people to conduct better requirement analysis.

\subsection{Experiment Setup}
\label{sec:expr_setup}
%In the experiments, we compared \emph{Desiree} with a \emph{Vanilla} RE approach, where a student participant uses the characteristics of a good software requirement specification (SRS) adapted from the IEEE standard 830-1998~\cite{committee_ieee_1998} and a set of guidelines for good requirements introduced in Wiegers et al.~\cite{wiegers_software_2013}. In the \emph{Vanilla} method, participants manually go through and improve stakeholder requirements using these desirable characteristics and guidelines. To prepare, we defined a set of requirements issues as in Table~\ref{tab:req_issues_kinds} by identifying the inverse of each characteristic introduced in the IEEE standard~\cite{committee_ieee_1998}.

In the experiments, we compared \emph{Desiree} with a \emph{Vanilla} RE approach, where a participant uses the characteristics of a good software requirement specification (SRS) adapted from the IEEE standard 830-1998~\cite{committee_ieee_1998} and a set of guidelines for writing good requirements introduced in Wiegers et al.~\cite{wiegers_software_2013}. In the \emph{Vanilla} method, participants manually go through and improve stakeholder requirements using these desirable characteristics and guidelines. This process approximates requirements walkthroughs using inspection checklists, and is used as a baseline representing how requirements are improved in practice.

%the IEEE standard~\cite{committee_ieee_1998} (i.e., contra
To prepare, we defined a set of requirements issues as in Table~\ref{tab:req_issues_kinds} by identifying the inverse of each characteristic introduced in Table~\ref{tab:req_desired_properties}. Our experiments check to see if people can identify and address more of these issues when refining stakeholder requirements with \emph{Desiree} or with the \emph{Vanilla} RE approach. We do not consider the ``Ranked'' characteristic in our experiments, as \emph{Desiree} currently does not support requirements prioritization. Similarly, we do not consider ``Formal'' as the \emph{Vanilla} RE method does not support it. We also do not compare \emph{Desiree} with the \emph{Vanilla} method on ``Traceability'' because \emph{Desiree} is a goal-oriented method, and as such it intrinsically supports requirements to requirements traceability (requirements to sources and requirements to design traceability are out of scope for our experiments). In addition, we introduce ``Unsatisfiable'', which is practically important but missing in the IEEE standard.
%(Inv) (Icmp) (Amb)(Vag) (Icns)(Umod) (Unsat)

\begin{table}[!htbp]
  \caption {Seven kinds of requirements issues}
  \label{tab:req_issues_kinds}
  \vspace {0.3 cm}
  \centering
  \small
  \setlength\tabcolsep{2pt}
  %\begin{tabular}{|c|p{0.8\textwidth}|}
  \begin{tabular}{|m{0.15\textwidth}|m{0.75\textwidth}|@{}m{0pt}@{}} %<{\centering}
  \hline
  \textbf{Issue} & \textbf{Definition} \\ \hline
  Invalid  & A requirement is invalid if it is not the one that stakeholders want. \\ \hline
  Incomplete & (1) incomplete requirement - a requirement is incomplete if necessary information is missing for implementation; (2) incomplete specification: a specification is incomplete if any significant requirement is missing.   \\ \hline
  Ambiguous  & A requirement is ambiguous if it has more than one interpretation. \\ \hline
  Unverifiable	& A requirement is vague if it specifies unclear or imprecise value regions. \\ \hline
  Inconsistent & A specification is inconsistent if there are: (1) conflicts between the specified properties of real-world objects; or (2) logical or temporal conflict; or (3) terms misuse or abuse. \\ \hline
  Unmodifiable & A specification is un-modifiable if: (1) it is not structurally organized; or (2) its requirements are redundant; or (3) it intermixes requirements  \\ \hline
  Unsatisfiable	& A requirement is practically unsatisfiable (unattainable) if it is impossible or too costly to fulfill.\\  \hline
  %Untraceable (Utrac) & An SRS is untraceable if: (1) some requirements do not have identifiers; (2) interrelations between requirements are not captured.  \\ \hline
  \end{tabular}
\end{table}

In general, these issues can be classified into two categories: (1) issues of individual requirements, including ``Invalid'', ``Ambiguous'', ``Unverifiable'', and ``Unsatisfiable''; (2) issues of requirements specifications, including ``Inconsistent'', and ``Unmodifiable''. Note that ``Incomplete'' can be further decomposed into ``Incomplete Requirement'' and ``Incomplete SRS'', which fall in these two categories, respectively.

\subsection{Research Question}
\label{sec:expr_rq}
Our research question can be stated as: \emph{\textbf{compared with the Vanilla method, can Desiree help people to identify and address more requirements issues when transforming stakeholder requirements to specifications}}?

%We define the null hypothesis, $H_0$, as: there is no statistical difference in the number of requirements issues found when using \emph{Desiree} vs. the \emph{Vanilla} method. The alternative hypothesis, $H_1$, is accordingly defined as: there is a statistical significance in the number of issues found using Desiree vs. the Vanilla method. These two hypotheses are formulated as in Eq.~\ref{eq:eq_hypo_test}, where ``$\mu_{Desiree}$'' and ``$\mu_{Vanilla}$'' represent the average number of identified issues when using the \emph{Desiree} and the \emph{Vanilla} approach, respectively.

We define the null hypothesis, $H_0$, as: there is no statistical difference in the number of requirements issues found when using \emph{Desiree} ($\mu_{D}$) vs. the \emph{Vanilla} method ($\mu_{V}$). The alternative hypothesis, $H_1$, is accordingly defined as: there is a positive statistical difference in the number of issues found using \emph{Desiree} vs. the \emph{Vanilla} method. These two hypotheses can be formulated as Eq.~\ref{eq:eq_hypo_test}.

\begin{equation}\label{eq:eq_hypo_test}
    \begin{aligned}
        H_0: \mu_{D} - \mu_{V} = 0; \\
        H_1: \mu_{D} - \mu_{V} > 0;
    \end{aligned}
\end{equation}

Similarly, we can define the null and alternative hypotheses for all the 7 kind of issues defined in Table~\ref{tab:req_issues_kinds}. For example, with regarding to ``Ambiguity'', the null hypothesis $H_{ambiguous\_0}$ can be stated as:  there is no statistical difference in the number of ambiguity issues found when using \emph{Desiree} vs. the \emph{Vanilla} method, and the alternative hypothesis $H_{ambiguous\_1}$ can be stated as: there is a positive statistical difference in the number of ambiguity issues found using \emph{Desiree} vs. the \emph{Vanilla} method.

%Considering that some participants may not finish modeling or analyzing all given requirements, we also measure the percentage of identified issues of those modeled/analyzed ones (symbolized as ``T$'$''). The null and alternative hypotheses for this measure can be defined similarly.

\subsection{Experiment Design}
\label{sec:expr_design}
The experimental task was to transform a set of given stakeholder requirements into a specification through refinements and operationalizations. To evaluate the differences in their performance, each participant was required to perform the task twice: s/he uses the \emph{Vanilla} method on a project X in the first session, and then uses \emph{Desiree} on another project Y in the second session. In both sessions, the participants discussed with stakeholders to elicit necessary information for addressing identified issues, and submitted a refined specification. For example, one probably need further information to quantify ``\emph{fast}'', which is vague. To keep independence between participants, they were asked to use texting to communicate with the stakeholder instead of speaking aloud. All experimental tasks were performed electronically and online.
%(Google or Zoho Docs)

The experiment was duplicated three times, the first at University of Trento, Italy, and the second and the third at Tsinghua University, China. In each experiment, we used two projects, \emph{Meeting Scheduler} (MS) and \emph{Realtor Buddy} (RB), which are selected from the PROMISE requirements set~\cite{menzies_promise_2012}, for the experimental task. We chose 10 requirements, which cover some typical functionalities and qualities (e.g., search, usability), from each project, and identified a list of issues for both projects. We also added some issues that are newly~\footnote{This means that participants have found issues that we failed to identify in advance for an experiment.} identified by participants into the reference issue lists in each experiment. We show the statistics of these issues in Table~\ref{tab:statistics_issues}, and refer interested readers to the detail in Appendix~\ref{sec:appendix_issues_in_MS} and~\ref{sec:appendix_issues_in_RB}.

\begin{table}[!htbp]
  \caption {The Statistics of the reference issues in the two testing projects}
  \label{tab:statistics_issues}
  \vspace {0.3 cm}
  \centering
  \small
  \setlength\tabcolsep{2pt}
  \begin{tabular}{|p{0.2\textwidth}|p{0.1\textwidth}<{\centering}|p{0.1\textwidth}<{\centering}
  |p{0.1\textwidth}<{\centering}|p{0.1\textwidth}<{\centering}|p{0.1\textwidth}<{\centering}|p{0.1\textwidth}<{\centering}|}
  \hline
   & \multicolumn{2}{c|}{\textbf{Experiment One}} & \multicolumn{2}{c|}{\textbf{Experiment Two}} & \multicolumn{2}{c|}{\textbf{Experiment Three}} \\ \hline
   & \textbf{MS} & \textbf{RB} & \textbf{MS} & \textbf{RB} & \textbf{MS} & \textbf{RB} \\ \hline
   Invalid	& - & -	& - & - & - & -\\ \hline
   Incomplete & 18 & 16 & 24 & 22 & 26 & 26\\ \hline
   Ambiguous & 2 & 4 & 4 & 6 & 4 & 5 \\ \hline
   Unverifiable & 10 & 6& 10 & 6 & 8 & 7 \\ \hline
   Inconsistency&  2& 3 & 3& 3 & 3 & 4\\ \hline
   Un-modifiable & 3 & 4& 4& 6 & 4 & 4\\ \hline
   Un-satisfiable & 4 & 6& 4& 6 & 4 & 4\\ \hline
   %Un-traceable & -	& - & - & - & - & -\\ \hline
   Total & 39 & 39 & 49 & 49 & 49 & 50\\ \hline
  \end{tabular}
\end{table}

There are three points to be noted. First, the number of ``Invalid'' issues is 0 since it is hard to justify which requirement is not desired by original stakeholders who provided the requirements set. Second, we have relatively more issues in experiment two since we have incorporated more issues that are newly identified by participants in this experiment. Third, observing that participants probably can not finish modeling or analyzing all given requirements, we have removed 2 requirements from each project in experiment three (i.e., we had only 8 requirements in the third experiment). These requirements are also presented in Appendix \ref{sec:appendix_issues_in_MS} and \ref{sec:appendix_issues_in_RB}.

We summarize the design for the three experiments in Table~\ref{tab:expr_design}. In experiment one, we had 17 participants: Master's students at the Department of Information Engineering and Computer Science, University of Trento, taking the RE course at the spring term of 2015. We also had 4 Ph.D. students or postdocs in the research group of Software Engineering and Formal Methods playing the role of stakeholder. We assigned two stakeholders to a project, and randomly separated the students into two teams, RG1 and RG2.

In the first session, we introduced the characteristics of a good SRS~\cite{committee_ieee_1998}and the textbook technique~\cite{wiegers_software_2013} in 30 minutes, presented the domain knowledge of the two projects in 10 minutes, and tested in 90 minutes. In this session, the group of students RG1 worked on MS and RG2 worked on RB. In the second session, we introduced the \emph{Desiree} framework and its supporting tool in 40 minutes, and tested in 90 minutes. In this session, the teams were given the other project.

%In the first session, the group of students RG1 worked on MS and RG2 worked on RB. In the second session, the teams were given the other project.
%(one at University of Trento, Trento, Italy, and the other two at Tsinghua University, Beijing, China)

%In the first session, we introduce the characteristics of a good SRS in 30 minutes, present the domain knowledge of the two projects in 10 minutes, and let students manually go through 10 requirements and write a refined requirement specification in 90 minutes. In this session, both teams use the vanilla approach, but work on different projects: RG1 on Meeting Scheduler and RG2 on Realtor Buddy. In the second session, we introduce the Desiree framework and its supporting tool (i.e., the Desiree tool) in 40 minutes, and then let students to refine and model 10 stakeholder require-ments using the tool in 90 minutes. In this session, the two teams work on the projects interchangeably: RG1 on Realtor Buddy and RG2 on Meeting Scheduler.

\begin{table}[!htbp]
  \caption {The design of the three controlled experiments }
  \label{tab:expr_design}
  \vspace {0.3 cm}
  \centering
  \small
  \setlength\tabcolsep{2pt}
  \begin{tabular}{|m{0.1\textwidth}|m{0.1\textwidth}<{\centering}|m{0.17\textwidth}<{\centering}|m{0.1\textwidth}<{\centering}
  |m{0.17\textwidth}<{\centering}|m{0.1\textwidth}<{\centering}|m{0.17\textwidth}<{\centering}|@{}m{0pt}@{}}
  %\begin{tabular}{|p{0.1\textwidth}|p{0.1\textwidth}<{\centering}|p{0.17\textwidth}<{\centering}|p{0.1\textwidth}<{\centering}|p{0.17\textwidth}<{\centering}|p{0.1\textwidth}<{\centering}|p{0.17\textwidth}<{\centering}|}
  %\begin{tabular}{|c|c|c|c|c|c|c|}

  \hline
   & \multicolumn{2}{c|}{\textbf{Experiment One}} & \multicolumn{2}{c|}{\textbf{Experiment Two}} & \multicolumn{2}{c|}{\textbf{Experiment Three}} \\ \hline
   Session	& One & Two	& One & Two & One & Two\\ \hline
   Approach	& \emph{Vanilla} & \emph{Desiree} & \emph{Vanilla} & \emph{Desiree} & \emph{Vanilla} & \emph{Desiree} \\ \hline
   Trainer & \multicolumn{6}{c|}{{Same}} \\ \hline

   \multicolumn{1}{|p{0.1\textwidth}<{\centering}|}{Language} & \multicolumn{2}{p{0.26\textwidth}<{\centering}|}{Teaching + Testing: English}&\multicolumn{2}{p{0.26\textwidth}<{\centering}|}{Teaching: Chinese; Testing: English} & \multicolumn{2}{p{0.26\textwidth}<{\centering}|}{Teaching + Testing: Chinese + English}\\ \hline

   Training  & 30 (min) & 30 + 10 (min) Method + Tool & 60 (min) & 60 + 60 (min) Method + Tool & 45 (min) & 45 + 60 (min) Method + Tool \\ \hline
   Testing &  90 (min) & 90 (min) & 90 (min) & 90 (min) & 90 (min) & 90 (min)\\ \hline
  \end{tabular}
\end{table}
%This experiment replicates the first, but with a few changes to the experimental design.
In the second experiment, we had 18 volunteer participants: Master's students at the Institute of Information System and Engineering, School of Software, Tsinghua University. Compared with experiment one, a few changes were made to the experimental design. First, to improve the consistency of stakeholders' answers, we randomly separated the 18 participants into 6 small teams of size 2 $-$ 4 (the size varied as participants changed their time schedules), and hired 1 constant stakeholder for all the 6 teams on the same project (note that the 6 teams conducted the experiment one by one, not concurrently). Second, based on our initial observations that the training time was too short, we increased the \emph{Desiree} training time from 40 minutes to 2 hours (1 hour for the method and 1 hour for the tool). In addition, we had also updated the \emph{Desiree} tool based on the feedback collected in the first experiment, mainly on the usability aspect (e.g., copy and paste).

In the third experiment, we had 30 participants: Master's students at the School of Software, Tsinghua University, taking the RE course at the fall term of 2015. This experiment replicates the first, with the only change on training time: 45 minutes training for the \emph{Desiree} method, and 60 minutes tutorial on the \emph{Desiree} tool. Also, we hired 6 students who have already participated experiment two as our stakeholders (7 in total, including the trainer). Similarly, stakeholders are randomly assigned to the two projects, and students are randomly separated into two teams according to the two projects. The teams were given different projects in the two sessions.

\subsection{Data Collection}
\label{sec:expr_data_collection}
The statistics of data collected from the three experiments are summarized in Table~\ref{tab:statistics_collected_data}. There are three points to noted. First, in each session of the three experiments, a participant was supposed to have a conversation, a refined requirements document and/or a requirements model. Second, in each experiment, the output of the \emph{Vanilla} session is just documents while that of the \emph{Desiree} session is a mix of models and documents: a participant was required to refine unmodelled requirements using natural language, but still following the \emph{Desiree} method. In a few cases, participants submitted only models or only texts in a \emph{Desiree} session. Third, in the experiments, a few participants (e.g., in the \emph{Vanilla} session of experiment three) have refined the given requirements without communicating to stakeholders.

In the first experiment, we collected 16 complete samples. On average, participants took 63 and 89.5 minutes to finish the task in session one and session two, respectively. In session one, we collected 17 online discussions in Google Doc and 17 requirement documents. In session two, we collected 16 online discussions, 11 \emph{Desiree} models, and 14 requirement documents. Specifically, in session two, 5 participants had only documents while 2 of them had only models.

%Specifically, in session two, 9 participants had both documents and models, 5 participants had only documents while 2 of them had only models.

%Note that in session two requirement documents are a complement of models: the participants need to refine unmolded requirements in natural language if they cannot model all of them with the \emph{Desiree} syntax, but still follow the \emph{Desiree} method.

\begin{table}[!htbp]
  \centering
  %\begin{threeparttable}
  \caption {Statistics of collected data: discussions, texts and models}
  \label{tab:statistics_collected_data}
  \vspace {0.3 cm}
  \small
  \setlength\tabcolsep{2pt}
  %\begin{tabular}{|p{0.2\textwidth}|p{0.1\textwidth}<{\centering}|p{0.1\textwidth}<{\centering}|p{0.1\textwidth}<{\centering}|p{0.1\textwidth}<{\centering}|p{0.1\textwidth}<{\centering}|p{0.1\textwidth}<{\centering}|}
  \begin{tabular}{|c|c|c|c|c|c|c|}
  \hline
   &  \multicolumn{2}{c|}{\textbf{Experiment One}} & \multicolumn{2}{c|}{\textbf{Experiment Two}}&  \multicolumn{2}{c|}{\textbf{Experiment Three}} \\ \hline
   & \emph{Vanilla} & \emph{Desiree} & \emph{Vanilla} & \emph{Desiree} & \emph{Vanilla} & \emph{Desiree} \\ \hline
   Time & 63 (min) & 89.5 (min) & 78 (min) & 94 (min) & 62 (min) &  97 (min) \\ \hline
   Discussion & 17 & 16 & 18 & 15 & 28 &  29 \\ \hline
   Requirements Document & 17 & 14 & 18 & 12 & 30 &  23 \\ \hline
   Requirements Models & - & 11& - & 15 & - &  29 \\ \hline
   Complete Sample & \multicolumn{2}{c|}{$16$ } & \multicolumn{2}{c|}{$15$ }&  \multicolumn{2}{c|}{$29$ } \\ \hline
  \end{tabular}
  \begin{comment}
  \begin{tablenotes}
     \footnotesize
      \item 1: in the \emph{Desiree} session of experiment one, 5 out of the 16 participants have only documents, and 2 out of the 16 have only models.
      \item 2: in the \emph{Desiree} session of experiment one, 3 out of the 15 participants have modeled all the given requirements using the \emph{Desiree} tool, and hence do not need to submit a requirements document.
      \item 3: in the \emph{Vanilla} session of experiment three, 2 out of the 30 participants have refined the given requirements without communicating with stakeholders; in the \emph{Desiree} session, 6 out of the 29 participants have modeled all the given requirements by using the \emph{Desiree} tool.
  \end{tablenotes}
  \end{comment}
  %\end{threeparttable}
\end{table}

In the second experiment, we had 15 complete samples: 18 participants finished session one, and 15 of them finished session two. On average, the participants took 78 and 94 minutes to finish their tasks in session one and two, respectively. For those who finished session one, we collected 18 online discussions in Zoho Doc and 18 requirement documents; for those who completed session two, we collected 15 online discussions, 15 \emph{Desiree} models, and 12 requirement documents.

In the third experiment, we had 29 complete samples: 30 participants finished session one, and 29 of them finished session two. On average, the participants took 62 and 97 minutes to finish their tasks in session one and two, respectively. For those who finished session one, we collected 28 discussions (24 online discussions in Wechat~\footnote{http://www.wechat.com/en/} and 4 face-to-face ones that are recorded on papers) and 30 requirement documents (16 electronic and 14 handwritten ones); for those who completed session two, we collected 29 online discussions, 29 \emph{Desiree} models, and 23 requirement documents. Note that we have allowed for face-to-face conversations and handwritten requirement documents in the \emph{Vanilla} session of experiment three because some students did not bring their laptops

%There are three points to noted. First, if not specified, discussions are defaulted as recorded online conversations, requirement documents and models are defaulted in an electronic form. Second, in the first session of experiment three, we have allowed for face-to-face conversation and handwritten requirement documents because some students did not bring their laptops and some students' laptops were running out of battery power. Third, in the second session of all the three experiments, requirement documents are a complement of requirements models: a participant needs to refine unmolded requirements in natural language if he/she cannot model all of them with the \emph{Desiree} tool, but still follow the \emph{Desiree} method.

%That is, one did not need to submit a refined requirement document if s/he has modeled all the given requirements. For example, among the 29 participants in session two of experiment three, we collected only 23 requirements documents because 6 of them have modeled all the 8 requirements in their \emph{Desiree} models.

%used Wechat as the communication tool, and

\subsection{Descriptive Statistics}
\label{sec:expr_descriptive statistics}
We carefully went through participants' discussions and refined requirements (including both documents and models) to check how many issues they have found in the experiments. We say a participant has identified an issue if either of the two conditions hold.

%We say a participant has identified an issue if: (1) s/he has asked a corresponding question, e.g., we gave a count of identified unverifiable issue if someone has asked ``how to measure fast?"''; or (2) s/he has eliminated an issue in his/her refined requirements specification, either documents or models, although s/he did not ask any related question. For example, a participant has eliminated a term inconsistency in the RB project by changing ``the product'' to ``the system'' without asking any questions. To keep consistency, the trainer performed the evaluation for all the three experiments. To keep consistency, the trainer performed the evaluation for all the three experiments.

\begin{enumerate}
  \item A participant has asked a corresponding question, e.g., we gave a count of identified unverifiable issue if someone has asked ``how to measure fast?''.
  \item A participant has eliminated an issue in his/her refined requirements specification, either documents or models, although s/he did not ask any related question. For example, a participant has eliminated a term inconsistency in the RB project by changing ``the product'' to ``the system'' without asking any questions.
\end{enumerate}

%and negative results are in red
%, where positive results are in bold

To keep consistency, the trainer performed the evaluation for all the three experiments. We show the average percentage of identified issues of participants with regarding to each kind of issue in Table~\ref{tab:req_issues_kinds}. We see that in experiment one, on average, a participant was able to find more issues with \emph{Desiree} than with the \emph{Vanilla} approach (33.49\% vs. 29.17\%), but discovered fewer issues with regarding to ambiguous and unverifiable issues. In experiment two, as the training time for \emph{Desiree} increased from 40 minutes to 2 hours, we can see that the participants performed better in general: they found more issues in total (45.71\% vs. 32.11\%). Experiment three has provided similar evidence as experiment two: with acceptable training time, participants are able to find more issues with the \emph{Desiree} approach (36.75\% vs. 25.89\%).

\begin{table}[!htbp]
  \caption {Statistics of issues identified by participants in the three experiments}
  \label{tab:statistics_identified_issues}
  \vspace {0.3 cm}
  \centering
  \small
  \setlength\tabcolsep{2pt}
  %\begin{tabular}{|p{0.2\textwidth}|p{0.1\textwidth}<{\centering}|p{0.1\textwidth}<{\centering}|p{0.1\textwidth}<{\centering}|p{0.1\textwidth}<{\centering}|p{0.1\textwidth}<{\centering}|p{0.1\textwidth}<{\centering}|}
  \begin{tabular}{|c|c|c|c|c|c|c|c|c|c|}
  \hline
   & \multicolumn{3}{c|}{\textbf{Experiment One}} & \multicolumn{3}{c|}{\textbf{Experiment Two}} & \multicolumn{3}{c|}{\textbf{Experiment Three}} \\ \hline
    & \emph{Vanilla} & \emph{Desiree} &	 Diff & \emph{Vanilla} & \emph{Desiree} &Diff &\emph{Vanilla} & \emph{Desiree} & Diff \\ \hline
Incomplete & 15.84\% & 19.49\% & 3.65\% & 27.30\% & 31.64\% & 4.34\% & 21.49\% & 30.77\% & 9.28\% \\ \hline
Ambiguous & 25.00\% & 20.31\% & -4.69\% & 32.22\% & 54.44\% & 22.22\% & 12.93\% & 38.45\% & 25.52\% \\ \hline
Inconsistent & 10.42\% & 12.50\% & 2.08\% & 6.67\% & 8.89\% & 2.22\% & 16.09\% & 16.38\% & 0.29\% \\ \hline
Unverifiable & 88.75\% & 81.46\% & -7.29\% & 81.33\% & 92.67\% & 11.34\% & 79.37\% & 90.27\% & 10.90\% \\ \hline
Unmodifiable & 20.83\% & 46.88\% & 26.05\% & 9.44\% & 43.33\% & 33.89\% & 3.45\% & 47.41\% & 43.96\% \\ \hline
Unsatisfiable & 3.65\% & 11.46\% & 7.81\% & 10.56\% & 50\% & 39.44\% & 3.45\% & 24.14\% & 20.69\% \\ \hline
Total & 29.17\% & 33.49\% & 4.32\% & 32.11\% & 45.71\% & 13.60\% & 25.89\% & 36.75\% & 10.86\% \\ \hline
  \end{tabular}
\end{table}

These statistics are calculated based on the number of issues identified by participants and the reference number of issues in the testing projects using Eq.~\ref{eq:eq_stat_calc}, where $Pct_{e, m}$ is the average percentage of identified issues of type $e$ with a method $m$ in an experiment (e.g., $Pct_{unverifiable, \; Desiree}$ represents the average percentage of identified unverifiable issues when using \emph{Desiree}, and has the value of 90.27\% in experiment three as shown in Table~\ref{tab:statistics_identified_issues}), $N_{i, \; e, \; m, \; p}$ is the number of issues of type $e$ identified by a participant $i$ with method $m$ on a project $p$ (e.g., $N_{1, \; unverifiable, \; Desiree, \; MS}$ represents the number of unverifiable issues identified by participant 1 when working on the MS project with the \emph{Desiree} method), $REF_{e, \; p}$ is the reference number of issue $e$ in a project $p$ (e.g., $REF_{unverifiable, \; MS}$ represents the reference number of unverifiable issues in the MS project), $K_p$ is the number of participants that work on a project $p$ (e.g.,  $K_{MS}$  represents the number of participants that work on the MS project). Meanwhile, we show the detailed numbers of issues identified by participants in experiment one, two, and three in Table~\ref{tab:data_detailed_issue_number_expr1}, Table~\ref{tab:data_detailed_issue_number_expr2}, Table~\ref{tab:data_detailed_issue_number_expr3}, respectively.
%and refer interested users to our online repository for more detail about those of experiment one and two at https://goo.gl/oeJ9Fi.
%, $P$ is the number of projects (i.e., 2 in our case).
\begin{equation}\label{eq:eq_stat_calc}
    \vspace {-0.3 cm}
    \begin{aligned}
        Pct_{e, m} = (\sum{p=\{MS, RB\}}\sum_{i=1}^{K_p} N_{i, e, m, p} / REF_{e, p} ) / ( \sum{p=\{MS, RB\}} K_p )
    \end{aligned}
\end{equation}
% + \sum_{i=1}^{K} N_{i, e, m, RB} / REF_{e, RB}

One may notice that the participants in experiment one have a poorer performance on two indicators: the percentage of identified ``ambiguous'' and ``unverifiable'' issues. In our observation, this is mainly because the training time of 40 minutes is too short for the participants to master the \emph{Desiree} framework, as attested by the fact that they have only modeled 2.08 requirements (out of 10) in the \emph{Desiree} tool on average. In fact, many students have spent a lot of time struggling with the syntax and the tool, and did not have enough time to analyze the requirements themselves. Given sufficient training time (2 hours in experiment two, 1 hour 45 minutes in experiment three), we can see the participants generally perform better when using \emph{Desiree} vs. the \emph{Vanilla} method.

One last thing to mention is that the learning of \emph{Desiree} varies from individual to individual. In the third experiment, we found that 24 out of the 29 (82.76\%) has better performance when using the \emph{Desiree} method, but the rest (5/29, 17.24\%) have slightly poorer performance (note that the overall performance of the 29 participants increased from 25.89\% to 36.75\%). This can be caused by individuals' learning ability (see further discussion in Section~\ref{sec:expr_threats}).

\begin{table}[!htbp]
  \begin{threeparttable}
  \caption {Detailed number of issues identified by the 16 participants in experiment one}
  \label{tab:data_detailed_issue_number_expr1}
  \vspace {0.3 cm}
  \centering
  \scriptsize
  \setlength\tabcolsep{2pt}
  %\begin{tabular}{|p{0.2\textwidth}|p{0.1\textwidth}<{\centering}|p{0.1\textwidth}<{\centering}|p{0.1\textwidth}<{\centering}|p{0.1\textwidth}<{\centering}|p{0.1\textwidth}<{\centering}|p{0.1\textwidth}<{\centering}|}
  \begin{tabular}{|c|c|c|c|c|c|c|c|c|c|c|c|c|c|c|c|c|}
  \hline

 \multirow{2}{*}{\textbf{}} & \multicolumn{2}{c|}{\textbf{Project}} & \multicolumn{2}{c|}{\textbf{Incomplete}} & \multicolumn{2}{c|}{\textbf{Ambiguous}} & \multicolumn{2}{c|}{\textbf{Inconsistent}} & \multicolumn{2}{c|}{\textbf{Unverifiable}} & \multicolumn{2}{c|}{\textbf{Unmodifiable}} & \multicolumn{2}{c|}{\textbf{Unsatisfiable}} & \multicolumn{2}{c|}{\textbf{Total}} \\ \hline
 & \emph{V} & \emph{D} & \emph{V} & \emph{D} & \emph{V} & \emph{D} & \emph{V} & \emph{D} & \emph{V} & \emph{D} & \emph{V} & \emph{D} & \emph{V} & \emph{D} & \emph{V} & \emph{D} \\ \hline
UT01 & MS & RB & 7/18 & 6/16 & 0/2 & 1/4 & 0/2 & 0/3 & 10/10 & 6/6 & 0/3 & 1/4 & 0/4 & 2/6 & 17/39 & 16/39\\ \hline
UT02 & MS & RB & 1/18 & 2/16 & 0/2 & 0/4 & 0/2 & 1/3 & 3/10 & 2/6 & 2/3 & 3/4 & 0/4 & 0/6 & 6/39 & 8/39\\ \hline
UT03 & MS & RB & 3/18 & 3/16 & 1/2 & 0/4 & 0/2 & 1/3 & 9/10 & 5/6 & 2/3 & 3/4 & 0/4 & 0/6 & 15/39 & 12/39\\ \hline
UT04 & MS & RB & 3/18 & 6/16 & 0/2 & 1/4 & 0/2 & 0/3 & 9/10 & 6/6 & 0/3 & 2/4 & 0/4 & 2/6 & 12/39 & 17/39\\ \hline
UT05 & MS & RB & 2/18 & 4/16 & 1/2 & 1/4 & 0/2 & 1/3 & 8/10 & 5/6 & 0/3 & 3/4 & 0/4 & 0/6 & 11/39 & 14/39\\ \hline
UT06 & MS & RB & 4/18 & 4/16 & 1/2 & 2/4 & 0/2 & 0/3 & 9/10 & 6/6 & 1/3 & 2/4 & 1/4 & 0/6 & 16/39 & 14/39\\ \hline
UT07 & MS & RB & 2/18 & 0/16 & 1/2 & 0/4 & 0/2 & 0/3 & 9/10 & 5/6 & 2/3 & 0/4 & 0/4 & 1/6 & 14/39 & 6/39\\ \hline
UT08 & RB & MS & 2/16 & 3/18 & 1/4 & 0/2 & 0/3 & 1/2 & 5/6 & 2/10 & 0/4 & 1/3 & 1/6 & 0/4 & 9/39 & 7/39\\ \hline
UT09 & RB & MS & 1/16 & 4/18 & 0/4 & 0/2 & 1/3 & 0/2 & 6/6 & 9/10 & 0/4 & 1/3 & 0/6 & 0/4 & 8/39 & 14/39\\ \hline
UT10 & RB & MS & 1/16 & 2/18 & 1/4 & 0/2 & 0/3 & 0/2 & 6/6 & 8/10 & 0/4 & 0/3 & 0/6 & 0/4 & 8/39 & 10/39\\ \hline
UT11 & RB & MS & 3/16 & 4/18 & 1/4 & 1/2 & 1/3 & 1/2 & 6/6 & 10/10 & 1/4 & 2/3 & 0/6 & 1/4 & 12/39 & 19/39\\ \hline
UT12 & RB & MS & 2/16 & 4/18 & 2/4 & 1/2 & 0/3 & 0/2 & 6/6 & 9/10 & 0/4 & 2/3 & 0/6 & 0/4 & 10/39 & 16/39\\ \hline
UT13 & RB & MS & 3/16 & 3/18 & 0/4 & 0/2 & 1/3 & 0/2 & 6/6 & 9/10 & 0/4 & 1/3 & 0/6 & 2/4 & 10/39 & 15/39\\ \hline
UT14 & RB & MS & 5/16 & 2/18 & 1/4 & 1/2 & 0/3 & 0/2 & 6/6 & 9/10 & 2/4 & 2/3 & 0/6 & 0/4 & 14/39 & 14/39\\ \hline
UT15 & RB & MS & 2/16 & 4/18 & 1/4 & 1/2 & 1/3 & 0/2 & 6/6 & 8/10 & 1/4 & 2/3 & 0/6 & 1/4 & 11/39 & 16/39\\ \hline
UT16 & RB & MS & 2/16 & 2/18 & 1/4 & 0/2 & 1/3 & 0/2 & 4/6 & 8/10 & 0/4 & 1/3 & 1/6 & 0/4 & 9/39 & 11/39\\ \hline
%AVG & - & - & 0.160218254 & 0.1980268959 & 0.253968254 & 0.2003968254 & 0.09259259259 & 0.126984127 & 0.8793650794 & 0.8166666667 & 0.2222222222 & 0.4722222222 & 0.03637566138 & 0.1150793651 & 0.2962962963 & 0.3331298331\\ \hline
AVG & - & - & 15.84\% & 19.49\% & 25\% & 20.31\% & 10.42\% & 12.5\% & 88.75\% & 81.46\% & 20.83\% & 46.88\% & 3.65\% & 11.46\% & 29.17\% & 33.49\%\\ \hline
  \end{tabular}
  \begin{tablenotes}
      \scriptsize
      \item AVG: average; \emph{V}: the \emph{Vanilla} session; \emph{D}: the \emph{Desiree} session; \emph{x}/\emph{y}: \emph{x} is the number of issues identified by a participant, \emph{y} is the reference number.
    \end{tablenotes}
  \end{threeparttable}
\end{table}

\begin{table}[!htbp]
  \begin{threeparttable}
  \caption {Detailed number of issues identified by the 15 participants in experiment two}
  \label{tab:data_detailed_issue_number_expr2}
  \vspace {0.3 cm}
  \centering
  \scriptsize
  \setlength\tabcolsep{2pt}
  %\begin{tabular}{|p{0.2\textwidth}|p{0.1\textwidth}<{\centering}|p{0.1\textwidth}<{\centering}|p{0.1\textwidth}<{\centering}|p{0.1\textwidth}<{\centering}|p{0.1\textwidth}<{\centering}|p{0.1\textwidth}<{\centering}|}
  \begin{tabular}{|c|c|c|c|c|c|c|c|c|c|c|c|c|c|c|c|c|}
  \hline

 \multirow{2}{*}{\textbf{}} & \multicolumn{2}{c|}{\textbf{Project}} & \multicolumn{2}{c|}{\textbf{Incomplete}} & \multicolumn{2}{c|}{\textbf{Ambiguous}} & \multicolumn{2}{c|}{\textbf{Inconsistent}} & \multicolumn{2}{c|}{\textbf{Unverifiable}} & \multicolumn{2}{c|}{\textbf{Unmodifiable}} & \multicolumn{2}{c|}{\textbf{Unsatisfiable}} & \multicolumn{2}{c|}{\textbf{Total}} \\ \hline
 & \emph{V} & \emph{D} & \emph{V} & \emph{D} & \emph{V} & \emph{D} & \emph{V} & \emph{D} & \emph{V} & \emph{D} & \emph{V} & \emph{D} & \emph{V} & \emph{D} & \emph{V} & \emph{D} \\ \hline
TS01 & RB & MS & 5/22 & 12/24 & 3/6 & 2/4 & 1/3 & 2/3 & 6/6 & 10/10 & 4/6 & 2/4 & 1/6 & 2/4 & 20/49 & 30/49\\ \hline
TS02 & RB & MS & 4/22 & 8/24 & 1/6 & 2/4 & 0/3 & 1/3 & 6/6 & 9/10 & 0/6 & 1/4 & 1/6 & 1/4 & 12/49 & 22/49\\ \hline
TS03 & RB & MS & 8/22 & 11/24 & 2/6 & 2/4 & 1/3 & 0/3 & 6/6 & 9/10 & 1/6 & 2/4 & 1/6 & 2/4 & 19/49 & 26/49\\ \hline
TS04 & MS & RB & 9/24 & 6/22 & 1/4 & 5/6 & 0/3 & 0/3 & 10/10 & 6/6 & 0/4 & 3/6 & 0/4 & 1/6 & 20/49 & 21/49\\ \hline
TS05 & MS & RB & 9/24 & 5/22 & 2/4 & 4/6 & 0/3 & 0/3 & 8/10 & 6/6 & 0/4 & 3/6 & 0/4 & 4/6 & 19/49 & 22/49\\ \hline
TS06 & MS & RB & 7/24 & 7/22 & 1/4 & 3/6 & 0/3 & 0/3 & 6/10 & 5/6 & 0/4 & 0/6 & 0/4 & 4/6 & 14/49 & 19/49\\ \hline
TS07 & MS & RB & 8/24 & 9/22 & 1/4 & 2/6 & 0/3 & 0/3 & 9/10 & 6/6 & 0/4 & 3/6 & 0/4 & 3/6 & 18/49 & 23/49\\ \hline
TS08 & RB & MS & 1/22 & 0/24 & 1/6 & 3/4 & 0/3 & 0/3 & 4/6 & 9/10 & 0/6 & 2/4 & 0/6 & 1/4 & 6/49 & 15/49\\ \hline
TS09 & RB & MS & 4/22 & 9/24 & 1/6 & 2/4 & 0/3 & 0/3 & 6/6 & 9/10 & 0/6 & 2/4 & 0/6 & 1/4 & 11/49 & 23/49\\ \hline
TS10 & MS & RB & 6/24 & 8/22 & 1/4 & 3/6 & 0/3 & 0/3 & 5/10 & 5/6 & 0/4 & 3/6 & 0/4 & 1/6 & 12/49 & 20/49\\ \hline
TS11 & MS & RB & 8/24 & 7/22 & 1/4 & 3/6 & 0/3 & 0/3 & 8/10 & 6/6 & 0/4 & 1/6 & 0/4 & 3/6 & 17/49 & 21/49\\ \hline
TS12 & MS & RB & 4/24 & 8/22 & 2/4 & 2/6 & 0/3 & 0/3 & 6/10 & 5/6 & 0/4 & 2/6 & 1/4 & 4/6 & 13/49 & 22/49\\ \hline
TS13 & MS & RB & 8/24 & 4/22 & 1/4 & 3/6 & 1/3 & 0/3 & 5/10 & 6/6 & 1/4 & 3/6 & 0/4 & 4/6 & 16/49 & 21/49\\ \hline
TS14 & RB & MS & 5/22 & 5/24 & 4/6 & 2/4 & 0/3 & 1/3 & 5/6 & 8/10 & 2/6 & 3/4 & 2/6 & 4/4 & 19/49 & 23/49\\ \hline
TS15 & RB & MS & 9/22 & 10/24 & 2/6 & 3/4 & 0/3 & 0/3 & 6/6 & 10/10 & 0/6 & 2/4 & 3/6 & 3/4 & 20/49 & 28/49\\ \hline
%AVG & - & - & 0.2705289502 & 0.3170995671 & 0.3229166667 & 0.5461309524 & 0.06845238095 & 0.09523809524 & 0.8205357143 & 0.9258928571 & 0.09895833333 & 0.4375 & 0.1108630952 & 0.5 & 0.3205174927 & 0.4590014577\\ \hline
AVG & - & - & 27.3\% & 31.64\% & 32.22\% & 54.44\% & 6.67\% & 8.89\% & 81.33\% & 92.67\% & 9.44\% & 43.33\% & 10.56\% & 50\% & 32.11\% & 45.71\%\\ \hline
  \end{tabular}
  \begin{tablenotes}
      \scriptsize
      \item AVG: average; \emph{V}: the \emph{Vanilla} session; \emph{D}: the \emph{Desiree} session; \emph{x}/\emph{y}: \emph{x} is the number of issues identified by a participant, \emph{y} is the reference number.
    \end{tablenotes}
  \end{threeparttable}
\end{table}

\begin{table}[!htbp]
  \begin{threeparttable}
  \caption {Detailed number of issues identified by the 29 participants in experiment three}
  \label{tab:data_detailed_issue_number_expr3}
  \vspace {0.3 cm}
  \centering
  \scriptsize
  \setlength\tabcolsep{2pt}
  %\begin{tabular}{|p{0.2\textwidth}|p{0.1\textwidth}<{\centering}|p{0.1\textwidth}<{\centering}|p{0.1\textwidth}<{\centering}|p{0.1\textwidth}<{\centering}|p{0.1\textwidth}<{\centering}|p{0.1\textwidth}<{\centering}|}
  \begin{tabular}{|c|c|c|c|c|c|c|c|c|c|c|c|c|c|c|c|c|}
  \hline

 \multirow{2}{*}{\textbf{}} & \multicolumn{2}{c|}{\textbf{Project}} & \multicolumn{2}{c|}{\textbf{Incomplete}} & \multicolumn{2}{c|}{\textbf{Ambiguous}} & \multicolumn{2}{c|}{\textbf{Inconsistent}} & \multicolumn{2}{c|}{\textbf{Unverifiable}} & \multicolumn{2}{c|}{\textbf{Unmodifiable}} & \multicolumn{2}{c|}{\textbf{Unsatisfiable}} & \multicolumn{2}{c|}{\textbf{Total}} \\ \hline
 & \emph{V} & \emph{D} & \emph{V} & \emph{D} & \emph{V} & \emph{D} & \emph{V} & \emph{D} & \emph{V} & \emph{D} & \emph{V} & \emph{D} & \emph{V} & \emph{D} & \emph{V} & \emph{D} \\ \hline
TP01 & MS & RB & 7/26 & 8/26 & 0/4 & 4/5 & 1/3 & 2/4 & 7/8 & 7/7 & 0/4 & 2/4 & 1/4 & 2/4 & 16/49 & 23/50\\ \hline
TP02 & MS & RB & 8/26 & 8/26 & 1/4 & 3/5 & 1/3 & 1/4 & 8/8 & 6/7 & 0/4 & 2/4 & 1/4 & 0/4 & 19/49 & 18/50\\ \hline
TP03 & MS & RB & 6/26 & 6/26 & 0/4 & 4/5 & 1/3 & 1/4 & 7/8 & 7/7 & 0/4 & 1/4 & 0/4 & 3/4 & 14/49 & 21/50\\ \hline
TP04 & MS & RB & 9/26 & 10/26 & 0/4 & 5/5 & 0/3 & 1/4 & 6/8 & 7/7 & 0/4 & 4/4 & 0/4 & 0/4 & 15/49 & 23/50\\ \hline
TP05 & MS & RB & 9/26 & 9/26 & 0/4 & 3/5 & 0/3 & 0/4 & 6/8 & 6/7 & 0/4 & 3/4 & 0/4 & 1/4 & 15/49 & 19/50\\ \hline
TP06 & MS & RB & 9/26 & 8/26 & 0/4 & 1/5 & 0/3 & 0/4 & 7/8 & 6/7 & 0/4 & 1/4 & 0/4 & 0/4 & 16/49 & 15/50\\ \hline
TP07 & MS & RB & 5/26 & 6/26 & 0/4 & 4/5 & 1/3 & 1/4 & 5/8 & 7/7 & 0/4 & 2/4 & 0/4 & 1/4 & 11/49 & 19/50\\ \hline
TP08 & MS & RB & 8/26 & 7/26 & 1/4 & 2/5 & 0/3 & 1/4 & 7/8 & 6/7 & 0/4 & 2/4 & 1/4 & 1/4 & 17/49 & 17/50\\ \hline
TP09 & MS & RB & 3/26 & 10/26 & 1/4 & 1/5 & 0/3 & 0/4 & 5/8 & 4/7 & 0/4 & 1/4 & 0/4 & 0/4 & 9/49 & 15/50\\ \hline
TP10 & MS & RB & 2/26 & 7/26 & 0/4 & 1/5 & 1/3 & 0/4 & 5/8 & 5/7 & 0/4 & 1/4 & 0/4 & 0/4 & 8/49 & 13/50\\ \hline
TP11 & MS & RB & 3/26 & 2/26 & 0/4 & 3/5 & 0/3 & 1/4 & 8/8 & 7/7 & 0/4 & 2/4 & 0/4 & 2/4 & 11/49 & 15/50\\ \hline
TP12 & MS & RB & 8/26 & 12/26 & 0/4 & 2/5 & 0/3 & 0/4 & 1/8 & 6/7 & 0/4 & 2/4 & 0/4 & 0/4 & 9/49 & 20/50\\ \hline
TP13 & MS & RB & 3/26 & 8/26 & 0/4 & 4/5 & 0/3 & 3/4 & 7/8 & 6/7 & 1/4 & 3/4 & 0/4 & 1/4 & 10/49 & 22/50\\ \hline
TP14 & RB & MS & 7/26 & 3/26 & 0/5 & 2/4 & 1/4 & 0/3 & 6/7 & 7/8 & 0/4 & 2/4 & 0/4 & 1/4 & 14/50 & 13/49\\ \hline
TP15 & RB & MS & 6/26 & 11/26 & 1/5 & 2/4 & 1/4 & 0/3 & 7/7 & 7/8 & 0/4 & 3/4 & 1/4 & 2/4 & 16/50 & 22/49\\ \hline
TP16 & RB & MS & 4/26 & 10/26 & 1/5 & 1/4 & 0/4 & 0/3 & 6/7 & 8/8 & 0/4 & 2/4 & 0/4 & 2/4 & 11/50 & 21/49\\ \hline
TP17 & RB & MS & 4/26 & 7/26 & 0/5 & 0/4 & 0/4 & 1/3 & 2/7 & 7/8 & 0/4 & 0/4 & 0/4 & 0/4 & 6/50 & 15/49\\ \hline
TP18 & RB & MS & 6/26 & 7/26 & 1/5 & 0/4 & 0/4 & 1/3 & 3/7 & 6/8 & 0/4 & 2/4 & 0/4 & 0/4 & 10/50 & 14/49\\ \hline
TP19 & RB & MS & 3/26 & 6/26 & 1/5 & 0/4 & 1/4 & 0/3 & 5/7 & 8/8 & 0/4 & 1/4 & 0/4 & 0/4 & 10/50 & 14/49\\ \hline
TP20 & RB & MS & 5/26 & 5/26 & 2/5 & 0/4 & 2/4 & 1/3 & 7/7 & 8/8 & 0/4 & 3/4 & 0/4 & 1/4 & 16/50 & 15/49\\ \hline
TP21 & RB & MS & 7/26 & 10/26 & 0/5 & 0/4 & 0/4 & 0/3 & 7/7 & 8/8 & 0/4 & 3/4 & 0/4 & 2/4 & 14/50 & 20/49\\ \hline
TP22 & RB & MS & 3/26 & 4/26 & 2/5 & 0/4 & 1/4 & 0/3 & 6/7 & 8/8 & 1/4 & 2/4 & 0/4 & 2/4 & 12/50 & 14/49\\ \hline
TP23 & RB & MS & 4/26 & 6/26 & 0/5 & 1/4 & 1/4 & 0/3 & 6/7 & 7/8 & 0/4 & 3/4 & 0/4 & 1/4 & 11/50 & 15/49\\ \hline
TP24 & RB & MS & 7/26 & 17/26 & 0/5 & 2/4 & 1/4 & 2/3 & 6/7 & 8/8 & 0/4 & 2/4 & 0/4 & 1/4 & 14/50 & 30/49\\ \hline
TP25 & RB & MS & 4/26 & 7/26 & 1/5 & 1/4 & 1/4 & 0/3 & 5/7 & 7/8 & 1/4 & 0/4 & 0/4 & 0/4 & 11/50 & 15/49\\ \hline
TP26 & RB & MS & 6/26 & 8/26 & 2/5 & 0/4 & 1/4 & 0/3 & 6/7 & 5/8 & 0/4 & 0/4 & 0/4 & 0/4 & 15/50 & 13/49\\ \hline
TP27 & RB & MS & 4/26 & 8/26 & 1/5 & 2/4 & 0/4 & 0/3 & 7/7 & 8/8 & 0/4 & 3/4 & 0/4 & 2/4 & 12/50 & 20/49\\ \hline
TP28 & RB & MS & 3/26 & 8/26 & 1/5 & 2/4 & 0/4 & 0/3 & 7/7 & 8/8 & 0/4 & 1/4 & 0/4 & 0/4 & 11/50 & 18/49\\ \hline
TP29 & RB & MS & 9/26 & 14/26 & 2/5 & 2/4 & 2/4 & 1/3 & 6/7 & 8/8 & 1/4 & 2/4 & 0/4 & 3/4 & 19/50 & 28/49\\ \hline
%AVG & - & - & 0.2169008876 & 0.3068602071 & 0.1225961538 & 0.4018028846 & 0.1578525641 & 0.1682692308 & 0.790521978 & 0.9004979396 & 0.03305288462 & 0.4765625 & 0.03665865385 & 0.2385817308 & 0.2596879906 & 0.3676510989\\ \hline
AVG & - & - & 21.49\% & 30.77\% & 12.93\% & 38.45\% & 16.09\% & 16.38\% & 79.37\% & 90.27\% & 3.45\% & 47.41\% & 3.45\% & 24.14\% & 25.89\% & 36.75\%\\ \hline
  \end{tabular}
  \begin{tablenotes}
      \scriptsize
      \item AVG: average; \emph{V}: the \emph{Vanilla} session; \emph{D}: the \emph{Desiree} session; \emph{x}/\emph{y}: \emph{x} is the number of issues identified by a participant, \emph{y} is the reference number.
    \end{tablenotes}
  \end{threeparttable}
\end{table}

\subsection{Hypothesis Testing}
\label{sec:expr_hypo_testing}

%We can statistically analyze the differences in terms of identified issues between the \emph{Desiree} and the \emph{Vanilla} approach. We first performed normality tests on the differences of all 10 indicators in Table~\ref{tab:req_issues_kinds} using the Shapiro-Wilk test, and found that only 4/10 of the indicators in experiment one and 3/10 of the indicators in experiments two are normally distributed. Hence, we decided to use Wilcoxon Signed-Rank tests (WSR), a non-parametric statistical test that is used when comparing repeated measurements on a single sample, for our hypothesis testing. As a complement, we also ran the paired t-test, which is robust to moderate violation of normality, on these indicators. Note that the WSR test tests the median while the paired T-test tests the mean of two groups of measurements, i.e., the percentage of issues identified by each student when using Desiree and the Vanilla method. We report the p-values in Table~\ref{tab:statistics_hypo_test}.

We statistically analyzed the participants' differences in terms of identified issues when using \emph{Desiree} vs. the \emph{Vanilla} approach. Since we have far less than 30 participants in experiment one and two, and our Shapiro-Wilk Normality~\cite{_shapiro-wilk_2015} tests showed that the participants' differences in experiment three are not normally distributed on 3/7 of the issue indicators, we employed both \emph{paired Student's t} test~\cite{_students_2015} and \emph{Wilcoxon Signed-Rank} test (WSR)~\cite{_wilcoxon_2015} for our one-tailed hypothesis testing.The paired T test assumes that the differences between pairs (repeated measurements on a sample, before and after a treatment) are normally distributed, and is robust to moderate violation of normality~\cite{mcdonald_handbook_2009}. As a complement, the WSR test is a non-parametric alternative to the paired T test if the differences between pairs are severely non-normal~\footnote{http://www.biostathandbook.com/pairedttest.html}~\cite{mcdonald_handbook_2009}.

\begin{table}[!htbp]
  %\vspace {-0.5 cm}
  \caption {Statistical p-value for issues identified by participants }
  \label{tab:statistics_hypo_test}
  \vspace {0.3 cm}
  \centering
  \small
  %\scriptsize
  \setlength\tabcolsep{2pt}
  %\begin{tabular}{|p{0.2\textwidth}|p{0.1\textwidth}<{\centering}|p{0.1\textwidth}<{\centering}|p{0.1\textwidth}<{\centering}|p{0.1\textwidth}<{\centering}|p{0.1\textwidth}<{\centering}|p{0.1\textwidth}<{\centering}|}
  \begin{tabular}{|c|c|c|c|c|c|c|}
  \hline
     & \multicolumn{2}{c|}{\textbf{Experiment One}} & \multicolumn{2}{c|}{\textbf{Experiment Two}} & \multicolumn{2}{c|}{\textbf{Experiment Three}} \\ \hline
     & \textbf{Paired T} & \textbf{WSR} & \textbf{Paired T} & \textbf{WSR} & \textbf{Paired T}  & \textbf{WSR} \\ \hline
        Incomplete & 0.08331 & \textbf{0.03717} & 0.10593 & 0.1221 & \textbf{0.00007} & \textbf{0.00021} \\ \hline
        Ambiguous & 0.75764 & 0.71403 & \textbf{0.00144} & \textbf{0.00288} & \textbf{0.00069} & \textbf{0.00145} \\ \hline
        Inconsistent & 0.37845 & 0.37097 & 0.33507 & 0.30345 & 0.47585 & 0.66204 \\ \hline
        Unverifiable & 0.91854 & 0.87966 & \textbf{0.01428} & \textbf{0.02747} & \textbf{0.00396} & \textbf{0.00519} \\ \hline
        Unmodifiable & \textbf{0.00300} & \textbf{0.003} & \textbf{0.00001} & \textbf{0.00052} & $<$\textbf{0.00001} & $<$\textbf{0.00001} \\ \hline
        Unsatisfiable & 0.07636 & 0.09583 & $<$\textbf{0.00001} & \textbf{0.00032} & \textbf{0.00006} & \textbf{0.00013} \\ \hline
        Total & 0.06068 & 0.05666 & $<$\textbf{0.00001} & \textbf{0.00032} & $<$\textbf{0.00001} & \textbf{0.00001} \\ \hline
  \end{tabular}
  %\vspace {-0.55 cm}
\end{table}

%From this table, we can see that there is strong evidence that we shall reject the null hypothesis $H_0$ as specified in Eq.~\ref{eq:eq_hypo_test}. That is, the experiment results provide very strong evidence that \emph{Desiree} can help people to identify more issues in general (the last row)

We report the p-values in Table~\ref{tab:statistics_hypo_test}~\footnote{Our hypothesis tests are conducted using the R software package (https://www.r-project.org/).}. We can see that there is strong evidence that \emph{Desiree} can help people to identify more issues in general (the last row): for both tests, p-value $\leq$ 0.00032 $<< \alpha = 0.05 $ (the common confidence level) in experiment two and three, and  p-value $\approx$ 0.05 in experiment one. Specifically, there are strong evidence that the \emph{Desiree} framework is able to help people to identify more ``Incomplete'', ``Ambiguous'', ``Unverifiable'', ``'Unmodifiable', and ``Unsatisfiable'' issues (their p-values $\leq$ 0.05 in at least two experiments). We also see that there is no evidence that \emph{Desiree} can help people to identify more ``Inconsistent'' issues (p-value $>>$ 0.05) in all the three experiments.

%Also, there is strong evidence in experiment two that Desiree can help people to identify more ambiguous, vague and unsatisfiable issues, and more issues in total (p-value $<$ 0.05). In both experiments, we see no evidence that \emph{Desiree} can be helpful on identifying incomplete (requirement), inconsistent, and un-traceable issues (p-value $>$ 0.05).

%To safely draw the conclusion of rejecting the null hypothesis $H_0$ in Eq.~\ref{eq:eq_hypo_test} (and other null hypotheses with p-value $<$ 0.05), there are two other things to be considered: (1) \emph{accumulated type I error}; (2) \emph{effect size}.

Note that there is a potential risk of \emph{accumulated type I error}~\cite{leblanc_statistics:_2004} since we are conducting multiple significance tests. A {type I error} means that we falsely reject the null hypothesis due merely to random sampling variation~\cite{leblanc_statistics:_2004}; that is, there is no significant difference. An accumulated type I error means the probability of making one or more type I errors out of the multiple tests. In our case we have 7 tests, the probability of making one or more type I errors out of the 7 tests is much larger that 0.05 (indeed, 0.3017~\footnote{This probability $\alpha_m$ is calculated by $\alpha_m = 1 - (1 - \alpha)^n$, where $\alpha = 0.05$, $n$ is the number of tests.}). Therefore, we applied the Bonferroni adjustment~\cite{_bonferroni_2015} to the p-values obtained in experiment three, and show them in Table~\ref{tab:statistics_analysis}.

\begin{table}[!htbp]
  \vspace {-0.1 cm}
  \caption {Analyzing experiment three: p-values, statistics, and effect}
  \label{tab:statistics_analysis}
  \vspace {0.3 cm}
  \centering
  \small
  \setlength\tabcolsep{2pt}
  %\begin{tabular}{|p{0.2\textwidth}|p{0.1\textwidth}<{\centering}|p{0.1\textwidth}<{\centering}|p{0.1\textwidth}<{\centering}|p{0.1\textwidth}<{\centering}|p{0.1\textwidth}<{\centering}|p{0.1\textwidth}<{\centering}|}
  \begin{tabular}{|c|c|c|c|c|c|c|c|c|}
  \hline
    & \multicolumn{4}{c|}{\textbf{Paired T Test}} & \multicolumn{4}{c|}{\textbf{Wilcoxon Signed-Rank (WSR)}}\\ \hline
    & Pvalue & Bonferroni & t(28) & Effect & Pvalue & Bonferroni & Zvalue & Effect \\ \hline
    Incomplete & 0.00007 & 0.00046 & 4.4302 & 0.82266 & 0.00021 & 0.0015 & 3.52263 & 0.64314 \\ \hline
    Ambiguous & 0.00069 & 0.0048 & 3.5536 & 0.6599 & 0.00145 & 0.01017 & 2.97754 & 0.54362 \\ \hline
    Inconsistent & 0.47585 & 1 & 0.06111 & 0.01135 & 0.66204 & 1 & -0.41805 & -0.07633 \\ \hline
    Unverifiable & 0.00396 & 0.02773 & 2.8598 & 0.53105 & 0.00519 & 0.03631 & 2.5631 & 0.46796 \\ \hline
    Unmodifiable & $<$0.00001 & $<$0.00001 & 8.6828 & 1.61236 & $<$0.00001 & 0.00002 & 4.57097 & 0.83454 \\ \hline
    Unsatisfiable & 0.00006 & 0.00044 & 4.4458 & 0.82556 & 0.00013 & 0.00094 & 3.64313 & 0.66514 \\ \hline
    Total & $<$0.00001 & $<$0.00001 & 6.5715 & 1.22030 & 0.00001 & 0.00008 & 4.24921 & 0.7758 \\ \hline
  \end{tabular}
\end{table}

%(i.e., no difference between the two groups being tested)

We also report here the related statistics, i.e., t values for the paired T tests, Z values for the Wilcoxon Signed-Rank tests. The p-values can be interpreted as the probability, under the assumption of the null hypothesis, of observing a result that is equal to or more extreme than an observed t-value (or Z-value). That is, the larger magnitude (positive or negative) a t value is of, the lesser likelihood it will be observed. For example, in our case, the probability of observing t(28) $\geq$  6.5715~\footnote{Here 28 is the degree of freedom, obtained from $n - 1$ ($n$ is the number of observations).} is less than 0.00001 (the adjusted p-value of the paired T test in the last row of Table~\ref{tab:statistics_analysis}) if we assume no difference between the \emph{Desiree} group and the \emph{Vanilla} group. The very small adjusted p-value (p-value $< 0.00001 << 0.05$) indicates a very strong evidence that the samples are not from the null distribution, and we can reject the null hypothesis at the confidence level $\alpha = 0.05$. The Z-values of the Wilcoxon Signed-Rank test can be interpreted similarly.

% (assuming that samples are from the distribution under the null hypothesis),
We also analyzed the \emph{effect sizes} (ES), which are shown Table~\ref{tab:statistics_analysis}. Effect size is the magnitude of the difference between two groups, i.e., the magnitude of a treatment effect~\cite{sullivan_using_2012}. We consider effect size because the significance analysis just says that the \emph{Desiree} method can help people to identify more issues when engineering requirements, but we do not know to what degree it affects the participants. We calculated the effect sizes for the hypothesis tests in experiment three using Eq.~\ref{eq:eq_es}~\cite{cohen_statistical_2013}, where $M_1$ and $M_2$ are the means of the two groups of samples, $S_1$ and $S_2$ are their standard deviations.
%are also interested in seeing
%\emph{d} is Cohen's \textbf{\emph{d}}~\cite{cohen_statistical_2013},
%\begin{comment}
\begin{equation}\label{eq:eq_es}
    \begin{aligned}
    ES = d / \sqrt{(d^2 + 4)}, \quad d = ( M_1 - M_2) / \sqrt{(S_1^2 + S_2^2)/2}
    \end{aligned}
\end{equation}
%\end{comment}

%We show the effect sizes also in Table~\ref{tab:statistics_analysis}.
According to Cohen's conventional criteria of ``small'' (effect size from 0.2 to 0.3 ), ``medium'' (around 0.5), or ``big'' (0.8 to infinity) effect~\cite{cohen_statistical_2013}, our effect sizes for ``Total'', ``Incomplete'', ``Ambiguous'', ``Unverifiable'', ``Unmodifiable'' and ``Unsatisfiable'' issues in the third experiment fall into either the ``medium'' or the ``large'' category.

More specifically, Coe~\cite{coe_its_2002} has presented the interpretations of effect sizes ranging from 0.1 to 3.0.  According to Coe~\cite{coe_its_2002}, an effect size of 0.8 means that the score of the average person in the experimental group is 0.8 standard deviations above the average person in the control group, and hence exceeds the scores of 79\% of the control group. Take ``Incomplete'' that has an associated effect size of 0.82266 for example, the average person in the experimental \emph{Desiree} group (ranked $15^{th}$, 29 in total) will score higher than $23 = 29 \times 79\%$ of the students in the control \emph{Vanilla} group (i.e., ranked $6^{th}$, also 29 in total). Checking the data carefully, we found the participant ranked $15^{th}$ in the \emph{Desiree} group was on a par with the participants ranked $5^{th} - 7^{th}$ in the \emph{Vanilla} group (these three participants have identified the same number of issues).

%According to Coe~\cite{coe_its_2002}, a effect size of 0.8 means that the average person in the experimental group would score higher than 79\% of a control group that was initially equivalent.

%using the effect size under the WSR test and paired T test
We calculated the predicated ranks within the \emph{Vanilla} group for the average person in the \emph{Desiree} group on each type of issue in experiment three, and checked their actual ranks. We show these results in Table~\ref{tab:es_analysis}. We can see that the predictions generally matches well with the actual ranks (some of the actual ranks are better than the predictions). We can also find that the predictions that are made based on the effect sizes provided by the paired T test match better with the real situation.

\begin{table}[!htbp]
  \vspace {-0.1 cm}
  \centering
  %\begin{threeparttable}
  \caption {Effect analysis for experiment three}
  \label{tab:es_analysis}
  \vspace {0.3 cm}
  \centering
  \small
  \setlength\tabcolsep{2pt}
  %\begin{tabular}{|p{0.15\textwidth}|p{0.1\textwidth}<{\centering}
  %|p{0.1\textwidth}<{\centering}|p{0.1\textwidth}<{\centering}|p{0.1\textwidth}<{\centering}|p{0.1\textwidth}<{\centering}|}
  \begin{tabular}{|c|c|c|c|c|c|c|c|}
  \hline
     & \multicolumn{3}{p{0.3\textwidth}<{\centering}|}{\textbf{Paired T Test}} &
     & \multicolumn{3}{p{0.3\textwidth}<{\centering}|}{\textbf{WSR Test}}\\ \hline
     & Effect  & Percentage & Predicted & \textbf{Actual} & Predicted & Percentage & Effect \\ \hline
    Incomplete & 0.82266 & 79\% & $6^{th}$ & \bm{ $5^{th} - 7^{th}$} & $8^{th}$ & 73\% & 0.64314 \\ \hline
    Ambiguous & 0.6599 & 76\% & $7^{th}$ & \bm{$1^{th} - 4 ^{th}$} & $9^{th}$ & 69\% & 0.54362 \\ \hline
    Inconsistent & 0.01135 & 50\% & $15^{th}$ & \bm{$16^{th}$} & $15^{th}$ & 50\% & -0.07633 \\ \hline
    Unverifiable & 0.53105 & 69\% & $9^{th}$ & \bm{$8^{th} - 12 ^{th}$} & $9^{th}$ & 69\% & 0.46796 \\ \hline
    Unmodifiable & 1.61236 & 95\% & $1^{th}$ & \bm{$1^{th}$} & $6^{th}$ & 79\% & 0.83454 \\ \hline
    Unsatisfiable & 0.82556 & 79\% & $6^{th}$ & \bm{$1^{th} - 4^{th}$} & $7^{th}$ & 76\% & 0.66514 \\ \hline
    Total & 1.22030 & 88\% & $3^{th}$ & \bm{$3^{th}$} & $6^{th}$ & 79\% & 0.7758 \\ \hline
  \end{tabular}
  %\begin{tablenotes}
  %    \small
  %    \item Percentage: the percentage of the \emph{Vanialla} group who would be below the $15^{th}$ person in the \emph{Desiree} group; Predict: the predicted rank of person in the \emph{Vanialla} group of 29 who would be equivalent to the $15^{th}$ person in \emph{Desiree} group.
  %  \end{tablenotes}
  %\end{threeparttable}
\end{table}

\subsection{Analysis}
\label{sec:expr_discussion}
In general, the results meet our expectations. In this subsection, we discuss the reasons why \emph{Desiree} can help people to find more requirements issues, and our observations in the experiments.

\emph{Incomplete}. In our observations, \emph{Desiree} is helpful in identifying incomplete requirements issues mainly because: (1) the description-based syntax drives analysts/engineers to think about the kinds of properties that shall be associated with the capability when specifying a function; (2) the syntax facilitates the consideration of ``\emph{which attributes shall be used to describe the description} (\emph{filler})?'' when specifying a slot-description pair. This helps to identify more missing slots/properties and missing content requirements, and thus contributes to improving the incompleteness of individual requirements and requirements specifications, respectively. Take ``\emph{the system shall be able to search meeting rooms records}'' as an example, with \emph{Desiree}, many participants were able to find the following missing information: who can search? what kinds of search parameters shall be used? Further, more participants have asked ``\emph{what kinds of information shall a meeting room record include}?'', identifying a missing content requirement.

%Moreover, the introduction of content goals contributes to find some missing data requirements.

%\emph{Ambiguous}. When interpreting a requirement, if there is more than one way to structure it, it indicates an ambiguity. For example, the simple requirement ``download contact info for client'' is ambiguous since ``for client'' can be attached to either the function ``download'' or the entity ``contact info'': ``Download $<$object: Contact\_info$>$ $<$for: Client$>>$'' or ``Download $<$object: Contact\_info $<$associated\_with: Client$>>$''. Moreover, the \emph{Desiree} framework works well on disambiguating the semantics of ``AND'' and ``OR'': in an ``AND-reduce/operationalization'', all the sub-elements need to be satisfied in order to fulfill the parent element; in an ``OR-reduce/operationalization'', the satisfaction of any of the sub-elements would fulfill the parent. This distinction inspires the participants to ask ``shall a meeting notification be sent (via Email and SMS) at the same time or can be sent separately?'' when interpreting the requirement ``the product shall be able to send meeting notifications via Email and SMS''.

\emph{Ambiguous}. \emph{Desiree} offers operational rules for identifying potential ambiguities: (1) checking the subject of a slot (property); (2) checking the cardinality of the description of a slot in a function description. These rules are shown to be useful in our experiments. For example, more students have identified the ambiguity in the requirement ``\emph{the system shall be able to download contact info for client}'': is ``for client'' attached to the function ``download'' or the entity ``contact info''?  More interestingly, for the requirement ``\emph{the system shall allow privileged users to view meeting schedules in multiple reporting views}'', after addressing the unverifiable issues of ``privileged user'' and ``multiple'', several participants have further asked ``\emph{Shall these reporting views be opened simultaneously or not}?'', identifying an implicit ambiguity issue.

\emph{Unverifiable}. We observed that the participants can easily find simple unverifiable issues in given requirements, but tend to miss ``deep'' vague issues in stakeholders' answers when using the \emph{Vanilla} method. With \emph{Desiree}, the structuring of each requirement could remind them about implicit unverifiable issues. For example, most of the participants were able to justify ``\emph{the product shall have good usability}'' as unverifiable, but few of them realized that ``\emph{the product shall be easy to learn for realtors}'', which was given by stakeholders as a refinement of the previous requirement, is still vague. With \emph{Desireee}, participants would keep asking ``\emph{how to measure easy}?''. That is, when using the \emph{Vanilla} method, participants were more likely to accept vague stakeholder answers, while using \emph{Desiree}, they were more likely to notice and correct vague responses.

%We carefully analyzed participants' online discussions and refined specifications with regard to vagueness and ambiguity, to check why the participants in experiment two are much better on the two aspects.

%\emph{Un-modifiable}. For a given requirement, with \emph{Desiree}, an analyst first needs to identify its concerns and separate the concerns if there are several. This helps to reduce ``unmodifiable'' issues (avoiding intermixed requirements). Note that \emph{Desiree} captures the interrelations between functions, qualities, and content (as discussed in Section ??), thus it offers much better modifiability and traceability. We do not compare these two approaches on this aspect because the \emph{Vanilla} approach does not possess such features.

\emph{Un-modifiable}. \emph{Desiree} requires analysts/engineers to identify the concerns of a requirement, and separate them if there are several. This helps to avoid intermixed requirements. With \emph{Desiree}, many participants were able to successfully decouple composite requirements into simple ones. For example, they decoupled ``\emph{the system shall be able to generate a CMA} (\emph{Comparative Market Analysis}) \emph{report in acceptable time}'' into ``\emph{generate a CMA report}'' ($F_1$ := Generate $<$object: CMA\_report$>$) and ``\emph{the generation shall be in acceptable time}'' ($QG_2$ := Processing\_time ($F_1$) :: Acceptable). Further, they were able to capture interrelations between requirements by utilizing the \emph{Desiree} tool. For example, in the above example, the two elements are interrelated through the use of $F_1$ as the subject of $QG_2$. This enables us to systematically identifying the requirements to be affected when updating a requirement.

%\emph{Un-satisfiable}. Our \emph{Desiree} has offered a ``de-Universalize'' operator, which is designed for weakening requirements in order to make them practically satisfiable. Moreover, the \emph{Desiree} tool is able to provide hints when applying the ``Observe'' operator, asking if de-universalization is needed. These can help analysts to find more potentially unattainable requirements.

\emph{Un-satisfiable}. \emph{Desiree} offers a ``de-Universalize'' operator for weakening requirements in order to make them practically satisfiable. The supporting tool also provides hints for relaxation when the ``Observe'' operator is applied. As such, the participants were able to identify more potentially un-satisfiable issues. For example, when operationalizing the QG ``\emph{the search of meeting rooms shall be intuitive}'' by assigning surveyed users, many of them have asked ``\emph{how many percentage of the surveyed users shall agree}?''

%\emph{Inconsistent}. Our framework assumes that conflicts are explicitly defined by analysts and provides an ``Resolve'' operator to resolve them. Also, if a requirements specification is written using our syntax, we can translate it to an OWL2 ontology and perform the inconsistency reasoning (to do this, we need relevant knowledge axioms, e.g., ``Open'' conflicts with ``Close'', ``Busy'' conflicts with ``Free''). Third, to keep the consistency of terminologies, the \emph{Desiree} tool has tried to be helpful on reusing defined terminologies/concepts through syntax-based content assistance.  However, its effect is not demonstrated in these experiments. One possible reason is that the participants's proficiency on \emph{Desiree} is insufficient since many students have pointed out that they need more time to get familiar with the syntax.

\emph{Inconsistent}. Our framework assumes that conflicts are explicitly defined by analysts and provides an ``Resolve'' operator to resolve them. As such, the framework does not as yet offer much help in identifying inconsistency issues. However, if a requirements specification is written using our syntax, we can translate it to an OWL2 ontology and perform the inconsistency reasoning (to do this, we need relevant knowledge axioms, e.g., ``Open'' conflicts with ``Close'', ``Busy'' conflicts with ``Free''). Moreover, to keep the consistency of terminologies, the \emph{Desiree} tool has tried to be helpful on reusing defined terminologies/concepts through syntax-based content assistance. However, its effect is not demonstrated in these experiments. One possible reason is that the participants's proficiency on \emph{Desiree} is insufficient since many students have pointed out that they need more time to get familiar with the syntax.

%Further, when specifying a slot restriction, \emph{Desiree} could drive people to think about ``what will be included in the filler'', identifying some possibly missing content requirements (``Incomplete'' issues). In the example requirement, a trained participant may find two missing CTGs: what does contact info include? What kind of info can be used to describe a client? Finally, the use of the ``de-universalize'' operator can help analysts to find potentially unattainable requirements.

%We can see strong evidence from the experiments that with sufficient training ($¡Ý$ 2 hours), Desiree can indeed help people to identify more issues when transforming stakeholder requirements into specifications. Hence, we can accordingly reject the null hypothesis, H0.
\subsection{Feedback}
\label{sec:expr_feedback}

We have also conducted a survey on the learnability, usability, usefulness and complexity of the two methods in each experiment. In general, the majority of the participants have reported that the \emph{Desiree} framework is useful or very useful, but is hard to learn.

\begin{enumerate}
  \item In the first experiment, we have collected 13 responses. In this survey, the participants reported that the \emph{Desiree} framework is complex (7/13), hard to learn (6/13), hard to use (5/11), but useful/very useful (9/13).
  \item In the second experiment, we have collected 14 responses. In this survey, the participants also reported that the \emph{Desiree} framework is complex (11/14) and useful/very useful (12/14). More participants have reported that \emph{Desiree} is neutral, instead of hard, on its learnability (7/14) and usability (7/14).
  \item In the third experiment, we have collected 24 responses. In the survey, the participants have reported that \emph{Desiree} is complex (4/24) or neutral (8/24), hard to learn (11/24), hard to use (7/24) or neutral to use (7/24), and many reported that \emph{Desiree} is useful/very useful (20/24),
\end{enumerate}

Specifically, the participants have reported that \emph{Desiree} is useful because it offers a structured way for classifying and representing requirements, and provides a systematic method for reducing complex requirements. They also pointed out that the framework is hard to learn mainly because of its grammar.

\begin{comment}
\begin{table}[!htbp]
  \centering
  \small
  \setlength\tabcolsep{2pt}
  \caption {Statistics of the Questionnaires}
  \label{tab:promise}
  \vspace {0.3 cm}
  \begin{tabular}{|c|c|c|c|c|c|c|c|}
  \hline
   & & \multicolumn{3}{p{0.2\textwidth}<{\centering}|}{\textbf{Complexity}} & \textbf{Learnability} & \textbf{Usability} & \textbf{Usefulness}\\ \hline
   & & (very) simple & neutral & (very) nard & & & \\ \hline
   \multirow{2}{*}{\textbf{E-One}} & Vanilla & & & & & &  \\
   & Desiree & & &  & & & \\ \hline
   \multirow{2}{*}{\textbf{E-Two}} & Vanilla & & & & & &  \\
   & Desiree & & & & & & \\ \hline
   \multirow{2}{*}{\textbf{E-Three}} & Vanilla & &&  & & &  \\
   & Desiree & & & & & & \\ \hline
  \end{tabular}
\end{table}
\end{comment}

\begin{comment}
\begin{table}[!htbp]
  \centering
  \small
  \setlength\tabcolsep{2pt}
  \caption {Statistics of the Questionnaires}
  \label{tab:promise}
  \vspace {0.3 cm}
  \begin{tabular}{|c|c|c|c|c|c|}
  \hline
   %& {\textbf{Complex}} & \textbf{Learnability} & \textbf{Usability} & \textbf{Usefulness}\\ \hline
   & {\textbf{Complex}} & \textbf{Learnability} & \textbf{Usability} & \textbf{Usefulness}\\ \hline
   {\textbf{Experiment One}} & & & &  \\ \hline
   {\textbf{Experiment Two}}  & & & &  \\ \hline
   {\textbf{Experiment Three}} & & & &  \\ \hline
  \end{tabular}
\end{table}
\end{comment}

In addition, we have got positive comments from the participants in each experiment.

\begin{table}[!htbp]
  \centering
  %\small
  %\setlength\tabcolsep{2pt}
  \begin{tabular}{|cp{0.85\textwidth}|} %<{\centering}
  \hline
  1. & ``\emph{Desiree} embodies correctness check. It enforces you to think if what you are doing is right, e.g., functional goals, quality goals ....'' \\
  2. &  ``The method helps a lot when reducing the complex requirements and help with the standard representation of those items. Nothing is useless. The method makes the analysis process clearer more or less. '' \\
  3. & ``The tool makes me thinking in the structural and the mind is more MECE (Mutually Exclusive, Collectively Exhaustive)'' \\
  4. & ``I will use the method. Like refine the goals and split it into 3 kind of goal: functional goal, quality goal and content goal. It is very useful for thinking overall.''\\
  5. & ``Maybe in the future I will use the \emph{Desiree} to pick functional requirement and non-functional (requirements) from a natural language (text).''\\
  \hline
  \end{tabular}
\end{table}

Interestingly, we have got feedback from the lecturer of the RE course, for which the participants of the first experiments were registered. He told us that many students became more analytical after the experience. When they were writing requirements, especially functional requirements, they followed a more structured approach. For example, students would ask questions that they did not ask before, e.g., who performs a function, and what is the purpose of a function. Interested readers can refer to Appendix.~\ref{cha:appendixes_feedback} for our questionnaire report.

%Interested readers can refer to our questionnaire and reports at https://goo.gl/oeJ9Fi.

\subsection{Threats to Validity}
\label{sec:expr_threats}
There are several threats to the validity of our evaluation.

  %\item \emph{Inconsistency between stakeholders' answers}. Stakeholders may give inconsistent answers to the same question. For example, for the question ``what is MLS (multiple listing service)'', a stakeholder may take it as ``what is the literal meaning of MLS'', and another one may understand it as ``what info shall a MLS include''. In this situation, we assume that the participants who ask this question have identified the corresponding issue, e.g., a missing content requirement about MLS in this example.
      % Also, a stakeholder may give extra information, which may prevent participants from asking further questions, and provide potential hints for dealing with other requirements.

\begin{enumerate}
  \item \emph{Independence between participants.} We have tried to minimize mutual interference between participants in each experiment by: (1) assuming an exam scenario and asking them to perform the experimental task individually; and (2) requiring them to use texting to communicate with stakeholders instead of speaking aloud (we had only 4 face-to-face conversations in the \emph{Vanilla} session of experiment three since these students did not bring their laptops).
  \item \emph{Assessment}. The experiment results were evaluated by only one person. We have used objective and consistent rules for making judgments, to minimize the impact of individual subjectivity. For example, stakeholders may give inconsistent answers to the same question ``\emph{what is MLS} (\emph{multiple listing service})'': a stakeholder may take it as ``\emph{what is the literal meaning of MLS}'', and another one may understand it as ``\emph{what info shall a MLS include}''. In this situation, we assume that all the participants who asked this question have identified the corresponding issue, e.g., a missing content requirement about MLS in this example.
  \item \emph{The nature of participants}. Most of the participants in our experiments are students, but some of the participants in the third experiment are part-time master students. The survey in the third experiment shows that 8 out of the 24 responded participants (29 participants in total) have more than 1 year's work experiences (5 out of the 8 are on RE). Compared to the participants who do not have work experiences, we observed in experiment three that the experienced participants were able to learn the \emph{Desiree} framework more quickly, and identify requirements issues in a more systematic way with \emph{Desiree}. Also, holding the studies in two different universities provides more confidence in the generalizability of our results. We could further minimize this threat by conducting our experiment in an realistic industrial setting.
  \item \emph{Sample size.} We have ran three experiments, and have collected 16, 15, and 29 complete samples, respectively. Especially, in the third experiment, the sample size of 29 is relatively safe to assume the normality for paired T test (in fact, our normality tests showed that the paired differences on 4/7 of the issue indicators are normally distributed). Moreover, we have ran the Wilcoxon Signed-Rank test, which does not assume any distribution of the population, as a complement of the paired T test. The threat of generalizing our conclusion is relative low.
  \item \emph{Training}. In our experiments, the \emph{Desiree} framework was taught by the designer; and how the \emph{Vanilla} RE method was taught may affect results. We have tried to be fair in teaching and not bias the results.
  \item \emph{Projects}. The \emph{Desiree} method can be more or less successful for different types of projects, e.g., larger or more realistic. We have tried to mitigate this by using more than one project.
  %\item \emph{Order}. \emph{Desiree} is used after the Vanilla method in each experiment. Although each participant applied the \emph{Desiree} approach to a different project in his/her second task, he/she may have learned from the \emph{Vanilla} application in his/her first task.
  \item \emph{Order}. The \emph{Desiree} method is used after the \emph{Vanilla} method in each experiment. Although each participant applied \emph{Desiree} to a different project in the second task, s/he may have learned from the \emph{Vanilla} application in her/his first task. We could have done counterbalancing: some groups apply \emph{Vanilla} then \emph{Desiree}, and others apply \emph{Desiree} then \emph{Vanilla}; however, this setup would have been difficult to implement as part of the course design, with alternating tutorials and exercises for different groups of students.
\end{enumerate}

% In the second and third experiments, the participants came from different areas. For example, in experiment two, we have 5 participants from areas such as workflow (2/15), data mining (2/15) and database (1/15) except software engineering (10/15). However, we did not include RE profes-sionals. This is a common threat in experiments.

\section{Chapter summary}
\label{cha:eval_summary}
In this chapter, we have presented a series of empirical evaluations conducted to evaluate our proposal, including: (1) assessing the coverage of our requirements ontology by applying it to all the 625 requirements in the PROMISE dataset~\cite{menzies_promise_2012}; (2) evaluating the expressiveness
of our description-based language by using it to rewrite all the 625 requirements in the dataset; (3) illustrating our methodology by performing a realistic \emph{Meeting Scheduler} case study; (4) evaluating the the effectiveness of the entire \emph{Desiree} framework through three controlled experiments. The evaluation results have suggested that our ontology, syntax and methodology are adequate in capturing requirements in practice, and have provided strong evidence that the \emph{Desiree} framework indeed can help people to perform better requirements engineering with a medium or big effect.

%\makeatletter
%\addcontentsline{toc}{chapter}{Bibliography}
%\bibliographystyle{plain}
%\bibliographystyle{re}
%\bibliography{REProb}

%\end{document}

%% file: conclusion.tex
%\vspace{4cm}
\chapter{Conclusion and Future Work}
\label{cha:conclusion}
In this thesis, we have proposed \emph{Desiree}, a requirements calculus for incrementally transforming stakeholder requirements into an eligible requirements specification, intending to effectively address the requirements problem, which was characterized by Jackson and Zave~\cite{jackson_deriving_1995} as $DA, S \models R$, i.e., finding the specification \emph{S} that for certain domain assumptions \emph{DA} entails given requirements \emph{R}.

The \emph{Desiree} framework includes a requirements ontology (a set of requirements concepts), a set of requirement operators, a description-based syntax for representing these concepts and operators, and a systematic methodology for applying the concepts and operators in order to transform stakeholder requirements into a formal, complete enough, unambiguous, consistent, measurable, satisfiable, modifiable and traceable requirements specification. Moreover, we have evaluated the coverage of our requirement ontology, the expressiveness of our requirements language, the applicability of our methodology, and the effectiveness of the entire framework, through a series of empirical evaluations.

In this chapter, we conclude the thesis by summarizing its contributions, pointing out its limitations, and sketching directions for further research.

\section{Contributions}
\label{sec:conclusion_controbution}
Our \emph{Desiree} framework is designed for requirements engineers, who play the role of bridging stakeholders and system developers. Users of our framework need to have necessary knowledge and/or need to be trained. Our \emph{Desiree} approach can be used in traditional software development processes, and can also be adapted into agile software development processes as our description-based language captures a rich set of interrelations between requirements and contributes to improve the modifiability of requirements specifications fragments (in an agile development process, each iteration has certain allocated user stories, which need to be refined into requirement specification fragments, see our discussion about the characteristics of agile RE in Section \ref{sec:RE_in_Agile}).

In general, the contributions of this work can be grasped from the research questions that we have presented in Section~\ref{sec:solution} (see also Section~\ref{sec:innovation}).
\begin{comment}
\begin{enumerate}
  \item What are the right concepts for requirements modeling?
  \item What are the proper representation for capturing requirements?
  \item How can we engineer a high quality specification from stakeholder requirements?
  \item How well is this approach when applied to realistic settings?
\end{enumerate}
\end{comment}

\vspace{6pt}
\textbf{RQ1:} What are the proper concepts for modeling requirements?
\vspace{6pt}

%CORE (Core Ontology for RE)~\cite{juretaa_core_2009}, the state of the art requirements ontology,

To answer the first research question, we have provided an ontological interpretation of requirements based on the ontological meaning of \emph{functions} and \emph{qualities}, and accordingly proposed a requirements ontology for classifying and modeling requirements. Our ontological interpretation of requirements contributes to addressing a set of limitations in the literature: (1) the narrow scope of existing definitions over functional requirements; (2) conceptual dispute of what NFRs are and how to capture them; (3) the inconsistency between quality hierarchies/models; (4) the flawed distinction between functional and non-functional requirements; (5) the difficulty on measuring the satisfaction of ``good-enough'' NFRs. Our requirements ontology also addresses the discovered deficiencies in CORE~\cite{juretaa_core_2009}, the state of the art requirements ontology, e.g., functional requirements refer to functions (capabilities) rather than their manifestations, requirements in practice can be a mix of concerns. Moreover, our requirements ontology has been evaluated by using realistic requirements in the public PROMISE requirements set~\cite{menzies_promise_2012}, and was proved to be adequate in capturing requirements in practice. To the best of our knowledge, our evaluation is the first attempt of applying a requirements ontology to realistic requirements data.
%the un-acknowledgement of requirements with mixed concerns

%Descriptions, in the form of ``$<s_i: D_i>$'', where $D_i$ restricts $s_i$, offers intuitive ways to refine requirements, e.g., adding a slot-description pair or specializing the description of a slot in a function description.
\vspace{6pt}
\textbf{RQ2:} What are the proper models for engineering requirements?
\vspace{6pt}

%What are the proper representation for capturing requirements?
To answer the second question, we have proposed a description-based representation for requirements. This description-based syntax offers several benefits: (1) it is able to capture both functional and non-functional requirements; (2) it facilitates the identification of requirements issues such as incompleteness, ambiguity and unverifiability, as shown in our evaluation in Section~\ref{sec:eval_framework}; (3) it offers intuitive ways to refine requirements, e.g., adding/removing a slot-description pair, generalizing/specializing the description of a slot in a function description; (4) it is able to capture the interrelation between requirements based on requirements detail rather than the whole requirement statement (e.g., we can specify an FC over the object of a function description, or define a QG/QC based on an F), this helps us to get an overview of NFRs that are associated with an F, specify crosscutting concerns of NFRs, and document NFRs (e.g., FRs and NFRs are not separated anymore, as they are in the IEEE 830-1998 standard~\cite{committee_ieee_1998}). Moreover, we have developed a prototype tool that is able to automatically translate the \emph{Desiree} syntax into Description Logics (DL) expressions, allowing users to query interrelations between requirements and detect inconsistencies among requirements.
%We have also evaluated the expressiveness of our language by using it to rewrite the 625 requirements in the PROMISE~\cite{menzies_promise_2012} data set. The results have shown the adequacy of our language in capturing requirements.
%high quality

\vspace{6pt}
\textbf{RQ3:} What are appropriate techniques and tools for transforming stake-
holder requirements into eligible specifications?
%\textbf{RQ3:} How to transform requirements into a specification?
\vspace{6pt}
%(including the requirements concepts, operators, the description-based syntax, and the methodology)
%How to engineer an eligible specification from stakeholder requirements?

To answer the third question, we have proposed a rich collection of requirements operators, and a methodology for applying the operators on requirements concepts (written in natural language or the \emph{Desiree} syntax) in order to transform stakeholder requirements into an eligible requirements specification. Compared with current goal modeling techniques such as KAOS~\cite{dardenne_goal-directed_1993}, Tropos~\cite{bresciani_tropos:_2004} and Techne~\cite{jureta_techne:_2010}, our framework offers two distinguishing features: (1) it allows us to weaken requirements, e.g., enlarging the expected quality region (``Scale''), or lowering the percentage over a set of individuals (``deUniversalize''); (2) it enables us to incrementally going from incomplete to complete (enough), unverifiable to verifiable, ambiguous to unambiguous, inconsistent/un-satisfiable to consistent/satisfiable, and informal to formal. In addition, with the supporting tool, we are able to translate requirements refinements/operationalization into DL axioms and perform entailment and ``what-if'' (fulfillment) analysis by using DL subsumption.
%the requirements concepts, operators, the description-based syntax, together with the methodology

\vspace{6pt}
\textbf{RQ4:} How well does our proposed framework work in realistic settings?
\vspace{6pt}
%How well is this approach when applied to realistic settings?

To answer the forth question, we have conducted a series of empirical evaluations to evaluate our proposal, including: (1) assessing the coverage of our requirements ontology by applying it to all the 625 requirements in the PROMISE dataset~\cite{menzies_promise_2012}; (2) evaluating the expressiveness
of our description-based language by using it to rewrite all the 625 requirements in the dataset; (3) illustrating our methodology by performing a realistic \emph{Meeting Scheduler} case study; (4) evaluating the the effectiveness of the entire \emph{Desiree} framework through three controlled experiments. The evaluation results show that our ontology and language are adequate in capturing requirements in practice, and provide strong evidence that the tool-supported \emph{Desiree} framework (including the ontology, operators, syntax and methodology) indeed can help people to perform better requirements engineering with a medium or big effect.

\section{Limitations}
\label{sec:conclusion_limitations}
Our approach also has several limitations on handling the following aspects.

%An another interesting way is ``slot mining'': statistically analyzing the requirements in specific application domains and eliciting a set of frequent slots.
\begin{itemize}
  \item \textbf{Built-in slots}. Our language currently does not have a built-in set of slots, and may result in different outputs when used by different users as they could use different words for the same relation. For example, when specifying the relation between ``students'' and ``clinical class'', one may use ``belong to'' while others could use ``associated with''. One possible solution to this is to develop ontologies of software systems and of application domains.
  \item \textbf{Instance-level constraints}. Our description-based language has difficulties on capturing some instance-level constraints. For example, ``\emph{Managers shall be able to move a student from one clinical lab section to another clinical lab section corresponding to} \emph{the same clinical class}''. We have used the ``\emph{same\_as}'' DL constructor~\cite{borgida_adding_1999}, which is more generally known as ``\emph{role value map}''~\cite{schmidt-schaus_s_subsumption_1988} to express these constraints (as discussed in Section~\ref{sec:eval_language_lessons}). However, the ``\emph{same\_as}'' constructor would cause the problem of undecidability on its general form~\cite{schmidt-schaus_s_subsumption_1988}, and thus imposing limitations on reasoning. This is mainly because the semantics our language is developed based on set-theory.
  \item \textbf{Temporal constraints}. Our language has limitations on handling temporal constraints. We currently represent temporal constraints with attributes such as ``before'', ``after'', and ``concurrent''. However, the reasoning part of such representations is severely limited. One possible way is to use temporal logic (e.g., use LTL as in KAOS~\cite{dardenne_goal-directed_1993}) for representing and reasoning about these kinds of requirements.
  \item \textbf{Algebraic constraints}. In addition, our language is unable to capture algebraic constraints such as ``given an initial balance $a$, after a withdrawal of $b$, the balance shall be $a - b = c$''. Our finding of temporal and algebraic constraints in realistic requirements implies that maybe hybrid-logic is needed for representing a complete requirements specification and reasoning over it.
  \item \textbf{Nested de-Universalization}. Our DL translation of refinement/operationalization is not able to capture the entailment semantics of nested $\bm{U}$, the semantics of which is expressed by using second-order (even higher-order) logic. For example, our DL translation is able to infer that $G_1 \models G_2$, where $G_1$ is ``\emph{the file search function shall take 30 seconds} (\emph{at least}) $80\%$ \emph{of the time}'', and $G_2$ is ``\emph{the file search function shall take 30 seconds} (\emph{at least}) $70\%$ \emph{of the time}''. However, it can note tell the entailment between $G_3$ ``$80\%$ \emph{of the system functions shall be fast at} $90\%$ \emph{of the time}'' and $G_4$ ``\emph{$90\%$ of the system functions shall be fast at $80\%$ of the time}''. This remains an open question even for human-being, as we need to consider the size of each set the the $\bm{U}$ operator is applied to, and also the importance of each set.
  \item \textbf{Simulation of entailment semantics}. We have used DL subsumption to simulate the entailment semantics of requirements operators. However, for some operators, the DL subsumption between the input and output elements does not conform to their entailment semantics. For example, when focusing $QG_1$ ``Security (\{the\_system\}) :: Good'' to $QG_2$ ``Security (\{the\_data\_module\}) :: Good'', we have $QG_1 \models QG_2$ (i.e., if the system is secure, then its date module is also secure); however, when translated into DL expressions, we will have $QG_2 \sqsubseteq QG_1$ if we replace the subject of $QG_1$, ``\{the\_system\}'',  with its part, ``\{the\_data\_module\}'', and specify a DA ``$\{the\_data\_module\} $:<$ \{the\_system\}$''. We used some tricks when applying such operators, e.g., expanding the subject of $QG_1$ to ``\{the\_data\_module\} $\lor$ \{the\_system\}'' instead of replacing it with its part, ``\{the\_data\_module\}''.
  \item \textbf{Prototype tool}. We have developed a prototype in support of the \emph{Desiree} framework.  However, the usability of the \emph{Desiree} tool, and the accessibility of the tutorial for the \emph{Desiree} approach (e.g., wiki, video, help manual) are still in need of improvement.
  \item \textbf{Evaluation}. We have conducted a series of evaluations to evaluate the framework (including its ontology, language, and methodology) using both case studies and controlled experiments. However, we have not assessed our framework in a realistic industrial setting (e.g., industrial case studies, sufficient professional participants). As such, there is still some risk of generalizing our conclusion.
\end{itemize}

%This is mainly because our language is developed based on sets.
%, and another interesting way is ``slot mining'': statistically analyzing the requirements in specific application domains and eliciting a set of frequent slots.

\section{Future Work}
\label{sec:conclusion_future_work}

There are many directions open for our future research. We discuss three most interesting ones in particular in this section.

%Several issues remain open, notably inconsistency handling.
%Inconsistency handling. The resolution of in-consistency may require one to prioritize, relax (e.g., relax the quality region, adding pre-condition) or even drop requirements. This interesting point will certainly be further explored within our framework.
%As we have mentioned in Section~\ref{sec:conclusion_limitations}, our framework currently does not have a built-in set of slots, and may result in different outputs when used by different users. Instead of developing ontologies of software systems and of application domains, another interesting idea is ``slot mining'': we can statistically analyze the requirements in specific application domains and elicit a set of frequent slots. This is inspired by the Berkeley Framenet project~\cite{baker_berkeley_1998}\cite{fillmore_background_2003}, which extracts information about the linked semantic and syntactic properties (frame elements) of English words (frames) from large electronic text corpora. A frame in FrameNet is akin to the desired function, and the associated frame elements serve as slots of that frame. That is, on one hand, we can reuse the frame elements of existing frames; on the other hand, we can collect realistic requirements, and accordingly adapt
\begin{itemize}
  \item \textbf{Slot mining}. As we have mentioned in Section~\ref{sec:conclusion_limitations}, our framework currently does not have a built-in set of slots, and may result in different outputs when used by different users. Instead of developing ontologies of software systems and of application domains, another interesting idea is to adapt the frame elements of existing frames in the Berkeley Framenet project~\cite{baker_berkeley_1998}\cite{fillmore_background_2003}, which extracts information about the linked semantic and syntactic properties (frame elements) of English words (frames) from large electronic text corpora. In general, a frame in FrameNet is akin to the ``FName'' in our function descriptions, and the associated frame elements act as slots of the function description. Based on this, we can statistically process corpora to elicit a set of common domain independent slots, or statistically analyze requirements texts in specific application domains to elicit a set of frequent domain dependent slots.
  \item \textbf{Requirements extraction from user feedback}. In our requirement ontology, we have specialized requirements into three categories: (1) function-related, including F, FG and FC; (2) quality-related, including QG and QC; (3) content-related, including CTG and SC. We have also provided a description-based syntax for structure them. The ``$<$slot: description$>$'' structures provide useful guidelines for extracting requirements from natural language texts (e.g., user feedback). Work has been done on eliciting functional requirements from texts, e.g., Casagrande et al.~\cite{casagrande_nlp-kaos_2014}; however, few attention has been paid to eliciting non-functional requirements (e.g., quality requirements and content requirements) from texts. The extraction of requirements (functional and non-functional) from texts, specifically, user feedback, is of importance to software/service evolution.
  %\item \textbf{Reasoning with quality goals}.
  \item \textbf{Requirements evolution}. Our description-based language is able to capture a rich set of interrelations between requirements, functional and non-funcitnoal. Therefore, an interesting research direction is to systematically and automatically detect the impact when changing a requirement. It will be very interesting to see how a requirements knowledge base evolves with changing requirements, a major topic in Software Engineering for the next decade.
  \item \textbf{Evaluation}. For evaluation, there are two interesting aspects that can be further explored: (1) the eligibility of produced requirements models (i.e., how many requirements issues does each produced requirements model have) when using the \emph{Desiree} approach; (2) how well is \emph{Desiree} when compared with classical goal-oriented techniques such as KAOS~\cite{dardenne_goal-directed_1993}, \emph{i}*~\cite{yu_modelling_2011} or  Techne~\cite{jureta_techne:_2010}?
\end{itemize}

%Therefore, an interesting research direction is how to effectively manage evolutionary changes to requirements.
%This interesting topic will be explored in the next steps of our work.

%% file: appendixs.tex
\chapter{\emph{Quality} Mapping}
\label{cha:appendixes_qtheory}
\input{qthoery}

\chapter{Graded Membership}
\label{cha:appendixes_membership}
\input{membership}

\chapter{The \emph{Desiree} Syntax}
\label{cha:appendixes_syntax}
\input{syntax}

\chapter{The Semantics of the U operator}
\label{cha:appendixes_semantics_u}
\input{usemantics}

\chapter{Detailed Requirements Issues}
\label{cha:appendixes_issues}
\input{ms}
\input{rb}

\chapter{\emph{Desiree} Feedback}
\label{cha:appendixes_feedback}
\input{questionnaire}

%% file: qthoery.tex
We adopt the quality theory introduced in the Unified Foundational Ontology (UFO)~\cite{guizzardi_ontological_2005}, which extends DOLCE~\cite{masolo_ontology_2003}. In this theory, a \emph{\textbf{quality}} is a basic perceivable or measurable entity that inheres in a particular entity and is unable to exist independently from it. A quality is a particular/ individual (e.g. ``$cost\#$'', the cost of a specific trip; by convention, we use ``\#'' to indicate individuals), and belongs to a \emph{\textbf{quality type}} QT (e.g., ``Cost''). Moreover, this theory differentiates a quality, e.g. ``$cost\#$'', from its value, e.g. ``1000 \euro'' which is called \emph{\textbf{quality value}} QV and is located in a \emph{\textbf{quality space}} QS with a specific structure (simply, a quality space consists of regions with sub-regions).

%Note that a quality can be intrinsic or relational. In the former case, it only depends on its inhering subject, and in the latter case it would also existentially depend on other entity(s) independent of its bearer. For example, the ``attractiveness'' of an application inheres in that application, and existentially depends on its users.

Originally, this theory associates a quality, say ``$cost\#$'', of type ``Cost'', to an individual subject, e.g., ``$trip\#$'', of type ``Trip''; and then associates with ``$cost\#$'' a specific value, say ``1000 \euro'', in the quality space ``EuroValues''. These associations are captured by the predicates \emph{qt}(\emph{cost}\#, \emph{trip}\#) and \emph{ql}(1000 \euro, \emph{cost}\#). Also, the quality type ``Cost'' and quality space ``EuroValues'' is associated through \emph{qs}(\emph{Cost}, \emph{EuroValues}).

To interpret NFRs, this original proposal needs to be adapted: (1) qualities discussed in RE are often not particulars, but rather quality types. For example, in the requirement ``\emph{the cost of trip shall be low}'', the quality type ``Cost'' is applied to a subject type ``Trip'', which implies a set of its instances in the future rather than a single one; (2) an NFR often constrains the desired quality value of a quality that inheres in a software-related subject, requiring us to associate the subject (type), quality type, quality, and quality value together through a unified characterization.

The key to these issues is to use functions that map from individuals to individuals, e.g., $\hat{qt}(Cost, trip\#) = cost\#$, instead of predicates that map individuals to Boolean values, e.g. $qt(cost\#, trip\#) = true$. We firstly define the necessary types: the subject/bearer of a particular quality is an individual subject, e.g., a trip, and of type \emph{\textbf{SubjectType}} (\emph{SubjT}), e.g., ``Trip''; a set of such individual subjects is of type \emph{\textbf{PowerSet}}(\emph{\textbf{SubjT}}), $\bm{\wp}(SubjT)$ for short; the type of a quality type is \emph{\textbf{QualityTypeSet}} (\emph{QTS}) , with each element being a different quality type, e.g., ``Cost'', ``Color'', ``Length'', etc. We then define in Table~\ref{tab:quality_functions} the basic functions, which are based on the three predicates in the original theory.

\begin{table}[!htbp]
  \caption {The basic quality-related functions}
  \label{tab:quality_functions}
  \vspace {0.3 cm}
  \centering
  \footnotesize
  \setlength\tabcolsep{2pt}
  %\begin{tabular}{|c|p{0.8\textwidth}|}
  %\begin{tabular}{|m{0.15\textwidth}|m{0.85\textwidth}|@{}m{0pt}@{}} %<{\centering} \raggedleft <{\centering}
  \begin{tabular}{|p{0.25\textwidth}|p{0.5\textwidth}|p{0.25\textwidth}<{\raggedright}|}
  \hline
  \textbf{Math Function} & \textbf{Meaning} & \textbf{Examples} \\ \hline

  $\hat{qs} : QTS \rightarrow QS$	& It associates a quality type QT in QTS with a relevant quality space QS	& $\hat{qs} (Cost) = EuroValues$ \\

  $\hat{qt}:QTS \rightarrow SubjT \rightarrow QT $ & It associates an individual subject of type SubjT with a particular quality of type QT in QTS	& $\hat{qt}(Cost)(trip\#) = cost\#$\\

  $\hat{ql}:QS \rightarrow QT  \rightarrow QV$ & It associates a particular quality of type QT with a quality value QV in quality space QS & $\hat{ql}(EuroValues) (cost\#) = 1000 \; (Euro) $ \\\hline
  \end{tabular}
\end{table}

Commonly, a function \emph{f} is denoted as $f: X  \rightarrow  Y$, in which $X$ and $Y$ are types. By the notation $f: X  \rightarrow  Y  \rightarrow  Z$, we are able to define a high-order function $f$ that takes as argument an element of $X$ and returns as value a function $f': Y  \rightarrow Z$. For example, the function $\hat{qt}: QTS \rightarrow SubjT \rightarrow QT$ in Table~\ref{tab:quality_functions} takes as argument a quality type ``Cost'' and returns as value a function $\hat{qt}(Cost):SubjT \rightarrow QT$. This returned function $\hat{qt}(Cost)$ in turn takes an individual subject of type \emph{SubjT} (e.g. $trip\#$ of \emph{Trip}), and returns a particular quality of type QT (e.g., $cost\#$ of \emph{Cost}).

Note that the function $\hat{qt}$ is partial -- it is undefined if the given quality type QT is not applicable to the individual subject of type \emph{SubjT}. For example, $\hat{qt}(Cost)(dream\#4)$ is undefined because dreams do not have costs. Similarly, $\hat{ql}$ is defined only if the given quality space QS is connected to the quality type QT of a particular quality.

Based on the defined types and functions, we now are able to define a high-order and partial function \emph{hasQV} as Eq.~\ref{eq:eq_hasqv_basic}, in which \emph{x} is an individual subject of type \emph{SubjT}. It can be used in our example as \emph{hasQV}(\emph{Cost})(\emph{trip}\#) to obtain the cost value 1000 \euro, see the reduction process in Eq.~\ref{eq:eq_hasqv_reduction}, where ``$=_{\mathrm{\beta}}$'' indicates a $\beta$ reduction~\cite{_lambda_2015}.
\begin{equation}\label{eq:eq_hasqv_basic}
    \begin{aligned}
        hasQV: QTS \rightarrow SubjT \rightarrow QV \equiv \lambda_{\mathrm{QT}}.\lambda_{\mathrm{\emph{x}}}.\hat{ql}(\hat{qs}(QT))(\hat{qt}(QT)(x))
    \end{aligned}
\end{equation}

This function deals with only individual subjects. Since a requirement usually concerns a set of subject, e.g., a set of trips, we shall extend this formalization to handle types. For this purpose, we apply $hasQV(QT)(x)$ to all instances of a subject type \emph{SubjT}, obtaining the set of quality values for all the individuals. By function overloading, we are able to introduce a new function but with the same name \emph{hasQV} as shown in Eq.~\ref{eq:eq_hasqv_map} , in which ``$\bm{\wp}(SubjT)$'' is the type of a set of individual subjects, ``$\bm{\wp}(QV)$'' is the type of a set of quality values, ``$\bm{MAP}$'' is a type of iteration in which a math function (e.g., ``$\lambda_{\mathrm{x}}.hasQV(QT)(x)$'') is successively applied to elements of sequence (e.g., ``$SubjT$'', we treat the set of individuals of type $SubjT$ as a sequence) in the same order as in Lisp~\footnote{http://jtra.cz/stuff/lisp/sclr/map.html}. It can be used as $hasQV(Cost)(Trip)$ to get the set of cost values of all trips. We can further overland $hasQV$ as shown in Eq.~\ref{eq:eq_hasqv_avg} to return a single QV, which can serve as the standard for comparison, e.g., what does ``\emph{low}'' mean to a trip?
\begin{equation}\label{eq:eq_hasqv_reduction}
    \begin{aligned}
       hasQV(Cost)(trip\#)
       & =((\lambda_{\mathrm{QT}}.\lambda_{\mathrm{\emph{x}}}.\hat{ql}(\hat{qs}(QT))(\hat{qt}(QT)(x)) \;\; Cost) \;\; trip\#) \\
       & =_{\mathrm{\beta}} (\lambda_{\mathrm{\emph{x}}}.\hat{ql}(\hat{qs}(Cost))(\hat{qt}(Cost)(x)) \;\; trip\#) \\
       & =_{\mathrm{\beta}} \hat{ql}(\hat{qs}(Cost))(\hat{qt}(Cost)(trip\#)) \\
       & =\hat{ql}(EuroValues)(\hat{qt}(Cost)(trip\#)) \\
       & =\hat{ql}(EuroValues)(cost\#) \\
       & = 1000 \; (Euro) \\
    \end{aligned}
\end{equation}
\vspace{-18pt}
\begin{equation}\label{eq:eq_hasqv_map}
    \begin{aligned}
        hasQV:QTS \rightarrow \bm{\wp}(SubjT) \rightarrow \bm{\wp}(QV) \equiv \bm{MAP}(\lambda_{\mathrm{\emph{x}}}.hasQV(QT)(x)  \;\; SubjT)
    \end{aligned}
\end{equation}
\vspace{-18pt}
\begin{equation}\label{eq:eq_hasqv_avg}
    \small
    \begin{aligned}
        hasQV:QTS \rightarrow \bm{\wp}(SubjT) \rightarrow QV \equiv
        average (\bm{MAP}(\lambda_{\mathrm{\emph{x}}}.hasQV(QT)(x) \;\; SubjT))
    \end{aligned}
\end{equation}

Note that we use function overloading instead of new functions for simplifying the syntax of requirements language. For example, the requirement ``\emph{the cost of trip shall be low}'' (``$hasQV(Cost)(Trip) :: Low$'') and ``\emph{the cost of this trip shall be low}'' (``$hasQV(Cost)(trip\#) :: Low$'') have the same language structure, but refer to different subjects (a set of trips versus and an individual trip), also different values. 

We simplify the function ``$hasQV$'' by treating a quality type $QT$ itself as a mapping from its subject of type $SubjT$ to a quality value. As such the two ``\emph{low cost}'' examples can be accordingly written as ``$Cost(Trip):: Low$'' and ``$Cost(trip\#):: Low$'', and can be generalized to the syntactic form ``$QT (SubjT) :: QRG$'', which is equal to the one shown in Eq.~\ref{eq:eq_qr_syntax}. There are two points to be noted: (1) we use ``$Q$'' instead of ``$QT$'' in Eq.~\ref{eq:eq_qr_syntax} because it this more acceptable to the RE community; (2) when the subject is an individual, e.g., ``$trip\#$'', we use curly brackets to indicate a singleton, e.g., ``$Cost(\{{trip}\}):: Low$''.

\begin{comment}
\begin{equation}\label{eq:eq_hasqv_simplified}
    \small
    \begin{aligned}
        Q (SubjT) :: QGG
    \end{aligned}
\end{equation}
\end{comment}

%% file: membership.tex
In this chapter, we use a simple example QR ``\emph{the cost of trip shall be low}'' to illustrate the calculation of the graded membership of a specific cost value with regarding to the desired quality region ``\emph{low}'', using the techniques of ``\emph{Graded Membership}'' proposed by Decock et al.~\cite{decock_what_2014}.

\vspace{-12pt}
\section{Prototype Points}
\label{sec:appendix_membership_points}
Suppose that the associated quality region ``\emph{low}'' of the example QR is represented by two prototype point values 500\euro \; and 700\euro. Similarly, we can use 800\euro \; and 1000\euro, and 1200\euro \; and 1500\euro \; to represent the region ``\emph{medium}'' and ``\emph{high}''. Given the three prototype regions, we will have three sets of prototype points: $Set_{low} = \{500, 700\}$, $Set_{medium} = \{800, 1000\}$, $Set_{high} = \{1200, 1500\}$. Since ``Cost'' has a one-dimensional structure that is isomorphic to the positive half-line of real numbers, a \emph{Voronoi} diagram can be constructed by selecting an element from each set (these elements form a completion) and calculating the median of each pair of prototype values (a pair of point values should be in \emph{adjacent} regions). In this example, we will have 8 completions. \\

\noindent 1: \{500, 800, 1200\} \\
2: \{500, 800, 1500\} \\
3: \{700, 800, 1200\} \\
4: \{700, 800, 1500\} \\
5: \{500, 1000, 1200\} \\
6: \{500, 1000, 1500\} \\
7: \{700, 1000, 1200\} \\
8: \{700, 1000, 1500\} \\

Take, for instance, the first completion \{500, 800, 1200\}, the corresponding diagram that partition the cost value space will be ((500 + 800) / 2, (800 + 1200) / 2) = (650, 1000). That is, the ``\emph{low}'', ``\emph{medium}'' and ``\emph{high}'' region will be (0, 650], (650, 1000], and (1000, 1000+), respectively. The 8 corresponding simple diagrams are listed as follows:\\

\noindent  1: 0 ... 650  ... 1000  ... \\
2: 0 ... 650  ... 1150  ... \\
3: 0 ... 750  ... 1000  ... \\
4: 0 ... 750  ... 1150  ... \\
5: 0 ... 750 ... 1100  ... \\
6: 0 ... 750  ... 1250  ... \\
7: 0 ... 850  ... 1100  ... \\
8: 0 ... 850  ... 1250  ... \\

Now if we have a cost value 740 \euro, we will have 6 out of 8 diagrams (the $3^{rd} \sim 8^{th}$ ones) classify it to the region low. Thus, the example QR will be satisfied to a degree of 0.75 in this case.

\vspace{-12pt}
\section{Prototype Intervals}
\label{sec:appendix_membership_intervals}
Clearly, a prototype region may have infinitely many points. Hence the approach introduced in Section~\ref{sec:appendix_membership_points} (with limited points) will not work in this situation. Hence, we introduce in this section how to calculate graded membership by using intervals (instead of points) to represent prototype regions.

Let \emph{QS} be an \emph{m}-dimensional space (e.g., the cost space is one-dimensional, the RGB color space is three-dimensional), with $R$ = \{$r_1$, ..., $r_n$\} being the set of \emph{n} prototype regions in \emph{QS}. By selecting a point $p_i$ (denoted as a \emph{m}-tuple $<$$x_{i1}$, ..., $x_{im}$$>$, $1 \leq i \leq n$) from each region $r_i$, we get a completion, which can be represented as a $n \times m$-tuple $<$$x_{11}$, ..., $x_{1m}$, ... $x_{n1}$, ... , $x_{nm}$$>$. We denote the set of all completions as $\prod (R)$.

We define $QS_{p,i}$ in Eq.~\ref{eq:eq_voronoi_in_region_i} as the set of completions that determine \emph{Voronoi} diagrams which locate a value \emph{m}-dimensional value \emph{p} in a prototypical region $r_i$. Note that a point \emph{p} is located in the region $r_i$ iff $d(p, <x_{i1}, ..., x_{im}>) < d(p,<x_{j1}, ..., x_{jm}>)$ for all $<x_{j1}, ..., x_{jm}> \; \in r_j \; (i \neq j)$ , where \emph{d} is the distance function.\\
\begin{equation}\label{eq:eq_voronoi_in_region_i}
    \small
    \begin{aligned}
            QS_{p,i} := \{ x_{11}, ..., x_{1m}, ... x_{n1}, ... , x_{nm} | d(p,x_{i1}, ..., x_{im}) < d(p,x_{j1}, ..., x_{jm}) \\ \forall <x_{j1}, ..., x_{jm}> \; \in \; r_j  \;\; (i \neq j)\}
    \end{aligned}
\end{equation}

%$\mathbb{R}^{n\times m}$
The membership of a \emph{m}-dimensional point \emph{p} in a region $r_i$ can be defined as a measure over a set of completions in terms of the volume occupied by the related set of coordinates in the space $\mathbb{R}^{n\times m}$.
Specifically, it can be calculated using Eq.~\ref{eq:eq_graded_membership}, in which $M$ is the membership function, $\mu$ is a function that calculates the volume of a set of completions.\\
\begin{equation}\label{eq:eq_graded_membership}
    \begin{aligned}
            M(QS_{p,i}) = \mu (QS_{p,i}) / \mu (\prod(R))
    \end{aligned}
\end{equation}

We consider a one-dimensional space (\emph{m} = 1) and two prototype regions $r_1$ and $r_2$. Without loss of generality, we assume that the two regions are represented by intervals [\emph{a}, \emph{b}] and [\emph{c}, \emph{d}] ($a < b < c < d$). To determine the graded membership functions for (concepts that are related to) $r_1$ and $r_2$, we need to look at the space $\mathbb{R}^2$ ($n$ = 2, $m$ = 1). In this space, we have all the completions $\prod(R)$ in the rectangle with vertices $A = (a, c)$, $B = (a, d)$, $C = (b, c)$, and $D = (b, d)$, being coordinates $<x, y>$ such that $x \in r_1$ and $y \in r_2$. To locate a point \emph{p} in the region $r_1$, we need $d(x - p) \leq d(y - p)$ to hold, or equivalently $p - x \leq y - p$.

Note that if $p - x = y - p$, then \emph{p} coincides with the \emph{Voronoi} points of the diagram generated by $<x, y> \; \in \; \prod(R)$. These set of points form a line defined by $p - x = y - p$. Thus the set $QS_{p,i}(<x, y>)$ consists of all the points in the rectangle ABCD that lie above this line. In this case, $\prod(R)$ is the area of ABCD and $QS_{p,i}$ is the area of ABCD that is above the line $y = 2p - x$. So, we have Eq.~\ref{eq:eq_graded_membership_example}.\\
\begin{equation}\label{eq:eq_graded_membership_example}
    \begin{aligned}
            M(QS_{p,i}) \; = & \; \mu(QS_{p,i}) / \mu(\prod(R)) \\ = & \; \mu(QS_{p,i}) / ((b - a)(d - c))	
    \end{aligned}
\end{equation}

Considering different situations $b - a < d - c$, $b - a = d - c$ and $b - a > d - c$, we are able to get the membership functions for the region $r_1$ as shown in Eq.~\ref{eq:eq_case_11} $\sim$~\ref{eq:eq_case_15}, Eq.~\ref{eq:eq_case_21} $\sim$~\ref{eq:eq_case_24}, Eq.~\ref{eq:eq_case_31} $\sim$~\ref{eq:eq_case_35}. Similarly, we can get the membership function for $r_2$.

Case 1: $b - a < d - c$
\begin{subnumcases}
\centering
\small
{M_{r_1}(p)=}
1 &  $p < (a + c)/2$ \label{eq:eq_case_11}\\
1-{(2p-a-c)^2} /2(b-a)(d-c) & $(a+c)/2 \leq p \leq (c+b)/2$ \label{eq:eq_case_12}\\
(2d+a+b-4p)/2(d-c) & $(c+b)/2 \leq p \leq (a+d)/2$ \label{eq:eq_case_13}\\
(b+d-2p)^2/2(b-a)(d-c) & $(a+d)/2 \leq p \leq (b+d)/2$ \label{eq:eq_case_14}\\
0 &  $p > (b+d)/2$ \label{eq:eq_case_15}
\end{subnumcases}

Case 2: $b - a = d - c$
\begin{subnumcases}
\centering
\small
{M_{r_1}(p)=}
1 & $p<(a+c)/2$ \label{eq:eq_case_21}\\
1-(2p-a-c)^2/2(b-a)(d-c) & $(a+c)/2 \leq p \leq (c+b)/2$ \label{eq:eq_case_22}\\
(b+d-2p)^2/2(b-a)(d-c) & $(a+d)/2\leq p\leq(b+d)/2 $ \label{eq:eq_case_23}\\
0 & $p > (b+d)/2$ \label{eq:eq_case_24}
\end{subnumcases}

Case 3: $b - a > d - c$
\begin{subnumcases}
\centering
\small
{M_{r_1}(p)=}
1 &  $p<((a+d))/2$ \label{eq:eq_case_31}\\
1-(2p-a-c)^2/2(b-a(d-c)  & $(a+c)/2 \leq p \leq (a+d)/2$ \label{eq:eq_case_32}\\
(2b+c+d-4p)/2(b-a) & $(a+d)/2 \leq p \leq (c+b)/2$ \label{eq:eq_case_33}\\
(b+d-2p)^2/2(b-a)(d-c) & $(c+b)/2 \leq p \leq ((b+d))/2$ \label{eq:eq_case_34}\\
0 &  $p > (b+d)/2$ \label{eq:eq_case_35}
\end{subnumcases}

Note that in a one-dimensional space with more than two prototype regions, the part of the graded membership function that classify points to one side of a prototype region depends on only that region and the adjacent one that lies in the same side. In such situation, we only need to apply the above method to each pair of adjacent prototype regions and then simply collate the obtained partial functions. For example, suppose that there is a region additional $r_3$. To get the membership function for $r_2$, we will need to consider the pair of $r_1$ and $r_2$, and $r_2$ and $r_3$. Once we get the two partial membership functions for $r_2$ separately, we could simply collate them to get the final one.

\begin{figure}[!htbp]
  \centering
  \vspace {-0.2 cm}
  \includegraphics[width=0.9\textwidth]{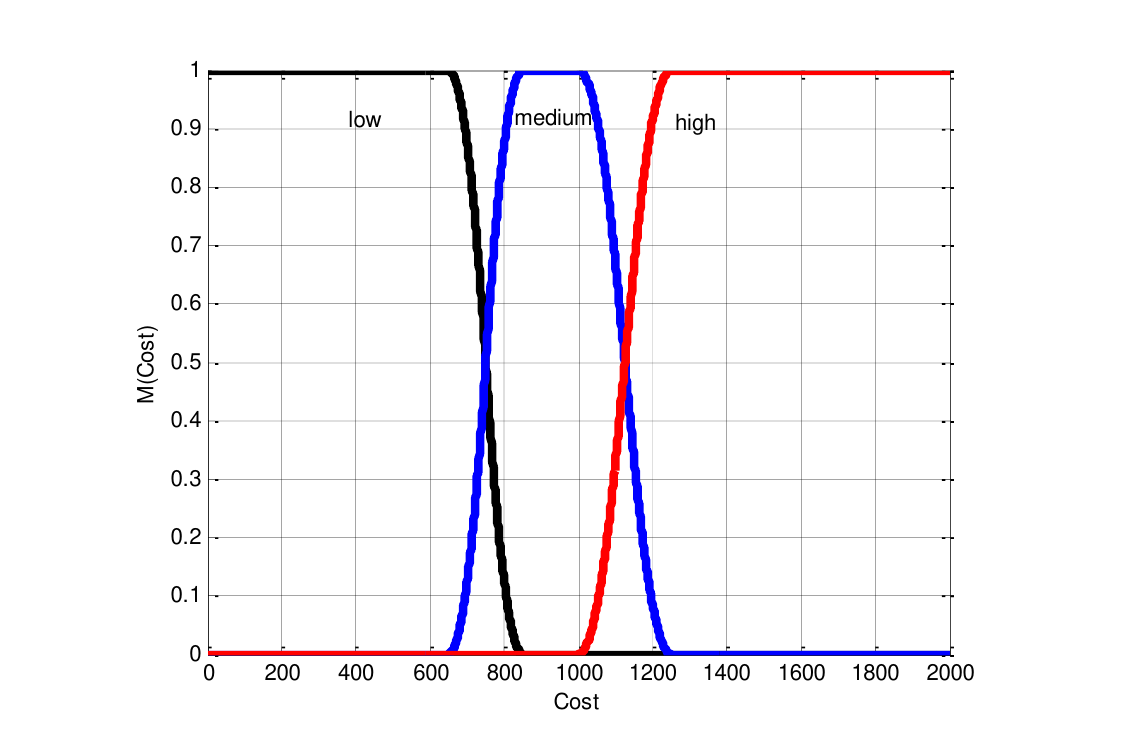}\\
  \vspace {-0.5 cm}
  \caption{The graded membership functions for \emph{Low}, \emph{Medium} and \emph{High}}\label{fig:membership}
\end{figure}

Now come back to our example, the cost space is one-dimensional (\emph{m} = 1), and we have \emph{n} = 3 prototype regions ``\emph{low}'', ``\emph{medium}'' and ``\emph{high}''. Suppose that the three regions are represented by interval [500\euro, 700\euro], [800\euro, 1000\euro] and [1200\euro, 1500\euro], respectively. By following the above introduced approach, we are able to get the membership function for ``\emph{low}'', ``\emph{medium}'', and ``\emph{high}'' as Eq.~\ref{eq:eq_low_11} $\sim$~\ref{eq:eq_low_14}, Eq.~\ref{eq:eq_medium_21} $\sim$~\ref{eq:eq_medium_28}, and Eq.~\ref{eq:eq_high_31} $\sim$~\ref{eq:eq_high_35} , which are depicted in Fig.~\ref{fig:membership}. Given a cost value 740\euro, we can obtain its membership: $M_{low}$(740) = 0.595, and $M_{medium}$(740) = 0.405.

%regarding ``low'' and ``medium''
\begin{subnumcases}
\centering
\scriptsize
{M_{low}(c)=}
1  &  $c<650$ \label{eq:eq_low_11}\\
1-(2c-1300)^2/80000 & $650 \leq c \leq 750$ \label{eq:eq_low_12}\\
(1700-2c)^2/80000 & $750 \leq c \leq 850$ \label{eq:eq_low_13}\\
0 & $c>850$ \label{eq:eq_low_14}
\end{subnumcases}

\begin{subnumcases}
\centering
\scriptsize
{M_{medium}(c)=}
0  &  $c<650$ \label{eq:eq_medium_21}\\
(2c-1300)^2/80000 & $650 \leq c \leq 750$ \label{eq:eq_medium_22}\\
1-(1700-2c)^2/80000 & $750 \leq c \leq 850$ \label{eq:eq_medium_23}\\
1 &  $850  \leq  c  \leq  1000$ \label{eq:eq_medium_24}\\
1-(c-1000)^2/30000 & $1000 \leq c \leq 1100$ \label{eq:eq_medium_25}\\
8-c/150 & $1100 \leq c \leq 1150$ \label{eq:eq_medium_26}\\
(2500-2c)^2/120000 & $1150 \leq c \leq 1250$ \label{eq:eq_medium_27}\\
0 &  $c>1250$ \label{eq:eq_medium_28}
\end{subnumcases}

\begin{subnumcases}
\centering
\scriptsize
{M_{high}(c)=}
0 & $c<1000$ \label{eq:eq_high_31}\\
(c-1000)^2/30000 & $1000 \leq c \leq 1100$ \label{eq:eq_high_33}\\
c/150-7 & $1100 \leq c \leq 1150$ \label{eq:eq_high_33}\\
1-(2500-2c)^2/120000 & $1150 \leq c \leq 1250$ \label{eq:eq_high_34}\\
1  &   $c>1250$ \label{eq:eq_high_35}
\end{subnumcases}

So far, we have discussed about one-dimensional conceptual space, like the ones for length, weight, processing time, and availability. For multi-dimensional spaces (e.g., the three-dimensional color space, the six-dimensional usability space), we need to first decompose the multi-dimensional spaces into single-dimensional spaces (e.g., refining ``Usability'' to its sub-qualities such as ``Learnability'', ``Operability'', which has a one-dimensional space such as ``learning time'' and ``operation time''; if not, we can keep refining it), and then use some mathematic rules (e.g., maximum, minimum, average, and others) for propagating up the degree of fulfillment.

%Its application to multiple-dimension space (e.g., the three-dimensional color space, the six-dimensional usability space) needs to be further investigated.

%% file: syntax.tex
%\section{The \emph{Desiree} Syntax}
%\label{sec:appendix_desiree_syntax}
%In this section, we present the full \emph{Desiree} syntax.

\begin{table}[!htbp]
  \begin{threeparttable}
  \caption {The BNF style \emph{Desiree} syntax}
  \label{tab:appendix_desiree_syntax}
  \centering
  \footnotesize
  %\vspace{-36pt}
  \setlength\tabcolsep{2pt}
  %\begin{alltt}
  \begin{tabular}{|rrp{0.65\textwidth}|}
      \hline
      \multicolumn{3}{c}{\textbf{\emph{Desiree} Syntax}} \\
      \hline
      \hline
      \multicolumn{3}{l}{\textbf{\emph{Descriptions:}}} \\ \hline
%(01) & SlotD $\rightarrow$ &   `$<$' Slot  `:' [(SOME $|$ [$\leq|\geq|=$] $n$)] Description `$>$'  $|$ SlotD SlotD \\
(01) & SlotD $\rightarrow$ &   `$<$' Slot  `:' [([$\leq|\geq$] $n$ $|$ SOME $|$ ONLY)] Description `$>$'\\
     & & $|$ SlotD SlotD \\
(02) & Description $\rightarrow$ &   Concept $|$ Val $|$ RegionExpr $|$ DataType \\
(03) & Concept $\rightarrow$ &   \emph{ConceptName} $|$ SlotD $|$ Concept `.' Slot $|$ Concept  Concept $|$ Concept `$\vee$' Concept $|$ Concept `$-$' Concept $|$ `\{' $ElemId_1$ ... `\}'  \\
(04) & RegionExpr $\rightarrow$ &   \emph{RegionName}  $|$ `[' Bound `,' Bound `]' $|$ `\{' $Val_1$, ... `\}'  \\ % /*enumeration*/
(05) & Bound $\rightarrow$ &   \emph{IntVal} $|$ \emph{RealVal}  $|$ MathExpression \\
(06) & MathExpression  $\rightarrow$ &   `\emph{count}' `(' Concept `)' $|$ ... \\
(07) & DataType $\rightarrow$ &   \emph{Integer} $|$ \emph{String} $|$ \emph{List} $|$ ... \\
\hline
\hline
\multicolumn{3}{l}{\textbf{\emph{Requirements:}}} \\ \hline

(08) & Req $\rightarrow$ &   Goal $|$ FG $|$ QG $|$ CTG \\
%(09) & Entity $\rightarrow$ &   `Entity' Concept \\
(09) & Goal $\rightarrow$ &   `Goal' Concept \\
%(11) & FG $\rightarrow$ &  Entity `$:<$' \emph{StateName} [SlotD] \\
(10) & FG $\rightarrow$ &  Concept `$:<$' Concept \\
(11) & QG $\rightarrow$ &  QualityName `(' Concept `)' `::' RegionExpr \\
(12) & CTG $\rightarrow$ &   Concept `$:<$' SlotD \\

\hline
\hline
\multicolumn{3}{l}{\textbf{\emph{Specifications:}}} \\ \hline
(13) & Spec $\rightarrow$ &   Function $|$ FC $|$ QC $|$ SC \\
(14) & Function $\rightarrow$ &   `Func' Concept \\
(15) & FC $\rightarrow$ &   Concept `$:<$' Concept \\
(16) & QC $\rightarrow$ &  QG [`$<$\emph{observed\_by}:' Concept `$>$'] \\
%(18) & SC $\rightarrow$ &   Entity `$:<$' SlotD \\
(17) & SC $\rightarrow$ &   Concept `$:<$' SlotD \\
\hline
\hline
\multicolumn{3}{l}{\textbf{\emph{Domain Assumptions:}}} \\ \hline
%(19) & DA $\rightarrow$ &  Concept  `$:<$ $|$ $\equiv$' Concept \\
(18) & DA $\rightarrow$ &  Concept  `$:<$' Concept \\

      \hline
  \end{tabular}
  \begin{tablenotes}
      \scriptsize
      \item[1:] ``[]'' means optional, ``$|$'' means alternatives.
    \end{tablenotes}
  \end{threeparttable}
\end{table}

%% file: usemantics.tex
%in Eq.~\ref{eq:eq_usemantics_examples}
\begin{comment}
\begin{equation}\label{eq:eq_usemantics_examples}
    \begin{aligned}
     QG_0 := & \; Response \; (SearchFnRuns) :: Fast \\
     QG_1 := & \; \bm{U} \; (QG_0, ?X, <inheres\_in : \; ?X>, 80\%) \\
     QG_2 := & \; \bm{O} \; (QG_0, <observed\_by : Surveyed\_user>) \\
     QG_3 := & \; \bm{O} \; (QG_1, <observed\_by : Surveyed\_user>) \\
     QG_4 := & \; \bm{U} \; (QG_3, ?O, <observed\_by : \; ?O >, 60\%)  \\
     QG_5 := & \; Response \; (Runs <run\_of: SystemFns>) :: fast \\
     QG_6 := & \; \bm{U} \; (QG_5, ?F, <inheres\_in : <run\_of : \; ?F >>, 80\%) \\
     QG_7 := & \; \bm{U} \; (QG_6, ?Y,   <inheres\_in: \; ?Y >, 90\%) \\
     QG_8 := & \; \bm{U} \; (QG_5, ?Y,   <inheres\_in: \; ?Y >, 90\%) \\
     QG_9 := & \; \bm{U} \; (QG_8, ?F, <inheres\_in : <run\_of : \; ?F>> \\
    \end{aligned}
\end{equation}
\end{comment}
\begin{comment}
\begin{equation}\label{eq:eq_usemantics_examples_searchruns}
    \small
    \begin{aligned}
        SearcFuncRuns := & \; Runs <run\_of: \{SearchFnIndividual\}> \\
    \end{aligned}
\end{equation}
\end{comment}
In this chapter, we use a set of QGCs to illustrate the semantics of combined $\bm{U}$ and $\bm{O_b}$ (e.g., ``\emph{$60\%$ of the observers shall agree that the search function is fast at $80\%$ of the time}''), and nested $\bm{U}$ (e.g., ``\emph{$90\%$ of the system functions shall be fast at $80\%$ of the time}''). We use a formula ``$AX_i$'' to specify the semantics of ``$QGC_i$'' ($0 \leq i \leq 9$).

\begin{equation}\label{eq:eq_usemantics_examples_QG0}
    %\small
    \begin{aligned}
        QG_0 := & \; Response \; (SearchFnRuns) :: Fast \\
        AX_0 \equiv & \; \forall r/SearchFnRuns.\forall q/Response.(inheres\_in(q, r) \\
                    &\rightarrow has\_value\_in(q, Fast))
    \end{aligned}
\end{equation}

\begin{equation}\label{eq:eq_usemantics_examples_QG1}
    %\small
    \begin{aligned}
        QG_0 := & \; Response \; (SearchFnRuns) :: Fast \\
        QG_1 := & \; \bm{U} \; (?X, QG_0, <inheres\_in : \; ?X>, 80\%) \\
        AX_1 \equiv & \; \exists ?X/\bm{\wp}(SearchFnRuns).(|?X|/|SearchFnRuns| > 0.8 \\
                    & \land \forall r/?X.\forall q/Response.(inheres\_in (q, r) \\
                    & \rightarrow has\_value\_in(q, Fast)))
    \end{aligned}
\end{equation}

\begin{equation}\label{eq:eq_usemantics_examples_QG2}
    %\small
    \begin{aligned}
     QG_0 := & \; Response \; (SearchFnRuns) :: Fast \\
     QG_2 := & \; \bm{O} \; (QG_0, <observed\_by : Surveyed\_user>) & \\
     AX_2 \equiv& \; \forall o/Surveyed\_user.\forall r/SearchFnRuns. \\
                & \; \forall q/Response.observed\_by(q, o) \land [inheres\_in(q, r) \\
                & \; \rightarrow has\_value\_in(q, Fast)]
    \end{aligned}
\end{equation}

\begin{equation}\label{eq:eq_usemantics_examples_QG3}
    %\small
    \begin{aligned}
     QG_0 := & \; Response \; (SearchFnRuns) :: Fast \\
     QG_1 := & \; \bm{U}\; (?X, QG_0, <inheres\_in : \; ?X>, 80\%) \\
     QG_3 := & \; \bm{O}\; (QG_1, <observed\_by : Surveyed\_user>)\\
     AX_3 \equiv& \; \forall o/Surveyed\_user.\exists ?X/\bm{\wp}(SearchFnRuns).\\
                & \; [|?X|/|SearchFnRuns|>0.8 \land \forall r/?X.\forall q/Response. \\
                & \; observed\_by(q, o) \land (inheres\_in(q, r)\\
                & \; \rightarrow has\_value\_in(q, Fast))]
     \end{aligned}
\end{equation}

\begin{equation}\label{eq:eq_usemantics_examples_QG4}
    %\small
    \begin{aligned}
     QG_0 := & \; Response \; (SearchFnRuns) :: Fast \\
     QG_1 := & \; \bm{U} \; (?X,  QG_0, <inheres\_in : \; ?X>, 80\%) \\
     QG_3 := & \; \bm{O} \; (QG_1, <observed\_by : Surveyed\_user>) \\
     QG_4 := & \; \bm{U} \; (?O,  QG_3, <observed\_by : \;?O >, 60\%)  \\
     AX_4 \equiv&  \; \exists ?O/\bm{\wp}(Surveyed\_user). \\
                & \; [|?O|/|Surveyed\_user|> 0.6 \\
                & \; \land \forall o/?O. \exists ?X/\bm{\wp}(SearchFnRuns).  \\
                & \; [|?X|/|SearchFnRuns| > 0.8 \\
                & \; \land \forall r/?X. \forall q/Response.observed\_by(q, o)  \\
                & \; \land (inheres\_in(q, r) \rightarrow has\_value\_in(q, Fast))]]
    \end{aligned}
\end{equation}

\begin{equation}\label{eq:eq_usemantics_examples_QC5}
    %\small
    \begin{aligned}
     QG_5 := \; & Response \; (Runs <run\_of: SystemFns>) :: Fast \\
     Foo(T) \equiv \; &[\lambda_{\mathrm{b}}.(Runs(b) \land \exists c. run\_of(b,c) \land T(c))] \\
     AX_5 \equiv \; & \; \forall r/Foo(SystemFns). \\
                    & \; \forall q/Response.inheres\_in(q,r) \\
                    & \; \rightarrow has\_value\_in(q,Fast)\\
          \equiv \; & \; \forall r/[\lambda_{\mathrm{b}}.(Runs(b) \land \exists c.run\_of(b,c) \land SystemFns(c))]. \\
                    & \; \forall q/Response.inheres\_in(q, r) \\
                    & \; \rightarrow has\_value\_in(q, Fast) \\
          \equiv \; & \; \forall r.(Runs(r) \land \exists c.run\_of(r, c) \land SystemFns(c)) \\
                    & \; \rightarrow  [ \forall q.Response(q).(inheres\_in(q, r) \\
                    & \; \rightarrow has\_value\_in(q, Fast))]
    \end{aligned}
\end{equation}

\begin{equation}\label{eq:eq_usemantics_examples_QC6}
    \small
    \begin{aligned}
     QG_5 := \; & Response \; (Runs <run\_of: SystemFns>) :: Fast \\
     QG_6 := \; & \bm{U} \; (?F, QG_5, <inheres\_in : <run\_of : \; ?F >>, 80\%) \\
     QG_7 := \; & \bm{U} \; (?Y, QG_6,  <inheres\_in: \; ?Y >, 90\%) \\
     AX_6 \equiv \; & \exists ?F/\bm{\wp}(SystemFns).|?F|/|SystemFns| > 0.8 \land \forall r/Foo(?F). \\
                    & \forall q/Response.inheres\_in(q, r) \rightarrow has\_value\_in(q, Fast) \\
          \equiv \; & \exists ?F/\bm{\wp}(SystemFns).|?F|/|SystemFns| > 0.8 \land \\
                    & \forall r.(Runs(r) \land \exists c.run\_of(r, c) \land ?F(c))\\
                    & \rightarrow [\forall q/Response.inheres\_in(q, r) \rightarrow has\_value\_in(q, Fast)] \\
     AX_7  \equiv \;& \exists ?F/\bm{\wp}(SystemFns).[|?F|/|SystemFns| > 0.8 \land \forall ?f/?F. \\
                    & \exists ?Y/\bm{\wp}(Foo(\{?f\}).[|?Y|/|Foo(\{?f\})| > 0.9  \\
                    & \land \forall r/?Y. \forall q/Response.inheres\_in(q, r) \rightarrow has\_value\_in(q,Fast)]] \\
           \equiv \;& \exists ?F/\bm{\wp}(SystemFns).[|?F|/|SystemFns| > 0.8 \land \forall ?f/?F .\\
                    & \exists ?Y/\bm{\wp}(\lambda_{\mathrm{b}}.(Runs(b) \land \exists c.run\_of(b,c) \land \{?f\}(c)).[|?Y|/|Foo(\{?f\})| > 0.9 \\
                    & \land \forall r/Y.\forall q/Response. inheres\_in(q, r) \rightarrow has\_value\_in(q,Fast)]] \\
           \equiv \;& \exists ?F/\bm{\wp}(SystemFns).[|?F|/|SystemFns| > 0.8 \land \forall ?f/?F. \\
                    &\exists ?Y/\bm{\wp}(\lambda_{\mathrm{b}}.(Runs(b) \land \exists c.run\_of(b,c) \land ?f=c).\\
                    &[|?Y|/|\{b|(Runs(b) \land \exists c. run\_of(b,c) \land ?f=c)\}| > 0.9 \\
                    &\land \forall r/Y. \forall q/Response. inheres\_in(q, r) \rightarrow has\_value\_in(q,Fast)] \\
         \equiv  \; & \exists ?F/\bm{\wp}(SystemFns).[|?F|/|SystemFns| > 0.8 \land \forall ?f/?F. \\
                    & \exists ?Y . [\forall y. (?Y(y) \rightarrow (Runs(y) \land run\_of(y, ?f) )) \\
                    & \land [|?Y|/|\{b|(Runs(b) \land  run\_of(b,?f))\}| > 0.9  \\
                    & \land \forall r/Y . \forall q / Response . inheres\_in(q, r) \rightarrow has\_value\_in(q,Fast)]]
    \end{aligned}
\end{equation}

\begin{equation}\label{eq:eq_usemantics_examples_QC8_9}
    \begin{aligned}
     QG_5 := \; & Response \; (Runs <run\_of: SystemFns>) :: Fast \\
     QG_8 := \; & \bm{U} \; (?Y, QG_5, <inheres\_in: \; ?Y >, 90\%) \\
     QG_9 := \; & \bm{U} \; (?F, QG_8, <inheres\_in : <run\_of : \; ?F>>, 80\%) \\
     AX_8 \equiv \; & \exists ?Y/\bm{\wp}( Foo(SystemFns) ) . |?Y| / |Foo(SystemFns)|>0.9 \\
                    & \land \forall r/?Y . \forall q/Response. inheres\_in (q, r) \rightarrow has\_value\_in(q,Fast)]\\
     AX_9 \equiv \; & \exists ?F / \bm{\wp}(SystemFns) . [|?F| / |SystemFns| > .8 \land \forall ?f/?F. \\
                    & \exists ?Y/\bm{\wp}(Foo(\{?f\})) . [ |?Y| / |Foo(\{?f\})| > 0.9 \\
                    & \land \forall r/?Y . \forall q/Response . inheres\_in(q, r) \rightarrow has\_value\_in(q,Fast)]]
    \end{aligned}
\end{equation}

%% file: ms.tex
\section{Meeting Scheduler}
\label{sec:appendix_issues_in_MS}
In this section, we show the requirements that are chosen from the \emph{Meeting Scheduler} project and used for our controlled experiments, as well as the requirements issues (both statistics and details) in them.

\subsection{Project Background}
The \emph{Meeting Scheduler} (MS) system is designed for university research groups to schedule meetings. In general, this system shall be able to create meeting entries, send meeting notifications to students and teachers in a research group. The system shall be also able to manage meeting room information. In addition, there is also a set of non-functional needs, such as usability, performance and interoperability.

\begin{itemize}
  \item \emph{\textbf{Meeting Notification}}: a notification message that tells people a meeting is happening, without requiring their feedback or response.
\end{itemize}

\subsection{The Chosen Stakeholder Requirements}

We have chosen a set of 11 stakeholder requirements, which cover some typical functionalities and qualities (e.g., search, usability), from the \emph{Meeting Scheduler} project. We have used 10/11, 10/11, 8/11 of the requirements (indicated by ``Y'' in Table~\ref{tab:number_issues_in_ms}) in experiment one, two, three, respectively.

\begin{enumerate}
   \item \emph{The product shall be able to manage meeting room records};
   \item \emph{The product shall allow for intuitive searching of conference rooms};
   \item \emph{The product shall be able to provide the search results of meeting rooms in an acceptable time};
   \item \emph{The product shall be able to send meeting notifications via different kinds of methods};
   \item \emph{The product will allow privileged users to view meeting schedules in multiple reporting views};
   \item \emph{The product shall work on a web server on multiple operating system};
   \item \emph{The product shall have good interoperability};
   \item \emph{The product shall have good usability};
   \item \emph{The product shall use standard buttons for navigation};
   \item \emph{The system shall be easy to maintain};
   \item \emph{The product shall display meeting rooms according to search parameters}.
\end{enumerate}

\subsection{Issues in the Chosen Requirements}
We present the statistics and details of requirements issues in these chosen requirements in Table~\ref{tab:number_issues_in_ms} and Table~\ref{tab:ms_detailed_issues}, respectively.

\begin{table}[h]
  \begin{threeparttable}
  \caption {Number of requirements issues in the \emph{Meeting Scheduler} project} 
  \label{tab:number_issues_in_ms}
  \vspace {0.3 cm}
  \centering
  %\small
  \scriptsize
  \setlength\tabcolsep{2pt}
  %\begin{tabular}{|c|p{0.8\textwidth}|}
  \begin{tabular}{|c|c|c|c|c|c|c|c|c|c|c|}
  \hline
\textbf{ID} & \textbf{EX-1 }& \textbf{EX-2} & \textbf{EX-3} & \textbf{Incomplete} & \textbf{Ambiguous} & \textbf{Inconsistent} & \textbf{Unverifiable} & \textbf{Unmodifiable} & \textbf{Unsatisfiable} & \textbf{Sum}\\ \hline
Req.1 & Y & Y & Y & 4 & - & - & - & 1 & - & 5\\ \hline
Req.2 & Y & Y & Y & 2 & 1 & 1 & 1 & 1 & 1 & 7\\ \hline
Req.3 & Y & Y & Y & 4 & - & 1 & 1 & - & 1 & 7\\ \hline
Req.4 & Y & Y & Y & 6 & 1 & - & 1 & 1 & - & 9\\ \hline
Req.5 & N & Y & N & 2 & 1 & - & 1 & - & - & 4\\ \hline
Req.6 & Y & Y & Y & 2 & 1 & - & 1 & - & - & 4\\ \hline
Req.7 & Y & Y & N & 1 & - & - & 2 & - & - & 3\\ \hline
Req.8 & Y & Y & Y & 2 & - & - & 2 & - & 1 & 5\\ \hline
Req.9 & Y & Y & Y & 3 & 1 & - & 1 & 1 & - & 6\\ \hline
Req.10 & Y & Y & Y & 3 & - & 1 & 1 & - & 1 & 6\\ \hline
Req.11 & Y & N & N & 1 & - & - & 1 & - & - & 2 \\ \hline
Sum & 10 & 10 & 8 & 30 & 5 & 3 & 12 & 4 & 4 & 58\\ \hline
  \end{tabular}
    \begin{tablenotes}
      \scriptsize
      \item[1:] EX-1: experiment one; EX-2: experiment two; EX-3: experiment three;
      \item[2:] We have used 10/11, 10/11, 8/11 of the requirements (indicated by ``Y'') in experiment one, two, three, respectively;
    \end{tablenotes}
  \end{threeparttable}
\end{table}

\vspace{1cm}
%\begin{landscape}
\begin{ThreePartTable}
\begin{TableNotes}
  \item[1:] The \emph{\textbf{In-Use}} column is used to represent the experiments that an issue (question) was counted in. For example, a value of ``NYY'' indicates that an issue was counted in experiment two and three, but not in experiment one. In addition, a value of ``---'' indicates that an issue was not counted as a requirement issue.
  \item[2:] The ``Q.R'' symbol indicates a repeated question.
  \end{TableNotes}

\begin{scriptsize}
\begin{longtable}{|c|c|p{0.4\textwidth}|p{0.45\textwidth}|}
\caption{Detailed requirements issues in the \emph{Meeting Scheduler} project}
\label{tab:ms_detailed_issues}
%\endfirsthead
\endhead

\endfoot
\insertTableNotes
\endlastfoot

        \hline
        \hline
        \multicolumn{4}{l}{|\textbf{General Questions}|} \\ \hline
\textbf{ID} & \textbf{In-Use} & \textbf{Issues} & \textbf{Stakeholder Answers}\\ \hline
Q.1 & --- & ``the system'' vs. ``the product'' (Inconsistent) & Yes, they are\\ \hline
Q.2 & --- & ``meeting room'' vs. ``conference room'' (Inconsistent) & Yes, they are\\ \hline
Q.3 & --- & ``search results of meeting room'' vs. ``search results of meeting room information'' (Inconsistent) & Yes, they are\\ \hline
Q.4 & --- & What shall be included for a user profile? (Incomplete Collection) & A user profile shall include: Id, name, gender, role, email, and phone.\\ \hline
Q.4.1 & --- & Are all of these attributes mandatory? & Yes, all these attributes are required\\ \hline
Q.5 & --- & What is the user of the system? (Incomplete Collection) & Secretaries, Teachers, Students (normal students and authorized students)\\ \hline
Q.5.1 & --- & What is the responsibility of secretary?  & manage meeting room records\\ \hline
Q.5.2 & --- & What is the responsibility of teacher? & create meetings, invite for participants\\ \hline
Q.5.3 & --- & What is the responsibility of authorized students? & create meetings, invite for participants\\ \hline
Q.5.4 & --- & What is the responsibility of normal students? & receive meeting notifications\\ \hline
Q.6 & --- & Who are surveyed users? & Surveyed secretaries, teachers and students\\ \hline
Q.6.1 & --- & The user that will do this usability test what kind of background will have?  & We assume that surveyed users are university secretaries, teachers and students.\\ \hline

\multicolumn{4}{l}{} \\ \hline
\hline
\multicolumn{4}{l}{|\textbf{Req.1: The product shall be able to manage meeting room records;}|} \\ \hline
Q.1 & YYY & Who can manage? (Incomplete Individual) & Secretaries\\ \hline
Q.2 & YYY & Who do you mean by ``Manage''? (Incomplete Abstract) & Add, update and delete meeting room records \\ \hline
Q.3 & YYY & What information shall a meeting room record include?  (Incomplete Collection) & Meeting room record shall include: room number, room type, room location, room capacity, busy period (date and time), and room equipment. \\ \hline
Q.4 & YYY & Busy period (Incomplete Collection) & Data and time\\ \hline
Q.5 & YYY & Shall we separate add, update and delete? (Unmodifiable) & Yes, they should be separated: the system shall add ...; the system shall update ...; the system shall delete.\\ \hline

\multicolumn{4}{l}{} \\ \hline
\hline
\multicolumn{4}{l}{|\textbf{Req.2: The product shall allow for intuitive searching of conference rooms;}|} \\ \hline
Q.1 & YYY & Who can search? (Individual Incompleteness) & Meeting organizers (authorized students and teachers)\\ \hline
Q.1.1 & --- & Does teacher needs to be authorized? & No, no need\\ \hline
Q.2 & YYY & What are the search parameters? (Individual Incompleteness) & I want to search by room type, by room capacity, by date and time\\ \hline
Q.3 & NYY & Can these parameters be combined? (Ambiguous) & Yes, they should be combined: room type + date and time; room capacity + date and time\\ \hline
Q.R & --- & What information will be included in a meeting room record? (Incomplete Collection) & See Q.3 in Req.1\\ \hline
Q.4 & NYY & Is ``conference room'' the same with ``meeting room''? (Inconsistent Term) & Yes, use ``meeting room''\\ \hline
Q.R & --- & What will be searched on? (Inconsistent Term) & Meeting room records (Q.3 in Req.3)\\ \hline
Q.5.3 & --- & Search on a list of meeting room records? & Yes, a list of meeting room records, not a single one\\ \hline
Q.5 & YYY & How to measure ``intuitive''? (Unverifiable) & Surveyed users in a usability test agree thatthe representation (UI design) of the search function are familiar to them\\ \hline
Q.5.1 & --- & What will be included in the representation? & Search box, search field, and search button,\\ \hline
Q.5.2 & --- & How to decide familiar or not? & Do not need to worry about this. A respondent will make his/her own judgment.\\ \hline
Q.6 & YYY & Do you mean all surveyed users? (Unsatisfiable) & 80\% of the surveyed users in a usability test\\ \hline
Q.7 & NYY & Search vs. the search shall be intuitive (Unmodifiable) & Yes, separate them\\ \hline

\multicolumn{4}{l}{} \\ \hline
\hline
\multicolumn{4}{l}{|\textbf{Req.3: The product shall be able to provide the search results of meeting rooms in an acceptable time;}|} \\ \hline
Q.1 & YYY & Acceptable is Unverifiable (Unverifiable) & Less than 30 seconds.\\ \hline
Q.1.1 & --- & Including 30 seconds? & Yes, inclusive\\ \hline
Q.2 & YYY & Do you mean all the search executions? (Unsatisfiable) & 90\% of the search executions\\ \hline
Q.2.1 & --- & Is it the average time? & No, I mean 90\% the time\\ \hline
Q.3 & YYY & Search results of meeting room or meeting room information ? (Inconsistent Term) & Meeting room records.\\ \hline
Q.4 & NYY & What are the search results? (Incomplete Abstract) & A set of meeting room records retrieved according to certain search conditions.\\ \hline
Q.4.1 & --- & Will you show the whole meeting room record or some of the attributes? What will be displayed? & Display a sub set of attributes (we call them ``meeting room items''):room number, room type, room location, and room capacity.\\ \hline
Q.5 & YYY & How to display the search results (Incomplete Individual) & Display them in a table, each meeting room item in a row\\ \hline
Q.6 & NYY & What shall be included for a user profile? (Incomplete Collection) & A user profile shall include: Id, name, gender, role, email, and phone.\\ \hline
Q.6.1 & --- & Are all of these attributes mandatory? & Yes, all these attributes are required\\ \hline
Q.7 & NNY & What is the user of the system? (Incomplete Collection) & Secretaries, Teachers, Students (normal students and authorized students)\\ \hline

\multicolumn{4}{l}{} \\ \hline
\hline
\multicolumn{4}{l}{|\textbf{Req.4: The product shall be able to send meeting notifications via different kinds of methods;}|} \\ \hline
Q.1 & YNY & Who do you mean by ``different''? (Unverifiable) & 2 methods, SMS (Short Messaging Service) and Email;\\ \hline
Q.2 & NYY & Send notification via Email and SMS at the same time? (Ambiguous) & No, separately.\\ \hline
Q.3 & YYY & Should the two methods be separated? (Unmodifiable) & Yes.\\ \hline
Q.4 & YYY & Who can send meeting notification?(Incomplete Individual) & Meeting organizer (authorized students and teachers)\\ \hline
Q.4.1 & --- & Does teacher needs to be authorized? & No, no need\\ \hline
Q.4.2 & --- & Need to send reminder? When to send reminder? & No need to send.\\ \hline
Q.5 & YYY & Who will receive meeting notification? (Incomplete Individual) & Meeting participants: team members, including students and teachers\\ \hline
Q.5.1 & --- & Can a meeting organizer send notification to him/her? & Yes, he/she can, because he/she is a team member\\ \hline
Q.6 & YYY & Send to how many participants at a time? (Incomplete individual) & a set, at least 2,\\ \hline
Q.6.1 & --- & At most? & At most 1000\\ \hline
Q.7 & NNY & When to send? (Incomplete individual) & After the meeting has been created\\ \hline
Q.8 & YYY & What information shall meeting notification include? (Incomplete collection) & Meeting notification shall include: meeting title, meeting time (including data and time), and meeting location.\\ \hline
Q.9 & YYY & Meeting time? (Incompleteness Collection) & Data and time\\ \hline

\multicolumn{4}{l}{} \\ \hline
\hline
\multicolumn{4}{l}{|\textbf{Req.5: The product will allow privileged users to view meeting schedules in multiple reporting views;}|} \\ \hline
Q.1 & NYN & Who are privileged users? (Incomplete Abstract) & Authorized students and teachers\\ \hline
Q.1.1 & --- & Does teacher needs to be authorized? & No, no need\\ \hline
Q.2 & NYN & What are meeting schedules? What shall a meeting schedule include? (Incomplete Collection) & It is a summary of meeting, it includes: meeting title, meeting time, and location.\\ \hline
Q.3 & NYN & What do you mean by ``multiple reporting view''? (Unverifiable) & Calendar View and List View\\ \hline
Q.4 & NYN & View at the same time or not? (Ambiguity) & No, separately\\ \hline

\multicolumn{4}{l}{} \\ \hline
\hline
\multicolumn{4}{l}{|\textbf{Req.6: The product shall work on a web server on multiple operating system;}|} \\ \hline
Q.1 & YYY & What kind of web server? (Incomplete Abstract) & Tomcat\\ \hline
Q.1.1 & --- & Is the web server in house or rented? & Rented\\ \hline
Q.2 & YYY & What do you mean by multiple? What specific operating systems are you referring to? (Unverifiable) & Windows\_XP, Windows\_7, Ubuntu\_15\_04, Mac\_OSX\_10\_8, Mac\_OSX\_10\_9, Mac\_OSX\_10\_10\\ \hline
Q.3 & YYY & Is it ambiguous? (Ambiguous) & The product shall be able to run on a web server that can run on the operating systems listed above\\ \hline
Q.4 & YYY & What does ``work on'' mean? (IncompleteAbstract) & Run on\\ \hline

\multicolumn{4}{l}{} \\ \hline
\hline
\multicolumn{4}{l}{|\textbf{Req.7: The product shall have good interoperability;}|} \\ \hline
Q.1 & YYN & What do you mean by ``good interoperability''? (Unverifiable) & I mean two aspects: (1) The product shall be accessible from different kinds of browsers; (2); as I stated in the requirement that talks about ``web server''.\\ \hline
Q.2 & YYN & What do you mean different kinds of browsers? (Unverifiable) & At least the following kinds: Chrome\_40\_0, Chrome\_41\_0, Chrome\_42\_0, Firefox\_35\_0, Firefox\_36\_0, Firefox\_37\_0, Safari\_8\_0.\\ \hline
Q.3 & NNN & Browsers on which kind of OS? Desktop or mobile? (Incomplete Individual) & Desktop and Mobile.\\ \hline

\multicolumn{4}{l}{} \\ \hline
\hline
\multicolumn{4}{l}{|\textbf{Req.8: The product shall have good usability;}|} \\ \hline
Q.1 & YYY & Good usability is Unverifiable (Unverifiable) & It is easy to learn for users ;\\ \hline
Q.2 & NNY & Easy to learn for what kind of users? (Incomplete Abstract) & Meeting organizers (authorized students and teachers).\\ \hline
Q.2.1 & --- & Does teacher needs to be authorized? & No, no need\\ \hline
Q.3 & YYY & What do you mean by easy to learn? (Unverifiable) & Users in a beta test shall be able to successfully create meeting entries within 10 minutes of learning;\\ \hline
Q.4 & YYY & Do you mean all users? (Unsatisfiable) & 90\% of users in a beta test\\ \hline
Q.5 & YYY & What information shall meeting entries include (Incomplete collection)? & It shall include meeting title, meeting time, meeting place, and a description\\ \hline

\multicolumn{4}{l}{} \\ \hline
\hline
\multicolumn{4}{l}{|\textbf{Req.9: The product shall use standard buttons for navigation;}|} \\ \hline
Q.1 & YYY & What is standard? (Unverifiable) & Standard style: dimension (length 30 pixels; width: 15 pixels) and color (grey background and black font).\\ \hline
Q.2 & YYY & Should all the buttons be standard or some of them be standard (Ambiguous)? & All buttons shall be standard;\\ \hline
Q.3 & YYY & Navigating whom? (Incomplete Individual) & Users shall be navigated; or navigate users; meeting organizers or creators\\ \hline
Q.4 & YYY & What kind of navigation? (Incomplete Individual) & Back and forth\\ \hline
Q.5 & NNY & Is there a navigation requirement (Incomplete Collection) & Yes, ``the system shall be able to navigate users''\\ \hline
Q.6 & YYY & Should it be separated? (Unmodifiable) & Yes\\ \hline

\multicolumn{4}{l}{} \\ \hline
\hline
\multicolumn{4}{l}{|\textbf{Req.10: The system shall be easy to maintain;}|} \\ \hline
Q.1 & NYY & What do you mean by maintain? (Incomplete) & Adding new features to the product.\\ \hline
Q.2 & NYY & For whom? (Incomplete Individual) & Software maintainers\\ \hline
Q.3 & YYY & Easy is Unverifiable (Unverifiable) & software maintainers are able to add new features into the product within 2 working days\\ \hline
Q.4 & YYY & Do you mean all the software maintainers? (Unsatisfiable) & No, I mean 90\% of software maintainers are able to do so\\ \hline
Q.5 & NYY & How to achieve good maintainability? (Incomplete Collection) & The product be designed using Design Patterns.\\ \hline
Q.5.1 & --- & Which design patterns? & MVC\\ \hline
Q.6 & NNY & Does ``the system'' and ``the product'' refer to the same thing? (Inconsistent) & Yes, use ``the product'' for consistency\\ \hline

\multicolumn{4}{l}{} \\ \hline
\hline
\multicolumn{4}{l}{|\textbf{Req.11: The product shall display meeting rooms according to search parameters;}|} \\ \hline
Q.1 & YNN & Display meeting room items, not meeting rooms, or meeting room detailed informaiton (Inconsistent Term) & Meeting room items (See Q.4.1 in Req.3)\\ \hline
Q.R & --- & What are the search parameter? & See Q.3 in Req.2\\ \hline
Q.R & --- & Who can search? & See Q.1 in Req.2\\ \hline
Q.R & --- & Display a list of meeting room items? not a single one? & See Q.5 in Req.3\\ \hline
Q.2 & YNN & What shall be included in meeing room item? (Incomplete Collection) & Id, name, type, capacity, location\\ \hline

\end{longtable}
\end{scriptsize}
%\end{table}
%\end{longtable}
\end{ThreePartTable}
%\end{landscape}

%% file: rb.tex
\section{Realtor Buddy}
\label{sec:appendix_issues_in_RB}
%In this section, we show the 10 chosen requirements of the \emph{Realtor Buddy} project, the statistics of issues in these 10 requirements, and the detailed issues in each requirement.

In this section, we show the requirements that are chosen from the \emph{Realtor Buddy} project and used for our controlled experiments, as well as the requirements issues (both statistics and details) in them.

\subsection{Project Background}
The \emph{Realtor Buddy} (RB) project is an application designed for helping real estate agents to sell properties. In general, the system shall allow the agent to manage property listing information in the multiple list service (MLS). There are also requirements on the non-functional aspects, such as look and feel, performance, and usability.

\begin{itemize}
  \item \emph{\textbf{Real Estate Agent}}: a real estate agent is a person who acts as an intermediary between sellers and buyers of real estate/real property and attempts to find sellers who wish to sell and buyers who wish to buy.
  \item \emph{\textbf{Multiple Listing Service}} (\emph{\textbf{MLS}}): a MLS is a suite of services that enables real estate brokers to establish contractual offers of compensation (among brokers), facilitate cooperation with other broker participants, and is a facility for the orderly correlation and dissemination of property/asset listing information to better serve real estate broker's clients, customers and the public.
  \item \emph{\textbf{Comparative Market Analysis}} (\emph{\textbf{CMA}}): a CMA report is an estimate of your home's value done by your real estate bro-ker to establish a listing/offer price when you decide that you want to sell or buy a home or property. This service is always offered free of charge and without obligation. A CMA should only be used as a reference for deciding at what price you should list or buy your home for.
  \item \emph{\textbf{Property Listing}}: a record of a property for lease or sale by an authorized real estate broker.
\end{itemize}

%The 10 chosen requirements of the \emph{Meeting Scheduler} project are as follows.
\subsection{The Chosen Stakeholder Requirements}

We have chosen a set of 10 typical stakeholder requirements from the \emph{Realtor Buddy} project. We have used 10/10, 10/10, 8/10 of the requirements (indicated by ``Y'' in Table~\ref{tab:statistics_issues_in_rb}) in experiment one, two, three, respectively.

\begin{enumerate}
   \item \emph{The system shall update or create new property listings in the MLS (Multiple Listing Service)};
   \item \emph{The system shall allow a real estate agent to query MLS information};
   \item \emph{The system shall display clear property images in the search results};
   \item \emph{The product shall generate a CMA (Comparative Market Analysis) report in an acceptable time};
   \item \emph{The user shall be able to download appointments and contact information for clients};
   \item \emph{The user interface shall have standard menus and buttons for navigation};
   \item \emph{The product shall have high availability};
   \item \emph{The system shall have a professional appearance};
   \item \emph{The system shall have good usability};
   \item \emph{The system shall be self-explanatory};
\end{enumerate}

\subsection{Issues in the Chosen Requirements}
We present the statistics and details of requirements issues in these chosen requirements in Table~\ref{tab:statistics_issues_in_rb} and Table~\ref{tab:rb_detailed_issues}, respectively.

\begin{table}[h]
  \begin{threeparttable}
  \caption {Number of requirements issues in the \emph{Realtor Buddy} project}
  \label{tab:statistics_issues_in_rb}
  \vspace {0.3 cm}
  \centering
  \scriptsize
  %\small
  \setlength\tabcolsep{2pt}
  %\begin{tabular}{|c|p{0.8\textwidth}|}
  \begin{tabular}{|c|c|c|c|c|c|c|c|c|c|c|}
  \hline
\textbf{ID} & \textbf{EX-1 }& \textbf{EX-2} & \textbf{EX-3} & \textbf{Incomplete} & \textbf{Ambiguous} & \textbf{Inconsistent} & \textbf{Unverifiable} & \textbf{Unmodifiable} & \textbf{Unsatisfiable} & \textbf{Sum}\\ \hline
Req.1 & Y & Y & Y & 7 & 1 & - & - & 1 & - & 9\\ \hline
Req.2 & Y & Y & Y & 4 & 1 & 3 & - & - & - & 8\\ \hline
Req.3 & Y & Y & Y & 5 & 1 & 1 & 1 & 1 & 1 & 10\\ \hline
Req.4 & Y & Y & Y & 5 & - & - & 1 & 1 & 1 & 8\\ \hline
Req.5 & Y & Y & N & 4 & 2 & - & - & 1 & - & 7\\ \hline
Req.6 & Y & Y & Y & 3 & 2 & - & 1 & 1 & - & 7\\ \hline
Req.7 & Y & Y & Y & 1 & - & - & 1 & - & 1 & 3\\ \hline
Req.8 & Y & Y & N & 1 & - & - & 1 & - & 1 & 3\\ \hline
Req.9 & Y & Y & Y & 1 & - & - & 2 & - & 1 & 4\\ \hline
Req.10 & Y & Y & Y & - & - & - & 1 & - & 1 & 2\\ \hline
Sum & 10 & 10 & 8 & 31 & 7 & 4 & 8 & 5 & 6 & 61\\ \hline
  \end{tabular}
  \begin{tablenotes}
      \scriptsize
      \item[1:] EX-1: experiment one; EX-2: experiment two; EX-3: experiment three;
      \item[2:] We have used 10/10, 10/10, 8/10 of the requirements (indicated by ``Y'') in experiment one, two, three, respectively;
    \end{tablenotes}
  \end{threeparttable}
\end{table}

\begin{ThreePartTable}
\begin{TableNotes}
  \item[1:] The \emph{\textbf{In-Use}} column is used to represent the experiments that an issue (question) was counted in. For example, a value of ``NYY'' indicates that an issue was counted in experiment two and three, but not in experiment one. In addition, a value of ``---'' indicates that an issue was not counted as a requirement issue.
  \item[2:] The ``Q.R'' symbol indicates a repeated question.
  \end{TableNotes}
\begin{scriptsize}
\begin{longtable}{|c|c|p{0.4\textwidth}|p{0.45\textwidth}|}
\caption{Detailed requirements issues in the \emph{Realtor Buddy} project}
\label{tab:rb_detailed_issues}
%\endfirsthead
\endhead

\endfoot
\insertTableNotes
\endlastfoot

        \hline
        \hline
        \multicolumn{4}{l}{|\textbf{General Questions}|} \\ \hline
\textbf{ID} & \textbf{Issues} & \textbf{Stakeholder Answers}\\ \hline
Q.R & --- & Are ``the system'' and ``the product'' the same thing? (Inconsistent Term) & Yes, they are\\ \hline
Q.R & --- & Are ``real estate agent'', ``realtor'', ``real estate agency'' and ``real estate broker'' the same concept? (Inconsistent Term) & Yes, they are.\\ \hline
Q.R & --- & ``Query'' vs. ``Search'' & Yes, they are.\\ \hline
Q.4 & --- & What is the query about? MLS information vs. Property listing information & Yes, they are.\\ \hline
Q.5 & --- & What is the user of the system? & Realtors. The system is designed for relators, not for internet users.\\ \hline
Q.6 & --- & Who are clients? & People who want to sell or buy homes; clients do not directly use the system.\\ \hline
Q.7 & --- & Who are surveyed users? & We assume that surveyed users are from the real estate industry (with at least 3 months experiences).\\ \hline
Q.R & --- & What shall be included for a user profile? (Incomplete Collection) & A user profile shall include: Id, name, gender, role, email, and phone number.\\ \hline
Q.R.1 & --- & All of the attributes mandatory? & Yes, all these attributes are required\\ \hline

\multicolumn{4}{l}{} \\ \hline
\hline
\multicolumn{4}{l}{|\textbf{Req.1: The system shall update or create new property listings in the MLS (Multiple Listing Service);}|} \\ \hline
Q.1 & YYY & Update what? New property listings? (Incomplete Individual) & Update property listings.\\ \hline
Q.2 & YYY & Who can update/create? (Incomplete Individual) & Realtors\\ \hline
Q.3 & NNY & Can a real estate agent update others' property listing records? (Incomplete Individual) & No, he/she cannot update others' records; he/she can only update his/her property listing records.\\ \hline
Q.4 & YYY & What information shall property listing includes?(Incomplete Collection) & Property listing information shall include location, listing price, property image, description, and property specification.\\ \hline
Q.5 & YYY & What is the currency for listing price? (Incomplete Collection) & Euro; Float (two digitals)\\ \hline
Q.6 & NYY & What is the format of property image? (Incomplete Collection) & JPEG; width: 2", length: 3"\\ \hline
Q.7 & NNY & Delete property listing info? (Incomplete Collection) & No, just give a mark if sold\\ \hline
Q.8 & NYY & Shall the two methods be supported or only one of them? (Ambiguous) & Both of them shall be supported by the system\\ \hline
Q.9 & YYY & Should be separated? (Unmodifiable) & Yes, it should be two requirements: update existing property listings, create new property listing.\\ \hline

\multicolumn{4}{l}{} \\ \hline
\hline
\multicolumn{4}{l}{|\textbf{Req.2: The system shall allow a real estate agent to query MLS information;}|} \\ \hline
Q.1 & YYY & What are the query parameters? What kind of queries? (Incomplete Individual) & The query parameters:location and price\\ \hline
Q.2 & NNY & Should the two parameters be separated or combined? (Ambiguous) & Query by location, query by price, query by location and price\\ \hline
Q.3 & NYY & Can anyone else query MLS information? Only realtors? (Incomplete Collection) & No, only real estate agents\\ \hline
Q.4 & NNY & Can a real estate agent query MLS information inserted by other agents? (Incomplete Individual) & Yes, he/she can search all the property listing info in the MLS \\ \hline
Q.R & --- & Do we need to delete a record when it has been sold out? & No deletion, just a marking ``Sold'' (see Q.7 in Req.1)\\ \hline
Q.5 & YYY & What is the query about? Or what is MLS information? (Inconsistent Term) & Hmm, it is property listing information. \\ \hline
Q.5.1 & --- & What exactly does query mean?  & Retrieve the information according to parameters and display them in a list\\ \hline
Q.6 & YYY & Are ``real estate agent'' and ``realtor'' the same concept? What is a ``real estate agent''? What is real? (Inconsistent) & Yes, they are.\\ \hline
Q.R & --- & What information shall a property listing includes? (Incomplete Collection) & See Q.4 in Req.1 \\ \hline
Q.7 & YYY & Are ``the system'' and ``the product'' the same thing? (Inconsistent Term) & Yes, they are the same thing.\\ \hline
Q.8 & NYY & User profile (user information)  (Incomplete Collection) & A user profile shall include id, name, user role, email, phone number\\ \hline

\multicolumn{4}{l}{} \\ \hline
\hline
\multicolumn{4}{l}{|\textbf{Req.3: The system shall display clear property images in the search results;}|} \\ \hline
Q.1 & YYY & What is ``clear''? (Unverifiable) & $>=$ 300 dpi (dot per inch), i.e., $>=$ 300 pixels per inch\\ \hline
Q.2 & NYY & What are search results? (Incomplete Abstract) & A set of property listings retrieved according to certain search parameters.\\ \hline
Q.3 & NNY & How to display the search results? (Incomplete Collection) & Display the retrieved property listings in a list, each property listing in a row.\\ \hline
Q.3.1 & --- & Display all the attributes of a property listing? & A sub set of attributes (we call them ``property listing items''): property location, listing price, property image, and description\\ \hline
Q.4 & NYY & Eachproperty listing shall have an image or some of them shall have an image? How many images (Incomplete individuals)? & In the search result, it is better for eachproperty listing to have an image, but it is possible that a property listing does not have an image.\\ \hline
Q.4.1 & --- & Does a property have how many images? & A property listing can have more than one image, but we only show one of them in the search results. To see all the images, you need to click ``view details''.\\ \hline
Q.5 & NNY & Which format of the image?(Incomplete Individual) & JPEG.\\ \hline
Q.6 & YNY & What is the size of the image?(Incomplete Individual) & width: 2", length: 3"\\ \hline
Q.6.1 & --- & How long is an inch? & 1 inch = 2.54 centimeter\\ \hline
Q.7 & YYY & Shall each image be clear (Ambiguous) & Yes, each images shown in the search result shall be clear\\ \hline
Q.8 & NNY & ``Query'' vs. ``Search'' & Yes, they are.\\ \hline
Q.9 & NYY & display image; displayed image shall be clear. (Unmodifiable) & Yes, they need to be separated\\ \hline
Q.10 & YYN & Shall all users agree that images are clear? (Unsatisfiable) & 90\% \\ \hline

\multicolumn{4}{l}{} \\ \hline
\hline
\multicolumn{4}{l}{|\textbf{Req.4: The product shall generate a CMA (Comparative Market Analysis) report in an acceptable time;}|} \\ \hline
Q.1 & YYY & How long is acceptable to you? (Unverifiable) & Less than 30 seconds .\\ \hline
Q.1.1 & --- & Including 30 seconds? & Yes, including 30, $<=$ 30 seconds.\\ \hline
Q.2 & YYY & Is 31 seconds OK? I mean, how about if sometimes it takes 31 seconds?  Or is ``30 seconds'' an average time? (Unsatisfiable) &  $<=$ 30 seconds 90\% of the time.\\ \hline
Q.3 & YYY & What should a CMA report include? (Incomplete collection) & In general, a CMA report shall include Id, location, and listing price\\ \hline
Q.3.1 & --- & What is a CMA report?  & Please refer to the domain description document.\\ \hline
Q.4 & YYY & Which format? (Incomplete individual) & PDF.\\ \hline
Q.5 & NYY & In which style? (Incomplete individual) & ``key: value'' pairs\\ \hline
Q.6 & NNY & Generate a CMA report at a time or a set of CMA reports at a time? (Incomplete Individual) & Generate one CMA report at a time\\ \hline
Q.7 & NNY & Who can generate? & Realtors\\ \hline
Q.8 & YYY & Should I separate the requirement into two? (Unmodifiable) & Yes, there are two requirements: ``the system shall generate CMA report'', ``the generation function shall take less than 30 seconds''.\\ \hline

\multicolumn{4}{l}{} \\ \hline
\hline
\multicolumn{4}{l}{|\textbf{Req.5:The user shall be able to download appointments and contact information for clients;}|} \\ \hline
Q.1 & YYN & What is ``user''? (Incomplete Abstract) & Realtors.\\ \hline
Q.2 & YYN & Download contact for the sake of client or download client contact?(Ambiguous) & Download client contact information.\\ \hline
Q.2.1 & --- & Who are clients? & People who want to buy or sell homes.\\ \hline
Q.3 & NYN & Download appointments and contact information at the same time?(Ambiguous) & No, download them separately.\\ \hline
Q.4 & YYN & What kind of contact information? What shall be included in contact information?(Incomplete Collection) & Name, Email, and Phone number.\\ \hline
Q.5 & YYN & What information shall be included in an appointment?(Incomplete Collection) & It shall include appointment time, appointment place, and the names of participants.\\ \hline
Q.6 & YYN & Which format? (Incomplete Individual) & PDF, for both.\\ \hline
Q.7 & YYN & Should it be separated into two requirements?(Unmodifiable) & Yes, it should be two requirements: ``download appointments'', and ``download client contact information''\\ \hline

\multicolumn{4}{l}{} \\ \hline
\hline
\multicolumn{4}{l}{|\textbf{Req.6: The user interface shall have standard menus and buttons for navigation;}|} \\ \hline
Q.1 & YYY & What do you mean by standard?  (Unverifiable) & Standard style: dimension (length 30 pixels; width: 15 pixels) and color (grey background and black font).\\ \hline
Q.2 & YYY & Should buttons be standard?  (Ambiguous) & Yes, use the same dimension for both menus and buttons\\ \hline
Q.3 & YYY & Should all menus be standard or some of them? (Ambiguous) & All of the menus shall be standard\\ \hline
Q.3.1 & --- & Should all buttons be standard or some of them? (Ambiguous) & All of the buttons shall be standard\\ \hline
Q.4 & NYY & Is there a navigation requirement?  Or shall I separate ``navigation'' from ``using menus and buttons''? (Unmodifiable) (Incomplete Collection) & Yes, the system shall be able to navigate users by using standard menus and buttons.\\ \hline
Q.5 & NYY & Navigation whom? (Incomplete Individual) & Navigate realtors\\ \hline
Q.6 & YYY & What kind of navigation? (Incomplete Individual) & Back and forth navigation \\ \hline
Q.7 & YYY & Shall menus and buttons be separated? (Unmodifiable) & Yes\\ \hline
Q.7.1 & --- & What is the standard color for menus?  & Gray background and black font\\ \hline
Q.7.2 & --- & What is the standard color for buttons?  & Gray background and black font\\ \hline
Q.7.3 & --- & What types of buttons? & Back, and forth\\ \hline
Q.7.4 & --- & Where will you put the menu? & The left of the page\\ \hline
Q.7.5 & --- & Where will you put the button? & The button of the page\\ \hline

\multicolumn{4}{l}{} \\ \hline
\hline
\multicolumn{4}{l}{|\textbf{Req.7: The product shall have high availability;}|} \\ \hline
Q.1 & NNY & What do you mean high availability? (Unverifiable) & The product shall be available for use 24 hours per day 365 days per year;\\ \hline
Q.2 & YYY & Do you mean available all the time? (Unsatisfiable) & It shall be available 99\% of the time in a year (time unit: hour)\\ \hline
Q.3 & YYY & What do you mean by ``available for use''? (IncompleteAbstract) & Available online for realtors\\ \hline

\multicolumn{4}{l}{} \\ \hline
\hline
\multicolumn{4}{l}{|\textbf{Req.8: The system shall have a professional appearance;}|} \\ \hline
Q.1 & YYN & What do you mean by professional? (Unverifiable) & Surveyed users in beta test agree that the appearance is professional\\ \hline
Q.2 & YYN & Do you mean all surveyed user in the beta test? (Unsatisfiable) & I mean 80\% of the surveyed users in beta test\\ \hline
Q.3 & YYN & How to make it professional? (Incomplete Collection) & Use standard navigation menus and buttons\\ \hline

\multicolumn{4}{l}{} \\ \hline
\hline
\multicolumn{4}{l}{|\textbf{Req.9: The system shall have good usability;}|} \\ \hline
Q.1 & NNY & What do you mean good usability? (Unverifiable) & The product shall be easy for a realtor to learn;\\ \hline
Q.2 & YYY & What do you mean by easy?  Easy is Unverifiable; too general (Unverifiable) & Realtors shall be able to manage property listing information within 10 minutes of learning.\\ \hline
Q.3 & YYY & What do you mean by ``manage''? (Incomplete Abstract) & add, delete, update and query property listing information .\\ \hline
Q.4 & YYY & Do you mean all the relators? (Unsatisfiable) & No, I mean 90\% of the relators.\\ \hline

\multicolumn{4}{l}{} \\ \hline
\hline
\multicolumn{4}{l}{|\textbf{Req.10: The system shall be self-explanatory;}|} \\ \hline
Q.1 & YYY & What do you mean by self explanatory? (Unverifiable) & The terms used in the system can be naturally understandable by relators in the realtor community.\\ \hline
Q.2 & YYY & Do you mean naturally understandable by all the relators? (Unsatisfiable) & No, I mean 90\% of the relators in the relator community.\\ \hline
Q.2.1 & --- & Do you mean all the words used in the system? & Yes, all words shall be in realtor's terminological dictionary.\\ \hline

\end{longtable}
\end{scriptsize}
%\end{table}
%\end{longtable}
\end{ThreePartTable}
%\end{landscape}

%% file: questionnaire.tex
\section{The \emph{Desiree} Questionnaire}
\label{sec:appendix_desiree_questionnaire}

This is a questionnaire about the first method (method 1, used in Session 1) and the second \emph{Desiree} method (method 2, used in Session 2)~\footnote{Note that the first 4 questions are only used in experiment three, the other 14 questions have been used in all the three experiments.}. Please let us know how you think about the two methods. We will keep improving our methodology and tool based on your feedback, comments and suggestions. Thanks very much!\\

%\includepdf[pages=-]{./pics/Desiree_Questionnaire.pdf}
%\includepdf[pages=1]{./pics/Desiree_Questionnaire.pdf}
%\includepdf[pages=1,pagecommand=h]{./pics/Desiree_Questionnaire.pdf}
%\includepdf[pages=2-]{./pics/Desiree_Questionnaire.pdf}

\begin{small}
\emph{\textbf{What is your education background}}? (only in experiment three)\\
\hspace*{0.8cm}
\begin{tabular}{rp{0.75\textwidth}}
    %\multicolumn{2}{l}{\emph{\textbf{What is your education background}}? (only in experiment three)} \\
    $\hspace{-.38cm}\Box$ & Bachelor \\
    $\hspace{-.38cm}\Box$ & Master  \\
    $\hspace{-.38cm}\Box$ & Ph.D. \\
    $\hspace{-.38cm}\Box$ & Post doctor \\
    $\hspace{-.38cm}\Box$ & Other \\
\end{tabular}

\vspace{12pt}
\emph{\textbf{Do you have any work experiences}}? (only in experiment three)\\
\hspace*{0.8cm}
\begin{tabular}{rp{0.75\textwidth}}
    %\multicolumn{2}{l}{\emph{\textbf{What is your education background}}? (only in experiment three)} \\
$\hspace{-.38cm}\Box$ & No or Little ($\leq$ 3 months) \\
$\hspace{-.38cm}\Box$ & $\leq$ 1 Year (e.g., having some internship) \\
$\hspace{-.38cm}\Box$ & $1 < \sim \leq 3$ Years \\
$\hspace{-.38cm}\Box$ & $3 < \sim \leq 5$ Years \\
$\hspace{-.38cm}\Box$ & $5 < \sim \leq 10$ Years \\
$\hspace{-.38cm}\Box$ & $> 10$ years \\
\end{tabular}

\vspace{24pt}
\emph{\textbf{What is your main working/researching/studying area}}? (only in experiment three)\\
\indent \emph{Can be multiple, e.g., System Developing, Business Modeling, Testing}

\vspace{36pt}
\emph{\textbf{Do you have work experiences on RE before}}? (only in experiment three)\\
\indent \emph{If yes, how many years?}

\vspace{36pt}
\emph{\textbf{1.How do you find the learnability of the first method}}? \\
\hspace*{0.8cm}
\begin{tabular}{rp{0.75\textwidth}}
    %\multicolumn{2}{l}{\emph{\textbf{What is your education background}}? (only in experiment three)} \\
$\hspace{-.38cm}\Box$ &   Very easy to learn \\
$\hspace{-.38cm}\Box$ &   Easy to learn \\
$\hspace{-.38cm}\Box$ &   Undecided / Neutral \\
$\hspace{-.38cm}\Box$ &   Hard to learn \\
$\hspace{-.38cm}\Box$ &   Very hard to Wlearn \\
\end{tabular}

\vspace{12pt}
\emph{\textbf{2. How do you find the learnability of the second method}}? \\
\hspace*{0.8cm}
\begin{tabular}{rp{0.75\textwidth}}
    %\multicolumn{2}{l}{\emph{\textbf{What is your education background}}? (only in experiment three)} \\
$\hspace{-.38cm}\Box$ &   Very easy to learn \\
$\hspace{-.38cm}\Box$ &   Easy to learn \\
$\hspace{-.38cm}\Box$ &   Undecided / Neutral \\
$\hspace{-.38cm}\Box$ &   Hard to learn \\
$\hspace{-.38cm}\Box$ &   Very hard to learn \\
\end{tabular}

\vspace{12pt}
\emph{\textbf{3. How do you find the usability of the first method}}? \\
\hspace*{0.8cm}
\begin{tabular}{rp{0.75\textwidth}}
    %\multicolumn{2}{l}{\emph{\textbf{What is your education background}}? (only in experiment three)} \\
$\hspace{-.38cm}\Box$ &   Very easy to use \\
$\hspace{-.38cm}\Box$ &   Easy to use \\
$\hspace{-.38cm}\Box$ &   Undecided / Neutral \\
$\hspace{-.38cm}\Box$ &   Hard to use \\
$\hspace{-.38cm}\Box$ &   Very hard to use \\
\end{tabular}

\vspace{12pt}
\emph{\textbf{4. How do you find the usability of the second method}}? \\
\hspace*{0.8cm}
\begin{tabular}{rp{0.75\textwidth}}
    %\multicolumn{2}{l}{\emph{\textbf{What is your education background}}? (only in experiment three)} \\
$\hspace{-.38cm}\Box$ &   Very easy to use \\
$\hspace{-.38cm}\Box$ &   Easy to use \\
$\hspace{-.38cm}\Box$ &   Undecided / Neutral \\
$\hspace{-.38cm}\Box$ &   Hard to use \\
$\hspace{-.38cm}\Box$ &   Very hard to use \\
\end{tabular}

\vspace{12pt}
\emph{\textbf{5. How do you find the usefulness of the first method for writing a good SRS}}? \\
\hspace*{0.8cm}
\begin{tabular}{rp{0.75\textwidth}}
    %\multicolumn{2}{l}{\emph{\textbf{What is your education background}}? (only in experiment three)} \\
$\hspace{-.38cm}\Box$ &   Very useful \\
$\hspace{-.38cm}\Box$ &   Useful \\
$\hspace{-.38cm}\Box$ &   Undecided / Neutral \\
$\hspace{-.38cm}\Box$ &   Not useful \\
$\hspace{-.38cm}\Box$ &   Useless \\
\end{tabular}

\vspace{12pt}
\emph{\textbf{6. How do you find the usefulness of the second method for writing a good SRS}}? \\
\hspace*{0.8cm}
\begin{tabular}{rp{0.75\textwidth}}
    %\multicolumn{2}{l}{\emph{\textbf{What is your education background}}? (only in experiment three)} \\
$\hspace{-.38cm}\Box$ &   Very useful \\
$\hspace{-.38cm}\Box$ &   Useful \\
$\hspace{-.38cm}\Box$ &   Undecided / Neutral \\
$\hspace{-.38cm}\Box$ &   Not useful \\
$\hspace{-.38cm}\Box$ &   Useless \\
\end{tabular}

\vspace{12pt}
\emph{\textbf{7. How do you find the complexity of the first method}}? \\
\hspace*{0.8cm}
\begin{tabular}{rp{0.75\textwidth}}
    %\multicolumn{2}{l}{\emph{\textbf{What is your education background}}? (only in experiment three)} \\
$\hspace{-.38cm}\Box$ &   Very simple  \\
$\hspace{-.38cm}\Box$ &   Simple \\
$\hspace{-.38cm}\Box$ &   Undecided / Neutral \\
$\hspace{-.38cm}\Box$ &   Complex \\
$\hspace{-.38cm}\Box$ &   Very complex \\
\end{tabular}

\vspace{12pt}
\emph{\textbf{8. How do you find the complexity of the second method}}? \\
\hspace*{0.8cm}
\begin{tabular}{rp{0.75\textwidth}}
    %\multicolumn{2}{l}{\emph{\textbf{What is your education background}}? (only in experiment three)} \\
$\hspace{-.38cm}\Box$ &   Very simple  \\
$\hspace{-.38cm}\Box$ &   Simple \\
$\hspace{-.38cm}\Box$ &   Undecided / Neutral \\
$\hspace{-.38cm}\Box$ &   Complex \\
$\hspace{-.38cm}\Box$ &   Very complex \\
\end{tabular}

\vspace{12pt}
\emph{\textbf{9. How do you find the usability of the Desiree tool}}? \\
\hspace*{0.8cm}
\begin{tabular}{rp{0.75\textwidth}}
    %\multicolumn{2}{l}{\emph{\textbf{What is your education background}}? (only in experiment three)} \\
$\hspace{-.38cm}\Box$ &   Very good \\
$\hspace{-.38cm}\Box$ &   Good \\
$\hspace{-.38cm}\Box$ &   Undecided / Neutral \\
$\hspace{-.38cm}\Box$ &   Bad \\
$\hspace{-.38cm}\Box$ &   Very bad \\
\end{tabular}

\vspace{12pt}
\emph{\textbf{10. Which parts of the Desiree method are the most and least hard to you}}?\\

\vspace{36pt}
\emph{\textbf{11. Which parts of the Desiree method are the most and least useful to you}}?\\

%\vspace{24pt}
%\emph{\textbf{12. Which parts of the Desiree method are the most and least hard to you}}?\\

\vspace{36pt}
\emph{\textbf{12. Will you use the Desiree method when you are writing SRS in the future}}?\\
\indent \indent \emph{\textbf{Parts of the method}}? \emph{\textbf{Which parts}}?

\vspace{36pt}
\emph{\textbf{13. Which parts of the Desiree tool work well}}?\\
\indent \indent \emph{\textbf{Which parts of the Desiree tool do we need to improve}}

\vspace{36pt}
\emph{\textbf{14. Do you have any other comments and suggestions}}?\\
\vspace{36 pt}
\end{small}

\section{Questionnaire Report}
\label{sec:appendix_questionnaire_reports}

We have collected 13 (out of 16), 14 (out of 15) and 24 (out of 29) responses in experiment one, experiment two, and experiment three. We report participants' feedback in each experiment on the learnability, usability, usefulness and complexity of the two methods (\emph{Vanilla} and \emph{Desiree}) in Table~\ref{tab:feedback_learnability}, Table~\ref{tab:feedback_usability}, Table~\ref{tab:feedback_usefulness}, Table~\ref{tab:feedback_complexity}, respectively. We also present participants' feedback on the usability of the \emph{Desiree} tool in Table.~\ref{tab:feedback_desiree_tool}. In these tables, the numbers indicate the number of participants.
%respondents that have participated in our experiments.

%``EX'' is a shorthand for ``Experiment''.

\begin{table}[!htbp]
  %\vspace{-0.5cm}
  \small
  \setlength\tabcolsep{2pt}
  \caption {Participants' feedback on the learnability of the two methods}
  \label{tab:feedback_learnability}
  \vspace {0.3 cm}
  \centering
  \begin{tabular}{|c|c|c|c|c|c|c|c|}
  \hline
   &  & Very Easy & Easy & Neutral & Hard & Very Hard & Total \\ \hline
 \multirow{2}{*}{\textbf{Experiment one}}& Method 1 & 3 & 10 & 0 & 0 & 0 & 13 \\ \cline{2-8}
 & Method 2 & 0 & 3 & 3 & 6 & 1 & 13 \\ \hline
 \multirow{2}{*}{\textbf{Experiment two}} & Method 1 & 1 & 10 & 2 & 1 & 0 & 14 \\ \cline{2-8}
 & Method 2 & 0 & 4 & 7 & 3 & 0 & 14 \\ \hline
 \multirow{2}{*}{\textbf{Experiment three}}& Method 1 & 2 & 13 & 8 & 1 & 0 & 24 \\ \cline{2-8}
 & Method 2 & 2 & 8 & 3 & 11 & 0 & 24 \\ \hline
  \end{tabular}
\end{table}

\begin{table}[!htbp]
  %\vspace{-0.5cm}
  \small
  \setlength\tabcolsep{2pt}
  \caption {Participants' feedback on the usability of the two methods}
  \label{tab:feedback_usability}
  \vspace {0.3 cm}
  \centering
  \begin{tabular}{|c|c|c|c|c|c|c|c|}
  \hline
   &  & Very Easy & Easy & Neutral & Hard & Very Hard & Total \\ \hline
 \multirow{2}{*}{\textbf{Experiment one}} & Method 1 & 5 & 6 & 1 & 0 & 0 & 12 \\ \cline{2-8}
 & Method 2 & 0 & 3 & 2 & 5 & 1 & 11 \\ \hline
 \multirow{2}{*}{\textbf{Experiment two}} & Method 1 & 0 & 7 & 4 & 3 & 0 & 14 \\ \cline{2-8}
 & Method 2 & 0 & 6 & 7 & 1 & 0 & 14 \\ \hline
 \multirow{2}{*}{\textbf{Experiment three}} & Method 1 & 2 & 12 & 10 & 0 & 0 & 24 \\ \cline{2-8}
 & Method 2 & 1 & 9 & 7 & 7 & 0 & 24 \\ \hline
  \end{tabular}
\end{table}

\begin{table}[!htbp]
  %\vspace{-0.5cm}
  \small
  \setlength\tabcolsep{2pt}
  \caption {Participants' feedback on the usefulness of the two methods}
  \label{tab:feedback_usefulness}
  \vspace {0.3 cm}
  \centering
  \begin{tabular}{|c|c|c|c|c|c|c|c|}
  \hline
 &  & Very Useful & Useful & Neutral & Not Useful & Useless & Total \\ \hline
 \multirow{2}{*}{\textbf{Experiment one}}& Method 1 & 2 & 8 & 3 & 0 & 0 & 13 \\ \cline{2-8}
 & Method 2 & 4 & 5 & 4 & 0 & 0 & 13 \\ \hline
 \multirow{2}{*}{\textbf{Experiment two}}& Method 1 & 1 & 5 & 4 & 4 & 0 & 14 \\ \cline{2-8}
 & Method 2 & 5 & 7 & 2 & 0 & 0 & 14 \\ \hline
 \multirow{2}{*}{\textbf{Experiment three}}& Method 1 & 4 & 10 & 8 & 2 & 0 & 24 \\ \cline{2-8}
 & Method 2 & 5 & 15 & 3 & 1 & 0 & 24 \\ \hline
  \end{tabular}
\end{table}

\begin{table}[!htbp]
  %\vspace{-0.5cm}
  \small
  \setlength\tabcolsep{2pt}
  \caption {Participants' feedback on the complexity of the two methods}
  \label{tab:feedback_complexity}
  \vspace {0.3 cm}
  \centering
  \begin{tabular}{|c|c|c|c|c|c|c|c|}
  \hline
 &  & Very Simple & Simple & Neutral & Complex & Very Complex & Total \\ \hline
 \multirow{2}{*}{\textbf{Experiment one}}& Method 1 & 2 & 8 & 1 & 1 & 0 & 12 \\ \cline{2-8}
 & Method 2 & 0 & 2 & 3 & 7 & 1 & 13 \\ \hline
 \multirow{2}{*}{\textbf{Experiment two}}& Method 1 & 0 & 9 & 3 & 1 & 1 & 14 \\ \cline{2-8}
 & Method 2 & 0 & 2 & 0 & 11 & 1 & 14 \\ \hline
 \multirow{2}{*}{\textbf{Experiment three}}& Method 1 & 0 & 4 & 13 & 7 & 0 & 24 \\ \cline{2-8}
 & Method 2 & 4 & 8 & 8 & 4 & 0 & 24 \\ \hline
  \end{tabular}
\end{table}

\begin{table}[!htbp]
  %\vspace{-0.5cm}
  \small
  \setlength\tabcolsep{2pt}
  \caption {Participants' feedback on the usability of the \emph{Desiree} tool}
  \label{tab:feedback_desiree_tool}
  \vspace {0.3 cm}
  \centering
  \begin{tabular}{|c|c|c|c|c|c|c|c|}
  \hline
 &  & Very Good & Good & Neutral & Bad & Very Bad & Total \\ \hline
 \textbf{Experiment one}& \emph{Desiree} tool & 1 & 3 & 6 & 2 & 1 & 13 \\ \hline
 \textbf{Experiment two}& \emph{Desiree} tool & 3 & 9 & 2 & 0 & 0 & 14 \\ \hline
 \textbf{Experiment three}& \emph{Desiree} tool & 2 & 11 & 8 & 3 & 0 & 24 \\ \hline
  \end{tabular}
\end{table}

Especially, we have investigated the work experience of the participants in experiment three. We show these statistics in Table.~\ref{tab:feedback_work_experience}.

\begin{table}[!htbp]
  %\vspace{-0.5cm}
  \small
  \setlength\tabcolsep{2pt}
  \caption {The statistics of the participants' work experience in experiment three}
  \label{tab:feedback_work_experience}
  \vspace {0.3 cm}
  \centering
  \begin{tabular}{|c|c|c|c|c|}
  \hline
  \multicolumn{2}{|c|}{\textbf{Work Experience}}& $\quad$ & \multicolumn{2}{c|}{\textbf{RE Experience}} \\ \hline
  No or Little ($\leq$ 3 months) & 11 &  & No RE Experience & 18 \\ \hline
$\leq$ 1 Year (e.g., having some internship) & 5 &  & RE Course & 1 \\ \hline
$1 < \sim \leq 3$ Years & 3 &  & 6 months & 1 \\ \hline
$3 < \sim \leq 5$ Years & 1 &  & 1 Year & 1 \\ \hline
$3 < \sim \leq 5$ Years & 3 &  & 2 Years & 2 \\ \hline
$3 < \sim \leq 5$ 10 years & 1 &  & 3 Years & 1 \\ \hline
  Total & 24 &  & Total & 24 \\ \hline
  \end{tabular}
\end{table}

We have also got many interesting (textual) feedback for the question Q.10 $\sim$ Q.14, we present them as below. Note that the numbers in the ``Id'' column are just sequencing numbers, not the identifiers of participants in our experiments.

\newpage


\section*{Supplementary material}

## Q.10: Which parts of the Desiree method are the most and least hard to you?

| | Id | Response |
|---|---|---|
| **Experiment One** | 1 | i didnt understand it well |
| | 2 | learn the syntax |
| | 3 | To understand the usage is the hardest part, like when to use function goal and when to use the function constrain. |
| | 4 | the distinction between the different quality goals was hard for me, but the goal and or link was easy |
| | 5 | the syntax is very challenging |
| | 6 | May be the links are abit strange, perhaps it would be better if you incorporate images as well on them. for a user to know which one to use if they are mistaken else the other parts are ok |
| | 7 | drawing the models |
| | 8 | modelling |
| **Experiment Two** | 1 | most hard part in the Desiree method is the grammar of the instructions<br>least hard part in the Desiree method is the representation of the relations between the concepts, the GUI helps a lot" |
| | 2 | "Drawing. Reading." |
| | 3 | To define the goal method between goal and functional goal … |
| | 4 | Learn to use the **tool** |
| | 5 | It is the most hard for me to remember the syntax of the Desiree method. |
| | 6 | every part is similar (**general**) |
| | 7 | Grammar |
| | 8 | The hardest part: How to recognize the type of goal. (**classification**)<br>     The least hard part: It is easy to use the tool. |
| | 9 | the most hard method: how to make sure my model is complete. |
| | 10 | Most: Memorizing the grammars and steps (reduce, interpret and etc.)<br>Least: Drawing the diagram |
| | 11 | The most hard part is the formula. The least hard is the interactive mode. |
| | 12 | **Tools** |
| | 13 | most hard: learning express method, applying new method<br>least hard: thinking in a more clear way |
| | 14 | The grammar of the language is the most hard, and how to recognize whether it's a quality good or QC is the least hard. |
| **Experiment Three** | 1 | Most: Syntax and grammar; Least: drawing |
| | 2 | How to distinguish the different type goals? Like Content goal, quality goal, and function goal, or more |
| | 3 | Desiree 画图作业 |
| | 4 | Usability |
| | 5 | "Reduce" operation's meaning and the using of DA |
| | 6 | Most: how organize the model when the quality constrains are too many (e.g., the availability of system is 24 X 365<br>Least: basic use of Desiree tool |
| | 7 | Goal definitions and splition, Goal reduce |
| | 8 | Most: 了解定义 goal 的语法 |
| | 9 | 分解需求，确定整体结构 |
| | 10 | Most: Grammar; Least: not sure. |
| | 11 | Most: the relationship between the node; least: the grammar |
| | 12 | Most: understand the grammar; least: operation of the tools |
| | 13 | I think setting goals is simple, but operationalization is difficult, especially the grammar of the codes. |
| | 14 | Most: Learning the tool. |
| | 15 | the most hard: 对各种 relation 的使用；the least hard: 创建项目等基本操作 |
| | 16 | Grammar |
| | 17 | Most: double "Ctrl + S" to save the files. Least: user habit likes eclipse |
| | 18 | Most: building the relationship among the nodes; least: I don't find anything not hard in it =_= |
| | 19 | 规范化的语言理解起来很困难，结果生成比较清晰 |
| | 20 | **Most:** 语法规则，Goal, FG, Func, QG, QC 之间的关系；**Least:** 结构化的过程，比如把一个 Goal 细化成 FG, QG, CG 等 |
| | 21 | None |
| | 22 | Defining functional constraints.<br>Deriving functions from functions. |
| | 23 | Dividing the requirement or goal into many and apply various method, makes much of the steps of RE clearer. |

## Q.11: Which parts of the Desiree method are the most and least useful to you?

| | Id | Response |
|---|---|---|
| **Experiment One** | 1 | The description is very useful for me. |
| | 2 | not explored enough to judge on this |
| | 3 | I hope the model is more useful, all the components, due to lack of experience it is difficult to judge me |
| | 4 | identifying the professional requirements is possible with the Desiree method as it poses questions to your requirement every time |
| **Experiment Two** | 1 | "The method helps a lot when **reducing** the complex requirements and help with the standard representation of those items. Nothing is useless. The method makes the analysis process clearer more or less. " |
| | 2 | every easy to know about the non-functional part... |
| | 3 | It is good. It is better if it is easier to learn the tool. (General) |
| | 4 | It is the most useful for me to use the method to **write** reasonable and unambiguous SRSs. |
| | 5 | every part is similar (General) |
| | 6 | Content Goal |
| | 7 | The most useful part: Practice drawing with the Desiree method. (refining) |
| | 8 | the most useful method: to write the function requirement (**structural description**) |
| | 9 | Most: Ensuring that all the details required in a **good SRS** will be considered. Least: The graph layout function for the diagram is not quite helpful at the moment |
| | 10 | The most useful part is the **hint**. |
| | 11 | The pattern of how to write requrements (**structural description**) |
| | 12 | most useful: clear thinking (**structural description**) least useful: no |
| | 13 | The method to **reduce** a complex requirement to several simple requirements is the most useful. I cannot figure the least useful part. |
| **Experiment Three** | 1 | Most: showing the graph; least: exporting the OWL |
| | 2 | To find more questions by a very useful thinking way |
| | 3 | Most: SRS |
| | 4 | Imagine |
| | 5 | Most: reduce; least: means |
| | 6 | Most: the tool makes me thinking in the structural and the mind is more MECE (Mutually Exclusive, Collectively Exhaustive) Least: the tool includes many grammar, it will take times to get similar with it |
| | 7 | Desiree uses specific grammar that is not openly defined, we need to set familiar with it first. |
| | 8 | Most: 能够帮助明晰需求 |
| | 9 | 可以更加明确需求的深层含义 |
| | 10 | Most: function, quality and content |
| | 11 | Most: functional goal |
| | 12 | Most: the way deal with requirements; least: not sure |
| | 13 | Interpret function is most useful. It is easy to use and efficient. I can't find a least useful method. |
| | 14 | Talking with the stakeholder can help us learn their real requirements. I think it's the most useful to me. |
| | 15 | the most useful: 会显示项目需求的结束; the least useful: 无 |
| | 16 | Help finding ambiguous descriptions in natural language. Too complicated... Not that standard. |
| | 17 | Most: reduce the goal into details; least: there is no guideline. |
| | 18 | Most: the structure of the requirement can help understand the requirement better; Least: everything is useful I think, but it's too hard to learn. >_< |
| | 19 | Most: 更清晰的理清思路; least: 生成的结果不是很直观 |
| | 20 | Most: 获取更具体，更准确，更全面的需求，通过结构化的方法。 |
| | 21 | Declaring & defining variables. |
| | 22 | Learning is not much simple. It takes much workshops (at least 3 ~ 4). In every workshop/class focus on some parts must be done. Focusing on quality, functionality and domain leads to better understanding of RE. |

**Q.12: Will you use the *Desiree* method when you are writing SRS in the future? Parts of the method? Which parts?**

| | Id | Response |
|---|---|---|
| **Experiment One** | 1 | the Three-Stage Process can be useful for the future |
| | 2 | Maybe, the function parts |
| | 3 | I think I will use most of them |
| | 4 | I would definitely use it is a good method.<br>mostly when writing functional requirements of the systems |
| | 5 | most definitely, so far all parts are likely important |
| | 6 | Maybe no. |
| | 7 | If fully functioning....yes i will. |
| **Experiment Two** | 1 | "I think I'll take the methodology in analysis process. If there is no time constraints, maybe I will use the Desiree tool as well. Use the standardized representation of the requirements, and the graph help with tracking the requirements." |
| | 2 | Maybe. I.D.K. (I don't know) |
| | 3 | i will try to use this method... on functional and non-functional part |
| | 4 | Yes. It is helpful. |
| | 5 | May be. Use the method to reduce complicate functional requirements and quality requirements. |
| | 6 | May be |
| | 7 | Yes. Drawing analysis tree. |
| | 8 | Yes. It helps me refine requirements.<br>Recognizing goal and drawing. |
| | 9 | yes! |
| | 10 | Yes. Especially when trying to cover all the details in functional / quality / content goals |
| | 11 | Yes, maybe in the future I will use the Desiree to pick functional requirement and non functional from a natural language. |
| | 12 | Not very sure |
| | 13 | Parts, thinking more methodically, but another way to express. |
| | 14 | I will try the whole method in future. |
| **Experiment Three** | 1 | Maybe not yet |
| | 2 | Yes! To find more details and make the requirements specification more measurable and believable. Use the tool and determine the goals and to be more details. |
| | 3 | Yes, I will use SRS |
| | 4 | No |
| | 5 | Yes, maybe I will use the Desiree tool to analyze vague requirements |
| | 6 | I will use the method. Like refine the goals and split it into 3 kind of goal: functional goal, quality goal and content goal. It is very useful for thinking overall. |
| | 7 | Maybe not, not used widely and publicly, but it's a good tool to help thinking. |
| | 8 | 因为至今在工作中没有发现使用该工具的工作环境，因此应该不会使用。 |
| | 9 | 在对一些较大规模的系统时 |
| | 10 | Function, quality and content |
| | 11 | Function, quality, content |
| | 12 | Maybe I will use the method, but not the tool … |
| | 13 | If I work in this area in future, I will spend more time to learn this tool, and use it as much as I can. I think it's professional and can help me to find more details. |
| | 14 | 会用一部分，使用画图功能，但是不会画全部，画简单的，简化形式的图 |
| | 15 | 应该不会 |
| | 16 | Maybe? |
| | 17 | I'll use the Desiree method. I think it's useful for RE. I mean I'll use all the methods. |
| | 18 | Of course, but I will use it only to help me understand the requirement, while I won't make any product or files from Desiree. |
| | 19 | 会使用它的分解功能和步骤来完成需求分析工作，但是不会凭借它来完成工作。 |
| | 20 | Yes, I will. Specifying the requirements by the structural method. |
| | 21 | I will use the Desiree tool to design. |
| | 22 | Not at all |
| | 23 | I would like to use Desiree method |
| | 24 | It depends.<br>Alternative 1: Will try to grip the tool then there will be a strong motivation to use.<br>Alternative 2: How to get requirements Important parts to focus while collecting requirements. |

## Q.13: Which parts of the Desiree tool work well?
## Which parts of the Desiree tool do we need to improve?

| | Id | Response |
|---|---|---|
| **Experiment One** | 1 | Some spellings should be revised. Error messages do not explain what is wrong, especially in the syntax checking |
| | 2 | The relationships work very well, but it is very hard to update the description, the scroll doesn't work and we have to manually save the description every time |
| | 3 | No comment, lack of experience |
| | 4 | so far the models I have tried to work on, do work very well |
| | 5 | The links, to a new user i simply understand the reduce the others are difficult to know when to implement. may be am quick to judge for i haven't explored enough |
| | 6 | -When i update a root node , the child nodes do not update. |
| **Experiment Two** | 1 | "Improve the interaction, for example the recognition of the joint point of the link. And the reference module should be link to each other automatically by name. The grammar should applied more flexible. The layout of the diagrams should be replace when it is near the boundary of the palette. Do well in the representation of the relations. The reduce, observe, deuniversazlize helps a lot when analyzing " |
| | 2 | "whole tool is work well... need to data entry too many place... if better have a copy function... :)" |
| | 3 | It can separate the question to the detail and gain a specific requirement in the end. It can be better if it is easier to learn. |
| | 4 | All parts of Desiree tool work well. No sugesstion. |
| | 5 | easy to handle |
| | 6 | Grammar tips. |
| | 7 | to be improved: (1) when i want to redo my operation, i can use "ctrl + z" |
| | 8 | The grammar is quite helpful and rigorous. However it would be better if more tips / hints could be provided for the users. Auto-complete will also be a good improvement for the future |
| | 9 | Guide for the next step work well. The layout and interactive mode need to improve. |
| | 10 | The idea of it. The learnability should be improved. |
| | 11 | Not very well, we have to do many operations to express a simple fact. |
| | 12 | I haven't used the tool for many times, I cannot answer this question |
| **Experiment Three** | 1 | No buttons to delete subpages, no method to change the ID when it's copied |
| | 2 | UI convenient. Sometimes there some bugs show up like the components miss when you delete on other pages. |
| | 3 | All, but I cannot use it very well |
| | 4 | Notification.<br>Improve the accelerator key (递增的 ID) |
| | 5 | The UI is friendly but it should be have more tooltips and hints, because the various numbers will be large, it is very hard to remember all. |
| | 6 | Interactions with requirements nodes are okay.<br>User experience needs to be improved. Always needs to copy from previous requirement node, should be better. |
| | 7 | 分解目标功能最好；继承目标描述方面需要优化 |
| | 8 | 画图比较清晰流畅。但是一些结构间的自动化构图做的还不够。 |
| | 9 | Make it easier to learn to use |
| | 10 | The tool should more easily to use. |
| | 11 | It create subject easily. But need more tips about the grammar. |
| | 12 | Most parts work well, but the grammar to describe the requirement is not flexible. It requires too much code, and brings about memory of memory. |
| | 13 | 我觉得这个工具本身是很好的。当用于处理一些较大型的项目，需要比较完善的文档时，这个工具将起到很大的作用。但是对于一些快速迭代开发的项目，这个工具的功能过于冗余，不适合使用。 |
| | 14 | 将需求明确分为了功能，质量，内容需求，针对不同需求分析。让画图更加简单，用 Desiree tool 耗费太多时间。 |
| | 15 | Hard to say.. |
| | 16 | Most parts of the Desiree tool work well. Only double "Ctrl + S" to save and cannot "Ctrl + Z" to revert (恢复) should be improved. |
| | 17 | The user interface is not friendly which is like being. Built ten years ago …… There should be some guide to teach users how to use it, as well as explaining terms such as "Func", "FG" … |
| | 18 | Work well: 大体上都可以<br>Need to improve: 删除途中元素太麻烦；没有 tutorial, introduction. |
| | 19 | 画图不错 （复制有一定的 Bug） |
| | 20 | User interface is not user friendly |
| | 21 | Improvements need in linking different functions and other entities manually |
| | 22 | It works good. The usability part must be focused. |

## Q.14: Do you have any other comments and suggestions?

| | Id | Response |
|---|---|---|
| **Experiment One** | 1 | I think i should should have had enough time to work with the tool in order to give meaningful feedback |
| | 2 | Improve the tool. |
| | 3 | the time was too small to adapt the model, but I have positive comment regarding the tool which is good for me |
| | 4 | providing user manuals for this tool so that one can use it easily. Could you think of providing an auto generated report of the requirements after one feeding in all the required information? |
| | 5 | Desiree embodies correctness check. before you even start to draw you caution your self if what you are doing is the right thing ie. functional goals, quality goals and chance of labeling functions and goals incorrectly are reduced which is not likely with the first method and to me this is really very important.<br><br>Thank you for the experience of the tool. its indeed helpful |
| | 6 | I think it will be a good tool to use when its fully functioning. |
| **Experiment Two** | 1 | How to simplified the interaction when drawing and raise the standardization of the representation at the same time |
| | 2 | this is a very good and useful tools... to define the requirement |
| | 3 | No. |
| | 4 | No. |
| | 5 | No. |
| | 6 | Disgree tool is very useful, maybe it's better to improve it,make it ever closer to coding. |
| | 7 | The test took too much time. |
| | 8 | I think the Desiree tool can be a online-collaboration application, then many people can edit the same time. |
| | 9 | A more friendly user interface will make this a wonderful tool for daily software requirement engineering! |
| | 10 | No |
| | 11 | I think I need more time to be familiar with the grammar and the tool. I think if we have time, maybe I will be more familiar with the grammar and the tool if I take the sessions about them after the 1st test. Because I can have more time to understand the grammar and practice with the tool |
| **Experiment Three** | 1 | No. |
| | 2 | 可以有份中文对照 |
| | 3 | Thanks for teaching. |
| | 4 | No, thanks for your work. |
| | 5 | NA. Thanks for all your valuable teaching and efforts. |
| | 6 | No |
| | 7 | 希望可以将 Desiree tool 开发的更加完善。 |
| | 8 | None |
| | 9 | None |
| | 10 | None .. |
| | 11 | I think this tool is useful but hard to learn. If the tool uses more graphical interfaces. It will be more friendly and easy to learn. |
| | 12 | 建议画图的时候可以选取 1 - 2 条覆盖面比较广的需求进行画图。不用全部都画出来。 |
| | 13 | 挺好的 |
| | 14 | Next time don't cost so much time please ... |
| | 15 | Teaching in double language is preferred. |
| | 16 | It is a very useful software which should be strengthened in its interface and guide. |
| | 17 | 一个系统中元素（如一个 Function）的复用方式需要更灵活 |
| | 18 | No |
| | 19 | Should provide help manual<br>Needs to provide tutorial<br>Auto links and manual links between different entities |
| | 20 | Good work & thanks for sharing your knowledge. Will be in contact with you. ☺ |

